\documentclass[final]{cvpr}

\usepackage{times}
\usepackage{epsfig}
\usepackage{graphicx}
\usepackage{amsmath}
\usepackage{amssymb}
\usepackage{subfigure}
\usepackage{amsmath,bm}
\usepackage{color}

\usepackage[pagebackref=true,breaklinks=true,colorlinks,bookmarks=false]{hyperref}

\def\confYear{CVPR 2021}

\makeatletter
\def\thanks#1{\protected@xdef\@thanks{\@thanks \protect\footnotetext{#1}}}
\makeatother

\begin{document}

%%%%%%%%% TITLE
\title{Robust Representation Learning with Feedback for Single Image Deraining}

\author{Chenghao Chen$^{1}$ and Hao Li$^{*1,2}$\\
\\
1. Department of Automation, Shanghai Jiao Tong University (SJTU), Shanghai, 200240, China.\\
2. École d’Ingénieurs SJTU-ParisTech (SPEIT), Shanghai, 200240, China.\\
\small{\textbf{Official CVPR 2021 paper: C. Chen and H. Li, ``Robust representation learning with feedback for single image deraining'',}} \\ 
\small{\textbf{IEEE/CVF Conf. on Computer Vision and Pattern Recognition (CVPR), 2021, pp.7742-7751}}
\thanks{Research supported by SJTU Young Talent Funding (WF220426002).}
\thanks{* Corresponding author: Hao Li ({Email: haoli@sjtu.edu.cn})}
}

\maketitle

%%%%%%%%% ABSTRACT
\begin{abstract}
A deraining network can be interpreted as a conditional generator that aims at removing rain streaks from image. Most existing image deraining methods ignore model errors caused by uncertainty that reduces embedding quality. Unlike existing image deraining methods that embed low-quality features into the model directly, we replace low-quality features by latent high-quality features. The spirit of closed-loop feedback in the automatic control field is borrowed to obtain latent high-quality features. A new method for error detection and feature compensation is proposed to address model errors. Extensive experiments on benchmark datasets as well as specific real datasets demonstrate that the proposed method outperforms recent state-of-the-art methods. Code is available at: \\ https://github.com/LI-Hao-SJTU/DerainRLNet
\end{abstract}
\vspace{-0.7cm}

\section{Introduction}
Outdoor vision systems are used widely such as on intelligent vehicles and for surveillance. They sometimes suffer from rain pollution, which is undesirable in practice. To handle this problem, study on image deraining has appeared, which aims at removing rain streaks from image.

Some methods exploit specific \textit{a priori} knowledge to clean images. For example, \cite{Alpher01} introduces the dark channel; \cite{Alpher02} maximizes the contrast among different target image regions. However, sparse rain streaks cannot be well removed by these methods. Some physical properties based deraining methods aim at separating the rain layer from the background layer via discriminative sparse coding \cite{Alpher06,Alpher18,Alpher21}, dictionary learning \cite{Alpher07}, and Gaussian mixture models \cite{Alpher22}.
However, specific \textit{a priori} knowledge based methods are susceptible to complex, diverse, and changeable scenarios.

\begin{figure}
\begin{center}
\begin{minipage}[b]{0.325\linewidth}
\includegraphics[width=1\linewidth,height=0.6\linewidth]{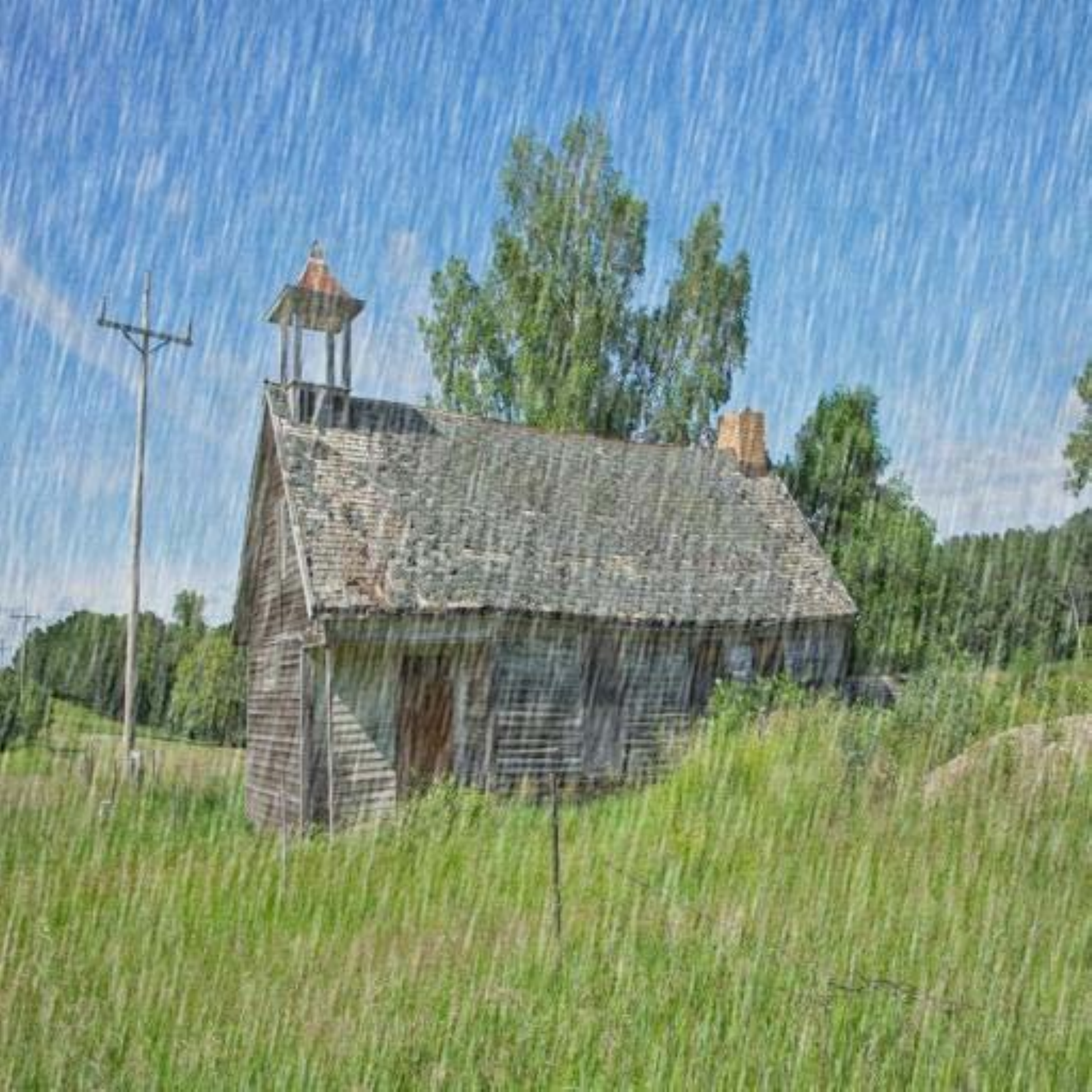}
\end{minipage}
\vspace{-0.15cm}
\begin{minipage}[b]{0.325\linewidth}
\includegraphics[width=1\linewidth,height=0.6\linewidth]{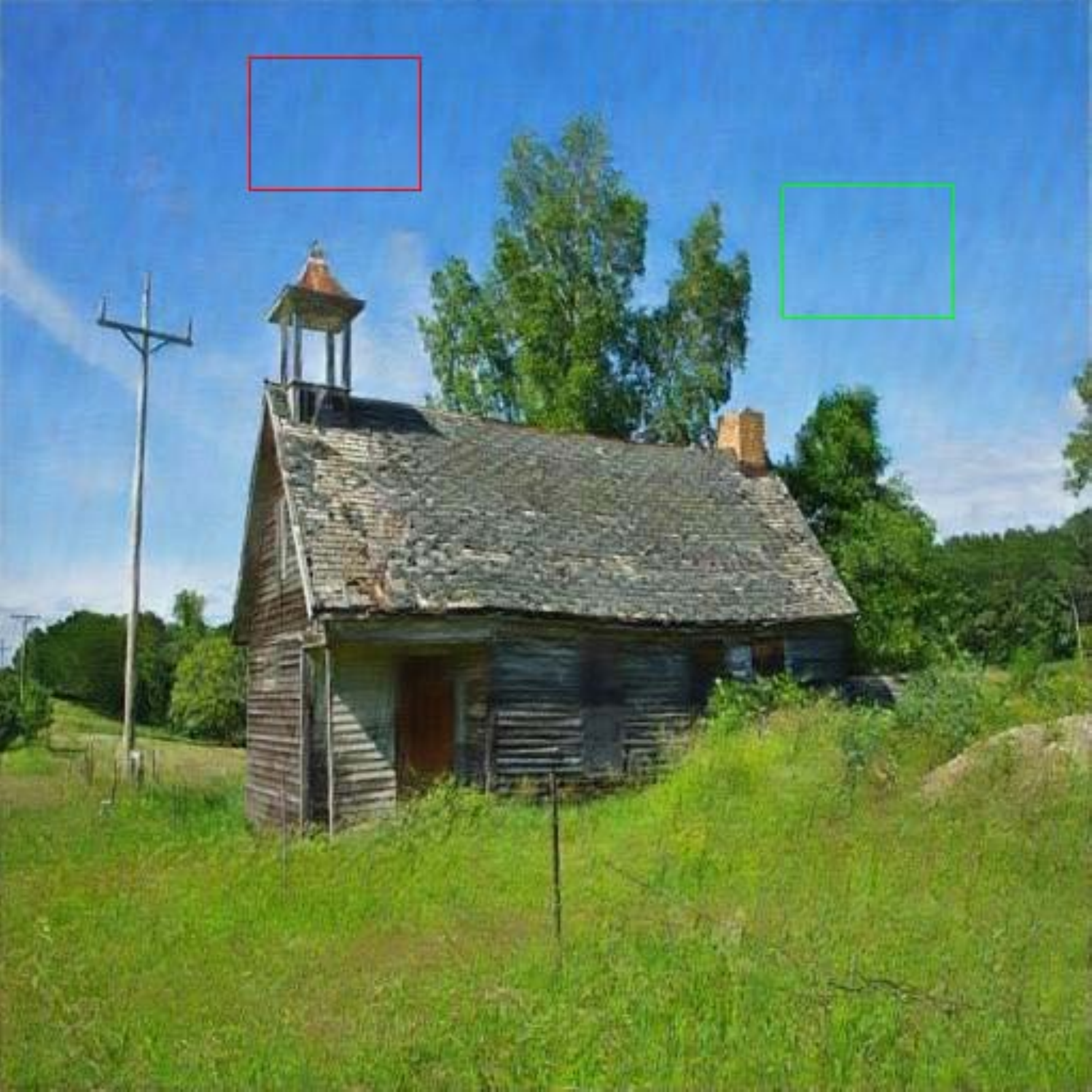}
\end{minipage}
\begin{minipage}[b]{0.325\linewidth}
\includegraphics[width=1\linewidth,height=0.6\linewidth]{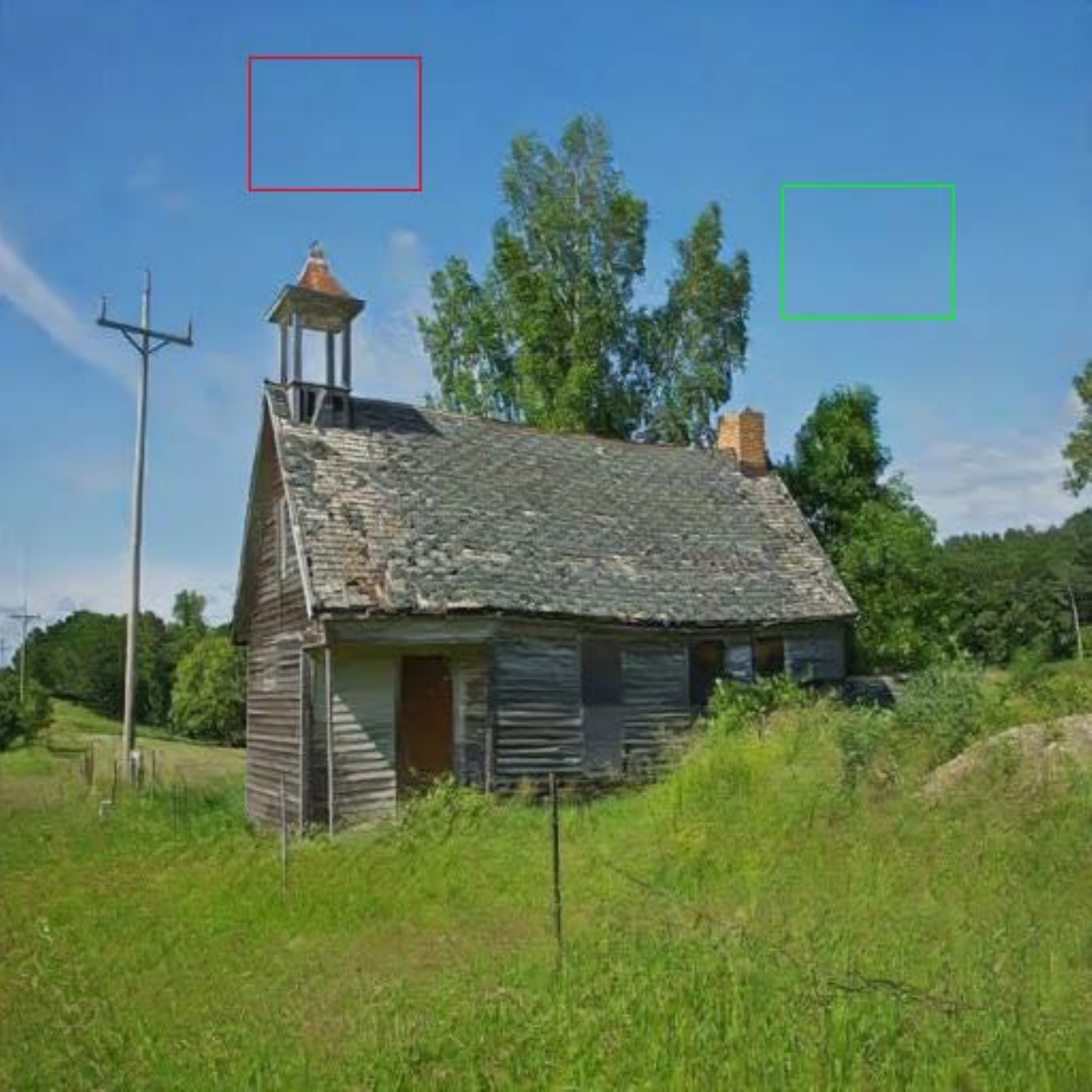}
\end{minipage}
\begin{minipage}[b]{0.325\linewidth}
\centering
\footnotesize{Rainy image}
\end{minipage}
\begin{minipage}[b]{0.325\linewidth}
\centering
\footnotesize{DID \cite{Alpher14}}
\end{minipage}
\begin{minipage}[b]{0.325\linewidth}
\centering
\footnotesize{Ours}
\end{minipage}
\begin{minipage}[b]{0.325\linewidth}
\includegraphics[width=1\linewidth,height=0.6\linewidth]{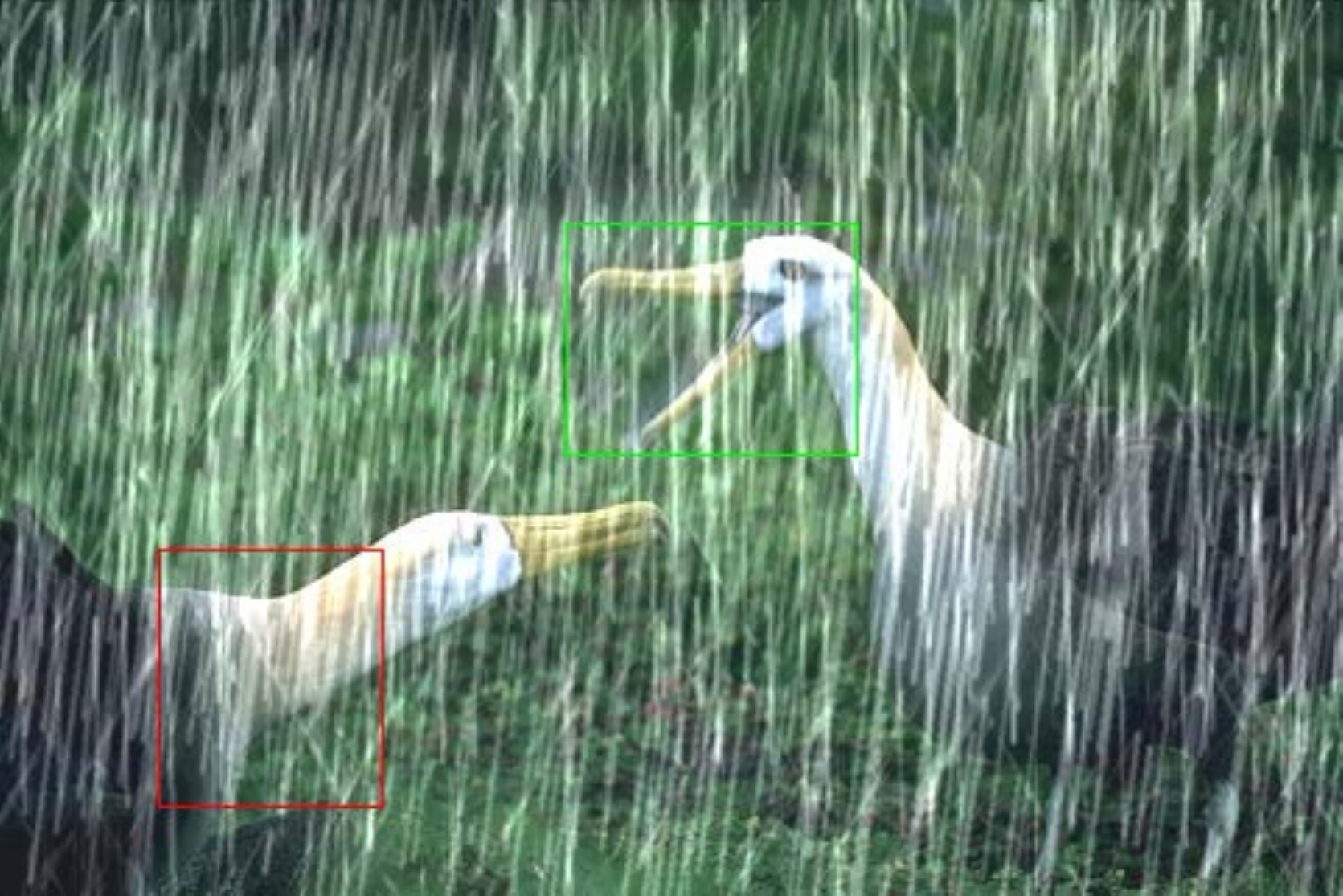}
\end{minipage}
\vspace{-0.15cm}
\begin{minipage}[b]{0.325\linewidth}
\includegraphics[width=1\linewidth,height=0.6\linewidth]{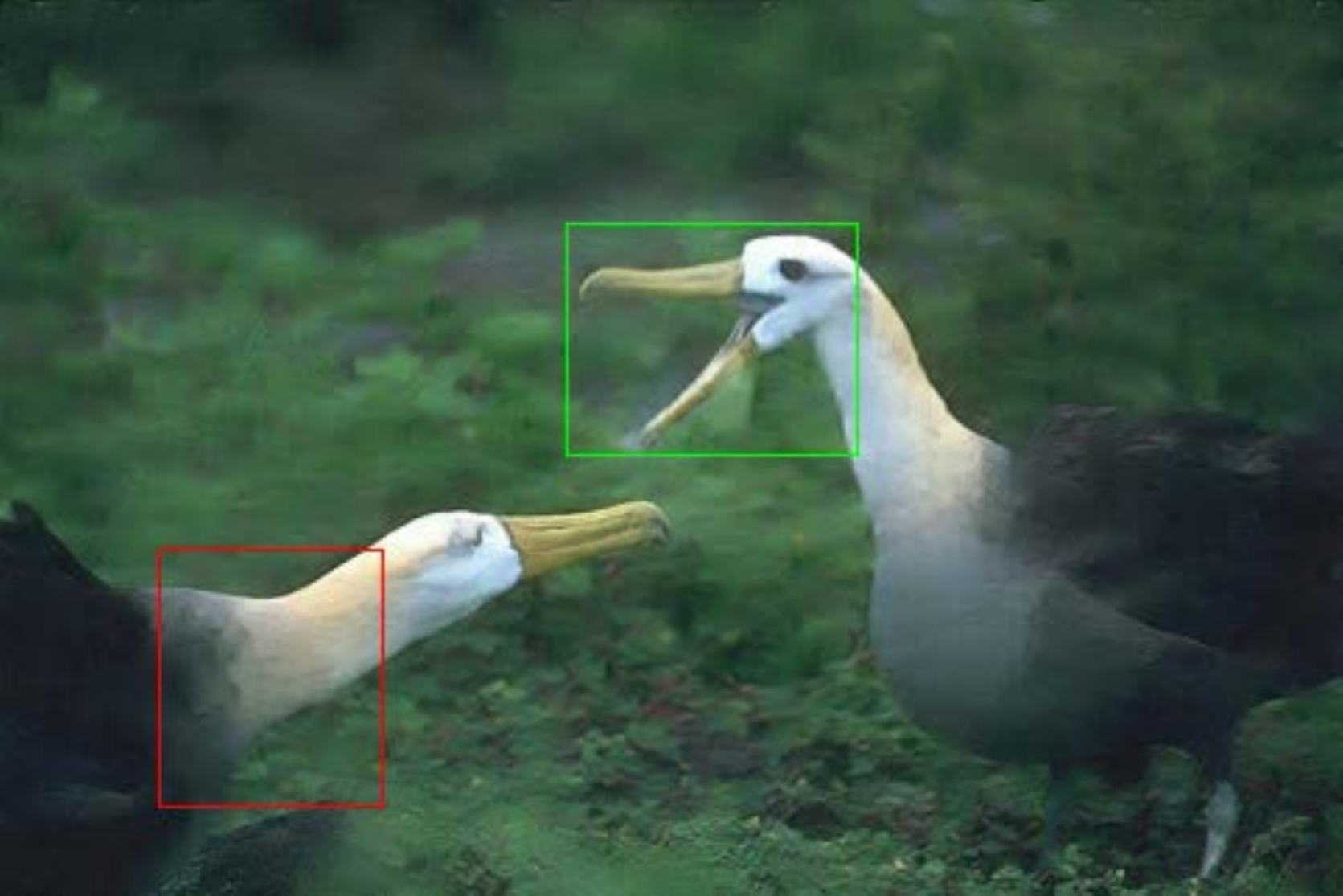}
\end{minipage}
\begin{minipage}[b]{0.325\linewidth}
\includegraphics[width=1\linewidth,height=0.6\linewidth]{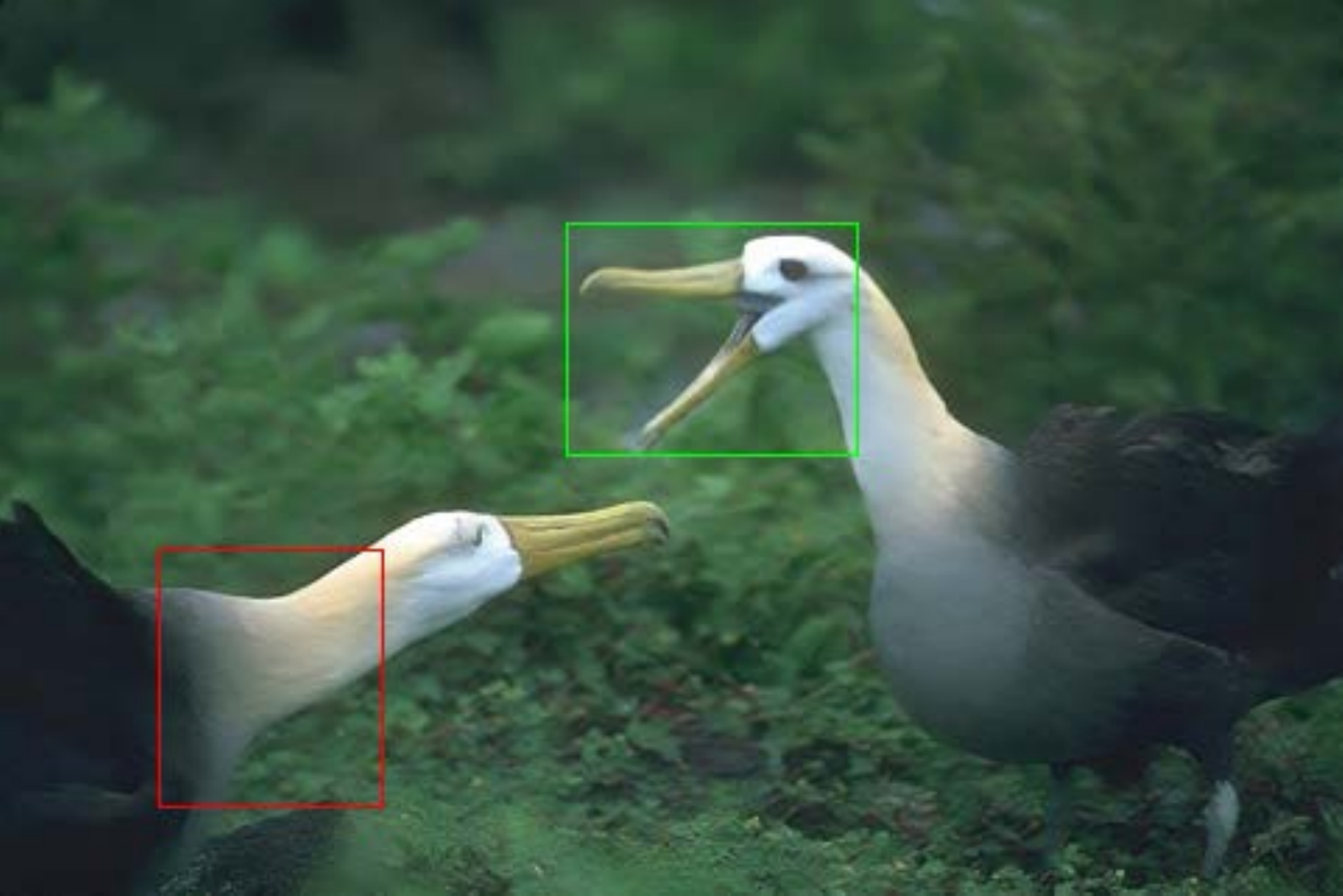}
\end{minipage}
\begin{minipage}[b]{0.325\linewidth}
\centering
\footnotesize{Rainy image}
\end{minipage}
\begin{minipage}[b]{0.325\linewidth}
\centering
\footnotesize{RESCAN \cite{Alpher26}}
\end{minipage}
\begin{minipage}[b]{0.325\linewidth}
\centering
\footnotesize{Ours}
\end{minipage}
\caption{Image deraining. DID \cite{Alpher14} tends to under-derain while RESCAN \cite{Alpher26} tends to remove details of clean image.}
\label{fig1}
\vspace{-0.9cm}
\end{center}
\end{figure}

In recent years, deep learning based image deraining methods are rising. A deraining network can be interpreted as a conditional generator, and high-quality output images can be generated if conditional embedding features can characterize target image contents \cite{Alpher16}. Most deep learning based deraining methods focus on designing novel network structures and guided features, such as residual based DDN \cite{Alpher13}, density based DID \cite{Alpher14}, recurrent structure based RESCAN \cite{Alpher26}, which can be regarded as strategies for embedding enhancement. Furthermore, \cite{Alpher16} embeds mixed feature layers into the model and decode it into a clean image. Methods that do not consider uncertainty-induced model errors tend to loss details and incur halo artefacts in the generated image (see Fig. \ref{fig1}).

\begin{figure*}
\begin{center}
\includegraphics[scale=0.45]{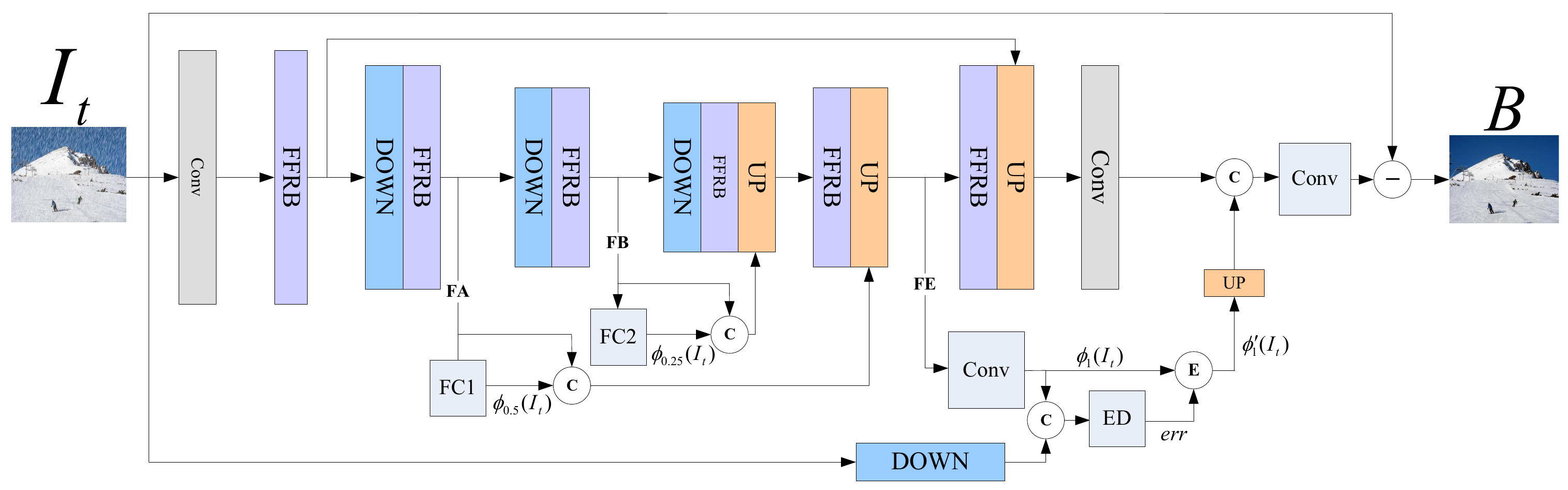}
\end{center}
\vspace{-0.4cm}
\caption{Compact overview of the proposed robust representation learning network structure (RLNet).}
\vspace{-0.2cm}
\label{overall}
\end{figure*}

To handle uncertainty-caused model errors, we propose a robust representation learning with feedback for image deraining. Given an image with rain streaks, the objective is to output the residual map. Then the clean image can be obtained by subtracting the residual map from the rainy image as illustrated in Fig. \ref{overall}. Since the embedding guides residual map generation, according to the image representation theory in \cite{Alpher39}, we try to find a functional relationship between basic embedding features and the optimal ones in the latent mapping space. The error detector and the feature compensator are designed to address model errors, for approximating the optimal embedding features.
The contributions of the paper are three-fold:

1) We analyzed the impact of uncertainty in the training process and the possibility of model error remapping via convolutional neural networks.

2) Based on the idea of closed-loop control, the error detector and the feature compensator are designed for addressing model error via feedback mechanism.

3) Unlike existing image deraining methods that embed low-quality features into the model directly, we replace low-quality features with latent high-quality features. This is a new perspective for improving the deraining performance.

%-------------------------------------------------------------------------

\section{Related Work}

\begin{figure*}
\begin{center}
\includegraphics[scale=0.45]{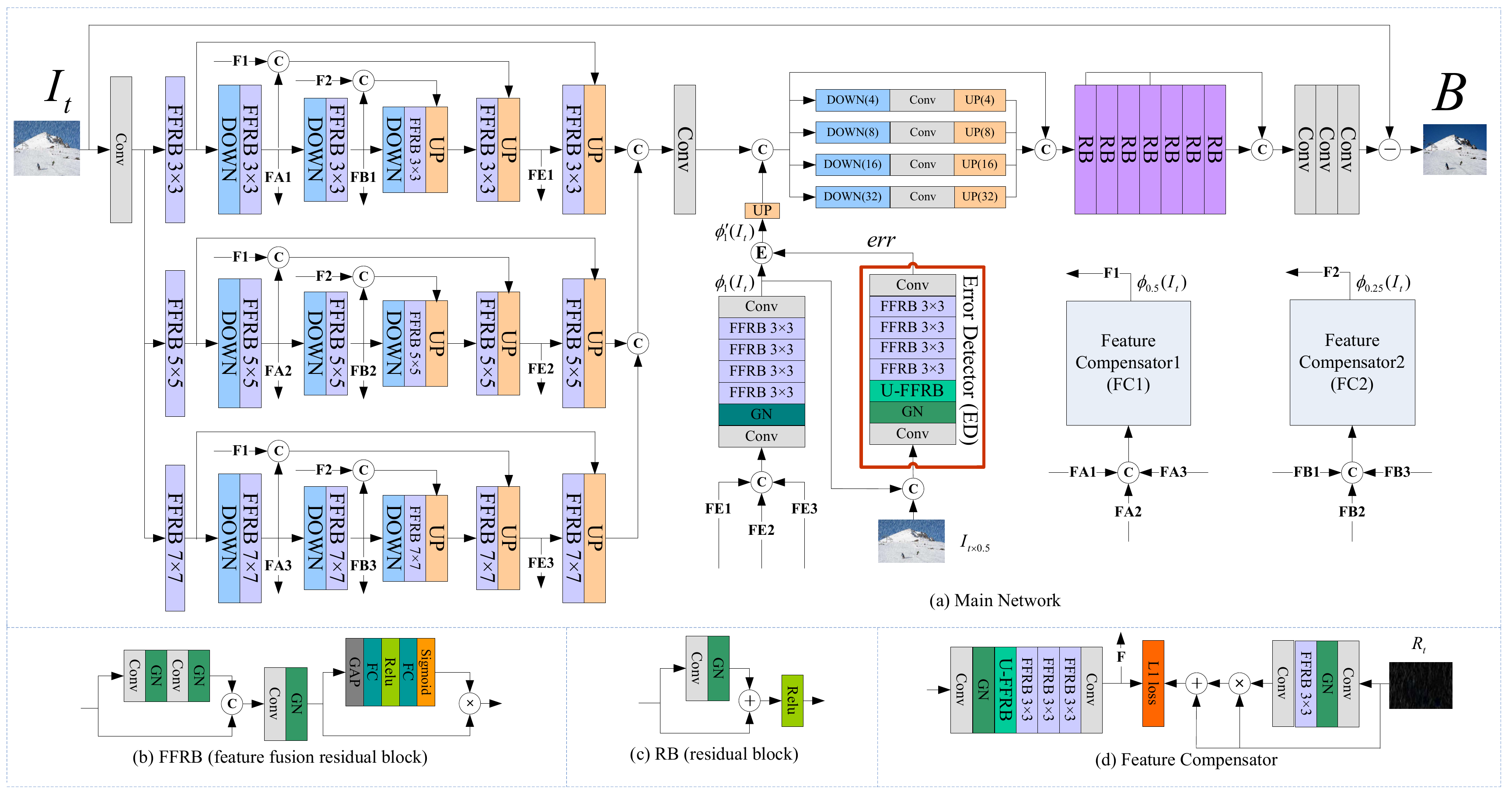}
\end{center}
\vspace{-0.4cm}
\caption{Full overview of the proposed robust representation learning network structure (RLNet). The ReLU functions after Conv and GN blocks are hidden for brevity. $-x \rightarrow $ represents information flow with index $x$, \textcircled{\scriptsize{\bf{E}}} is error compensation, \textcircled{\bf{c}} is concatenation operation, \textcircled{\footnotesize{\bm{$\times$}}} is pixel-wise multiplication, \textcircled{\footnotesize{\bm{$+$}}} is pixel-wise addition and \textcircled{\footnotesize{\bm{$-$}}} is pixel-wise subtraction. Zoom in to see small blocks better.}
\vspace{-0.5cm}
\label{fig2}
\end{figure*}

\subsection{Single Image Deraining}
Single image deraining plays a basic role for deraining. Unlike video based methods which analyse the difference between adjacent images, it is hard to remove rain streaks from a single image for lack of temporal information and ill-posed nature. For the more challenging single image deraining, traditional methods based on specific optimisition methods \cite{Alpher17,Alpher05,Alpher20,Alpher21,Alpher07} often tend to produce degraded images due to the limited mapping transformation. Recently, with the aid of the CNN, \cite{Alpher13} first focused on high-frequency rainy information for rain streak removal and demonstrated impressive restoration performance of the negative rain streak mapping network. Motivated by the deep residual neural network \cite{Alpher12}, \cite{Alpher03} proposes a deep detail network that is also trained in high-frequency domain to reduce range of intermediate variables from input to output. To handle heavy rain streaks, Li et al. decomposes a rainy image into a background layer and several rainy layers at different depth, and a recurrent CNN is proposed to remove rain streaks at state-wisely different depths \cite{Alpher26}. Similarly, In \cite{Alpher15}, a context aggregation network combined with memory units is used to remove rain streaks stage by stage.

\subsection{Representation Learning}
A high-quality residual map can be generated if conditional embedding features are able to depict contents of the residual map. Yang et al. \cite{Alpher25} decomposed a rainy image into a background layer and a rain streak layer, and located rain streaks via the binary map. However, both of removed details and remaining rain streaks on the clean image reflected the limitations of simple binary map guidance. Different from \cite{Alpher25}, Zhang et al. \cite{Alpher14} took the rain density into account to guide the network learning. Compared with the inadequacy of the rain density label that only represents image-level rain features, Qian et al. \cite{Alpher27} proposed to use the attention map to guide the residual map generation. We interpret that effectiveness mainly comes from the introduction of the rain streak features that guides the network to focus more on the rain streak regions. Later, Wei et al. \cite{Alpher28} introduced that the combined multi-stream convolutional structure can better describe context information. Despite the improved performance, these methods learned the image-level or pixel-level representation without considering the detail loss caused by the uncertainty during training. Different from the structural improvement, a confidence guided method \cite{Alpher29} studied the benefits from the residual map and its confidence. However, \cite{Alpher29} simply used weight coefficients as confidence properties, which failed to consider the suitable confidence representation, error distribution complexity and error compensation. Thus, this method tends to miss details or introduce halo artifacts.

\section{Feedback based Representation Learning}
\subsection{Problem Formulation}
In \cite{Alpher30,Alpher31}, the authors explained two types of uncertainty i.e. epistemic uncertainty a.k.a model uncertainty or systematic uncertainty, and aleatoric uncertainty a.k.a. statistical uncertainty that captures noise inherent in the observations. The variational inference can formulate epistemic uncertainty to compute variance. Maximum-aposterior or maximum-likelihood inference can formulate aleatoric uncertainty. To concisely describe our conditional optimization task, we model two output (conditional output and target output) and assume that probability distribution of each satisfies the Gaussian distribution. The minimisation objective, $\mathcal{P}=-\log p\left(y_{1}, y_{2} | f^{W}(x)\right)$, is given as:
\begin{equation}
\hspace{-0.2cm}
\begin{aligned}
\mathcal{P} \!&=\!-\log\! p\!\left(y_{1}\!\mid\! f^{W_{1}}\!\left(\!x, \!f^{W_{2}}\!(x)\right)\right)\!-\!\log\! p\left(y_{2} \!\mid \!f^{W_{2}}(x)\right) \\
& \propto \!\frac{1}{2 \sigma_{1}^{2}}\left\|y_{1}\!-\!f^{W_{1}}\!\left(x, f^{W_{2}}\!(x)\right)\right\|^{2} \\
&+\!\frac{1}{2 \sigma_{2}^{2}}\left\|y_{2}\!-\!f^{W_{2}}\!(x)\right\|^{2}\!+\!\log \sigma_{1} \sigma_{2}.
\end{aligned}\label{eq1}
\vspace{-0.1cm}
\end{equation}
where $p(\cdot)$ represents the probability function, $f^{W_{i}}(\cdot) (i=1 \text { or } 2)$ is the function of the corresponding network, $x$ is the input rainy image and $y_{i} (i=1 \text { or } 2)$ is the output. We denote  the mean of distribution $p\left(y_{i} | f^{W_{i}}(\cdot)\right) (i=1 \text { or } 2)$ as $y_{i}$ and the variance as $\sigma_{i}^{2}$. The operations with subscript label 2 are introduced to generate the embedding residual map to depict the contents of the residual map truth. The operations with subscript label 1 are introduced to generate the final residual map with the aid of the embedding residual map, and the clean image can be obtained by subtracting the residual map from the rainy image. Due to the uncertainty during training, the effect of changing the various modules in the network is limited. Thus we improve the deraining performance based on finding a functional relationship between the basic embedding residual map and the optimal one in the latent mapping space.

\noindent
{\bf Feedback mechanism.} After model training, some embeddings with large errors often increase uncertainty to degrade the deraining performance \cite{Alpher30}. To obtain high-quality features, the spirit of closed-loop feedback is incorporated into the CNN.
%As shown in Fig. \ref{fig3} (a), $R_{t}(t)$ is the setting value, and $R(t)$ as the output value needs to keep approaching $R_{t}(t)$. If there is an error between $R(t)$ and $R_{t}(t)$, the module of integral control $(1/S)$ with the error as input will rectify the output value $R(t)$.
In the automatic control system, the closed-loop control is capable of reducing the error through the integral function and the feedback. 
To apply the idea of the closed-loop control, we treat the training datasets as discrete inputs.
As shown in Fig. \ref{overall},
after generating embedding residual map $\phi_{1}(I_{t})$, we introduce the error detector (ED) with embedding residual map $\phi_{1}(I_{t})$ and rainy image $I_{t}$ as inputs to learn the error between embedding residual map $\phi_{1}(I_{t})$ and corresponding residual map truth $R_{t}$. The resulting error map is used to compensate the embedding residual map for the better feature representation.

\noindent
{\bf Effect analysis.} The role of the error detector is reflected in the training process and the results. For the training process, the effect of error compensation on the embedding residual map enables the R-to-R module (connecting this embedding residual map and the final output residual map) to be learned more accurately. With the training convergence of the R-to-R module, the final output loss function have a smaller impact on the O-to-R module (connecting the input rainy image and this embedding residual map), so that O-to-R module learns mainly based on the embedding feature loss function and acts as a feature guidance module. The decoupled learning is helpful to obtain appealing results \cite{Alpher40}. To this end, the model error induced by uncertainty can be reduced by error compensation. For the results, the rectified embedding residual map are always more conducive to generate better deraining results as shown in Fig. \ref{fig5}.

\subsection{CNN-based feedback process}
\begin{figure}
\begin{center}
\subfigure[]{
\begin{minipage}[b]{0.43\linewidth}
\includegraphics[width=1\linewidth]{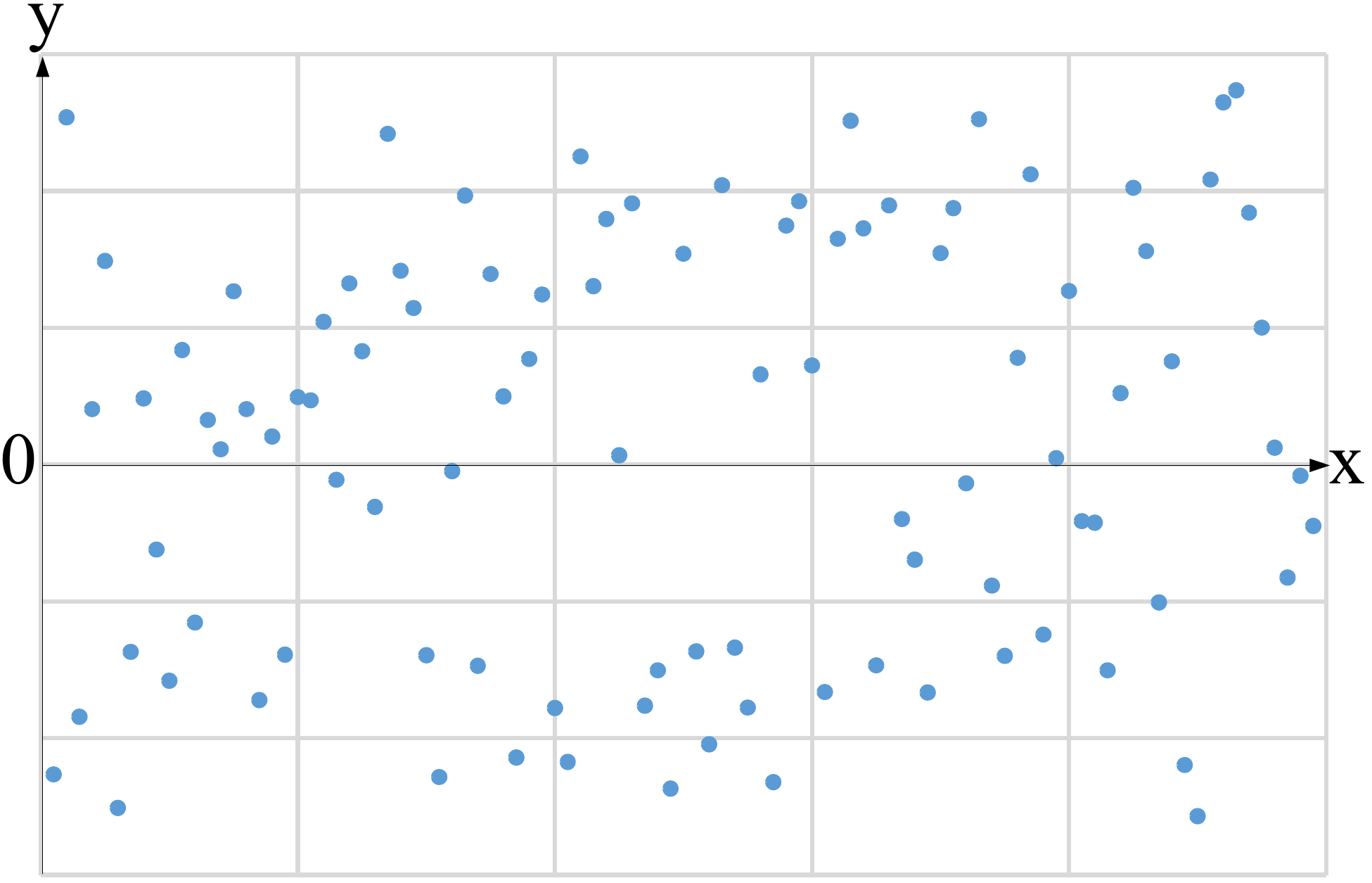}
\end{minipage}}
\subfigure[]{
\begin{minipage}[b]{0.43\linewidth}
\includegraphics[width=1\linewidth]{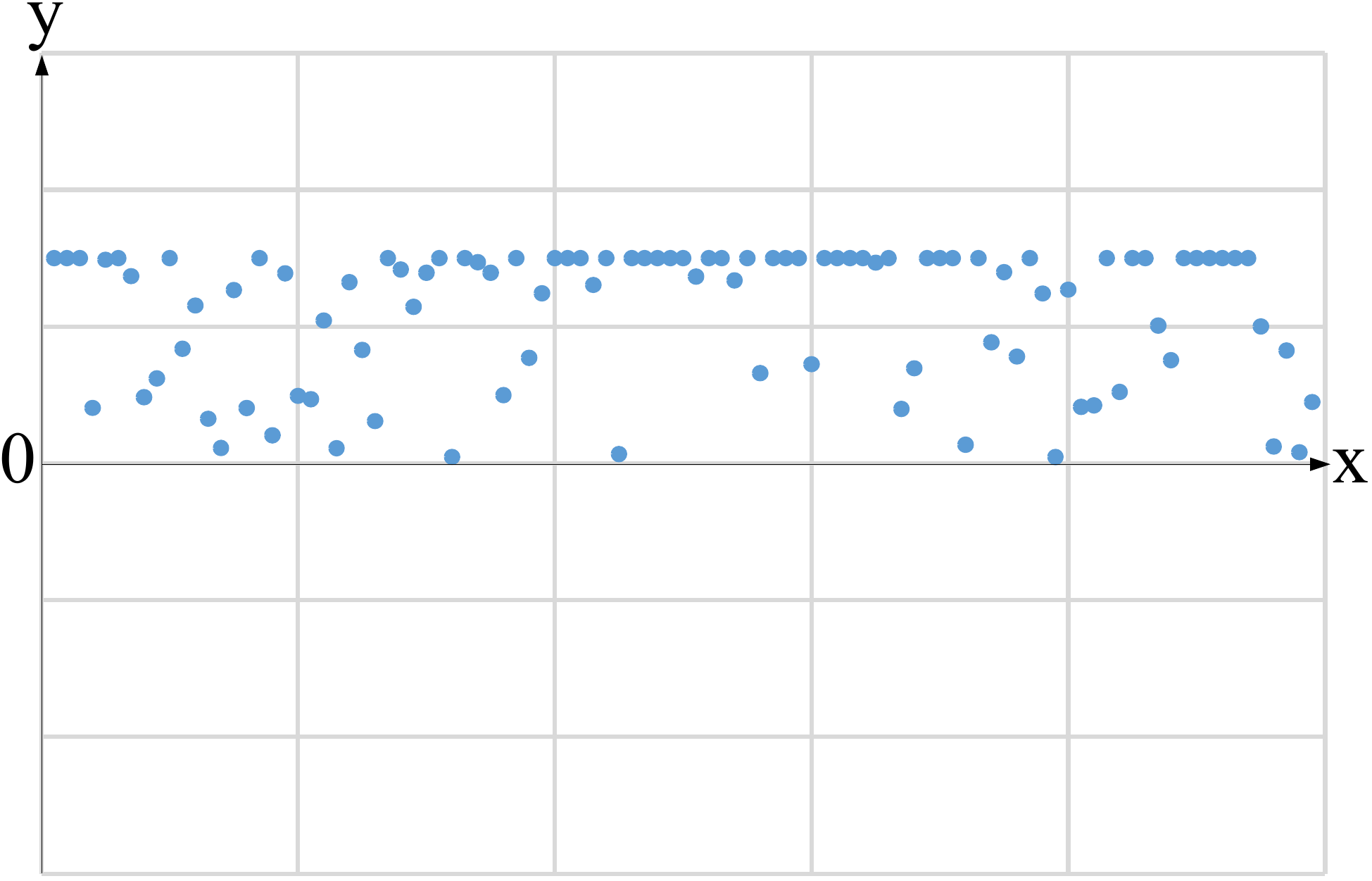}
\end{minipage}}
\caption{Abstract mapping from feature points (X-axis) to error reciprocals (Y-axis). (a) Distribution of 99 random error reciprocals within a certain range. (b) Distribution of random error reciprocals after taking the absolute value and upper limit truncation.}
\label{fig4}
\vspace{-0.8cm}
\end{center}
\end{figure}

\noindent {\bf Error distribution complexity.} In general, the variable errors caused by the uncertainty during training can hardly be remapped by CNN due to the complexity of the error distribution. By abstracting error reciprocals as points that fluctuate above and below the zero value, Fig. \ref{fig4} abstractly shows that operation of taking the absolute value of the error reciprocals can reduce the complexity of the error reciprocals distribution. Furthermore, the upper limit further simplifies the complexity of the error reciprocals distribution. Small errors (corresponding to the large error reciprocals) that have little effect on the results are truncated by the upper limit, which is conducive to remap the error map.

\begin{figure}[htbp]
\begin{center}
\begin{minipage}[b]{0.325\linewidth}
\includegraphics[width=1\linewidth,height=0.6\linewidth]{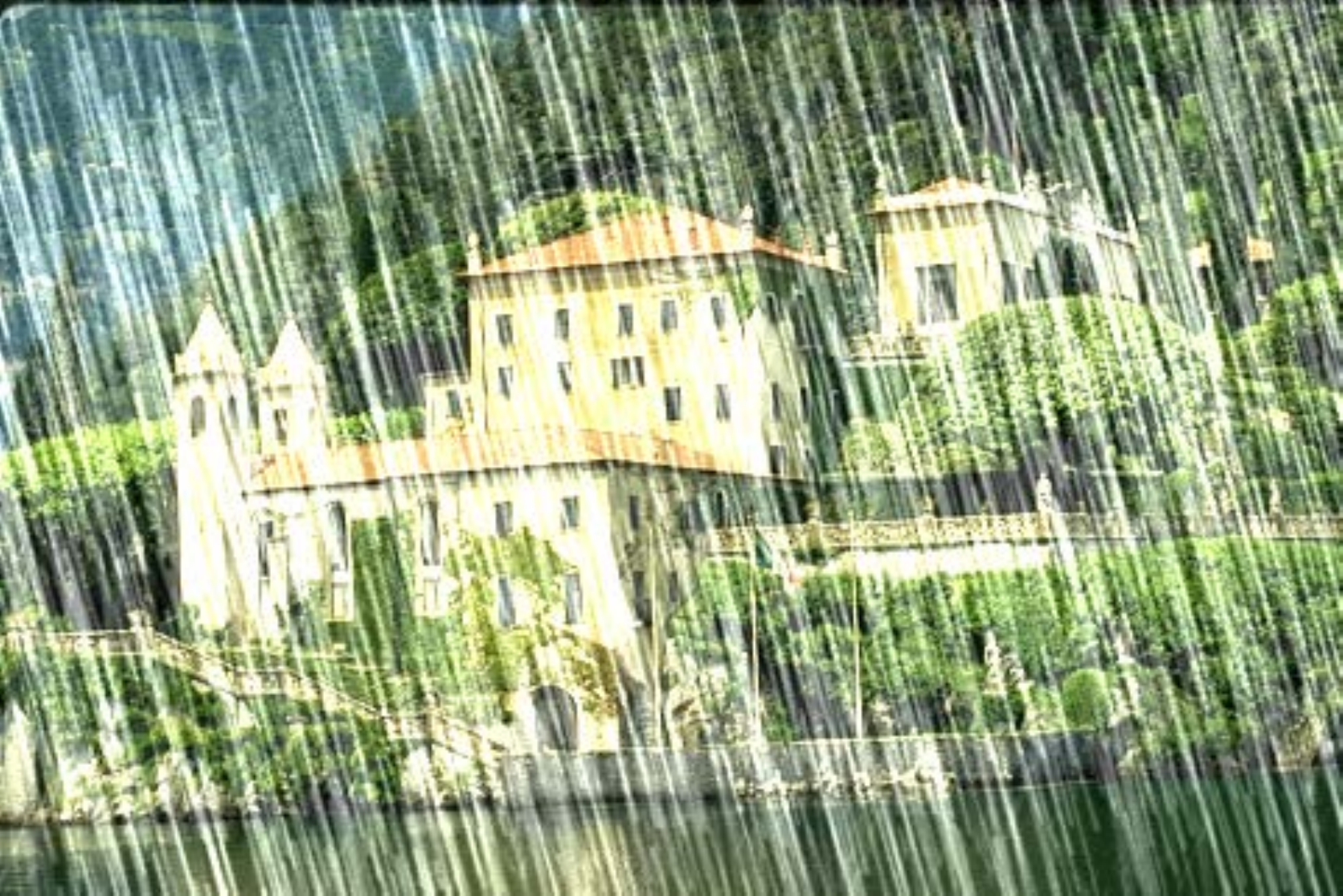}
\end{minipage}
\vspace{-0.15cm}
\begin{minipage}[b]{0.325\linewidth}
\includegraphics[width=1\linewidth,height=0.6\linewidth]{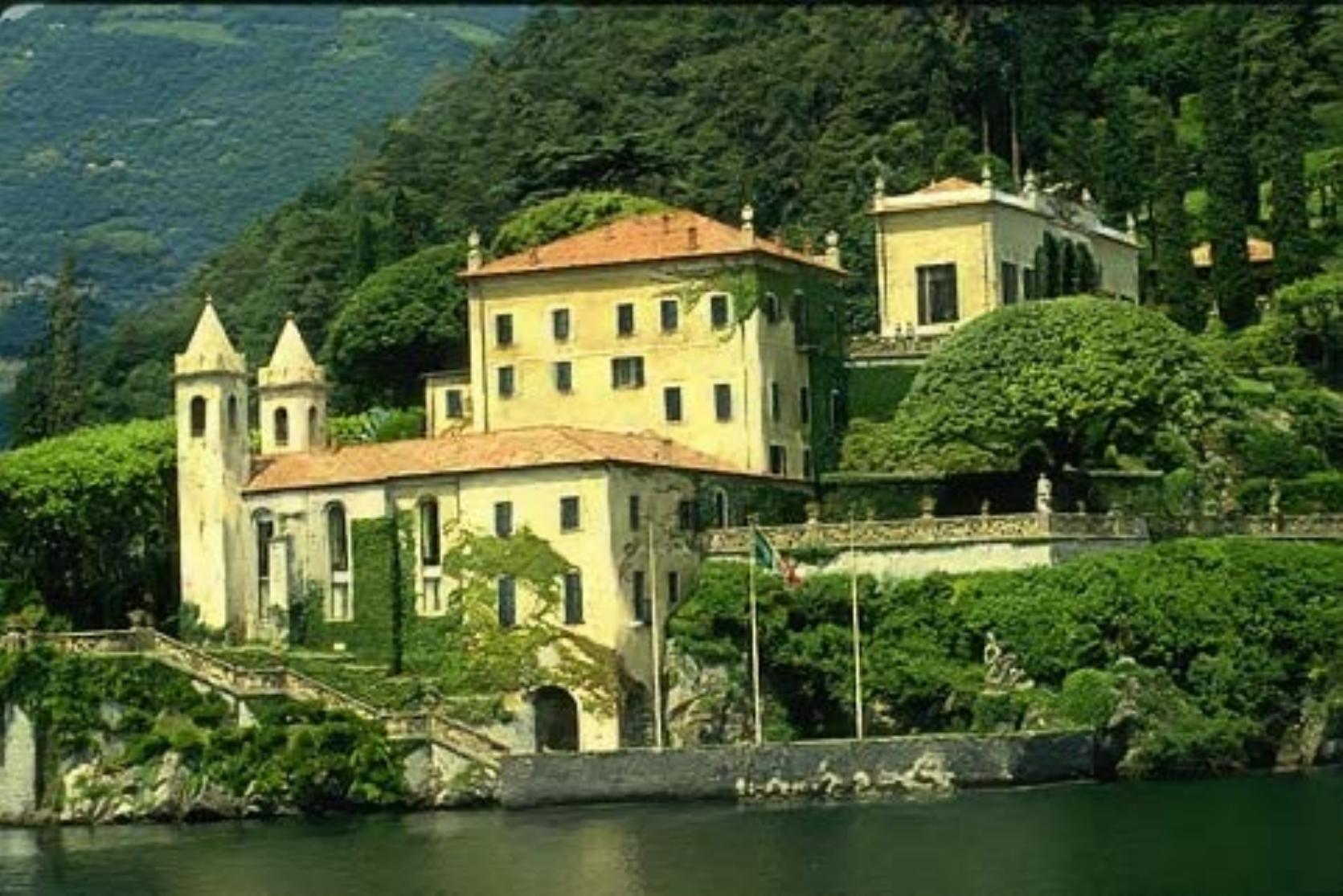}
\end{minipage}
\begin{minipage}[b]{0.325\linewidth}
\includegraphics[width=1\linewidth,height=0.6\linewidth]{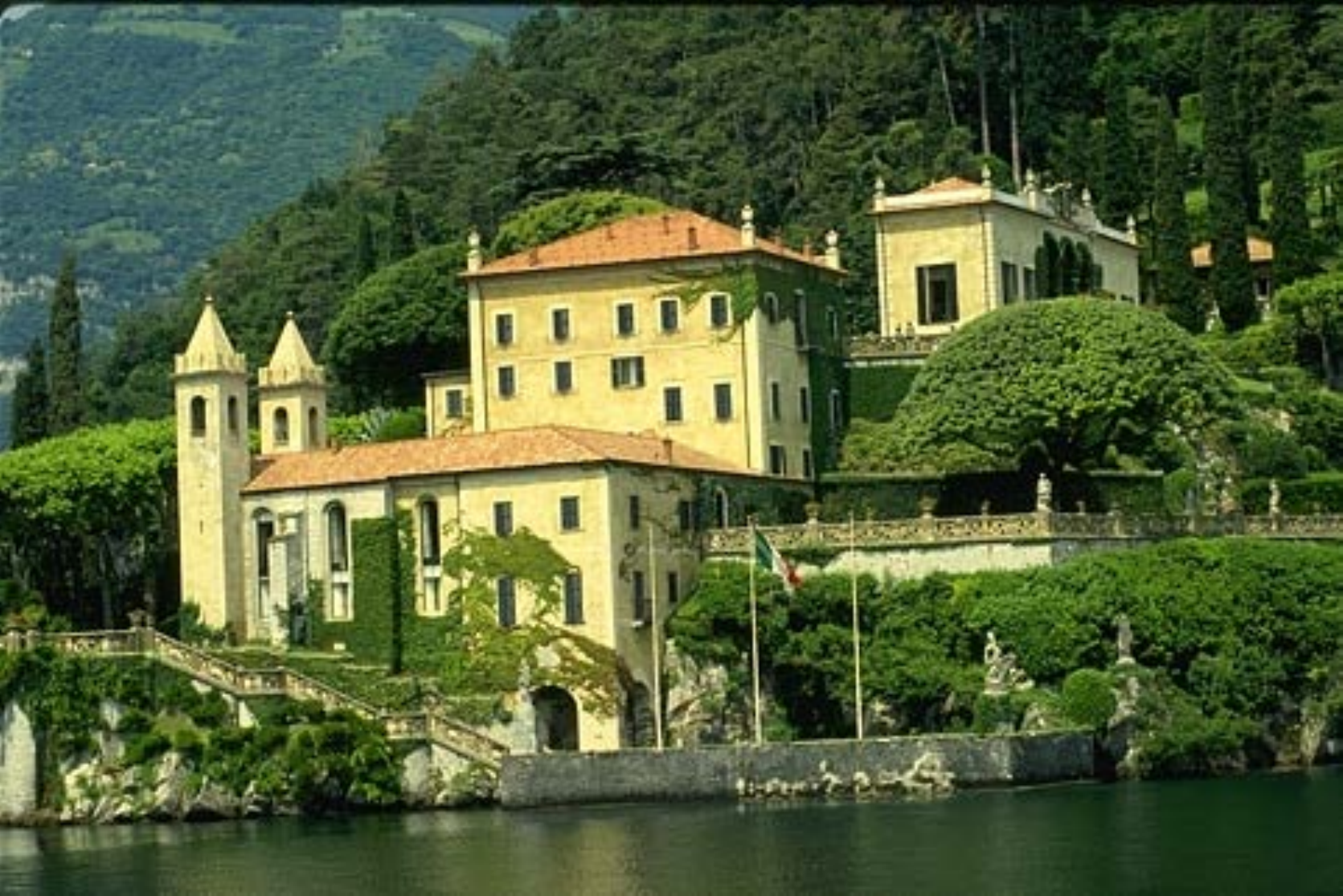}
\end{minipage}
\begin{minipage}[b]{0.325\linewidth}
\centering
\small{(a)}
\end{minipage}
\begin{minipage}[b]{0.325\linewidth}
\centering
\small{(b)}
\end{minipage}
\begin{minipage}[b]{0.325\linewidth}
\centering
\small{(c)}
\end{minipage}
\begin{minipage}[b]{0.325\linewidth}
\includegraphics[width=1\linewidth,height=0.6\linewidth]{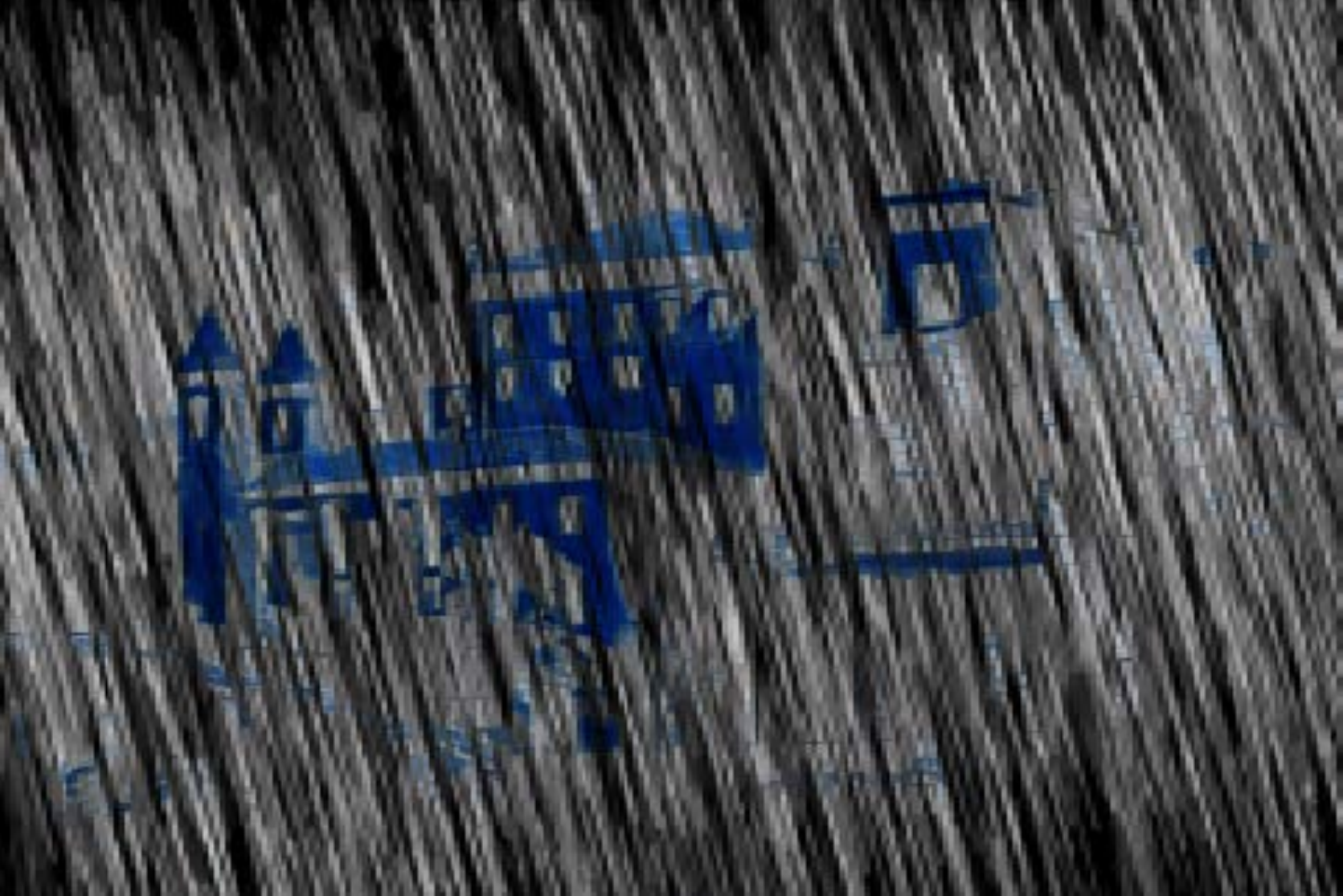}
\end{minipage}
\vspace{-0.15cm}
\begin{minipage}[b]{0.325\linewidth}
\includegraphics[width=1\linewidth,height=0.6\linewidth]{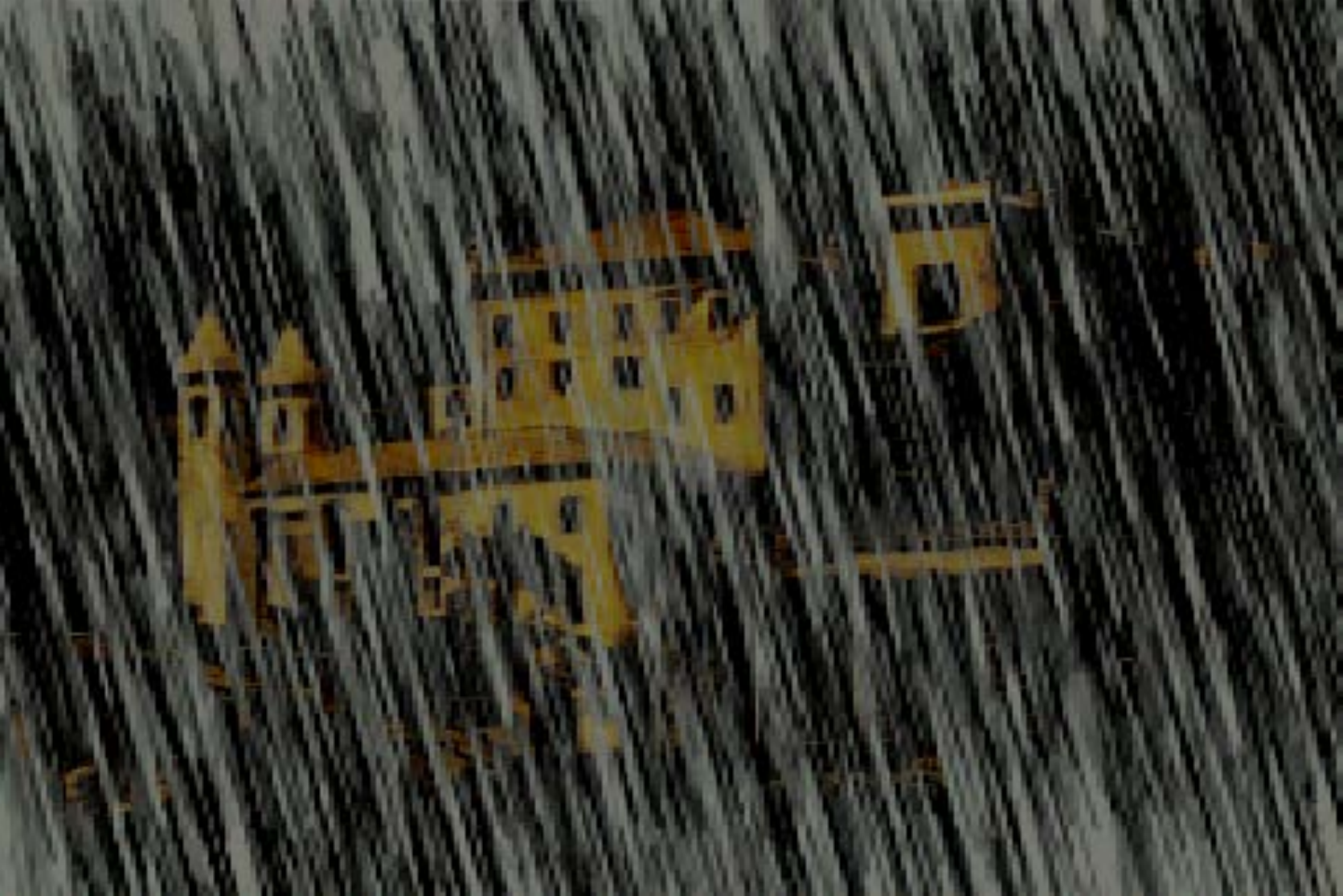}
\end{minipage}
\begin{minipage}[b]{0.325\linewidth}
\includegraphics[width=1\linewidth,height=0.6\linewidth]{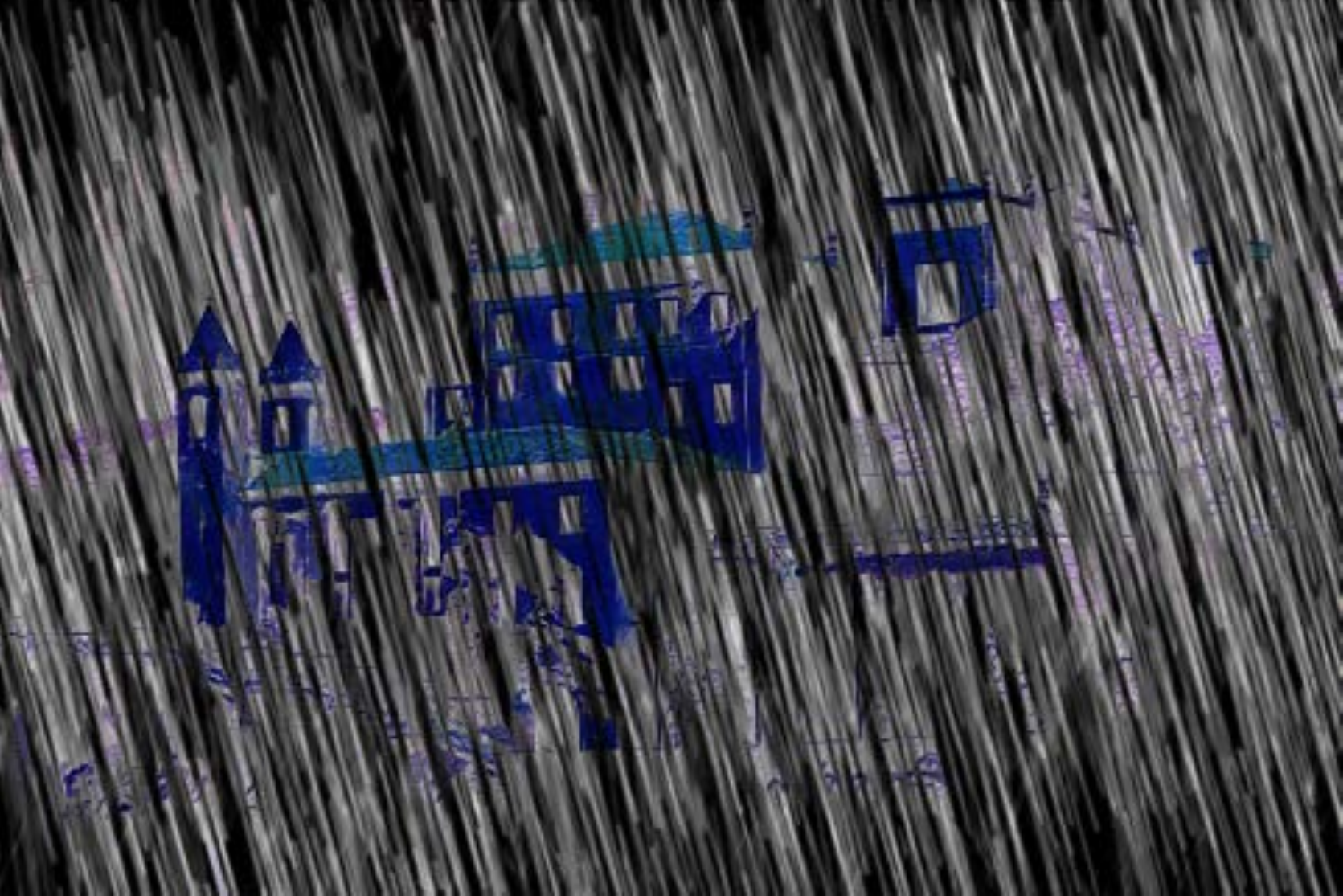}
\end{minipage}
\begin{minipage}[b]{0.325\linewidth}
\centering
\small{(d)}
\end{minipage}
\begin{minipage}[b]{0.325\linewidth}
\centering
\small{(e)}
\end{minipage}
\begin{minipage}[b]{0.325\linewidth}
\centering
\small{(f)}
\end{minipage}
\caption{(a) Rainy image. (b) Derained result of RLNet. (c) Groundtruth. (d) Embedding residual map ($\times 0.5$) not recrified by the error detector. (e) Error map multiplied by 10 for visualization. (f) Final generated residual map.}
\label{fig5}
\vspace{-0.6cm}
\end{center}
\end{figure}

\noindent
{\bf Error detector.} Specifically, we adopt the reciprocal of absolute errors multiplied by the threshold parameter $\theta_{1}$ as training value to train the error detector, and use the sigmoid function at the end of the error detector. In such a case, the upper limit of the error detector output is 1, and the error detector output is closer to 0 as long as the error is larger. For the embedding residual map and the error detector, the corresponding optimization problems are:
\begin{equation}\min _{\phi_{1}(\cdot)}\left\|R_{t \times 0.5}-\phi_{1}\left(I_{t}\right)\right\|_{1},\end{equation}
\begin{equation}\min _{\varphi(\cdot)} \| \frac{\theta_{1}}{\left|R_{t \times 0.5}-\phi_{1}\left(I_{t}\right)\right|}-\varphi\left(I_{t \times 0.5}, \phi_{1}\left(I_{t}\right)\right)\|_{1},\label{eq4}\end{equation}
where $I_{t}$ is the rainy image, $I_{t \times 0.5}$ is the rainy image at 0.5 scale size of $I_{t}$, $R_{t}$ represents the residual map truth, $R_{t \times 0.5}$ is the residual map truth at 0.5 scale size of $R_{t}$, $\phi_{1}(\cdot)$ is the function trained to map $I_{t}$ as the residual map $(\times 0.5)$, $\varphi(\cdot)$ is the function of the error detector, and $\theta_{1}$ represents the threshold parameter. Due to the sparseness of rain streaks, reducing the size of the residual map by half can well represent the original residual map (see Fig. \ref{fig5}(d)(f)), and can greatly reduce the amount of calculation. Note that the last part of $\varphi(\cdot)$ is the sigmoid funtion and small errors are truncated by the upper limit. In addition, It should be emphasized that $\left|R_{t \times 0.5}-\phi_{1}\left(I_{t}\right)\right|$ in the optimization problem (\ref{eq4}) is fixed, and the operation of taking the absolute value of the error reciprocal is very important for reducing the error distribution complexity.
The absolute error map (named err in Eq. \ref{eq5}) and the rectified embedding residual map is calculated by the following rectification process:
\begin{equation}e r r=\frac{\theta_{1}}{\varphi\left(I_{t \times 0.5}, \phi_{1}\left(I_{t}\right)\right)}-\theta_{1},\label{eq5}\end{equation}
\begin{equation}\phi_{1}^{\prime}\left(I_{t}\right)=\phi_{1}\left(I_{t}\right)-\operatorname{err}\left(\mathbf{1}-2 \phi_{1}\left(I_{t}\right)\right).\label{eq6}\end{equation}
where $\mathbf{1}$ is all-one matrix and $\phi_{1}^{\prime}\left(I_{t}\right)$ represents the rectified embedding residual map. In general, absolute errors (see Eq. \ref{eq5}) can not compensate features well. For this reason, we introduce a method to reasonably use absolute errors.

For the generated clean image, regions that should be rain-free always exist rain streaks due to heavy rain regions involved in the training process. Specifically, for the local optimal solution with locally similar inputs composed of rain and backgrounds, small pixel values and large pixel values of the residual map continuously fit the network to each of them during training, which indicates that smaller pixel values of the residual map are more likely to be generated by the network with larger values due to the pull of heavy rain pixels during training, especially when backgrounds of input rainy images are similar such that inputs for the network convolution are similar. Hence, the embedding residual map $\phi_{1}\left(I_{t}\right),\left( 0 \leq \phi_{1}\left(I_{t}\right) \leq \mathbf{1}\right)$ can describe the trend of corresponding errors. Note that the magnitude of embedding residual map values is much larger than error values. We multiply the absolute error map by the coefficient $\left(\mathbf{1}-2\phi_{1}\left(I_{t}\right)\right)$ that represents the confidence map of the absolute error map. To this end, the transformed error map (see Fig. \ref{fig5}(e)) are used to compensate the embedding residual map as shown in Eq. \ref{eq6}. The embedding residual map value from small to large corresponds to the error map value from negative to positive in terms of probability.

\begin{figure}[htbp]
\begin{center}
\begin{minipage}[b]{0.325\linewidth}
\includegraphics[width=1\linewidth,height=0.6\linewidth]{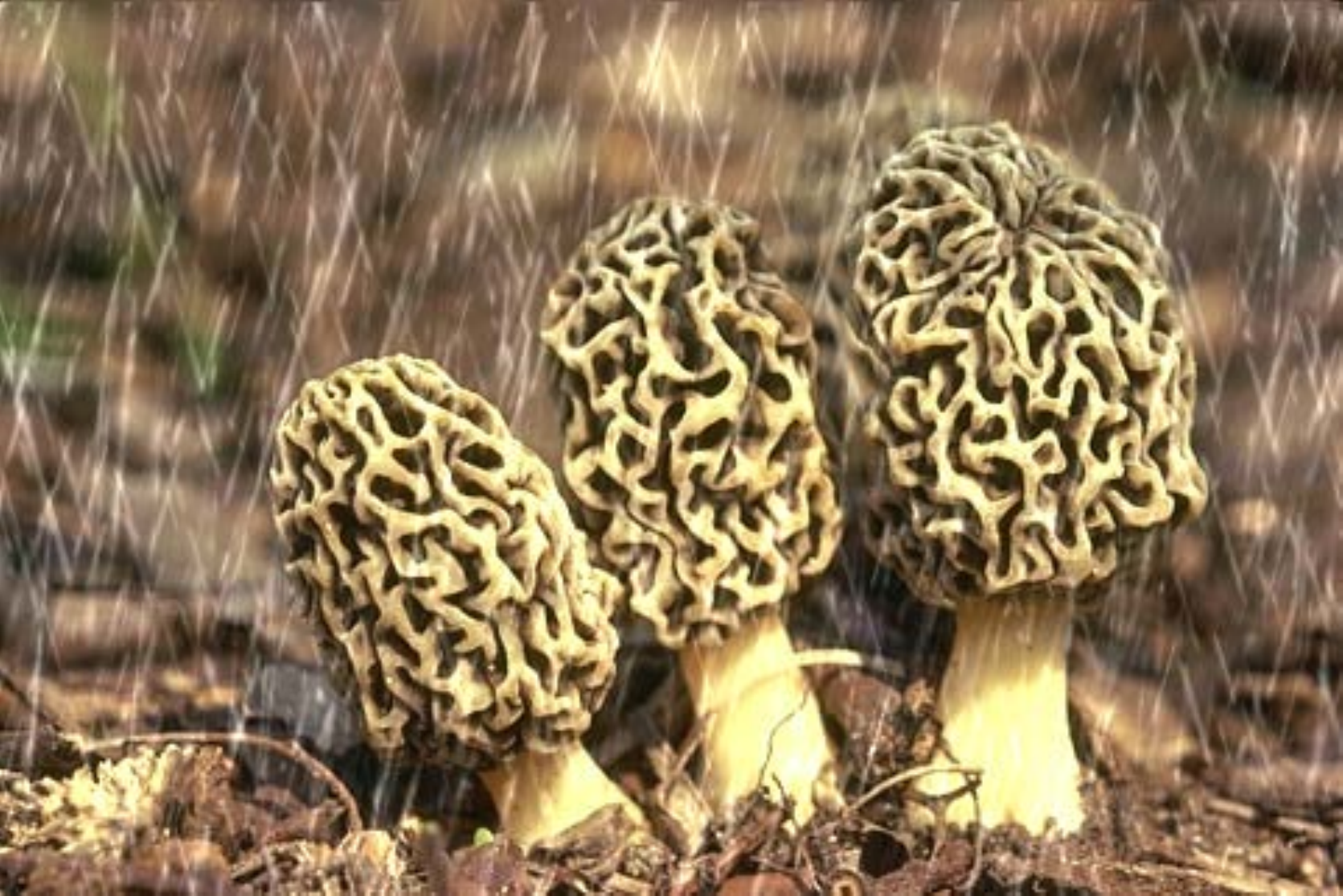}
\end{minipage}
\vspace{-0.15cm}
\begin{minipage}[b]{0.325\linewidth}
\includegraphics[width=1\linewidth,height=0.6\linewidth]{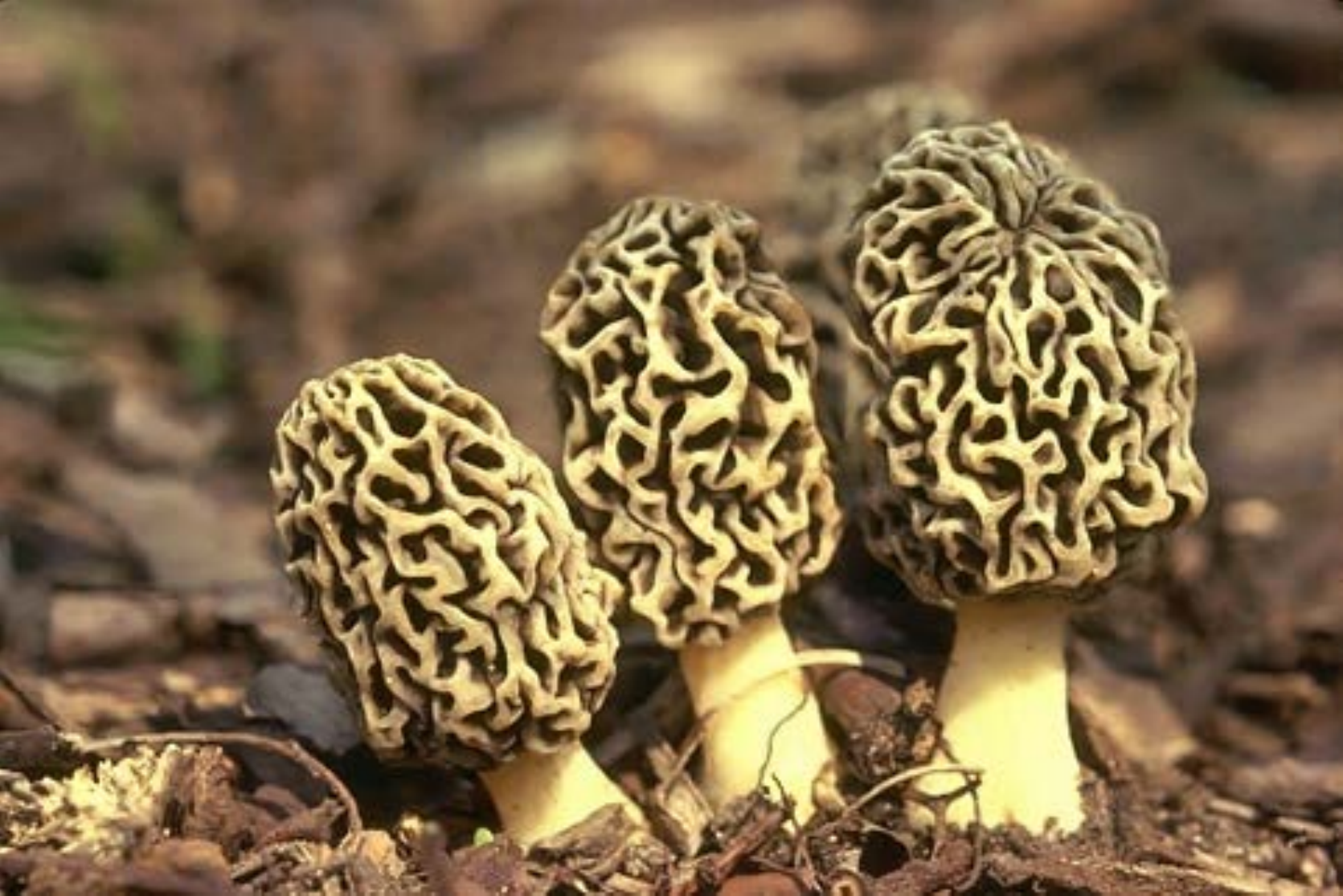}
\end{minipage}
\begin{minipage}[b]{0.325\linewidth}
\includegraphics[width=1\linewidth,height=0.6\linewidth]{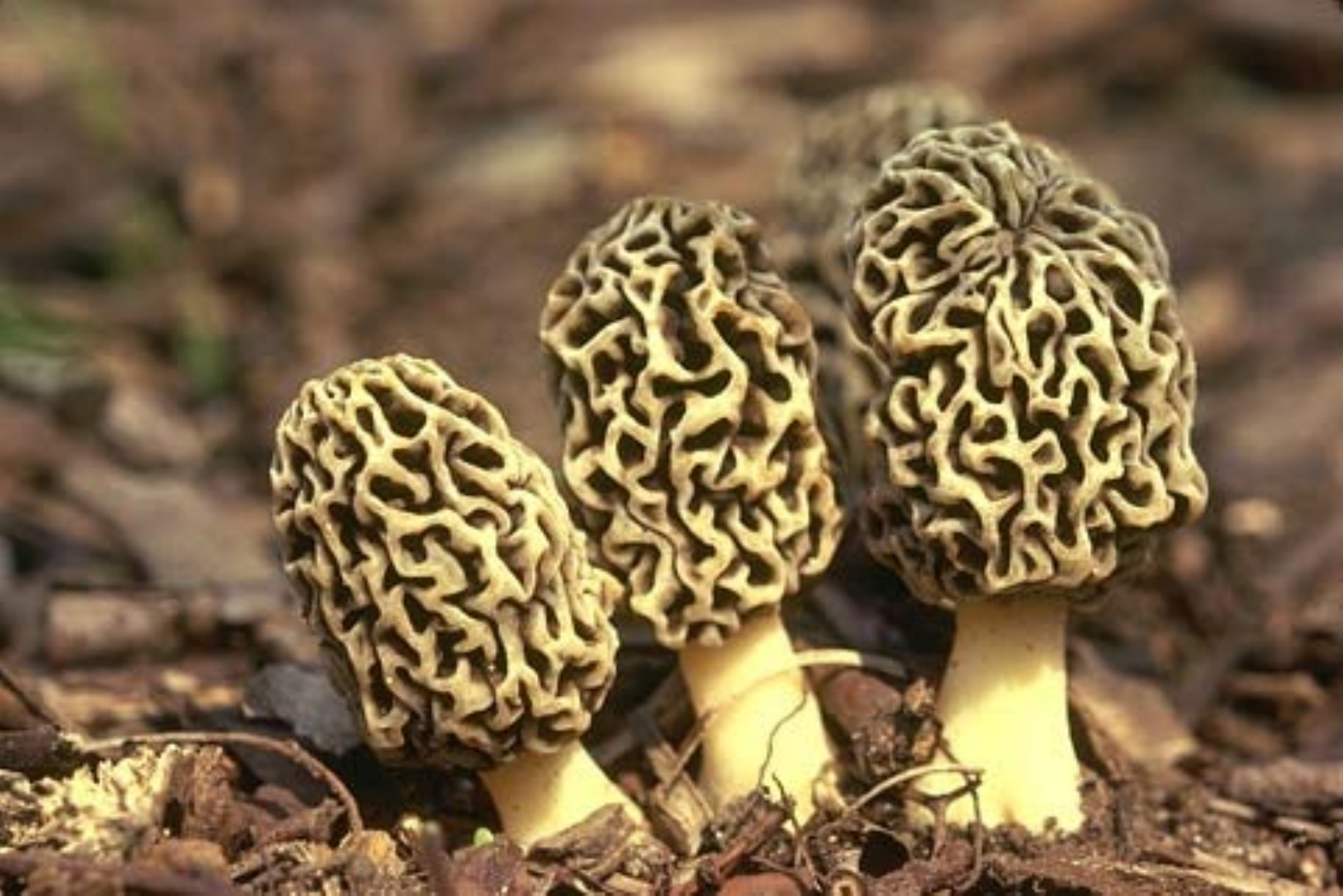}
\end{minipage}
\begin{minipage}[b]{0.325\linewidth}
\centering
\small{(a)}
\end{minipage}
\begin{minipage}[b]{0.325\linewidth}
\centering
\small{(b)}
\end{minipage}
\begin{minipage}[b]{0.325\linewidth}
\centering
\small{(c)}
\end{minipage}
\begin{minipage}[b]{0.325\linewidth}
\includegraphics[width=1\linewidth,height=0.6\linewidth]{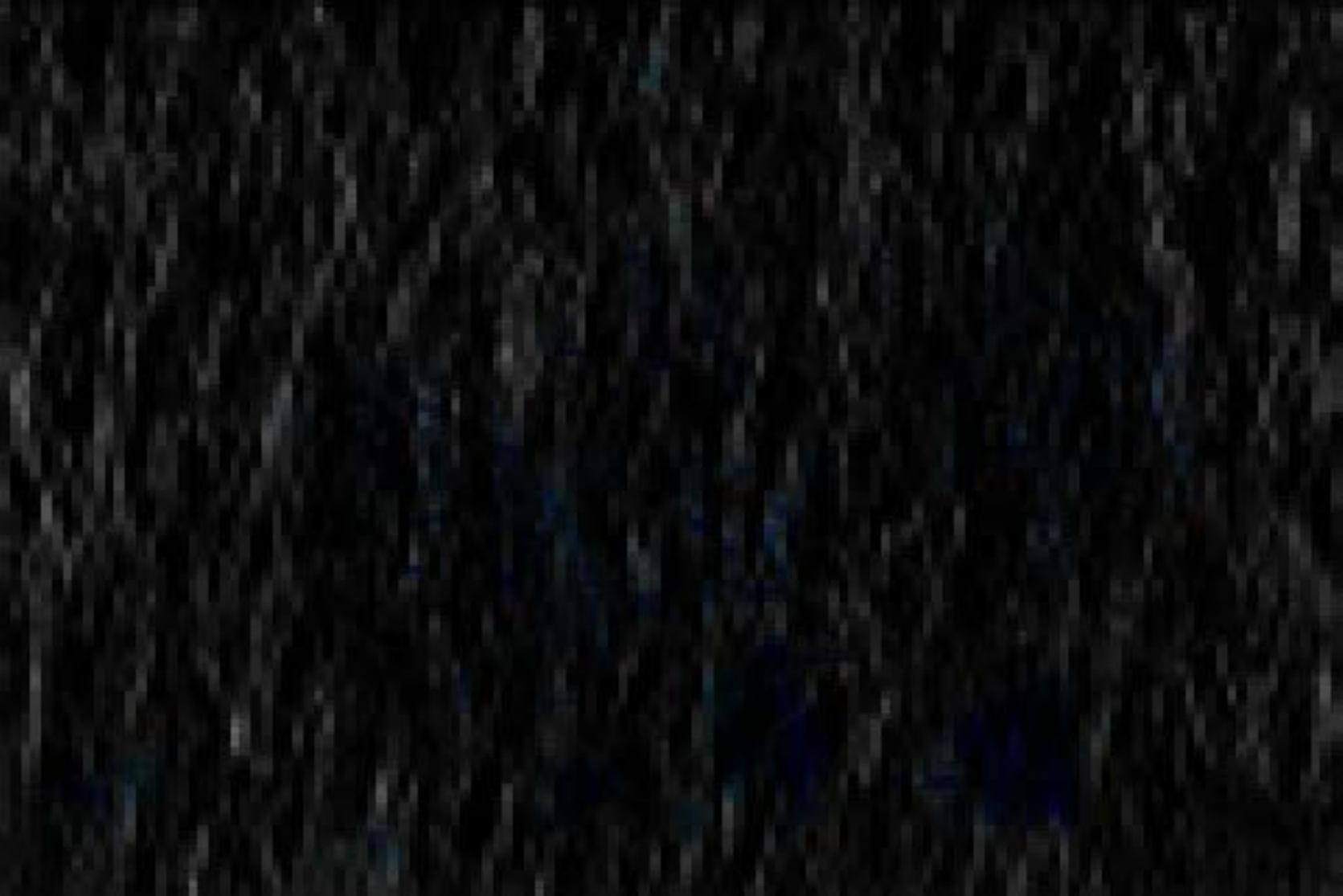}
\end{minipage}
\vspace{-0.15cm}
\begin{minipage}[b]{0.325\linewidth}
\includegraphics[width=1\linewidth,height=0.6\linewidth]{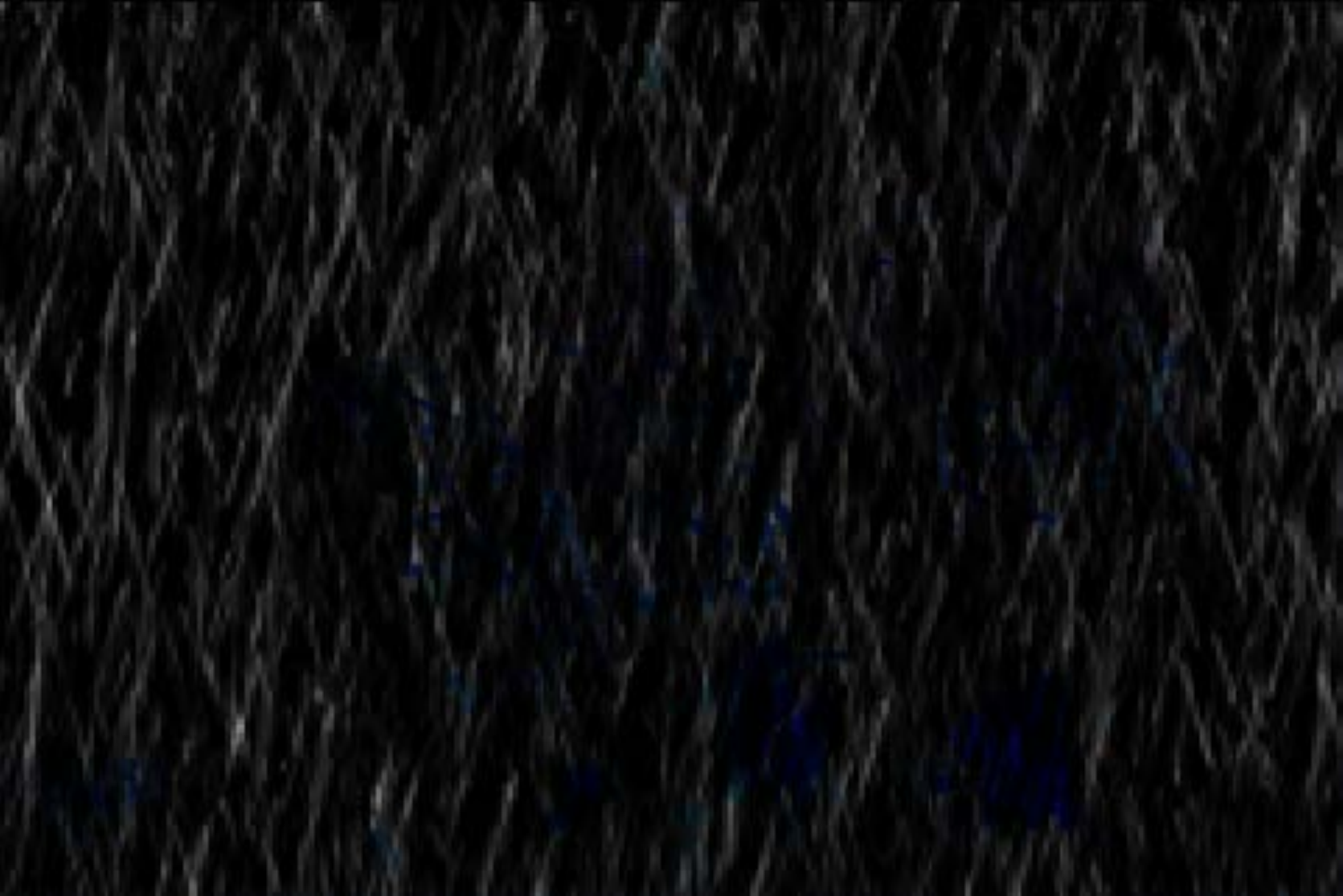}
\end{minipage}
\begin{minipage}[b]{0.325\linewidth}
\includegraphics[width=1\linewidth,height=0.6\linewidth]{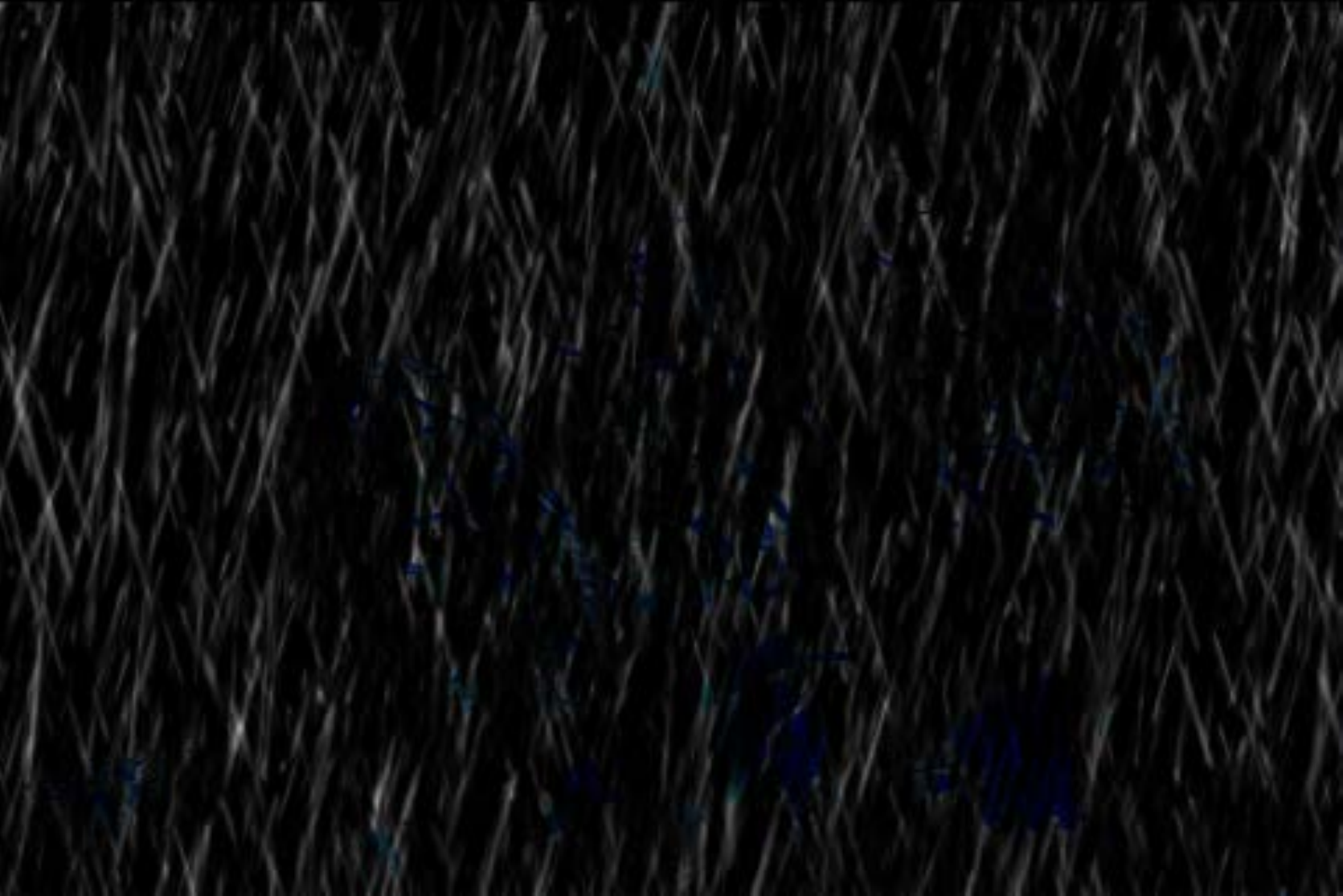}
\end{minipage}
\begin{minipage}[b]{0.325\linewidth}
\centering
\small{(d)}
\end{minipage}
\begin{minipage}[b]{0.325\linewidth}
\centering
\small{(e)}
\end{minipage}
\begin{minipage}[b]{0.325\linewidth}
\centering
\small{(f)}
\end{minipage}
\caption{(a) Rainy image. (b) Derained result of RLNet. (c) Groundtruth. (d) Embedding residual map ($\times 0.25$) output by the feature compensator. (e) Embedding residual map ($\times 0.5$) output by the feature compensator. (f) Final generated residual map.}
\label{fig6}
\vspace{-0.7cm}
\end{center}
\end{figure}

\noindent
{\bf Feature compensator.} Similarly, our goal in the feature compensator is to design a light-weight module to obtain the high-quality embedding residual map to guide other embedding residual map. In this case, the embedding residual map that needs to be rectified by the error detector becomes more accurate to reduce the difficulty of error remapping. According to Eq. \ref{eq1}, we wish to obtain a embedding residual map less affected by uncertainty. Recalling that the CNN is of the complex nonlinear function, we adopt the CNN-based module (see Fig. \ref{fig2}(d)) to transform residual map truth $R_{t}$ for error compensation, which can be learned as follows:
\begin{equation}\min _{\phi_{i}(\cdot), \omega_{i}(\cdot)}\left\|R_{t \times i}+\theta_{2} \omega_{i}\left(R_{t \times i}\right) R_{t \times i}-\phi_{i}\left(I_{t}\right)\right\|_{1}+\zeta,\end{equation}
where $\zeta=P\left(\omega_{i}\left(R_{t \times i}\right)\right)=\lambda\times\left\|\omega_{i}\left(R_{t \times i}\right)\right\|_{2}^{2}$, $P(\cdot)$ means the regularizer, $\lambda$ is the setup parameter, $R_{t \times i}$ represents the residual map at $i$ $(i=0.25 \text { or } 0.5)$ scale size of $R_{t}$. $\phi_{i}(\cdot)$ is the function trained to map $I_{t}$ as the transformed residual map at $i$ scale size (as shown in Fig. \ref{fig6}(d)(e)), $\omega_{i}(\cdot)$ represents the transformation function and $\theta_{2}$ is the transformation parameter that determines the level of transformation. As $\omega_{i}(\cdot)$ constrains the error by transforming features, the loss of details caused by uncertainty will be reduced.
\subsection{Network Architecture}
\noindent
{\bf Feature Fusion Residual Block.} We adopt the group normalization \cite{Alpher33} and the SE block \cite{Alpher26} for the better feature fusion. As shown in Fig \ref{fig2}(b), the feature fusion residual block (FFRB) is formulated as:
\begin{equation}F F R B(x)=S E(G N(\operatorname{Conv}(\operatorname{Res}(x)))),\end{equation}
where $x$ is the input signal and $\operatorname{Res}(\cdot)$ represents residual block containing group normalization (GN). Since the better performance of group normalization than batch normalization and instance normalization when the batch size is small \cite{Alpher33}, the group normalization is used to reduce internal covariate shift. By using SE block \cite{Alpher26}, the feature channels with more contextual information will be intensified by the larger weight coefficient offered by SE block. Mathematically, SE block can be expressed as:
\begin{equation}S(x)=\operatorname{Sigmoid}(F C(\operatorname{Relu}(F C(G A P(x))))) \otimes x,\end{equation}
where $\operatorname{GAP}(\cdot)$ is the global average pooling and $\operatorname{FC}(\cdot)$ is the fully connected layer.

\noindent
{\bf Multi-stream Residual Architecture.} The proposed deraining network is built upon the encoder-decoder architecture that is widely adopted in image deraining \cite{Alpher16,Alpher29}. A large receptive field induced by encoder-decoder module U-FFRB is able to obtain context information. Since rain streaks commonly at defferent shape, density and scale, only utilizing one-stream CNN \cite{Alpher15,Alpher16,Alpher29} may lose some useful information. Motivated by the effectiveness of multi-scale convolutional kernels \cite{Alpher14,Alpher28}, we use the multi-stream residual network to extract multi-scale concentrated features. The kernel sizes of the FFRB belonging to their respective streams are set to 3, 5, and 7, respectively, to find back the lost details. From Fig \ref{fig2}(a), the multi-stream architecture can be described as:
\begin{equation}\mathrm{M}(x)=\left[\mathrm{U}_{3 \times 3}(\mathrm{x}), \mathrm{U}_{5 \times 5}(\mathrm{x}), \mathrm{U}_{7 \times 7}(\mathrm{x})\right],\end{equation}
where $\operatorname{U}_{i \times i}(\cdot)$ represents U-FFRB module with the kernel size $i$, and $[\cdot]$ is the concatenation operation.

\noindent
{\bf Error detector and feature compensator.} For the feature compensator, an additional feature transformation module is introduced to adaptively transform residual map truth, and a concise encoder-decoder branch is constructed as shown in Fig \ref{fig2}(d) to learn the transformed residual map. For the correlation between different streams, the learned transformed residual map is copied into three copies and respectively embedded into three streams with skip-connection as shown in Fig \ref{fig2}(a). For the error detector, two concise branches containing FFRB are constructed to learn the residual map and the error map. Noted that one of the branches constructed for error map generation contains the encoder-decoder module. The error map is used to compensate corresponding embedding residual map (see Fig \ref{fig2}(a)).

\noindent
{\bf Refinement module.} The main goal of the refinement module is to finely adjust the feature maps clustered together. We use the spatial pyramid pooling \cite{Alpher04} to further obtain multi-scale features. The scale factors are set to 4, 8, 16 and 32, respectively. For the feature maps with different sizes, the point-wise convolution is utilized to reduce their channels and the up-sample operation adopting the nearest interpolation is utilized to restore original size. As shown in Fig \ref{fig2}(a), the down-up structure can be formulated as:
\begin{equation}\begin{array}{c}
D(x)=\left[\left(\operatorname{Conv}\left(x \downarrow_{4}\right)\right) \uparrow_{4},\left(\operatorname{Conv}\left(x \downarrow_{8}\right)\right) \uparrow_{8},\right. \\
\left.\left(\operatorname{Conv}\left(x \downarrow_{16}\right)\right) \uparrow_{16},\left(\operatorname{Conv}\left(x \downarrow_{32}\right)\right) \uparrow_{32}, x\right].
\end{array}\end{equation}
The next seven resblocks with the group normalization \cite{Alpher33} are designed as shown in Fig \ref{fig2}(c).
\subsection{Loss Function}
\begin{figure*}
\begin{center}
\begin{minipage}[b]{0.16\linewidth}
\includegraphics[width=1\linewidth,height=0.6\linewidth]{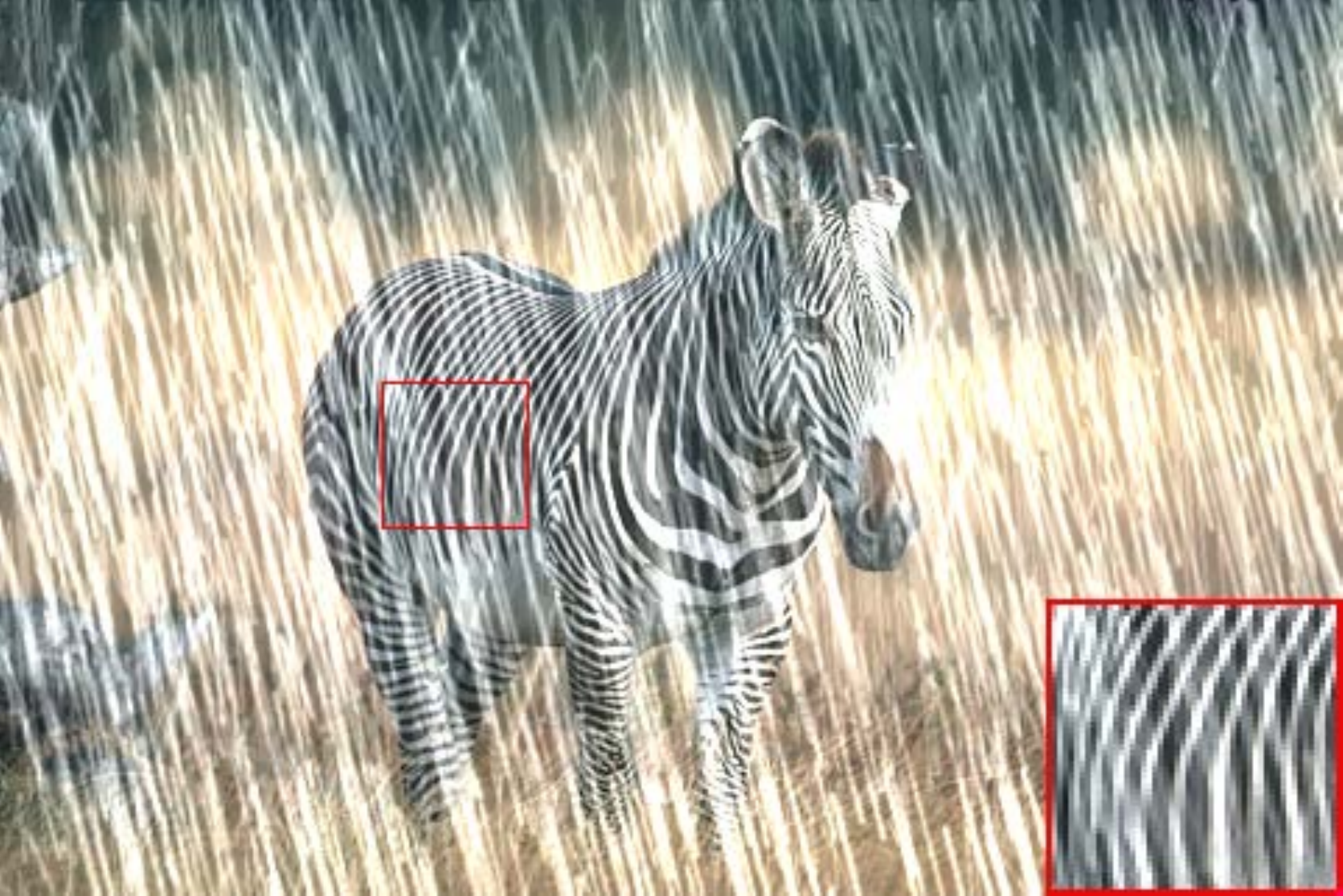}
\end{minipage}
\vspace{-0.15cm}
\begin{minipage}[b]{0.16\linewidth}
\includegraphics[width=1\linewidth,height=0.6\linewidth]{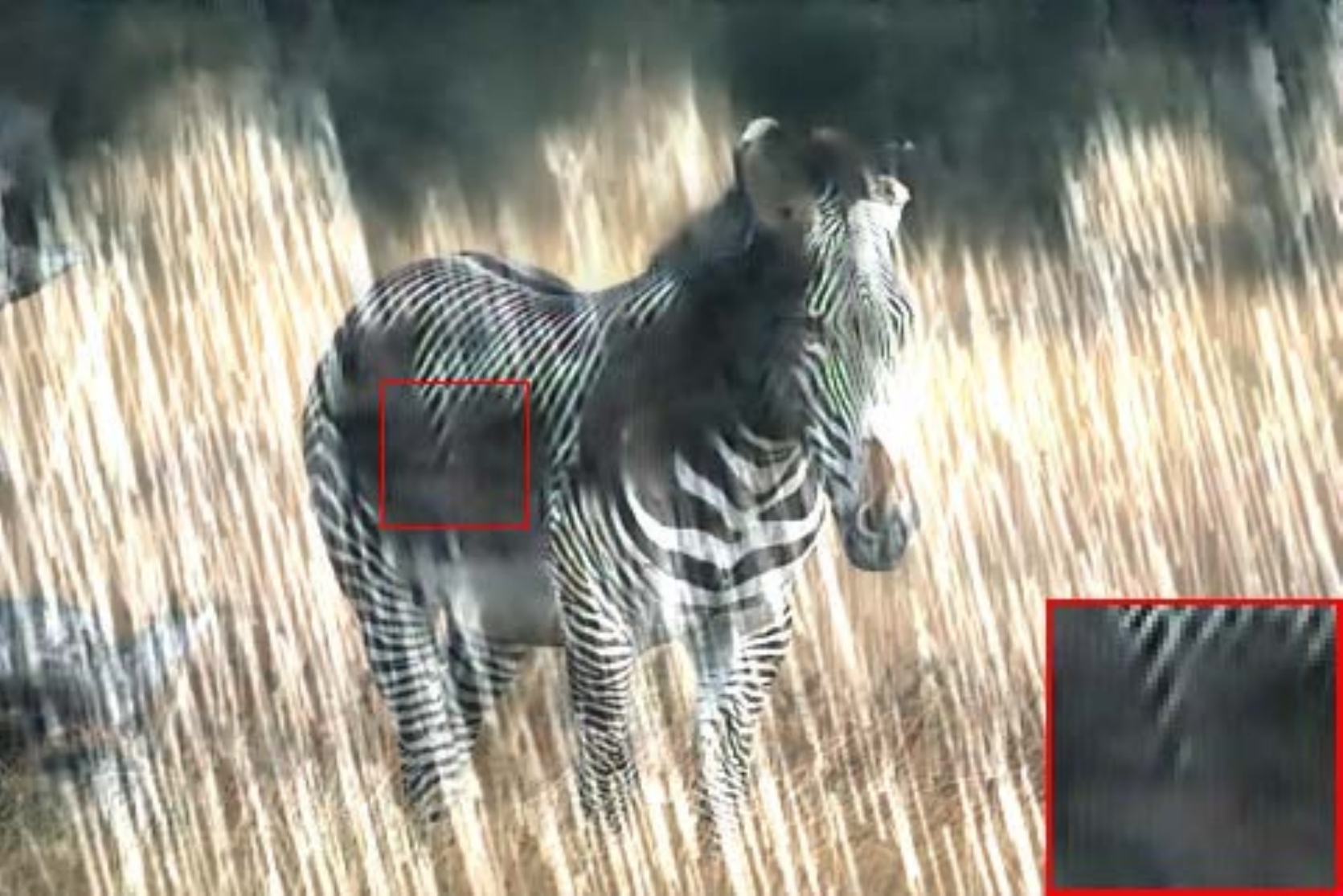}
\end{minipage}
\begin{minipage}[b]{0.16\linewidth}
\includegraphics[width=1\linewidth,height=0.6\linewidth]{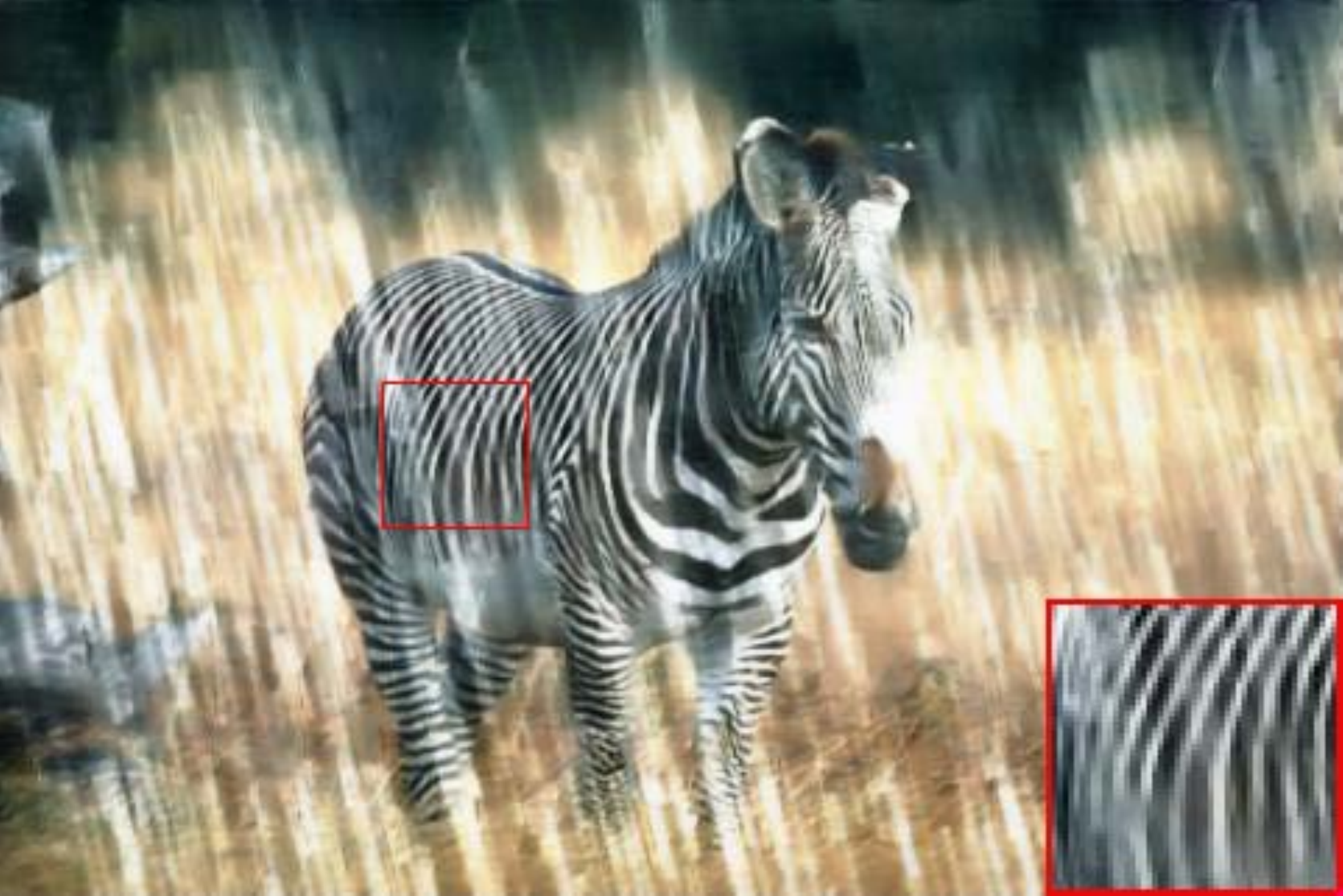}
\end{minipage}
\begin{minipage}[b]{0.16\linewidth}
\includegraphics[width=1\linewidth,height=0.6\linewidth]{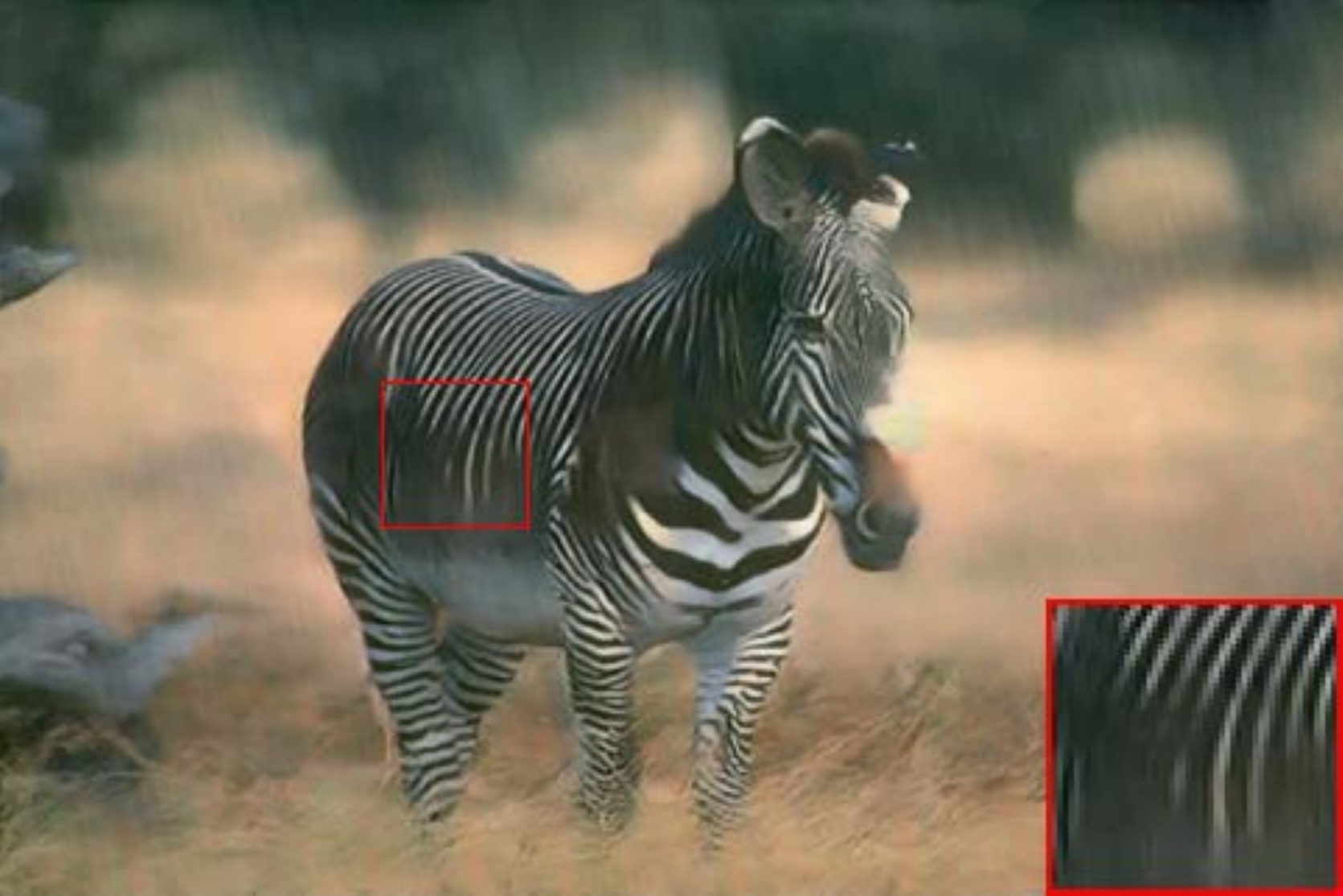}
\end{minipage}
\begin{minipage}[b]{0.16\linewidth}
\includegraphics[width=1\linewidth,height=0.6\linewidth]{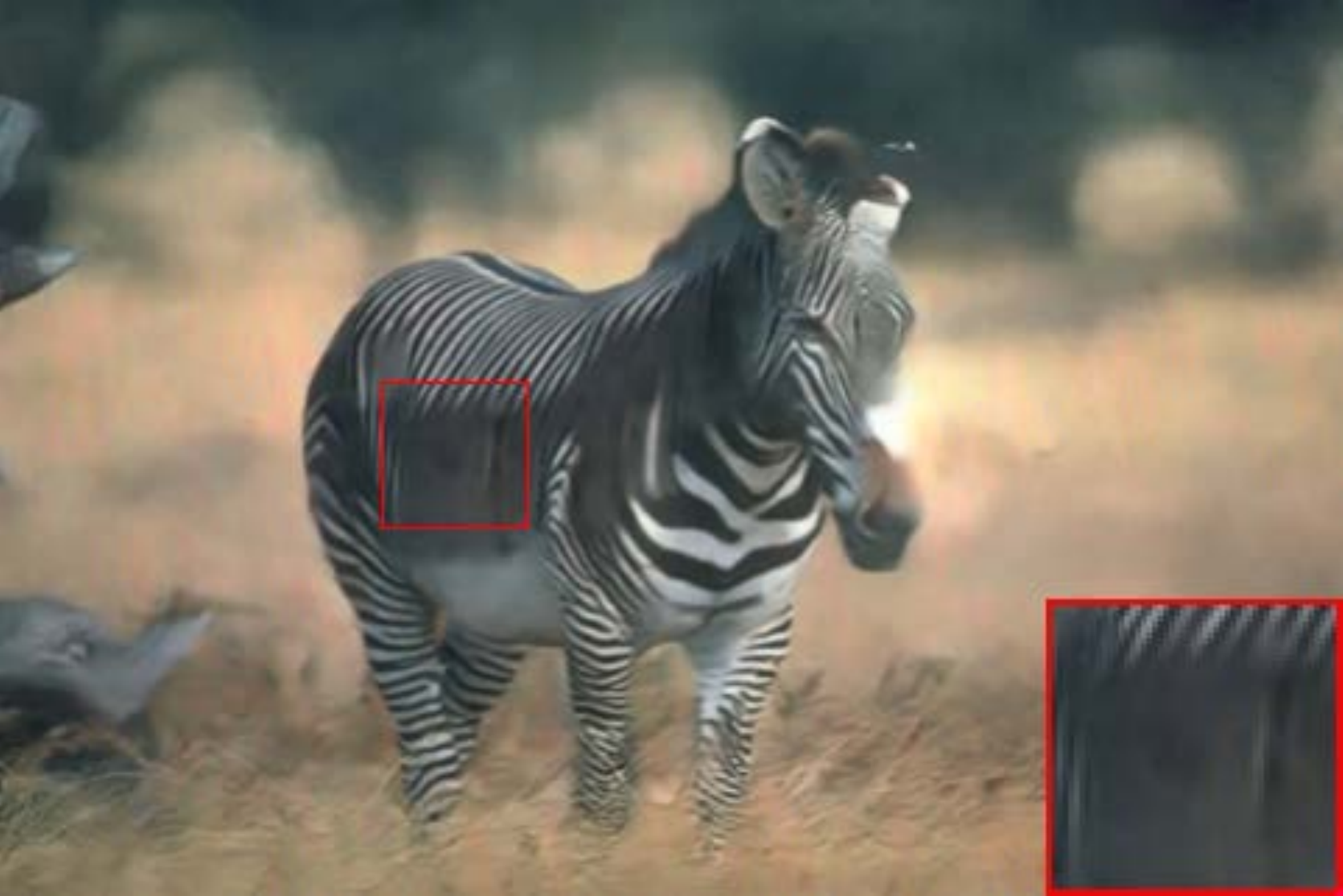}
\end{minipage}
\begin{minipage}[b]{0.16\linewidth}
\includegraphics[width=1\linewidth,height=0.6\linewidth]{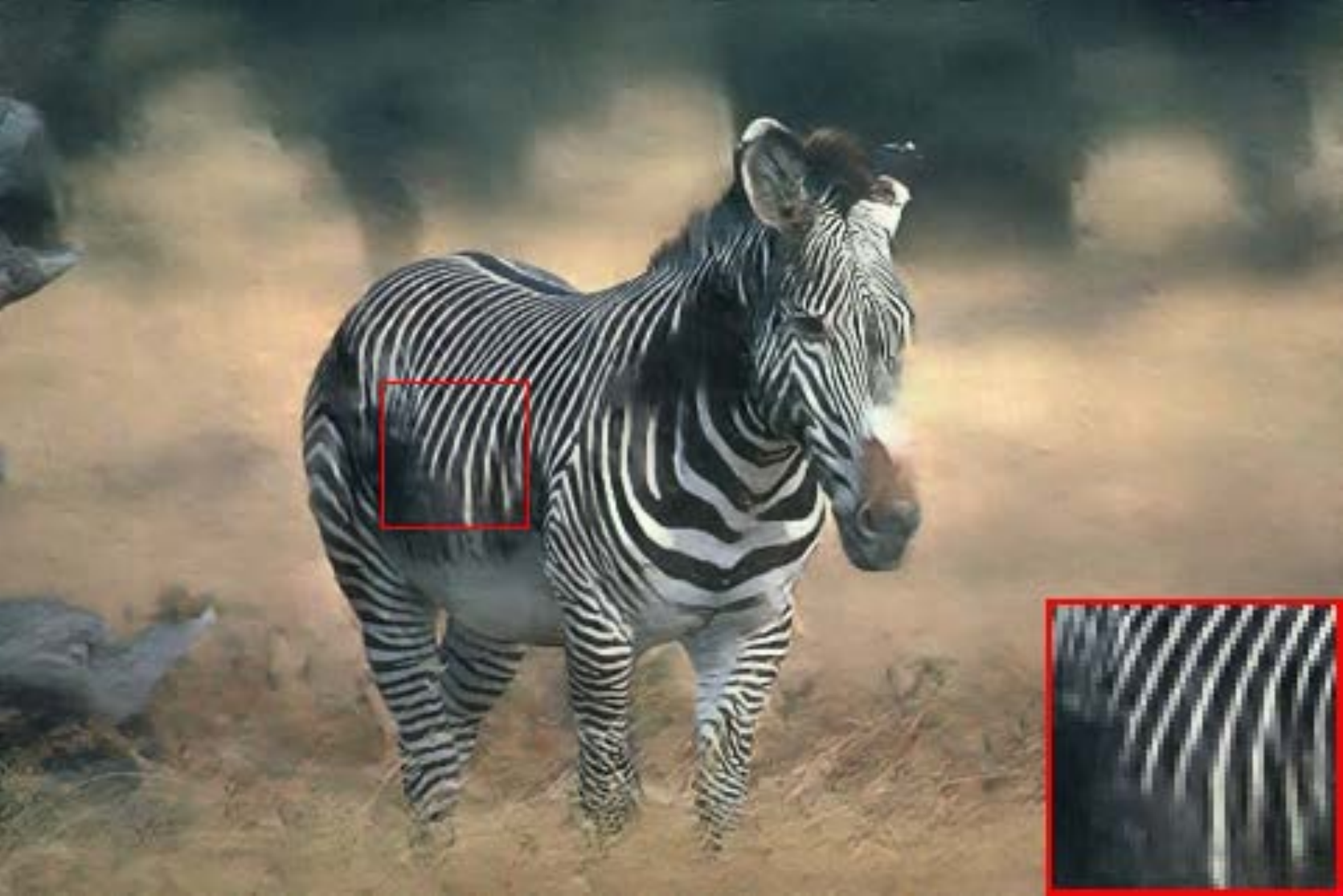}
\end{minipage}
\begin{minipage}[b]{0.16\linewidth}
\centering
\footnotesize{Rainy image}
\end{minipage}
\begin{minipage}[b]{0.16\linewidth}
\centering
\footnotesize{SPANet \cite{Alpher37}}
\end{minipage}
\begin{minipage}[b]{0.16\linewidth}
\centering
\footnotesize{DID \cite{Alpher14}}
\end{minipage}
\begin{minipage}[b]{0.16\linewidth}
\centering
\footnotesize{UMRL \cite{Alpher29}}
\end{minipage}
\begin{minipage}[b]{0.16\linewidth}
\centering
\footnotesize{MSPFN \cite{Alpher41}}
\end{minipage}
\begin{minipage}[b]{0.16\linewidth}
\centering
\footnotesize{RESCAN \cite{Alpher26}}
\end{minipage}
\begin{minipage}[b]{0.16\linewidth}
\includegraphics[width=1\linewidth,height=0.6\linewidth]{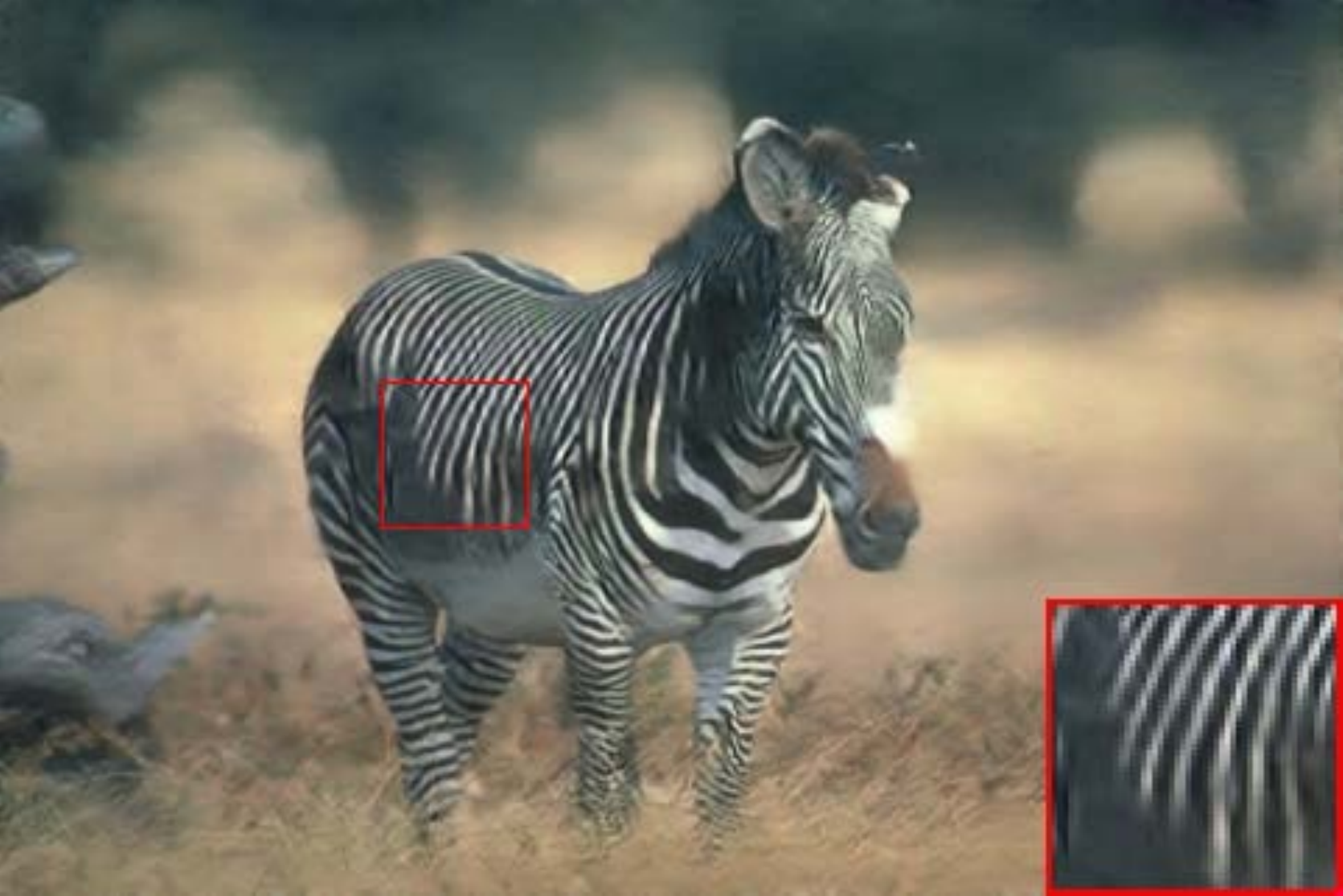}
\end{minipage}
\vspace{-0.15cm}
\begin{minipage}[b]{0.16\linewidth}
\includegraphics[width=1\linewidth,height=0.6\linewidth]{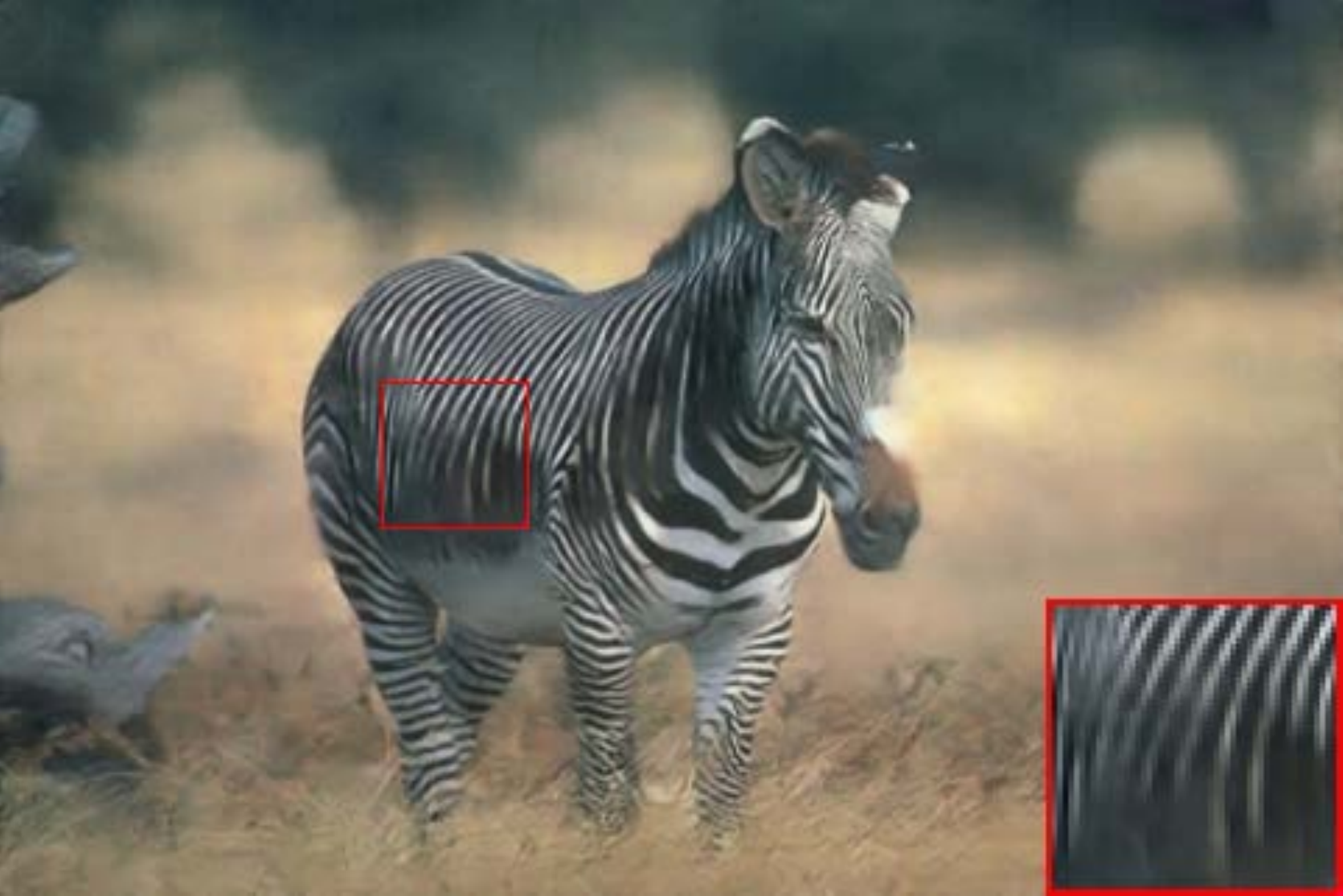}
\end{minipage}
\begin{minipage}[b]{0.16\linewidth}
\includegraphics[width=1\linewidth,height=0.6\linewidth]{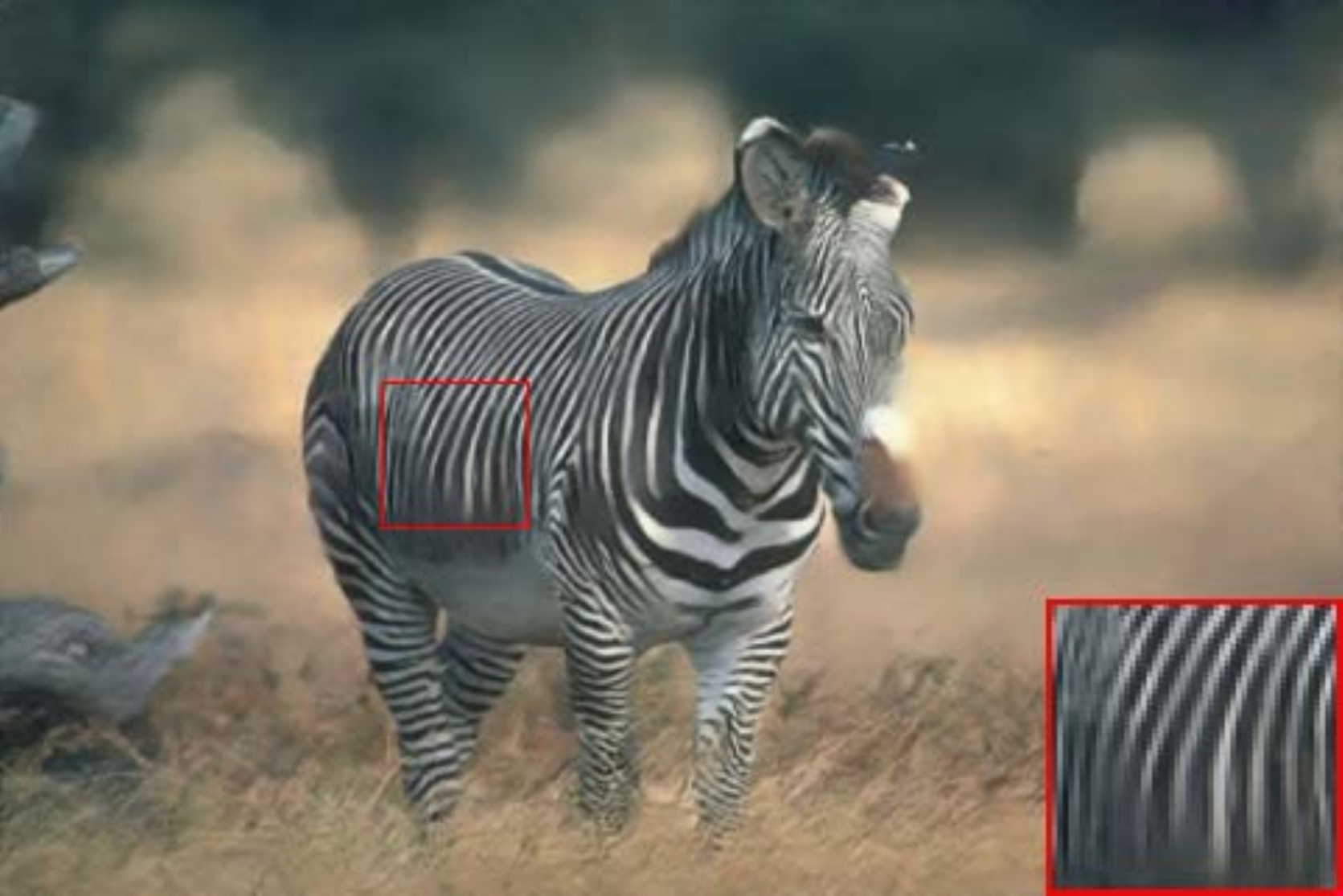}
\end{minipage}
\begin{minipage}[b]{0.16\linewidth}
\includegraphics[width=1\linewidth,height=0.6\linewidth]{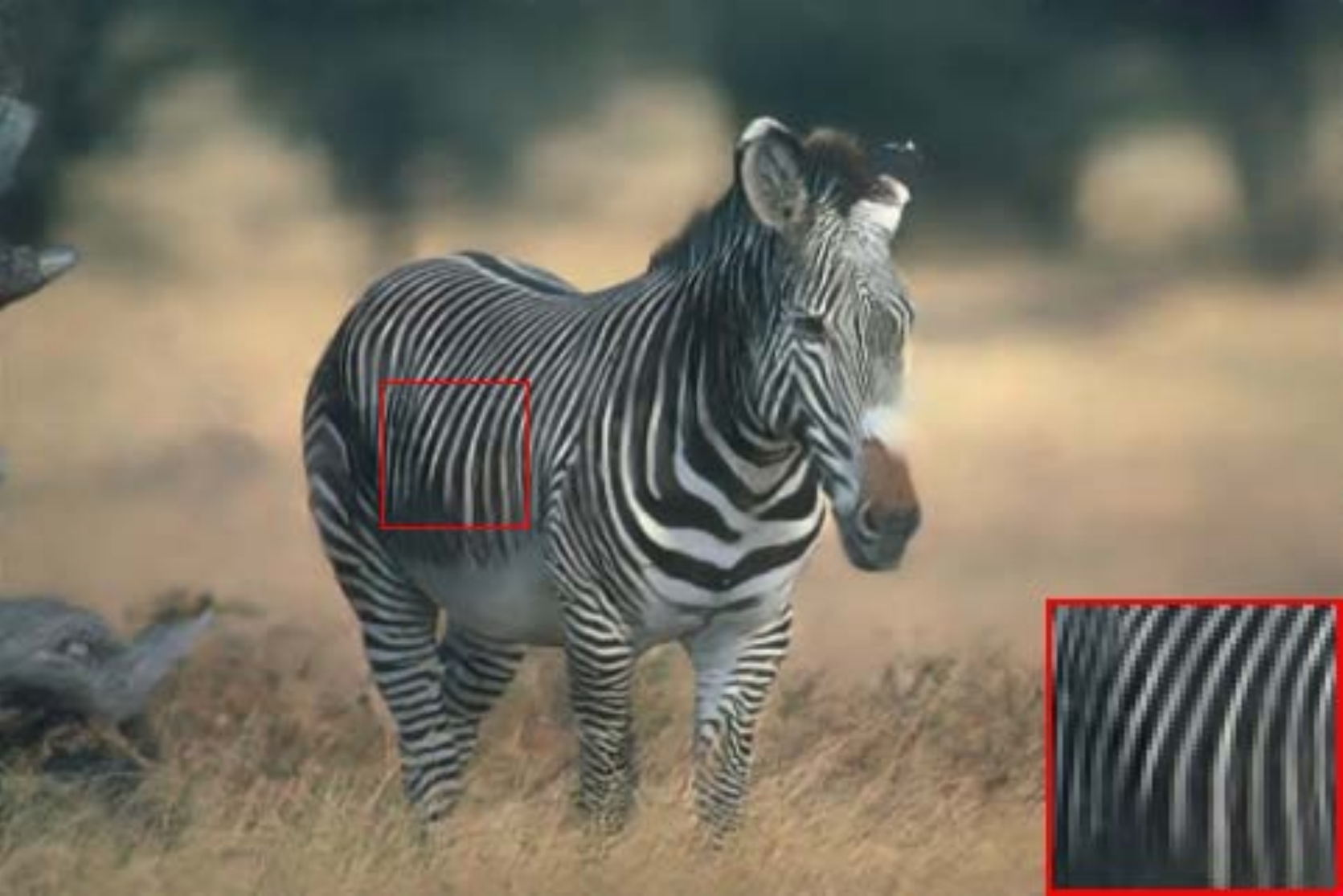}
\end{minipage}
\begin{minipage}[b]{0.16\linewidth}
\includegraphics[width=1\linewidth,height=0.6\linewidth]{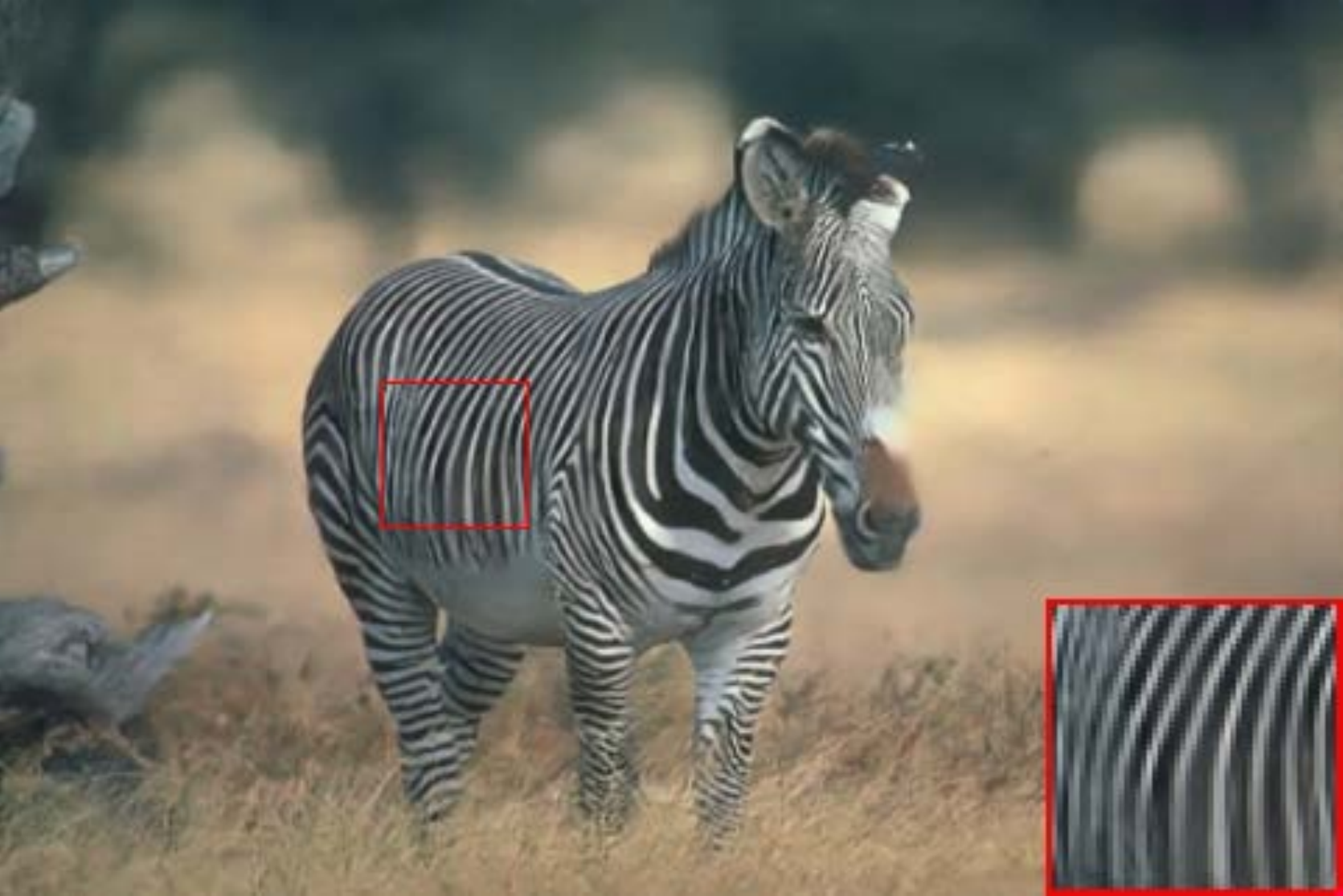}
\end{minipage}
\begin{minipage}[b]{0.16\linewidth}
\includegraphics[width=1\linewidth,height=0.6\linewidth]{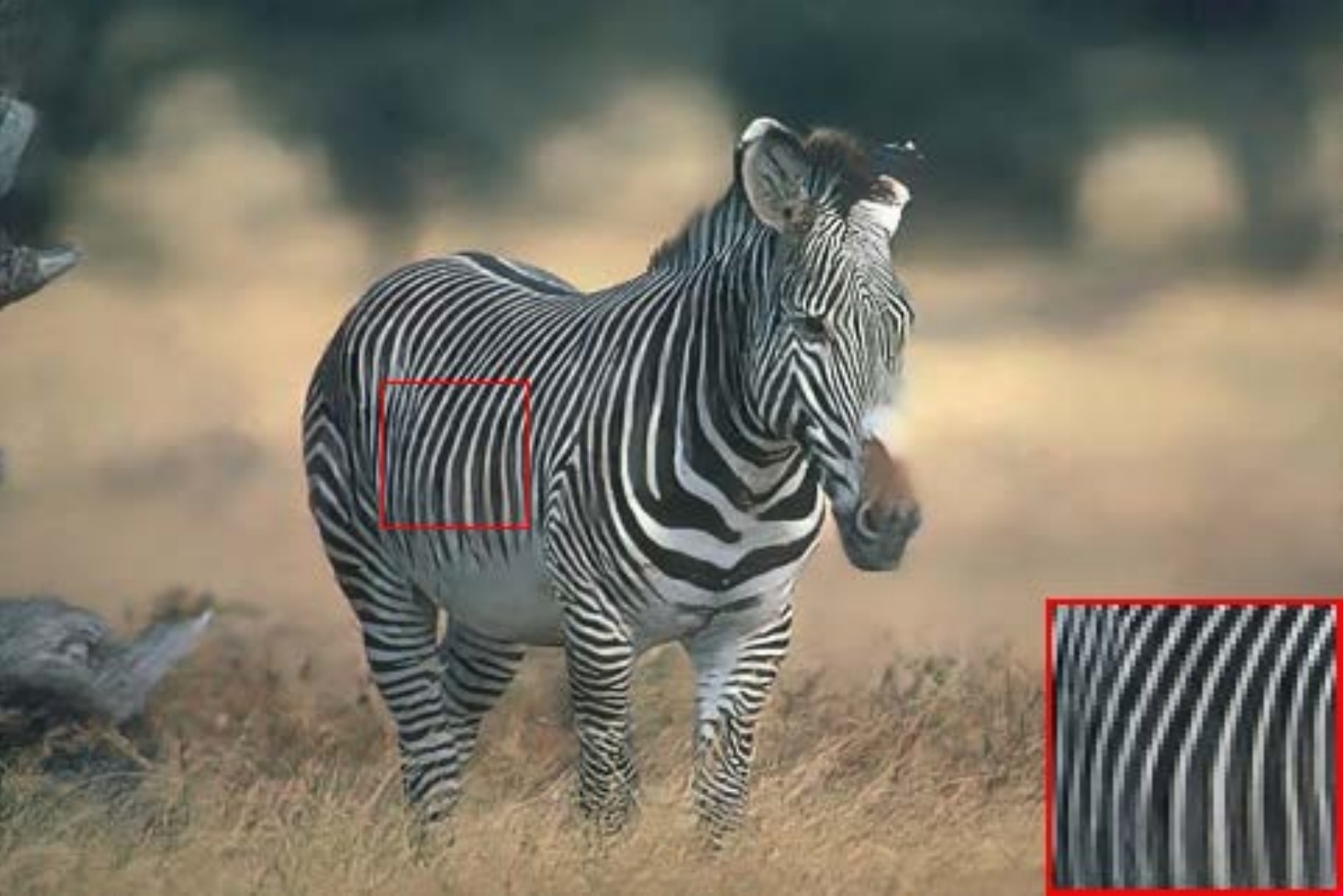}
\end{minipage}
\begin{minipage}[b]{0.16\linewidth}
\centering
\footnotesize{M1}
\end{minipage}
\vspace{-0.1cm}
\begin{minipage}[b]{0.16\linewidth}
\centering
\footnotesize{M2}
\end{minipage}
\begin{minipage}[b]{0.16\linewidth}
\centering
\footnotesize{M3}
\end{minipage}
\begin{minipage}[b]{0.16\linewidth}
\centering
\footnotesize{M4}
\end{minipage}
\begin{minipage}[b]{0.16\linewidth}
\centering
\footnotesize{M6}
\end{minipage}
\begin{minipage}[b]{0.16\linewidth}
\centering
\footnotesize{M7}
\end{minipage}
\caption{$1^{st}$ column: rainy image. $2^{nd}$-$12^{th}$ column: derained images. M1 to M7 (excluding M5) are visualizations of the ablation study.}
\label{fig7}
\vspace{-0.9cm}
\end{center}
\end{figure*}
The squared penalty of Mean Square Error (MSE) penalizes large errors and tolerates small errors, which tends to produce the over-smoothed image. Thus, Mean Absolute Error (MAE) is used to obtain better robustness. For the error detector rectifying the embedding residual map, we minimize the following two loss functions:
\begin{equation}L_{e 1}=\left\|R_{t \times 0.5}-\phi_{1}\left(I_{t}\right)\right\|_{1},\end{equation}
\begin{equation}L_{e 2}=\left\|\frac{\theta_{1}}{\left|R_{t \times 0.5}-\phi_{1}\left(I_{t}\right)\right|}-\varphi\left(I_{t \times 0.5}, \phi_{1}\left(I_{t}\right)\right)\right\|_{1},\label{eq9}\end{equation}
Note that $\left|R_{t \times 0.5}-\phi_{1}\left(I_{t}\right)\right|$ in Eq. \ref{eq9} is fixed. For the feature compensator, the loss function and the regularizer are formulated as follow:
\begin{equation}\begin{aligned}
L_{c} =\sum_{i \in\{0.25,0.5\}}\left\|R_{t \times i}+\theta_{2} \omega_{i}\left(R_{t \times i}\right) R_{t \times i}-\phi_{i}\left(I_{t}\right)\right\|_{1},
\end{aligned}\end{equation}
\begin{equation}\begin{aligned}
L_{p} =\left\|\omega_{i}\left(R_{t \times i}\right)\right\|_{2}^{2},
\end{aligned}\end{equation}
In the proposed model, the final output residual map is refined via using the following loss function:
\begin{equation}L_{f}=\left\|R_{t}-f\left(I_{t}\right)\right\|_{1},\end{equation}
where $f(\cdot)$ represents the overall network function. Furthermore, motivated by the goal of generating derained image to approximate its ground-truth image, we utilize the SSIM function \cite{Alpher35} as the additional evaluation metric for the generated clean image. It is formulated as follows:
\begin{equation}L_{S S I M}=-S S I M\left(B_{t}, I_{t}-f\left(I_{t}\right)\right),\end{equation}
where $B_{t}$ is the clean image truth.
The combination of the MAE based loss function and SSIM loss function can preserve the per-pixel similarity as well as preserving the global structure. The overall loss function used to train proposed RLNet is formulated as follows:
\begin{equation}L_{a l l}=L_{f}+L_{SSIM}+\lambda L_{p}+\lambda_{1} L_{e 1}+\lambda_{2} L_{e 2}+\lambda_{3} L_{c}, \end{equation}
where $\lambda$, $\lambda_{1}$, $\lambda_{2}$ and $\lambda_{3}$ are setup parameters.

%------------------------------------------------------------------------
\section{Experiments and Results}
%-------------------------------------------------------------------------
\subsection{Experiment Settings}
\noindent
{\bf Datasets.} For fairness, we use the same benchmark datasets as other methods. {\bf DID-data} with different rain magnitudes provided by Zhang et al. \cite{Alpher14} contains 12000 training images and 1200 testing images, {\bf Rain800} collected by Zhang et al. \cite{Alpher36} contains 700 training images and 100 testing images. Two datasets are synthesized by Yang et al. \cite{Alpher25}, namely {\bf Rain200H} and {\bf Rain 100L}. Rain200H (with heavy rain streaks) consists of 1800 training images and 200 testing images, and Rain100L (with light rain streaks) consists of 200 training images and 100 testing images. In addition, \cite{Alpher15,Alpher37} supply some real-world rainy images as a test set. Deraining methods are trained on the synthetic dataset Rain200H but are tested on real-world images.

\noindent
{\bf Training Details.} During training, a image pair (or its horizontal flip) is resized as $512\times512$ for training. We implemented all experiments on a Titan Xp GPU and use a batch size of 1. The first stage is used to train the error detector and the feature compensator with the initial value $\lambda=0.01$, $\lambda_{1}=0.6$, $\lambda_{2}=0$, $\lambda_{3}=0.6$, $\theta_{1}=0$, $\theta_{2}=0.05$: Adam is used as the optimizer to train the model and ends after 90 epochs. The initial learning rate is 0.0002 and divided by 5 when reaching 50, 65, 80 epochs. When reaching 20 epochs, $\theta_{2}$ is set as 0.15. When reaching 30 epochs, $\lambda_{2}$ is set as 6. By using the pre-trained weights obtained in the first stage, the fine-turning training process with the initial value $\lambda=0.01$, $\lambda_{1}=0.6$, $\lambda_{2}=0$, $\lambda_{3}=0.6$, $\theta_{1}=0.15$, $\theta_{2}=0.05$ is as follows: The initial learning rate is 0.0002 and divided by 2 every 30 epochs. The 240 epochs in total. When reaching $30 \times K (K=1,2,3,4,5,6)$ epochs, $\lambda_{2}$ is set as 0. When reaching $30 \times K +15 (K=0,1,2,3,4,5)$ epochs, $\lambda_{2}$ is set as 0.6. 

\noindent
{\bf Quality Comparisons.} Peak Signal to Noise Ratio (PSNR) \cite{Alpher43} and Structural Similarity Index (SSIM) \cite{Alpher35} are adopted to assess the performance of different methods on benchmark datasets. Since there is no ground-truth image in real-world datasets, NIQE \cite{Alpher44} is adopted to evaluate the generalization ability of different methods on real datasets.
%-------------------------------------------------------------------------
\subsection{Ablation Study}
\begin{table*}
\begin{center}
\caption{PSNR and SSIM comparisons on four benchmark datasets. Red and blue colors are used to indicate $1^{st}$ and $2^{nd}$ rank, respectively. $^{\triangleright}$ denotes some metrics of this method are copied from \cite{Alpher38}. $^{\circ}$ denotes the model is re-trained due to no pre-trained wight provided by the author. To be fair, UMRL \cite{Alpher29} and MSPFN \cite{Alpher41} are fine-tuned with the Rain100L training set when evaluated on the Rain100L test set.}
\vspace{-0.2cm}
\scalebox{0.655}{
\begin{tabular}{c|c|c|c|c|c|c|c|c|c|c|c}
\hline
Method & LP \cite{Alpher22} &DSC \cite{Alpher18}$^{\triangleright}$ & DDN \cite{Alpher13}& DID \cite{Alpher14}&SPANet \cite{Alpher37}&RESCAN \cite{Alpher26}$^{\circ}$& UMRL \cite{Alpher29}& PReNet \cite{Alpher15}& MSPFN \cite{Alpher41} &RCDNet \cite{Alpher45} &RLNet\\
\hline\hline
DID-data \cite{Alpher14} & 22.46/0.801 & 21.44/0.789& 27.33/0.853& 27.93/0.861&22.96/0.720& 29.12/0.880& 30.35/0.891&\textcolor{blue}{30.40}/\textcolor{blue}{0.891}&30.34/0.881& 29.81/0.859 &\textcolor{red}{32.62}/\textcolor{red}{0.917}\\
\hline
Rain200H \cite{Alpher25} & 14.26/0.420 & 15.66/0.544& 20.12/0.635& 15.54/0.520&13.27/0.412& 25.92/0.823& 23.01/0.744& 27.64/0.884 &24.30/0.748&\textcolor{blue}{28.83}/\textcolor{blue}{0.886} &\textcolor{red}{28.87}/\textcolor{red}{0.895}\\
\hline
Rain100L \cite{Alpher25} & 29.11/0.881 & 24.16/0.866& 33.50/0.944& 23.79/0.773&27.85/0.881& 36.58/0.970& 32.39/0.921& 36.28/0.979 &33.50/0.948& \textcolor{red}{38.60}/\textcolor{red}{0.983} &\textcolor{blue}{37.38}/\textcolor{blue}{0.980}\\
\hline
Rain800 \cite{Alpher36} & 20.46/0.729 & 18.56/0.599& 21.16/0.732& 21.22/0.750&21.22/0.687& 23.90/0.828& 23.24/0.808&22.83/0.790 &\textcolor{blue}{25.52}/\textcolor{blue}{0.830}& 24.59/0.821 &\textcolor{red}{27.95}/\textcolor{red}{0.870}\\
\hline
\end{tabular}\label{table4}}
\end{center}
\vspace{-0.8cm}
\end{table*}
We conduct all ablation experiments on Rain200H for its heavy rain streaks distributed heterogeneously.

\noindent
{\bf Absolute value operation.} Using the non-absolute error map to directly add to the embedding residual map, a network termed as RLNet- with simple error compensation is constructed. Since the error map as a variable target is difficult to fit, even if we double the parameters of the error detector for RLNet-, the resulting 27.70 dB only exceeding the original size model RLNet- in Table \ref{table1} by 0.13 dB.

\noindent
{\bf Threshold parameter $\bm{\theta_{1}}$.} In order to obtain a light-weight error detector, the absolute value operation is adopted and the threshold parameter $\theta_{1}$ is introduced to dynamically adjust upper limit of error reciprocals for better error map remapping as shown in Eq. \ref{eq9}. We keep a certain $\theta_{2}=0.15$ fixed and change the $\theta_{1}$ to find a better parameter setting for the error detector. Table \ref{table1} lists the PSNR and SSIM values of four RLNet models with $\theta_{1}=0.03,0.04,0.05,0.06$. When $\theta_{1}=0.03,0.04, 0.06$, the RLNet performs a little inferior to RLNet with $\theta_{1}=0.05$. It can be interpreted as the $\theta_{1}$ that is too large will weaken the role of the error detector, and the $\theta_{1}$ that is too small will increase the difficulty of error map remapping. Then we set the better parameter setting ($\theta_{1}=0.05$) for the proposed RLNet.

The results in Table \ref{table5} verify the effect of the error detector in the training process and results. In addition, we apply the embedding residual map and the error detector to the base network UMRL$^*$ \cite{Alpher29}. The obtained UMRL$^*+$E achieves 2.14 dB performance improvement.
\begin{table}
\begin{center}
\caption{Effect of threshold parameter $\theta_{1}$.}
\scalebox{0.79}{
\begin{tabular}{c|c|c|c|c|c}
\hline
Method &RLNet-& $\theta_{1}=0.03$ &$\theta_{1}=0.04$ & $\theta_{1}=0.05$& $\theta_{1}=0.06$\\
\hline\hline
PSNR &27.57&28.48&28.57 &{\bf 28.87} &28.61   \\
\hline
SSIM & 0.856& 0.877&0.882 &{\bf 0.895} &0.881  \\
\hline
\end{tabular}\label{table1}}
\end{center}
\vspace{-0.6cm}
\end{table}
\begin{table}
\begin{center}
\caption{Effect of the error detector. $-$E means that RLNet does not contain the error detector. $\pm$E denotes that RLNet uses the error detector in training but removes it after training. RLNet is our default model containing the error detector.}
\scalebox{0.72}{
\begin{tabular}{c|c|c|c|c|c}
\hline
Method & UMRL$^*$\cite{Alpher29}& RLNet$-$E &RLNet$\pm$E & UMRL$^*+$E& RLNet\\
\hline\hline
PSNR &22.31 &27.51 &28.05&24.45&{\bf 28.87}   \\
\hline
SSIM & 0.767&0.851&0.880&0.791 &{\bf 0.895}  \\
\hline
\end{tabular}\label{table5}}
\end{center}
\vspace{-0.8cm}
\end{table}

\noindent
{\bf Transformation Parameter $\bm{\theta_{2}}$.} We keep a certain $\theta_{1}=0.05$ fixed and change the $\theta_{2}$ to find a better parameter setting for the feature compensator. Table \ref{table2} lists the PSNR and SSIM values of four RLNet models with $\theta_{2}=0.05,0.1,0.15,0.2$. RLNet with $\theta_{2}=0.05,0.1,0.2$ performs a little inferior to RLNet with $\theta_{2}=0.15$. For the too small value of $\theta_{2}$ weakening the role of the feature compensator and the too large value of $\theta_{2}$ increasing the complexity of feature distribution, it is resonable to see those results from Table \ref{table2}. Hence, we set $\theta_{2}=0.15$ for the RLNet.

\noindent
{\bf Network Architecture.} As shown in the Table \ref{table3}), M1 denotes a single-stream U-net (kernel size is 3) without other modules. The base network, M1 itself enables the PSNR and SSIM to reach 26.91 dB and 0.830\% respectively. M2 replaces the residual block with FFRB for M1. The FFRB increases the PSNR by 0.22 dB and the SSIM by 0.3\%. Moreover, M3 modifies single-stream mechanism of M2 to multi-stream mechanism, with which M3 surpasses M2 by 0.29 dB and 1.4\%. When compared with M3, M4 with the unrectified embedding residual map hardly improves performance. M6 adds the error detector to M4. By comparing M6 and M4, it can be seen that error detector by itself contributes 1.29 dB and 3.4\%. M5 in Table \ref{table3}) is obtained from M6 by removing the error detection oriented loss function $L_{e 2}$ but keeping all other conditions unchanged. Experimental results show that M5 no longer achieves the same performance as M6 but has degraded performance as M4. This implies that the proposed error detection mechanism does play an important role in enhancing the model performance. Besides, the RLNet$\pm$E in Table \ref{table5} that uses the error detector in training but removes it after training will finally have the same model size as the RLNet without using error detection at all, yet the former outperforms the latter. M7 adds $L_{c}$+$L_{p}$ to M6 (i.e. $\theta_{2}$ is changed from 0 to 0.15). The results show that M7 with the feature compensation obtains the better performance over M6 by 0.18 dB and 1.4\%. The LPIPS \cite{Alpher46} values of M4 (without error detector), M5 (without $L_{e 2}$) and M6 (with error detector) are 0.053, 0.052 and 0.041 respectively when tested on Rain200H. From Fig. \ref{fig7}, one can see that the error detector and the feature compensator further improve deraining performance.
\begin{table}
\begin{center}
\caption{Effect of transformation parameter $\theta_{2}$.}
\scalebox{0.9}{
\begin{tabular}{c|c|c|c|c}
\hline
Method & $\theta_{2}=0.05$ &$\theta_{2}=0.1$ & $\theta_{2}=0.15$&$\theta_{2}=0.2$\\
\hline\hline
PSNR &28.74 &28.81 &{\bf 28.87} &28.85   \\
\hline
SSIM & 0.889&0.890 &{\bf 0.895} &0.893  \\
\hline
\end{tabular}\label{table2}}
\end{center}
\vspace{-0.6cm}
\end{table}
\begin{table}
\begin{center}
\caption{Ablation study on different modules.}
%\vspace{-0.35cm}
\scalebox{0.75}{
\begin{tabular}{c|c|c|c|c|c|c|c}
\hline
Model & M1 &M2 & M3& M4& M5& M6& M7\\
\hline\hline
Base Network & \checkmark & \checkmark& \checkmark& \checkmark& \checkmark & \checkmark& \checkmark\\
FFRB & & \checkmark& \checkmark& \checkmark& \checkmark& \checkmark  & \checkmark\\
Multi-stream & & & \checkmark& \checkmark& \checkmark& \checkmark & \checkmark\\
Embedding & & & &\checkmark &\checkmark & \checkmark & \checkmark\\
E-Detector & & & & &\checkmark & \checkmark& \checkmark\\
$L_{e 2}$ & & & & & & \checkmark& \checkmark\\
$L_{c}$+$L_{p}$ & & & & & & & \checkmark\\
\hline
PSNR &26.91 &27.13 &27.42 &27.39 &27.40&28.69&{\bf 28.87}   \\
\hline
SSIM & 0.830&0.833 &0.847 &0.846 &0.847&0.881&{\bf 0.895} \\
\hline
\end{tabular}\label{table3}}
\end{center}
\vspace{-0.8cm}
\end{table}

\subsection{Experiments on Benchmark Datasets}
\begin{figure*}
\begin{center}
%\vspace{-0.1cm}
\begin{minipage}[b]{0.12\linewidth}
\includegraphics[width=1\linewidth,height=0.6\linewidth]{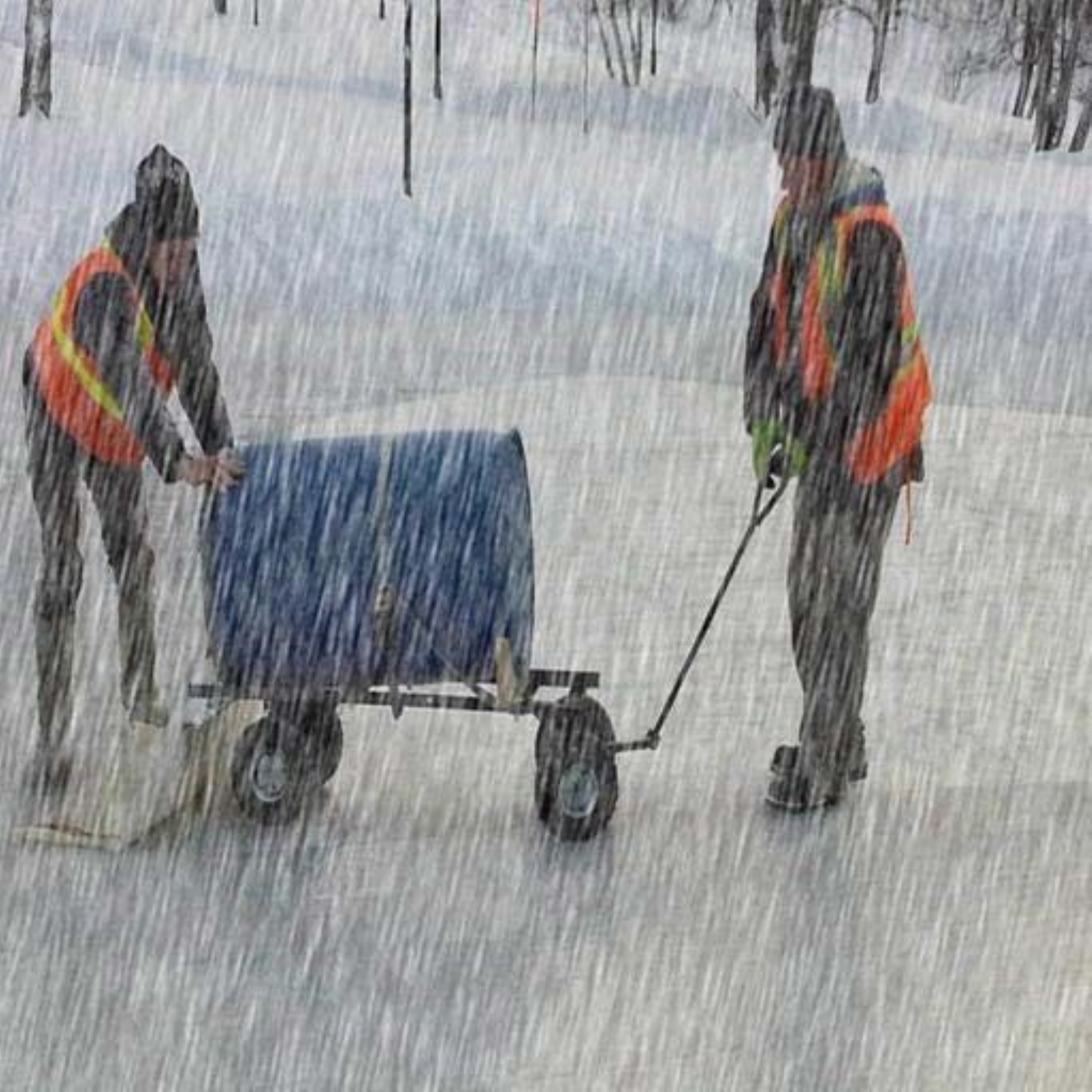}
\end{minipage}
\begin{minipage}[b]{0.12\linewidth}
\includegraphics[width=1\linewidth,height=0.6\linewidth]{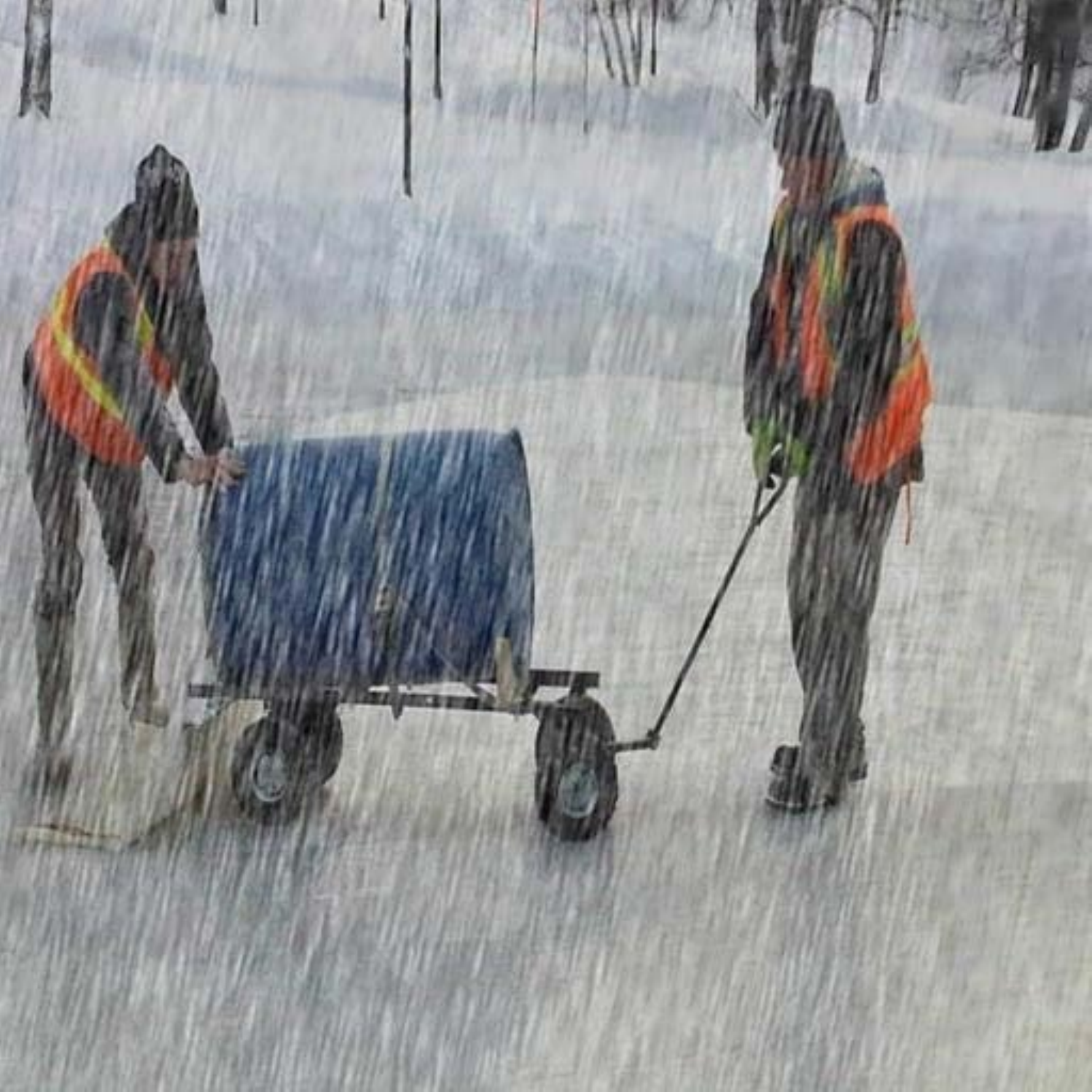}
\end{minipage}
\begin{minipage}[b]{0.12\linewidth}
\includegraphics[width=1\linewidth,height=0.6\linewidth]{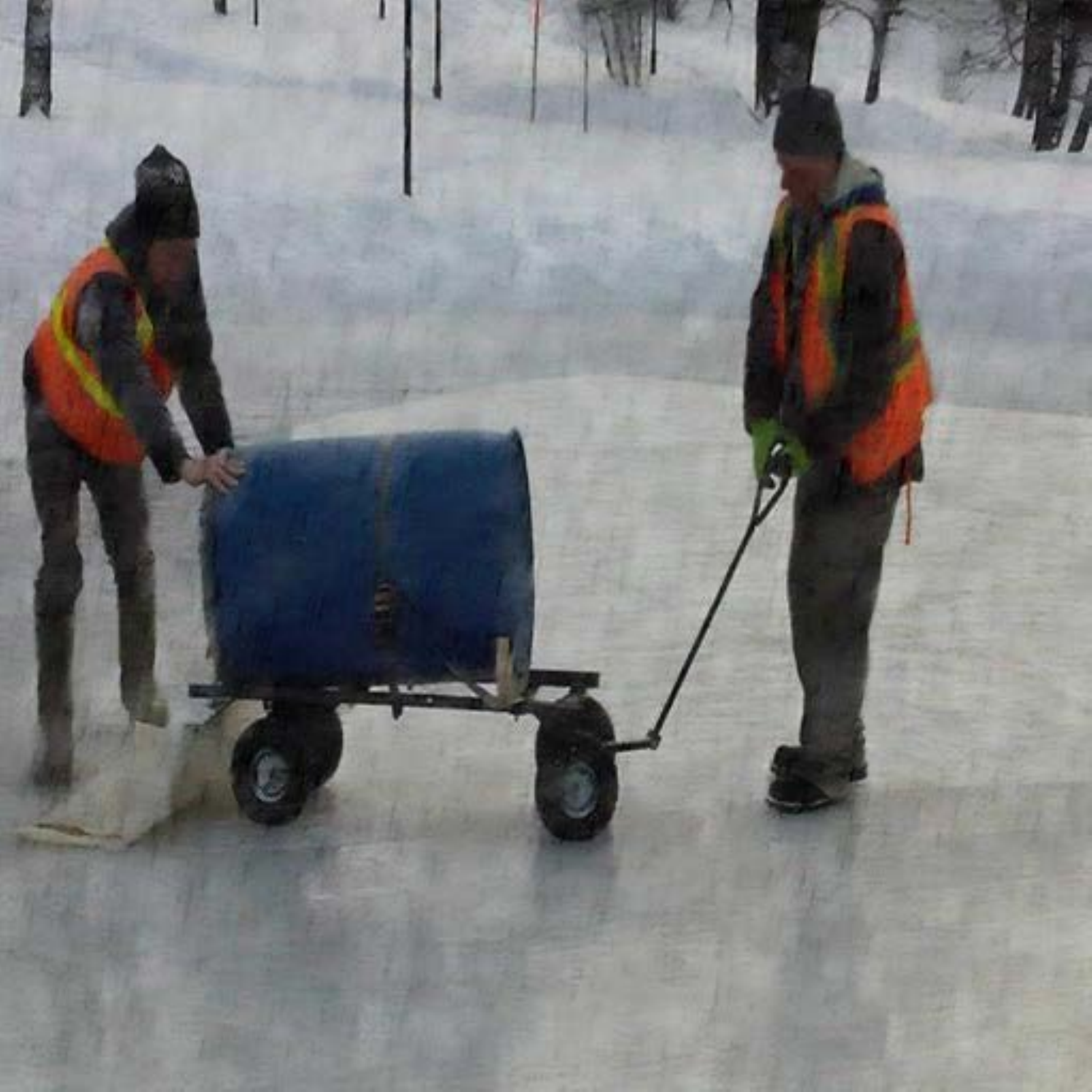}
\end{minipage}
\begin{minipage}[b]{0.12\linewidth}
\includegraphics[width=1\linewidth,height=0.6\linewidth]{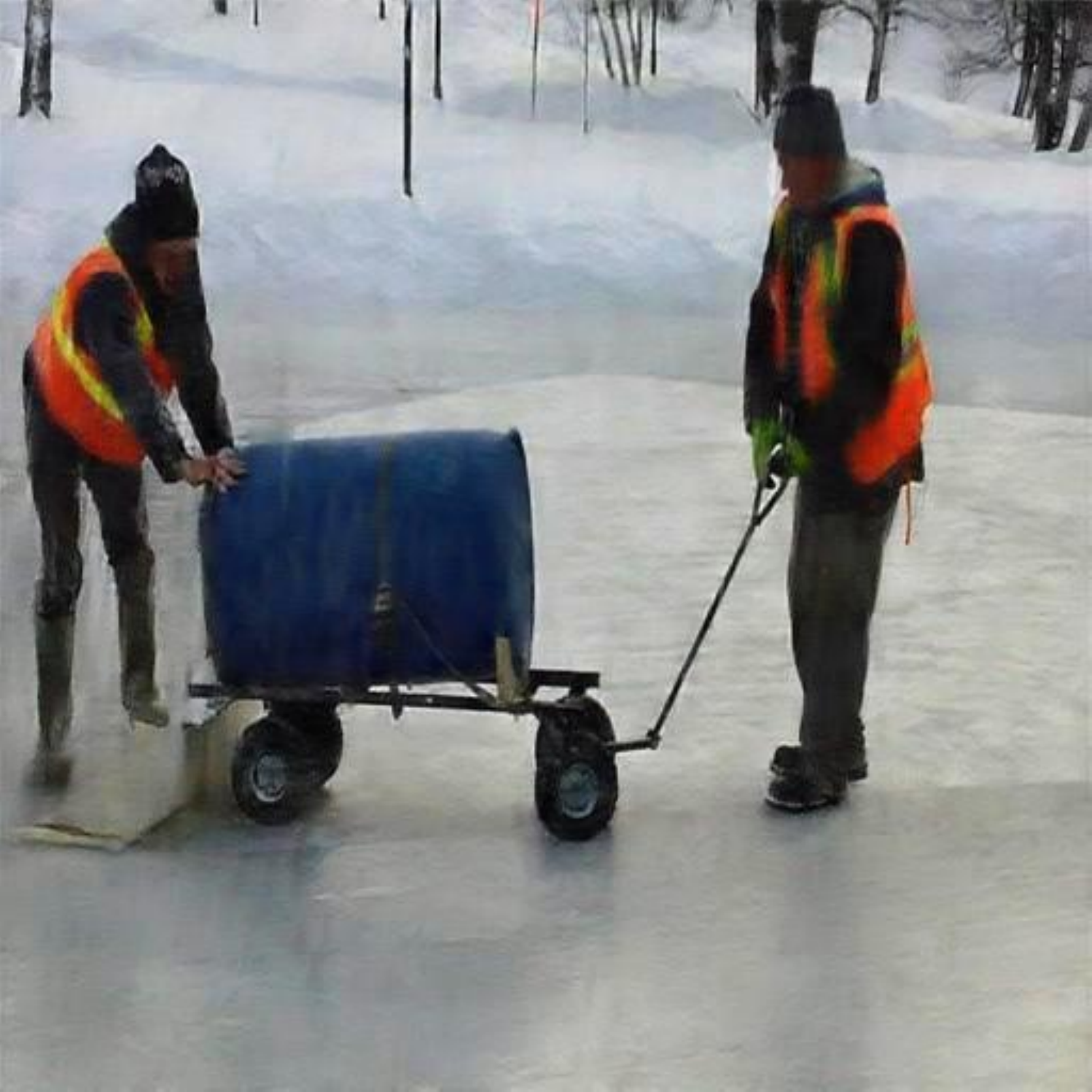}
\end{minipage}
\begin{minipage}[b]{0.12\linewidth}
\includegraphics[width=1\linewidth,height=0.6\linewidth]{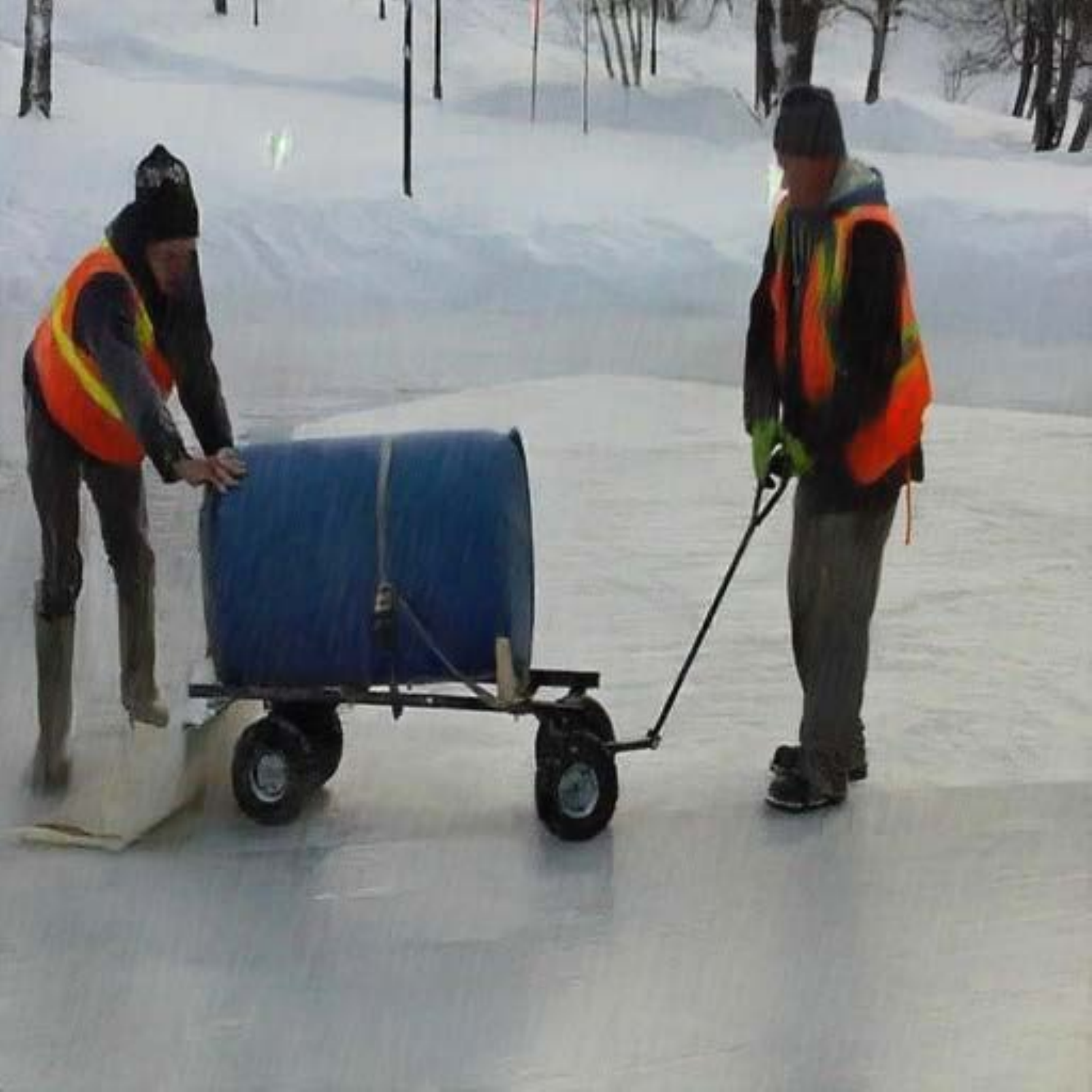}
\end{minipage}
\begin{minipage}[b]{0.12\linewidth}
\includegraphics[width=1\linewidth,height=0.6\linewidth]{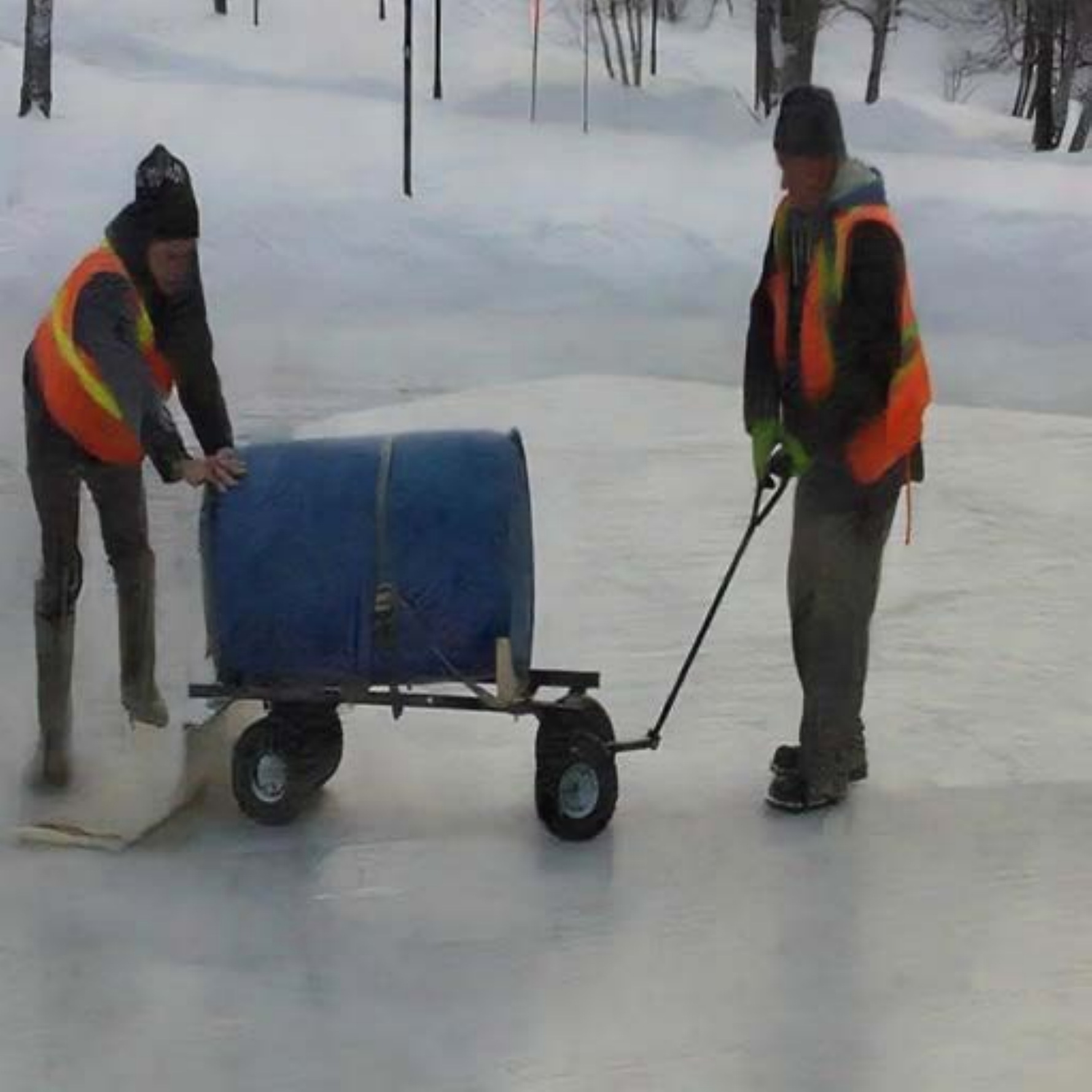}
\end{minipage}
\begin{minipage}[b]{0.12\linewidth}
\includegraphics[width=1\linewidth,height=0.6\linewidth]{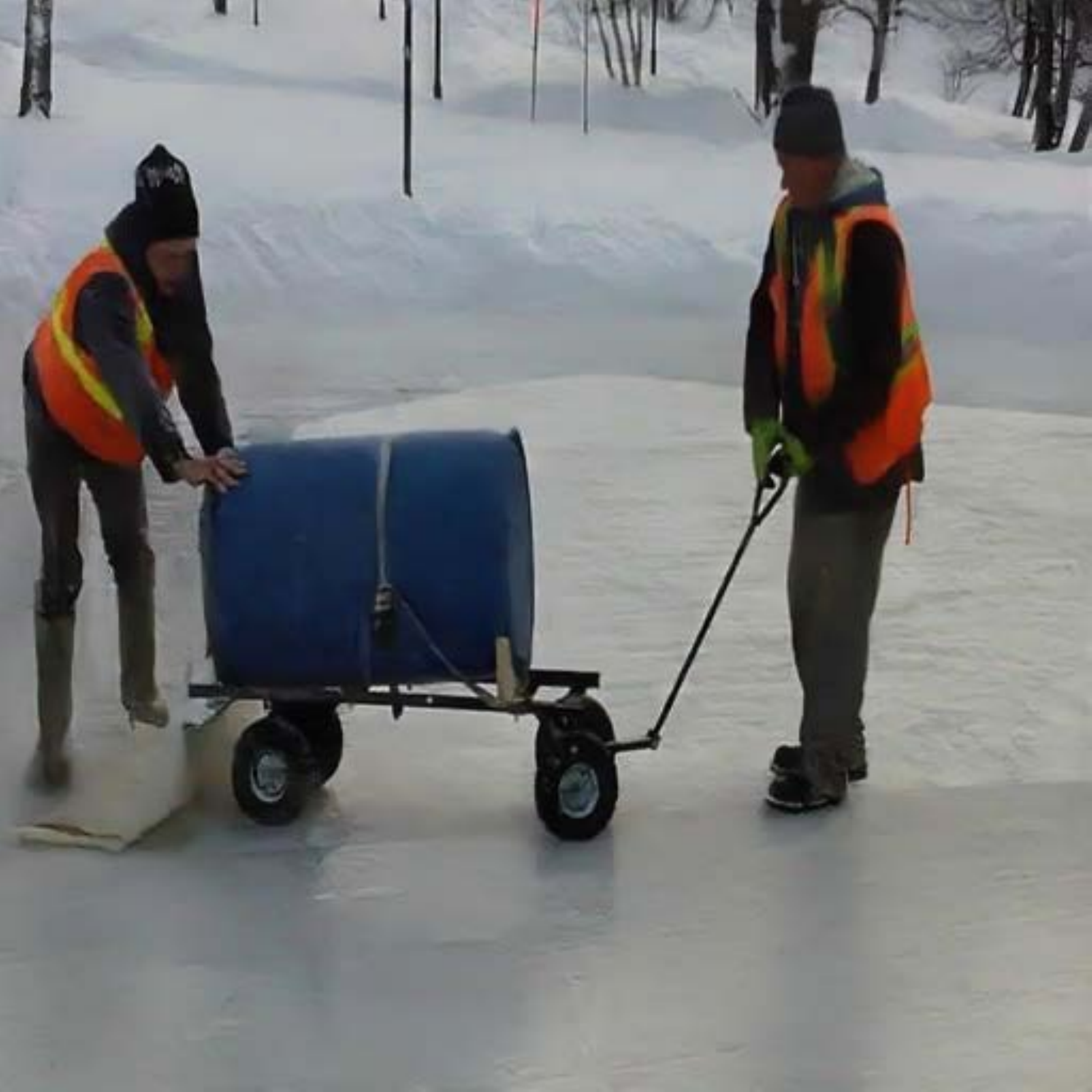}
\end{minipage}
\begin{minipage}[b]{0.12\linewidth}
\includegraphics[width=1\linewidth,height=0.6\linewidth]{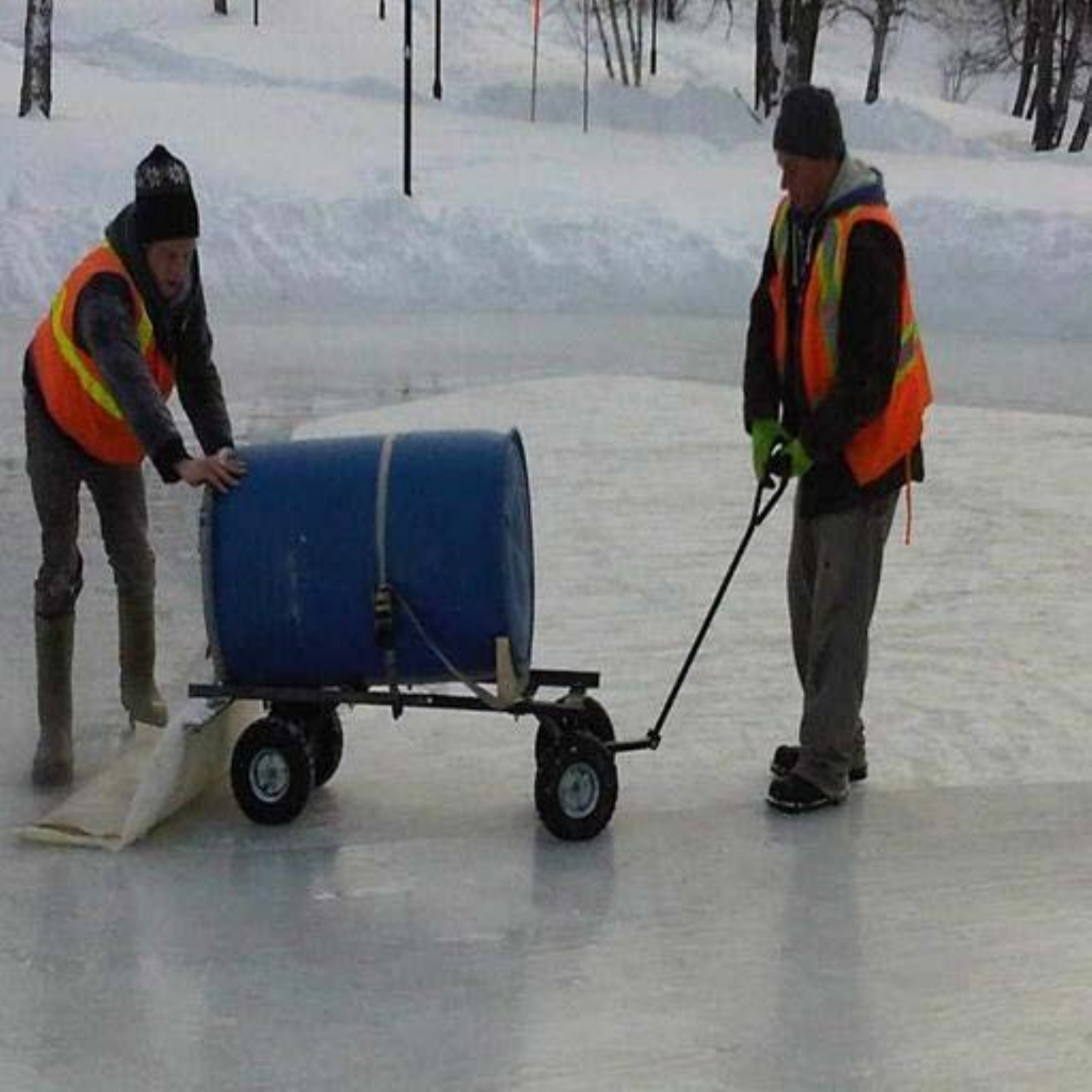}
\end{minipage}
\begin{minipage}[b]{0.12\linewidth}
\includegraphics[width=1\linewidth,height=0.6\linewidth]{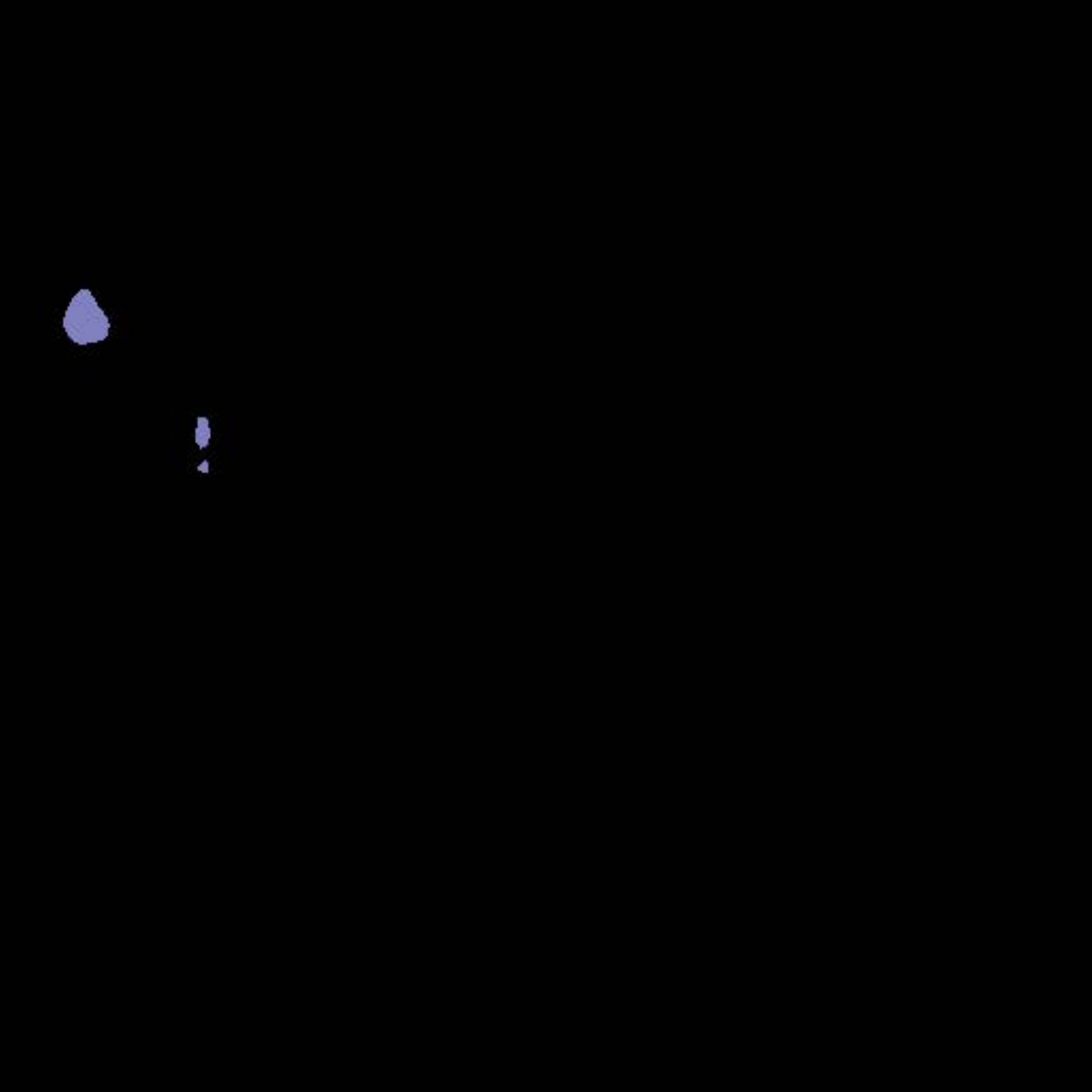}
\end{minipage}
\begin{minipage}[b]{0.12\linewidth}
\includegraphics[width=1\linewidth,height=0.6\linewidth]{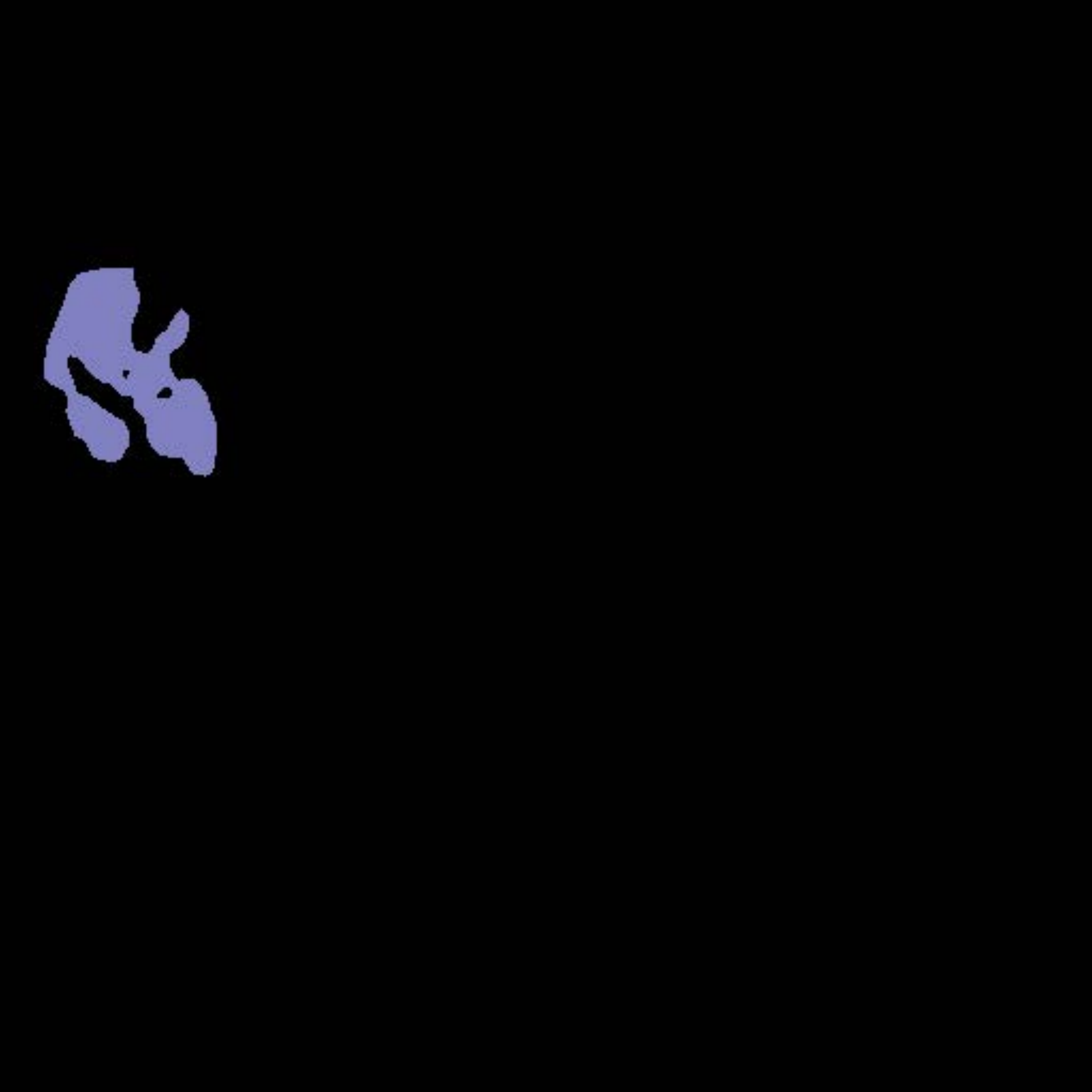}
\end{minipage}
\begin{minipage}[b]{0.12\linewidth}
\includegraphics[width=1\linewidth,height=0.6\linewidth]{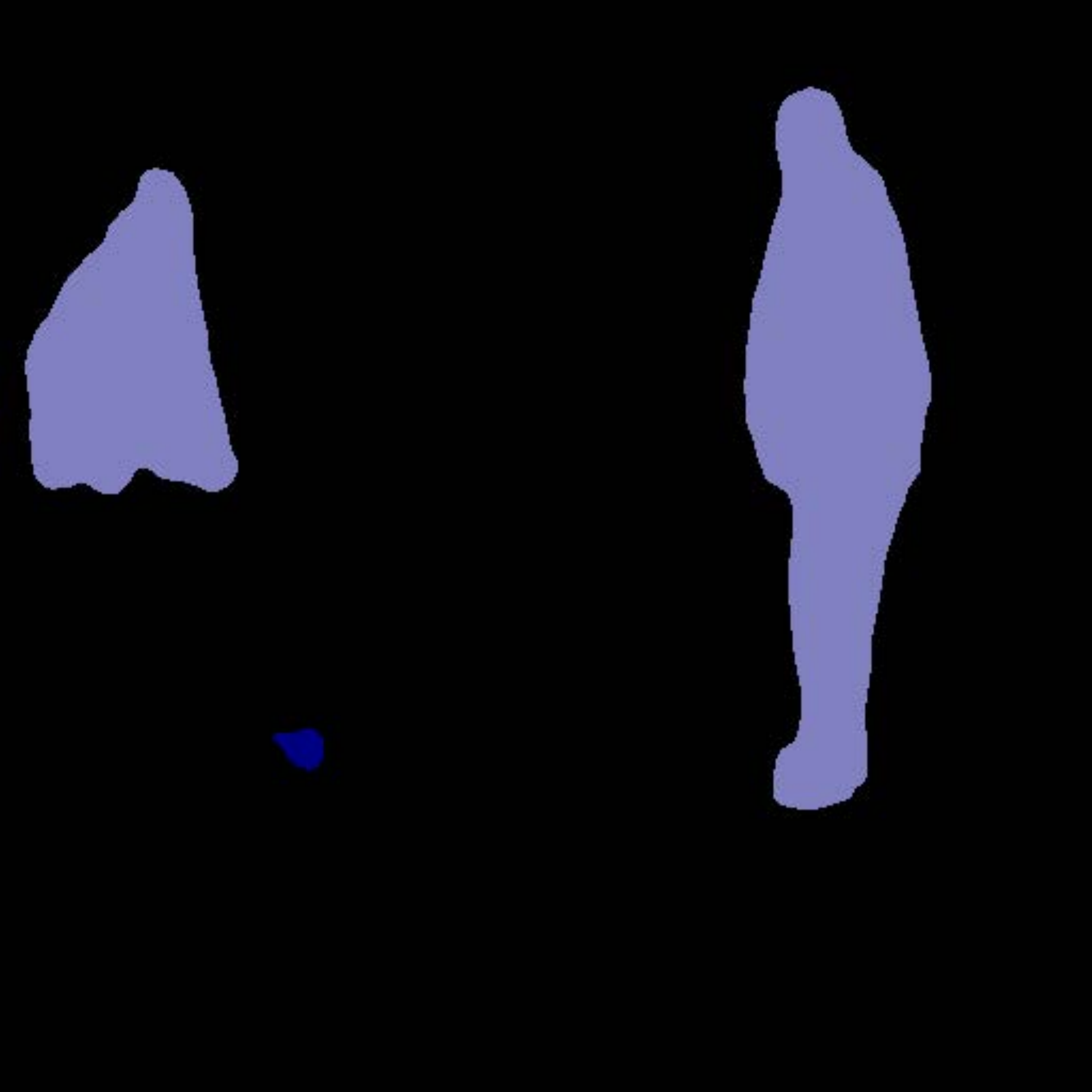}
\end{minipage}
\begin{minipage}[b]{0.12\linewidth}
\includegraphics[width=1\linewidth,height=0.6\linewidth]{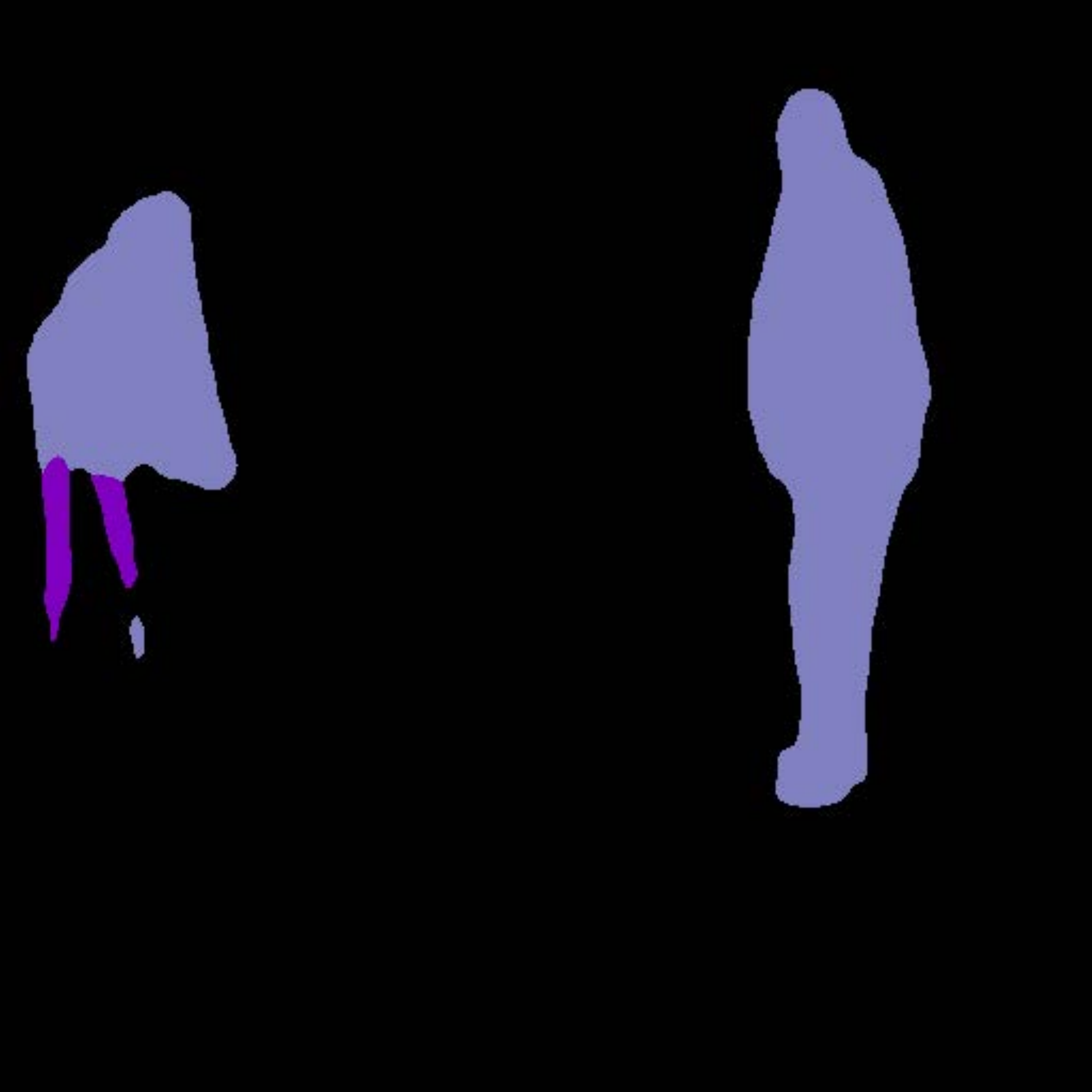}
\end{minipage}
\begin{minipage}[b]{0.12\linewidth}
\includegraphics[width=1\linewidth,height=0.6\linewidth]{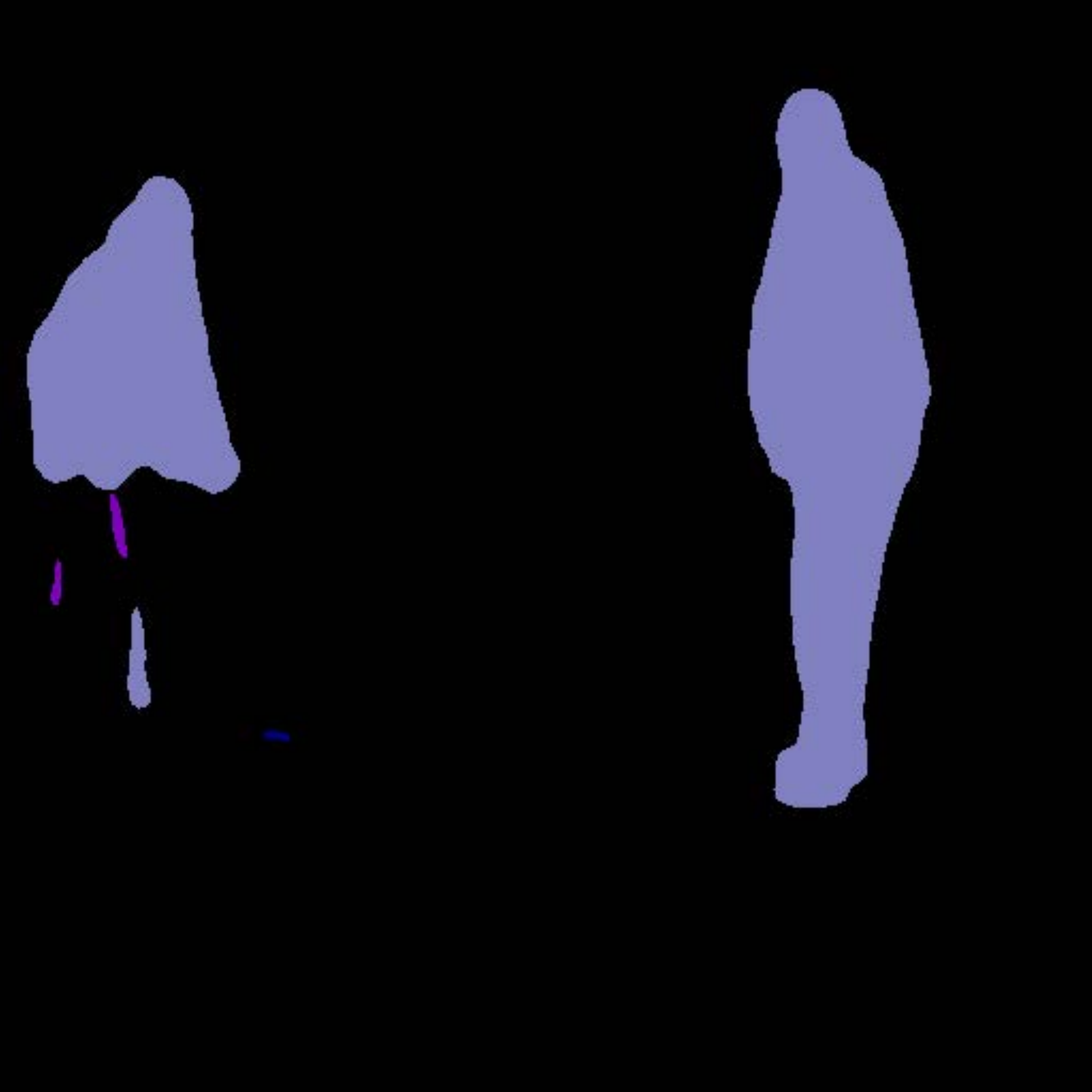}
\end{minipage}
\begin{minipage}[b]{0.12\linewidth}
\includegraphics[width=1\linewidth,height=0.6\linewidth]{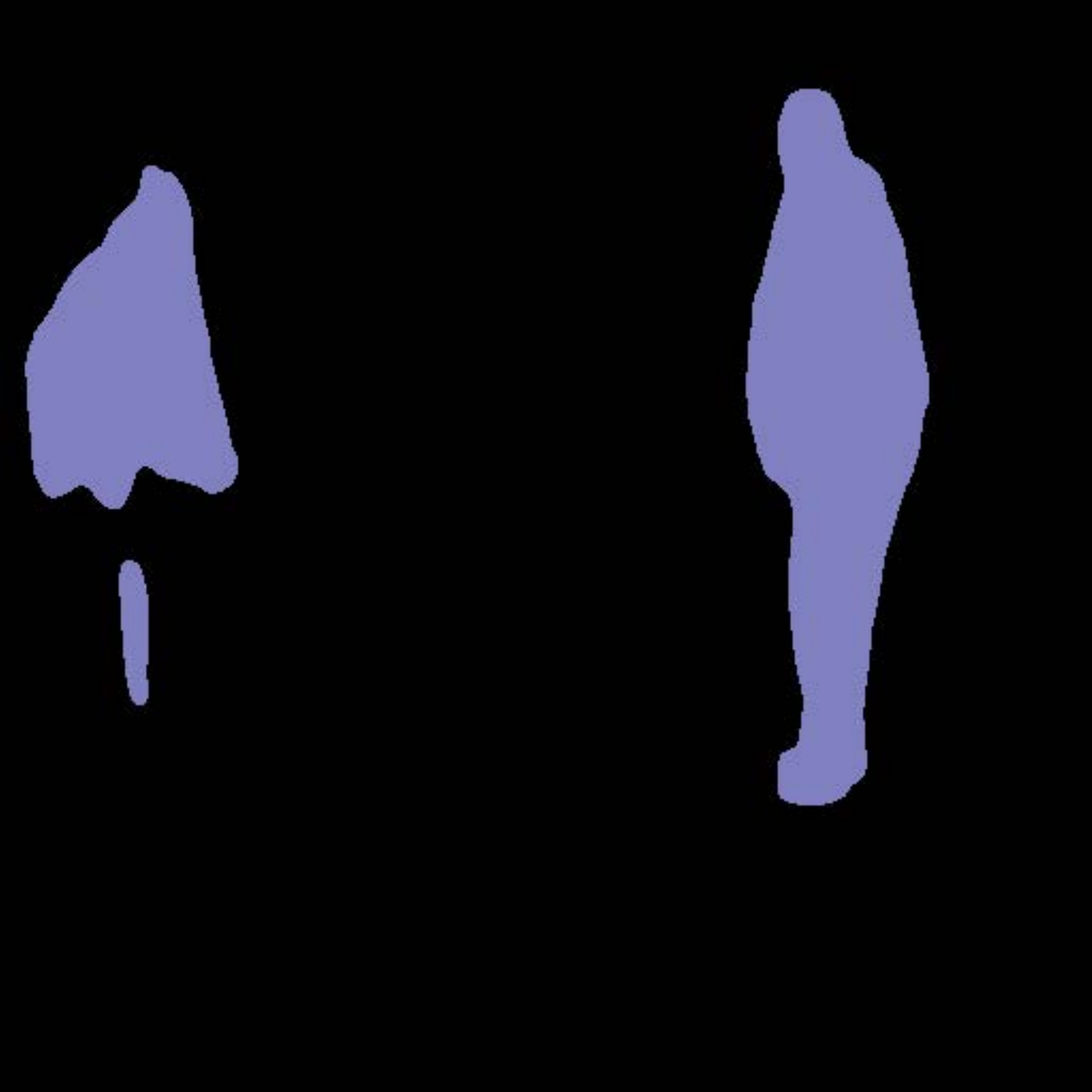}
\end{minipage}
\begin{minipage}[b]{0.12\linewidth}
\includegraphics[width=1\linewidth,height=0.6\linewidth]{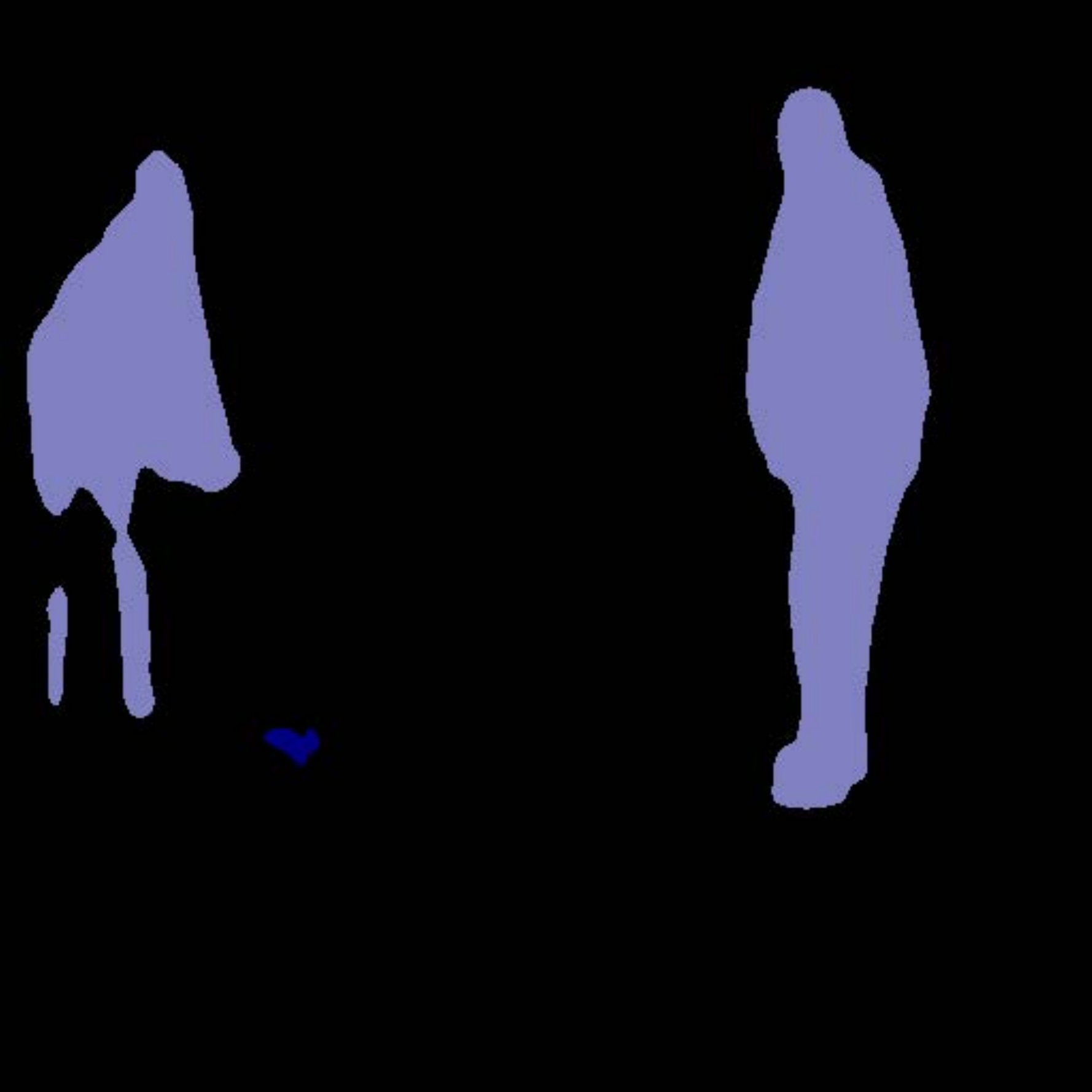}
\end{minipage}
\begin{minipage}[b]{0.12\linewidth}
\includegraphics[width=1\linewidth,height=0.6\linewidth]{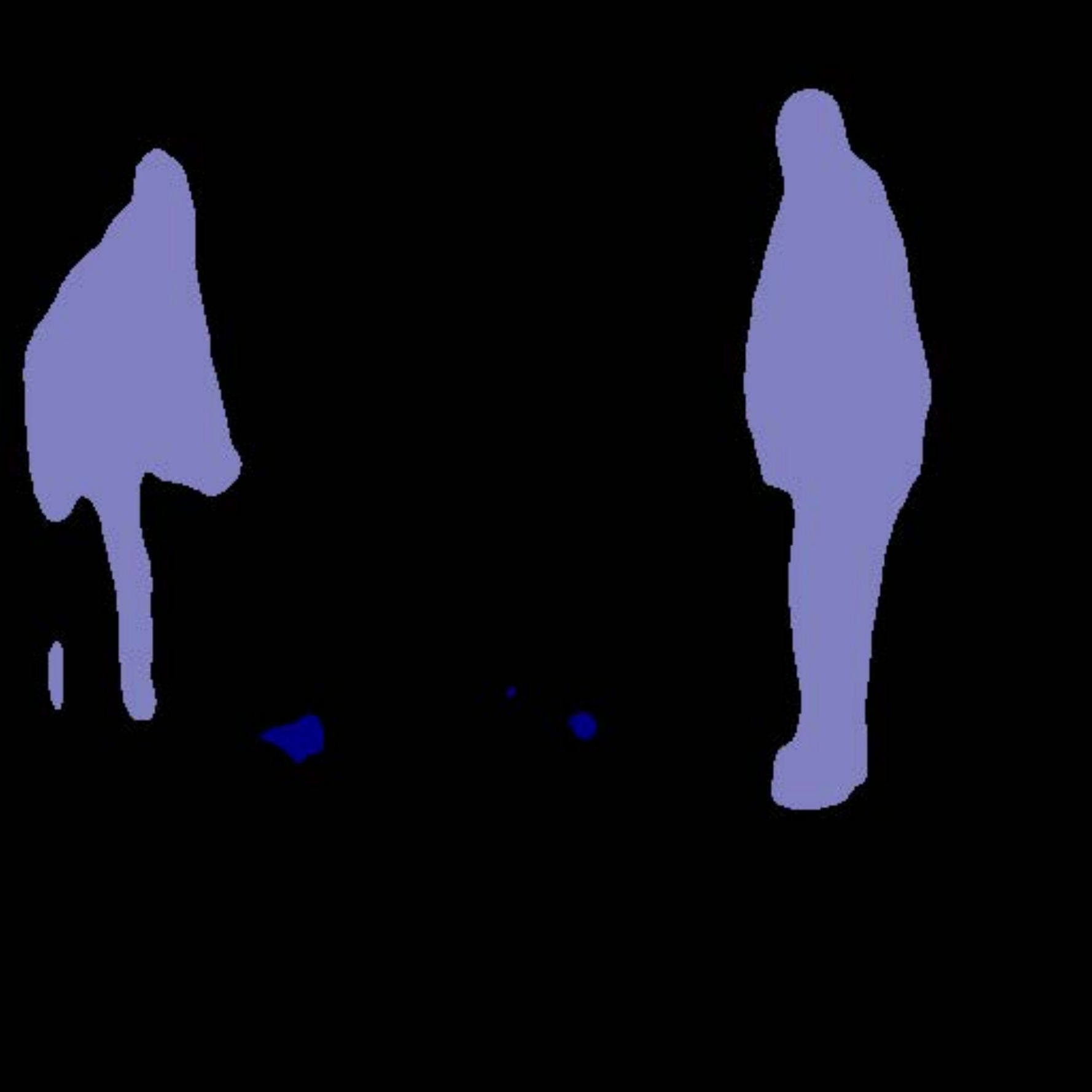}
\end{minipage}
\begin{minipage}[b]{0.12\linewidth}
\centering
\footnotesize{(a) Rainy images}
\end{minipage}
\begin{minipage}[b]{0.12\linewidth}
\centering
\footnotesize{(b) SPANet \cite{Alpher37}}
\end{minipage}
\begin{minipage}[b]{0.12\linewidth}
\centering
\footnotesize{(c) RESCAN \cite{Alpher26}}
\end{minipage}
\begin{minipage}[b]{0.12\linewidth}
\centering
\footnotesize{(d) DID \cite{Alpher14}}
\end{minipage}
\begin{minipage}[b]{0.12\linewidth}
\centering
\footnotesize{(e) UMRL \cite{Alpher29}}
\end{minipage}
\begin{minipage}[b]{0.12\linewidth}
\centering
\footnotesize{(f) MSPFN \cite{Alpher41}}
\end{minipage}
\begin{minipage}[b]{0.12\linewidth}
\centering
\footnotesize{(g) Ours}
\end{minipage}
\begin{minipage}[b]{0.12\linewidth}
\centering
\footnotesize{(h) Groundtruth}
\end{minipage}
\caption{Examples of joint deraining and segmentation. DeepLabv3+ \cite{Alpher42} is adopted for segmentation. Zoom in to see the details.}
\label{fig10}
\vspace{-0.7cm}
\end{center}
\end{figure*}
\begin{figure*}
\centering
\subfigure[Rainy images]{
\begin{minipage}[b]{0.137\linewidth}
\includegraphics[width=1\linewidth,height=0.6\linewidth]{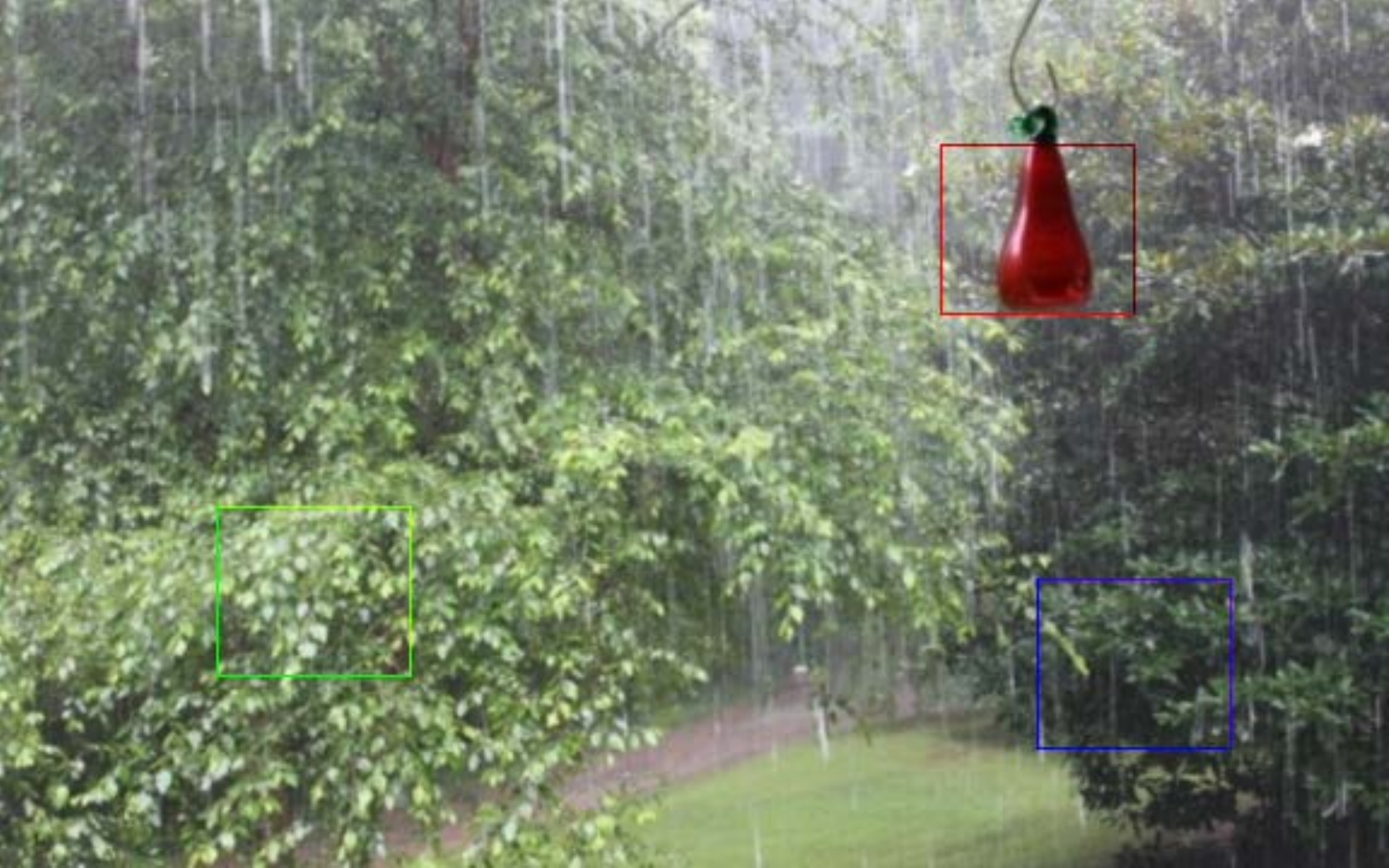}
\vspace{0.05cm}
\begin{minipage}[b]{0.33\linewidth}
\includegraphics[width=1\linewidth,height=0.6\linewidth]{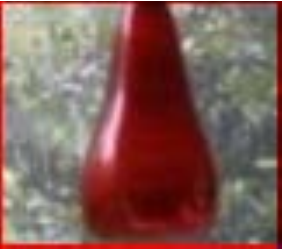}
\end{minipage}
\hspace{-0.15cm}
\begin{minipage}[b]{0.33\linewidth}
\includegraphics[width=1\linewidth,height=0.6\linewidth]{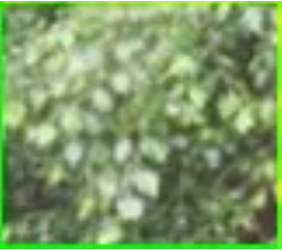}
\end{minipage}
\hspace{-0.15cm}
\begin{minipage}[b]{0.33\linewidth}
\includegraphics[width=1\linewidth,height=0.6\linewidth]{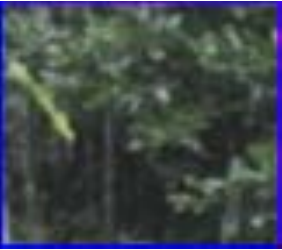}
\end{minipage}
\includegraphics[width=1\linewidth,height=0.6\linewidth]{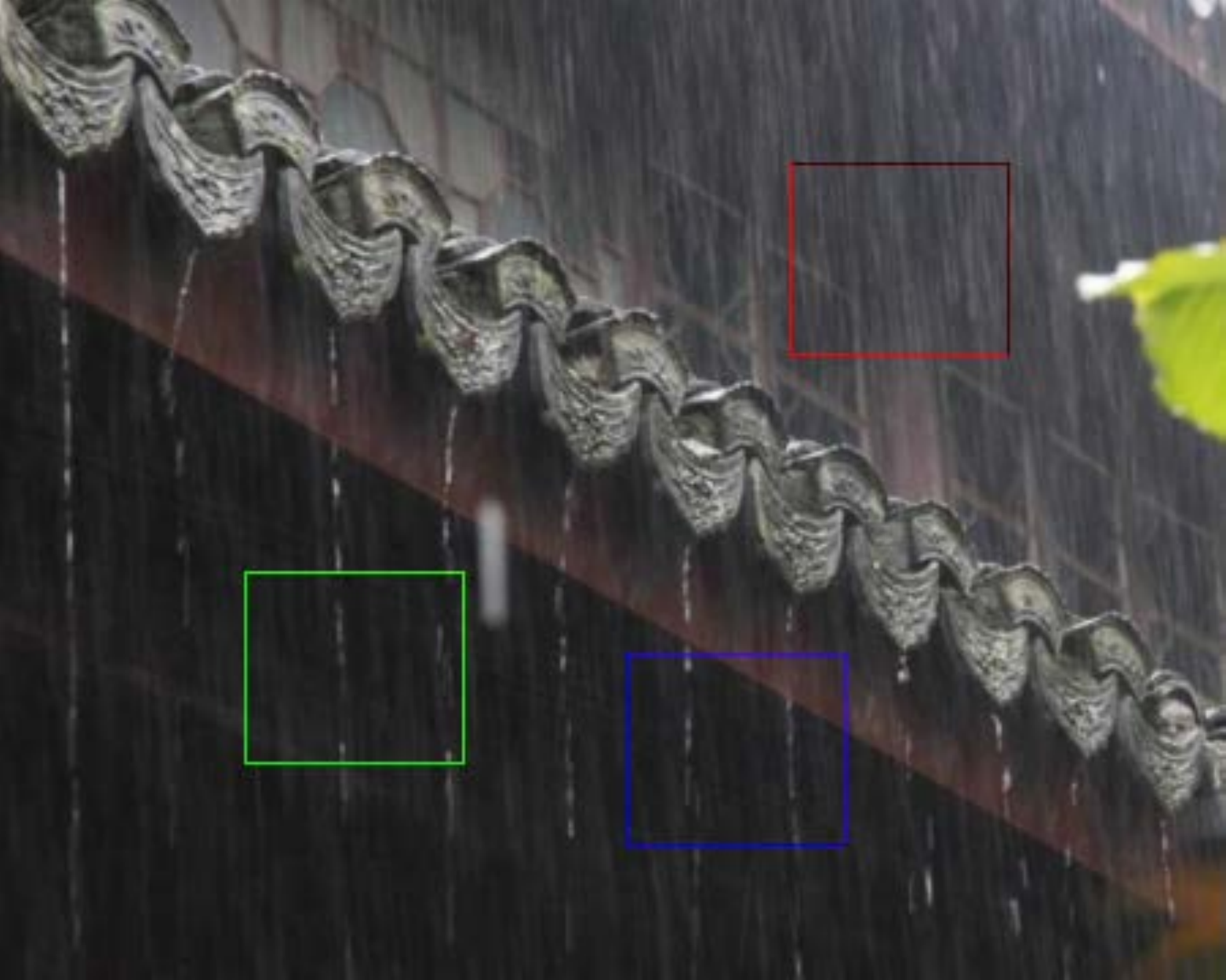}
\begin{minipage}[b]{0.33\linewidth}
\includegraphics[width=1\linewidth,height=0.6\linewidth]{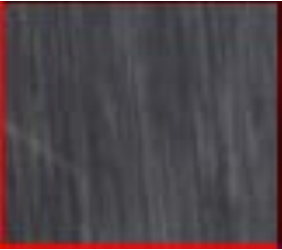}
\end{minipage}
\hspace{-0.15cm}
\begin{minipage}[b]{0.33\linewidth}
\includegraphics[width=1\linewidth,height=0.6\linewidth]{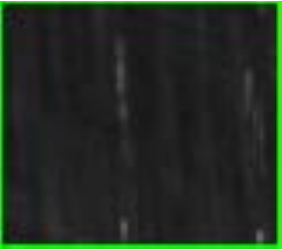}
\end{minipage}
\hspace{-0.15cm}
\begin{minipage}[b]{0.33\linewidth}
\includegraphics[width=1\linewidth,height=0.6\linewidth]{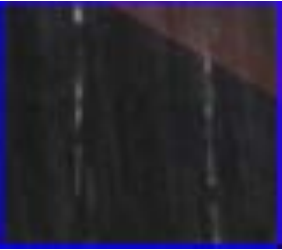}
\end{minipage}
\end{minipage}}
\hspace{-0.15cm}
\subfigure[DID \cite{Alpher14}]{
\begin{minipage}[b]{0.137\linewidth}
\includegraphics[width=1\linewidth,height=0.6\linewidth]{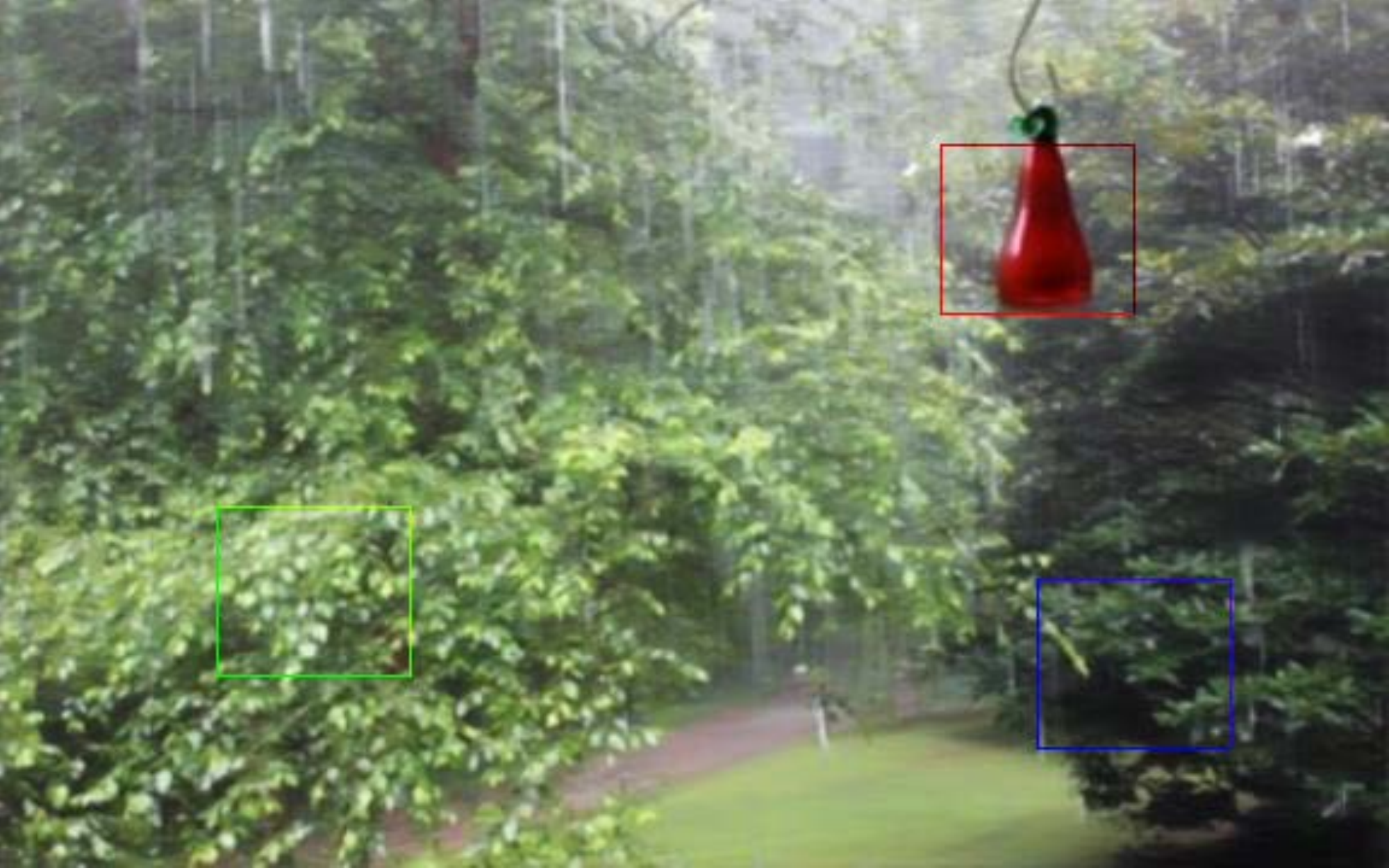}
\vspace{0.05cm}
\begin{minipage}[b]{0.33\linewidth}
\includegraphics[width=1\linewidth,height=0.6\linewidth]{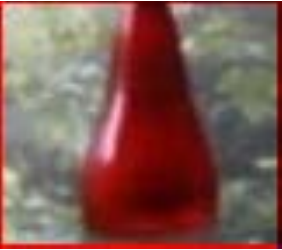}
\end{minipage}
\hspace{-0.15cm}
\begin{minipage}[b]{0.33\linewidth}
\includegraphics[width=1\linewidth,height=0.6\linewidth]{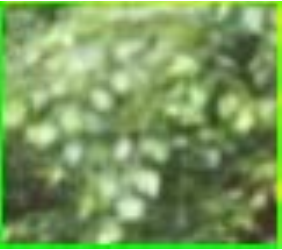}
\end{minipage}
\hspace{-0.15cm}
\begin{minipage}[b]{0.33\linewidth}
\includegraphics[width=1\linewidth,height=0.6\linewidth]{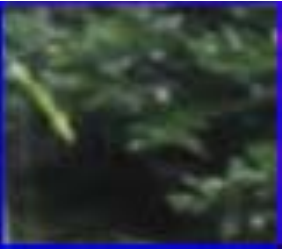}
\end{minipage}
\includegraphics[width=1\linewidth,height=0.6\linewidth]{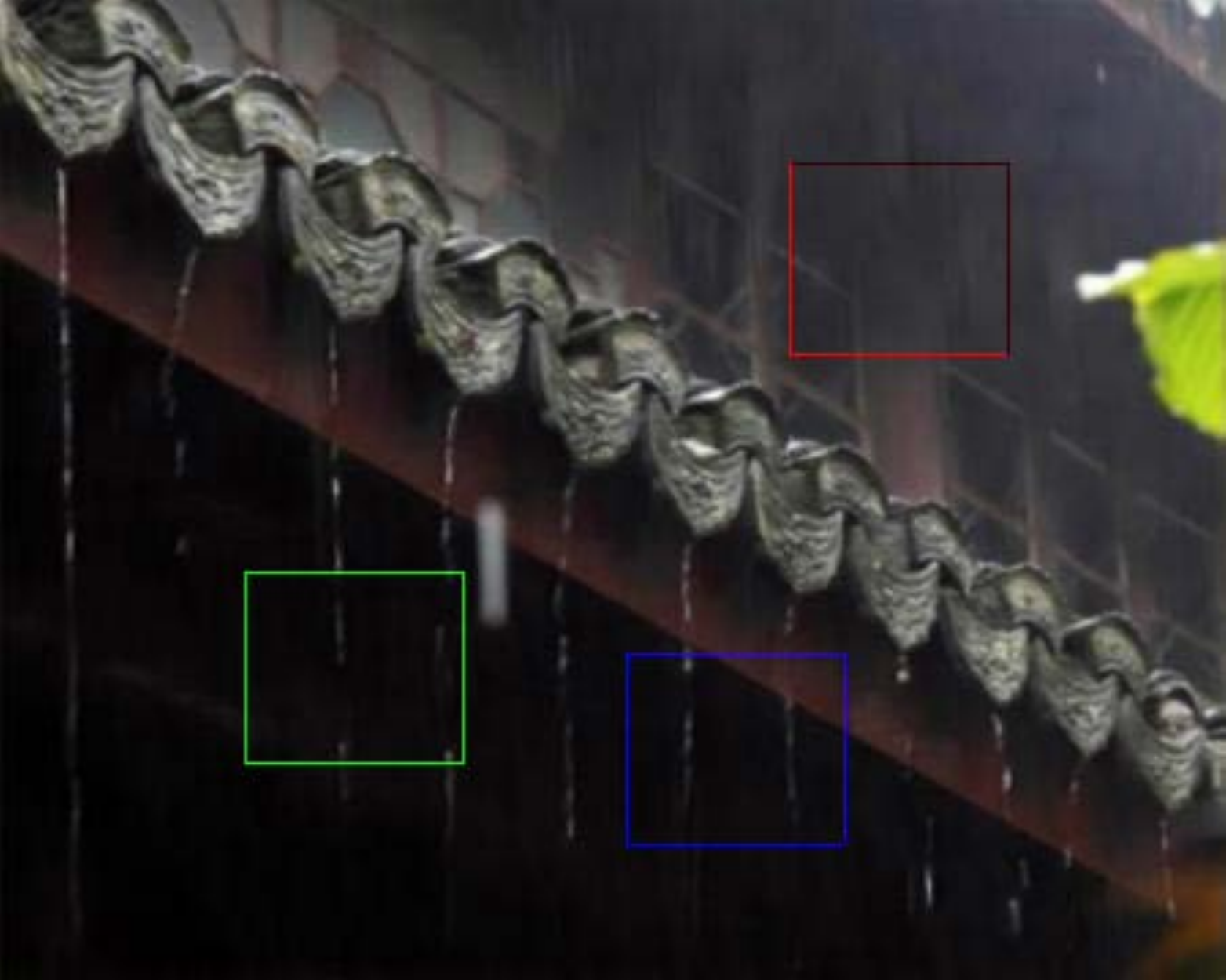}
\begin{minipage}[b]{0.33\linewidth}
\includegraphics[width=1\linewidth,height=0.6\linewidth]{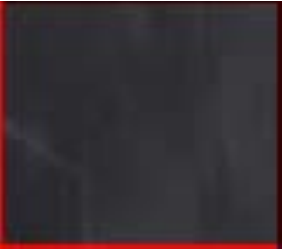}
\end{minipage}
\hspace{-0.15cm}
\begin{minipage}[b]{0.33\linewidth}
\includegraphics[width=1\linewidth,height=0.6\linewidth]{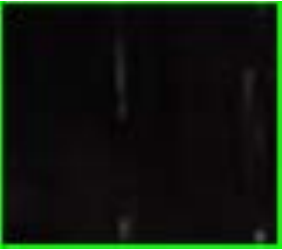}
\end{minipage}
\hspace{-0.15cm}
\begin{minipage}[b]{0.33\linewidth}
\includegraphics[width=1\linewidth,height=0.6\linewidth]{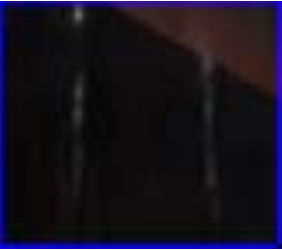}
\end{minipage}
\end{minipage}}
\hspace{-0.15cm}
\subfigure[MSPFN \cite{Alpher41}]{
\begin{minipage}[b]{0.137\linewidth}
\includegraphics[width=1\linewidth,height=0.6\linewidth]{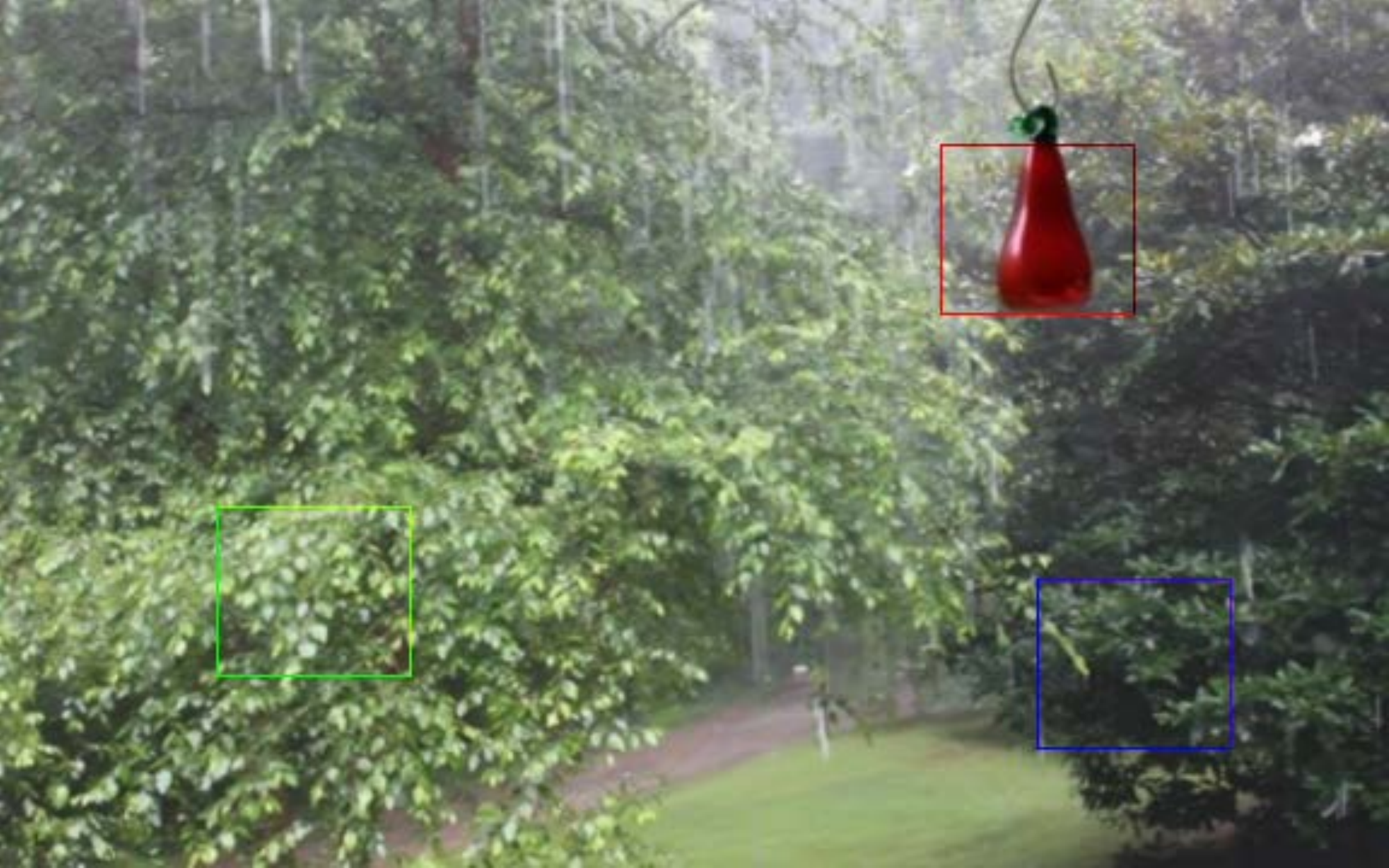}
\vspace{0.05cm}
\begin{minipage}[b]{0.33\linewidth}
\includegraphics[width=1\linewidth,height=0.6\linewidth]{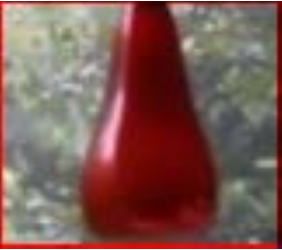}
\end{minipage}
\hspace{-0.15cm}
\begin{minipage}[b]{0.33\linewidth}
\includegraphics[width=1\linewidth,height=0.6\linewidth]{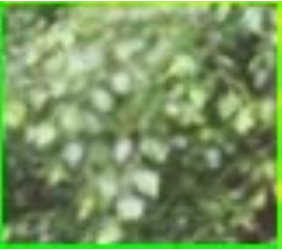}
\end{minipage}
\hspace{-0.15cm}
\begin{minipage}[b]{0.33\linewidth}
\includegraphics[width=1\linewidth,height=0.6\linewidth]{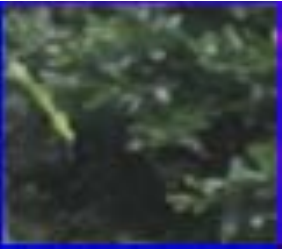}
\end{minipage}
\includegraphics[width=1\linewidth,height=0.6\linewidth]{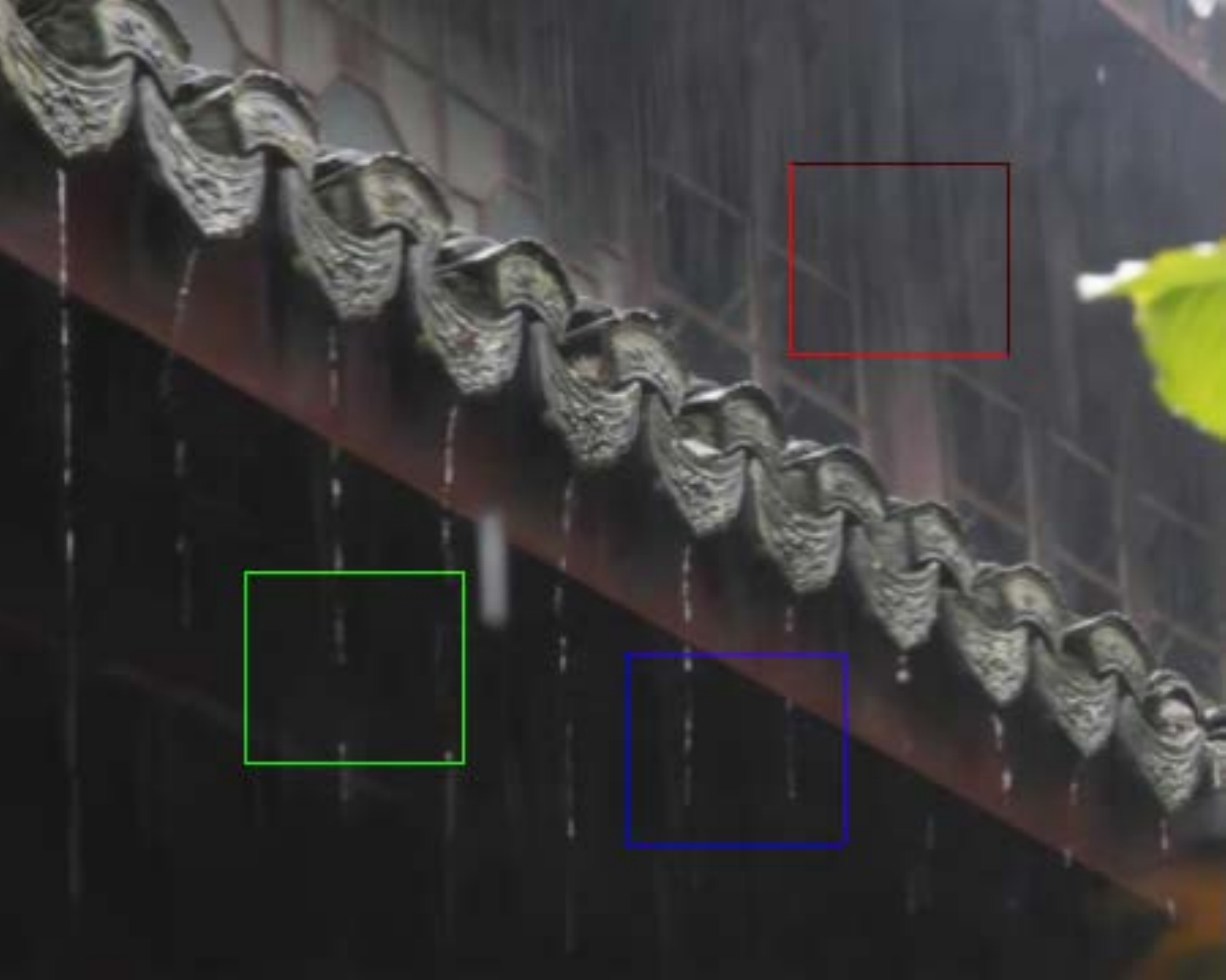}
\begin{minipage}[b]{0.33\linewidth}
\includegraphics[width=1\linewidth,height=0.6\linewidth]{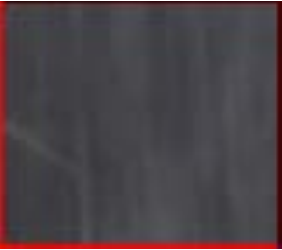}
\end{minipage}
\hspace{-0.15cm}
\begin{minipage}[b]{0.33\linewidth}
\includegraphics[width=1\linewidth,height=0.6\linewidth]{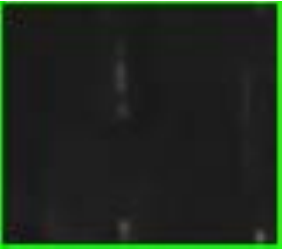}
\end{minipage}
\hspace{-0.15cm}
\begin{minipage}[b]{0.33\linewidth}
\includegraphics[width=1\linewidth,height=0.6\linewidth]{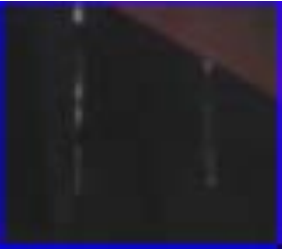}
\end{minipage}
\end{minipage}}
\hspace{-0.15cm}
\subfigure[RESCAN \cite{Alpher26}]{
\begin{minipage}[b]{0.137\linewidth}
\includegraphics[width=1\linewidth,height=0.6\linewidth]{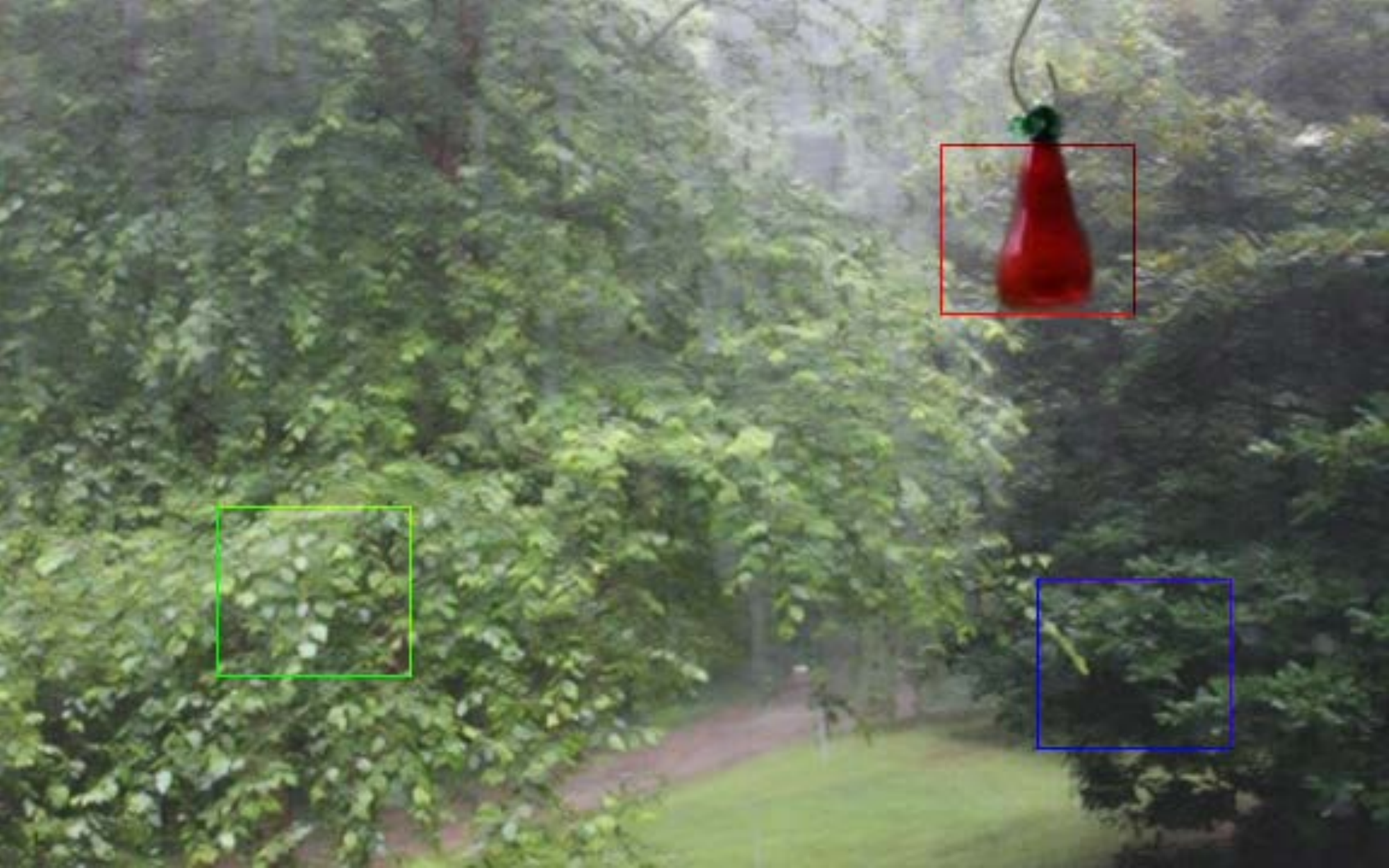}
\vspace{0.05cm}
\begin{minipage}[b]{0.33\linewidth}
\includegraphics[width=1\linewidth,height=0.6\linewidth]{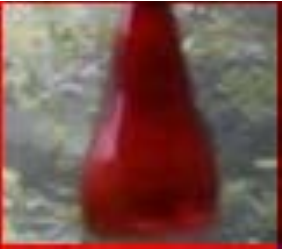}
\end{minipage}
\hspace{-0.15cm}
\begin{minipage}[b]{0.33\linewidth}
\includegraphics[width=1\linewidth,height=0.6\linewidth]{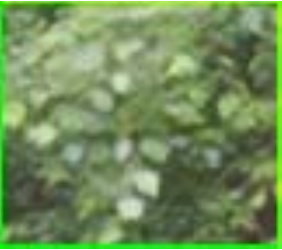}
\end{minipage}
\hspace{-0.15cm}
\begin{minipage}[b]{0.33\linewidth}
\includegraphics[width=1\linewidth,height=0.6\linewidth]{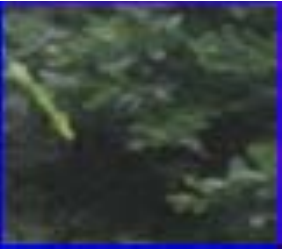}
\end{minipage}
\includegraphics[width=1\linewidth,height=0.6\linewidth]{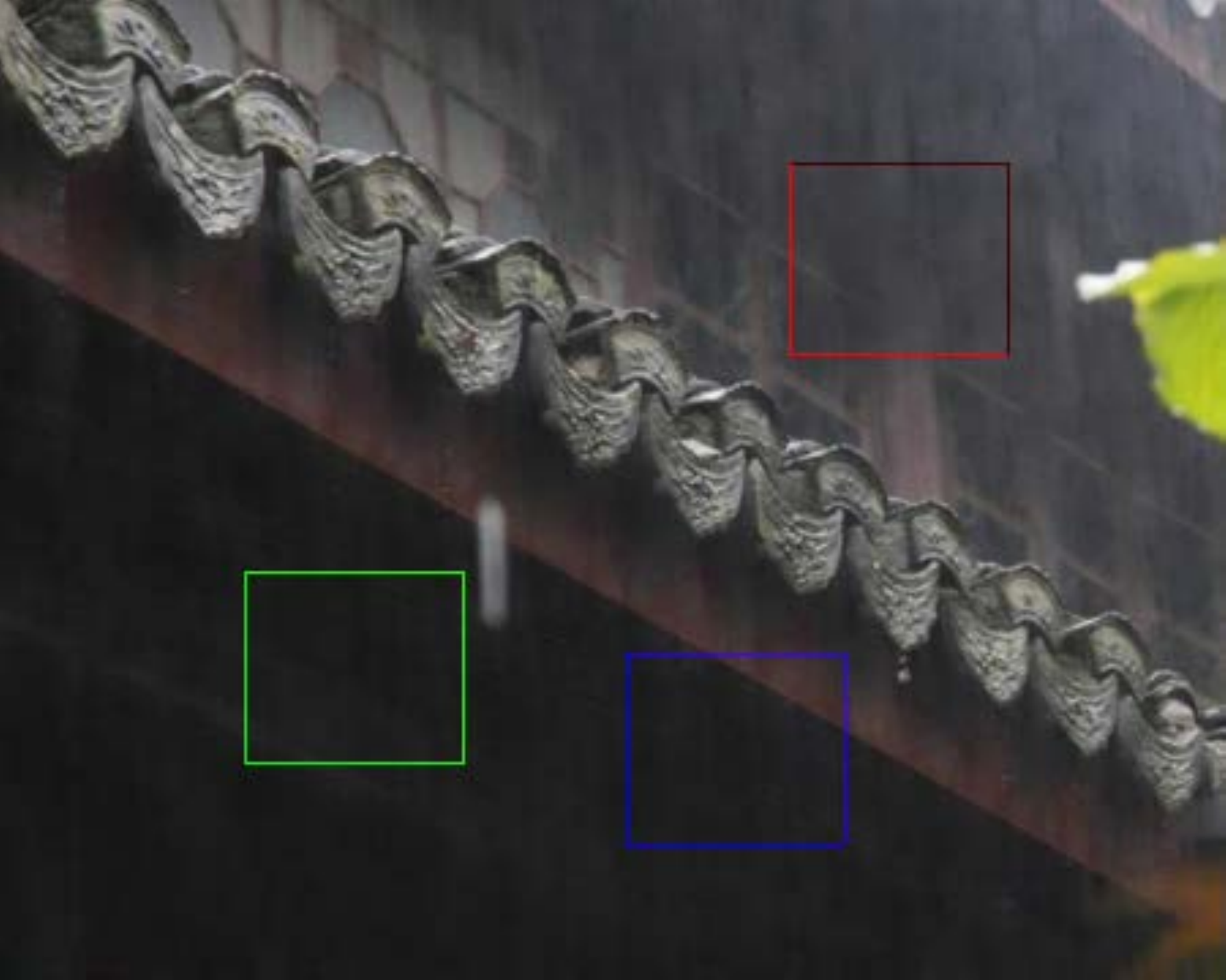}
\begin{minipage}[b]{0.33\linewidth}
\includegraphics[width=1\linewidth,height=0.6\linewidth]{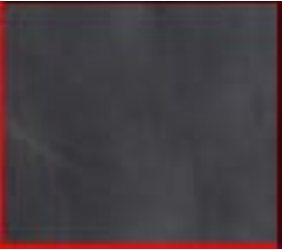}
\end{minipage}
\hspace{-0.15cm}
\begin{minipage}[b]{0.33\linewidth}
\includegraphics[width=1\linewidth,height=0.6\linewidth]{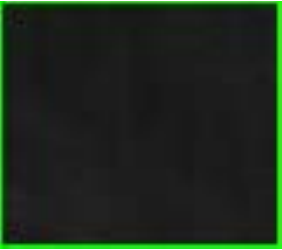}
\end{minipage}
\hspace{-0.15cm}
\begin{minipage}[b]{0.33\linewidth}
\includegraphics[width=1\linewidth,height=0.6\linewidth]{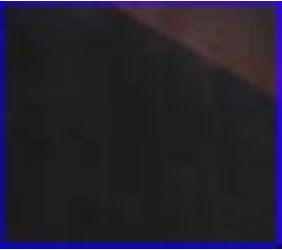}
\end{minipage}
\end{minipage}}
\hspace{-0.15cm}
\subfigure[UMRL \cite{Alpher29}]{
\begin{minipage}[b]{0.137\linewidth}
\includegraphics[width=1\linewidth,height=0.6\linewidth]{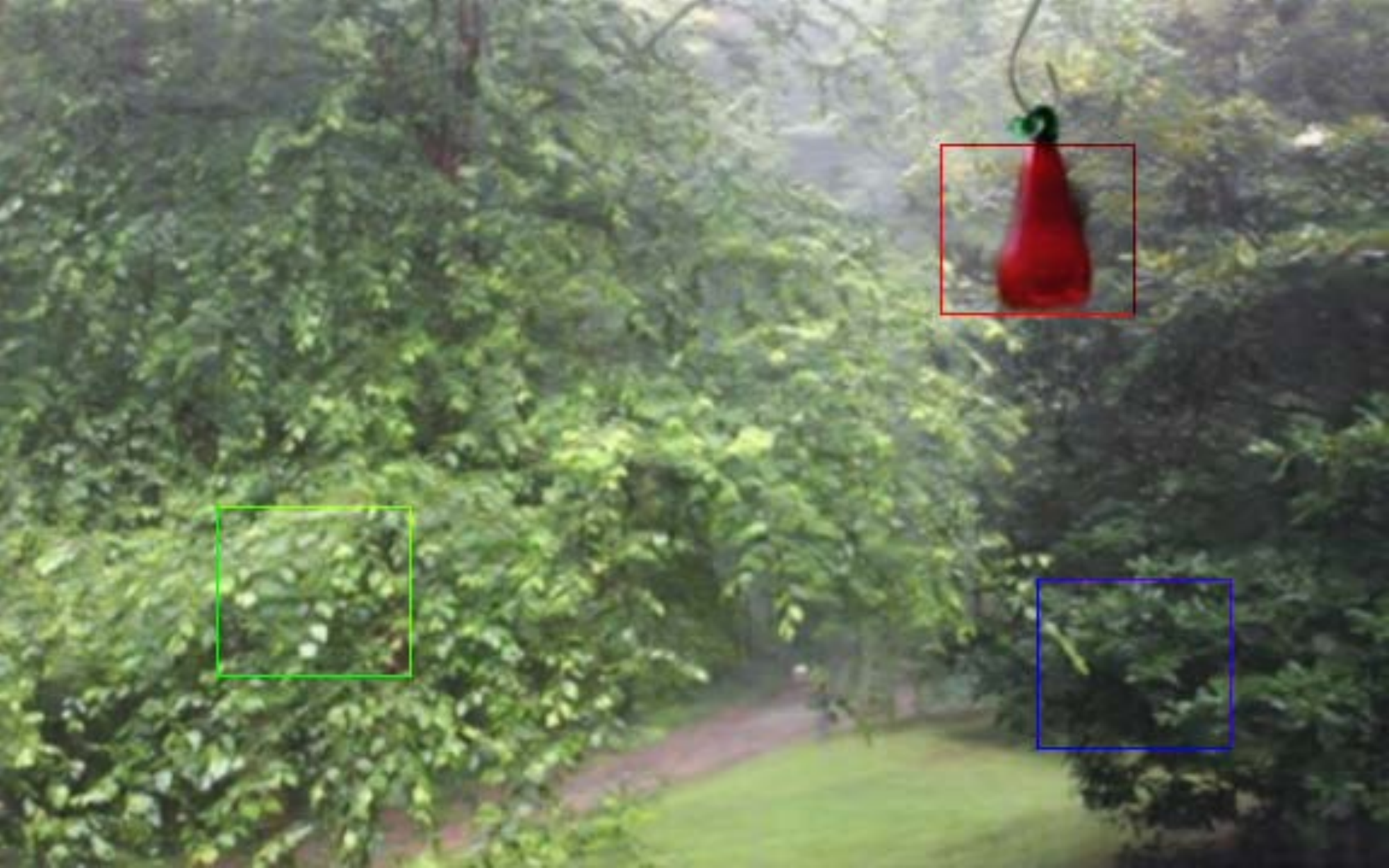}
\vspace{0.05cm}
\begin{minipage}[b]{0.33\linewidth}
\includegraphics[width=1\linewidth,height=0.6\linewidth]{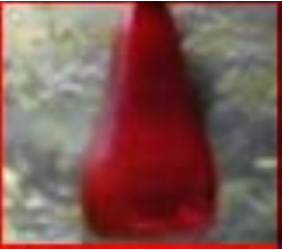}
\end{minipage}
\hspace{-0.15cm}
\begin{minipage}[b]{0.33\linewidth}
\includegraphics[width=1\linewidth,height=0.6\linewidth]{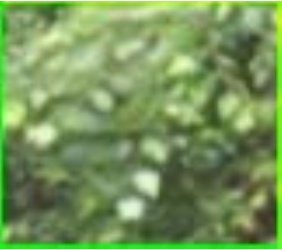}
\end{minipage}
\hspace{-0.15cm}
\begin{minipage}[b]{0.33\linewidth}
\includegraphics[width=1\linewidth,height=0.6\linewidth]{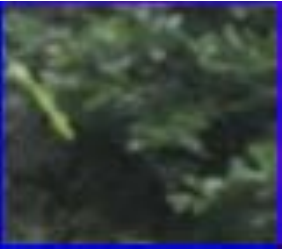}
\end{minipage}
\includegraphics[width=1\linewidth,height=0.6\linewidth]{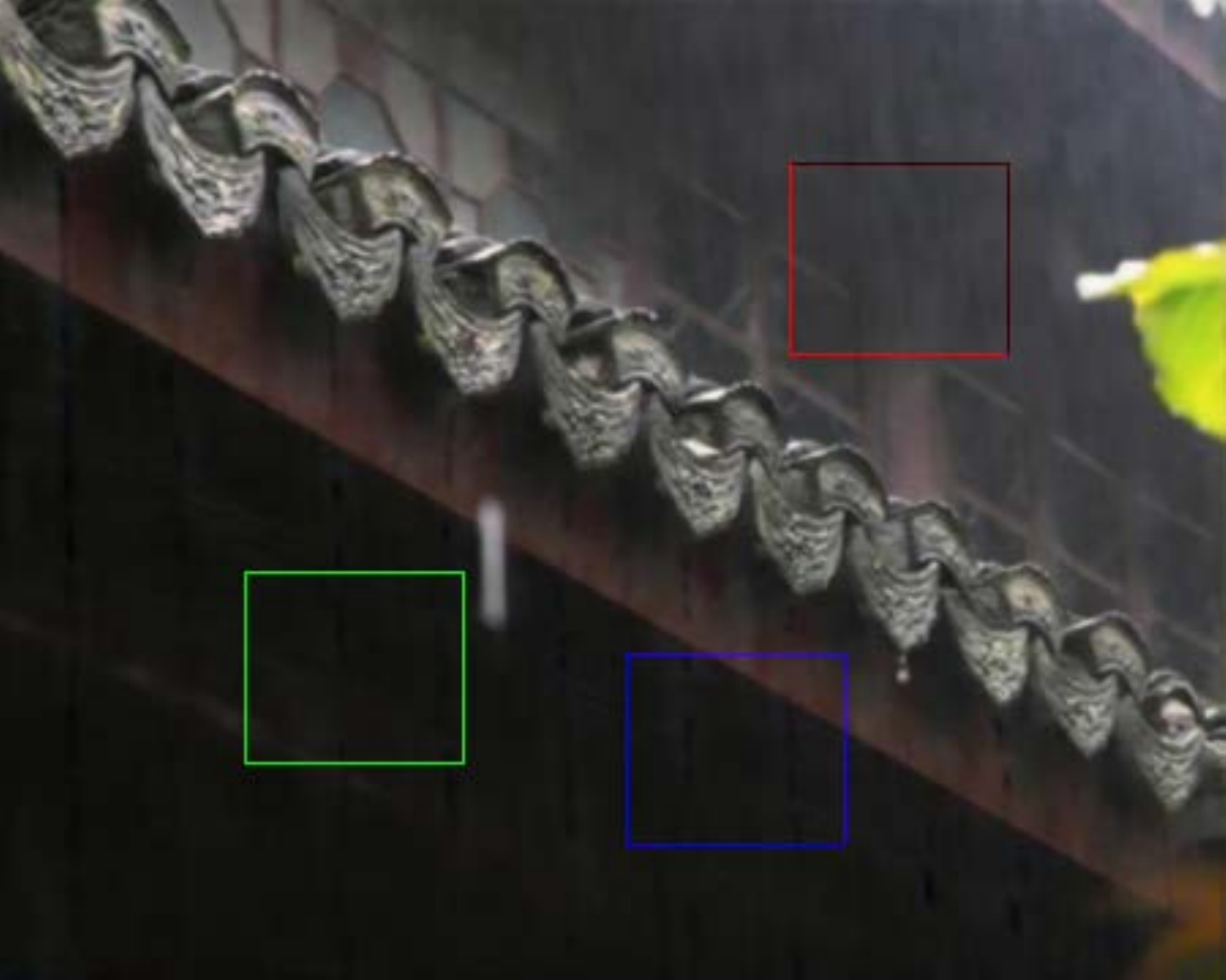}
\begin{minipage}[b]{0.33\linewidth}
\includegraphics[width=1\linewidth,height=0.6\linewidth]{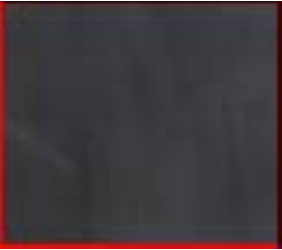}
\end{minipage}
\hspace{-0.15cm}
\begin{minipage}[b]{0.33\linewidth}
\includegraphics[width=1\linewidth,height=0.6\linewidth]{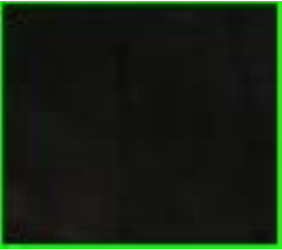}
\end{minipage}
\hspace{-0.15cm}
\begin{minipage}[b]{0.33\linewidth}
\includegraphics[width=1\linewidth,height=0.6\linewidth]{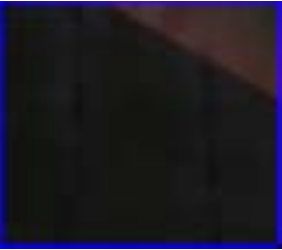}
\end{minipage}
\end{minipage}}
\hspace{-0.15cm}
\subfigure[RCDNet \cite{Alpher45}]{
\begin{minipage}[b]{0.137\linewidth}
\includegraphics[width=1\linewidth,height=0.6\linewidth]{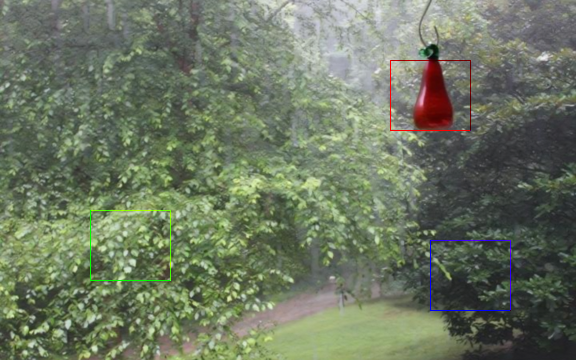}
\vspace{0.05cm}
\begin{minipage}[b]{0.33\linewidth}
\includegraphics[width=1\linewidth,height=0.6\linewidth]{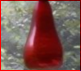}
\end{minipage}
\hspace{-0.15cm}
\begin{minipage}[b]{0.33\linewidth}
\includegraphics[width=1\linewidth,height=0.6\linewidth]{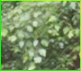}
\end{minipage}
\hspace{-0.15cm}
\begin{minipage}[b]{0.33\linewidth}
\includegraphics[width=1\linewidth,height=0.6\linewidth]{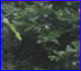}
\end{minipage}
\includegraphics[width=1\linewidth,height=0.6\linewidth]{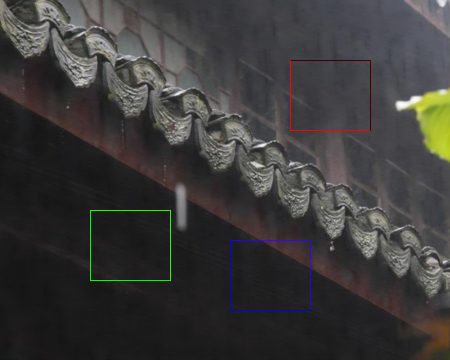}
\begin{minipage}[b]{0.33\linewidth}
\includegraphics[width=1\linewidth,height=0.6\linewidth]{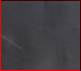}
\end{minipage}
\hspace{-0.15cm}
\begin{minipage}[b]{0.33\linewidth}
\includegraphics[width=1\linewidth,height=0.6\linewidth]{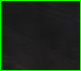}
\end{minipage}
\hspace{-0.15cm}
\begin{minipage}[b]{0.33\linewidth}
\includegraphics[width=1\linewidth,height=0.6\linewidth]{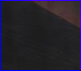}
\end{minipage}
\end{minipage}}
\hspace{-0.15cm}
\subfigure[Ours]{
\begin{minipage}[b]{0.137\linewidth}
\includegraphics[width=1\linewidth,height=0.6\linewidth]{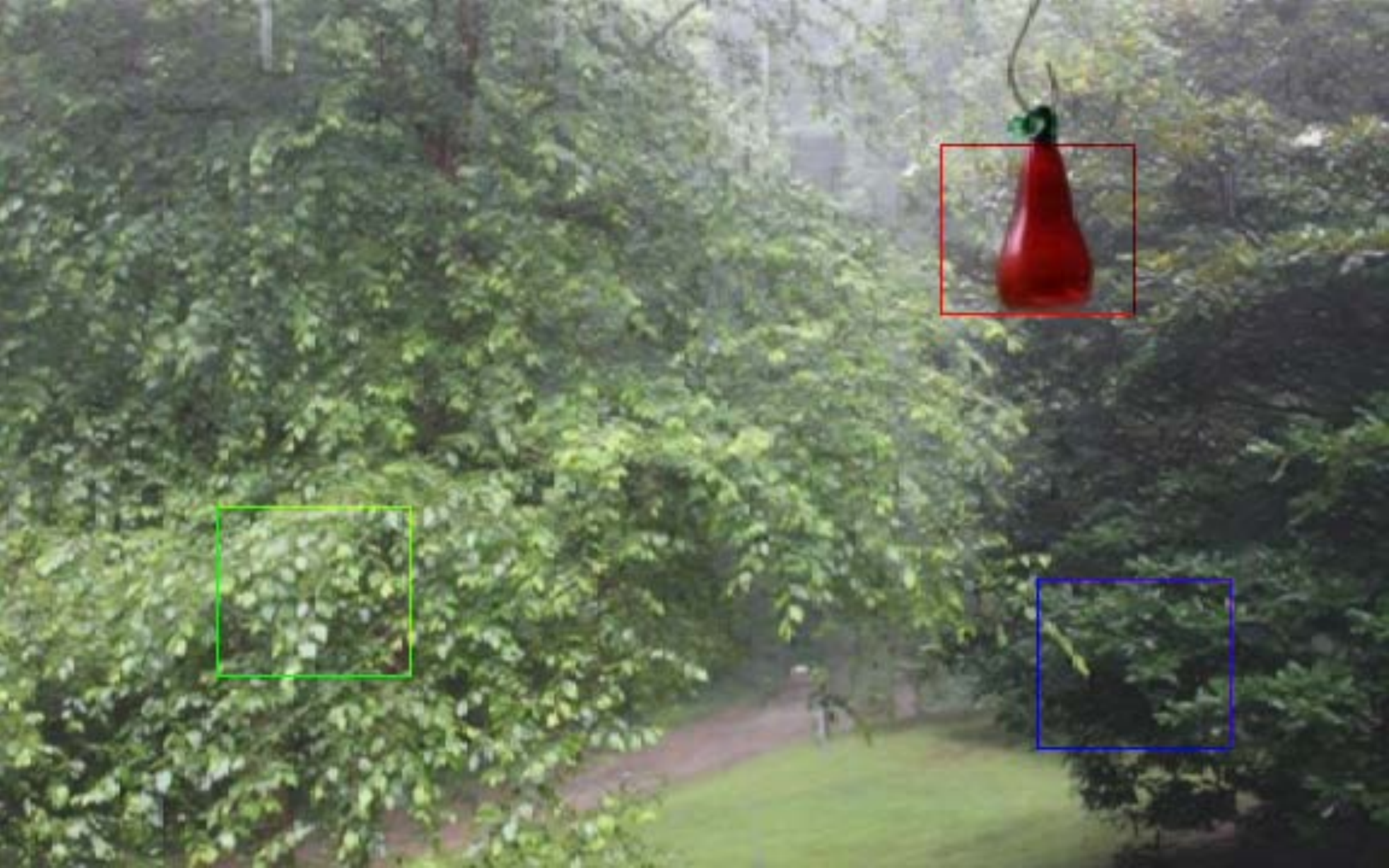}
\vspace{0.05cm}
\begin{minipage}[b]{0.33\linewidth}
\includegraphics[width=1\linewidth,height=0.6\linewidth]{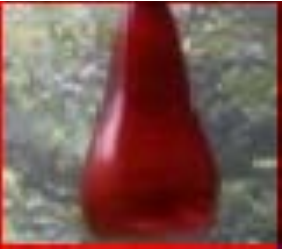}
\end{minipage}
\hspace{-0.15cm}
\begin{minipage}[b]{0.33\linewidth}
\includegraphics[width=1\linewidth,height=0.6\linewidth]{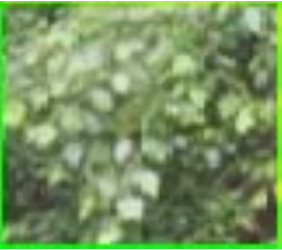}
\end{minipage}
\hspace{-0.15cm}
\begin{minipage}[b]{0.33\linewidth}
\includegraphics[width=1\linewidth,height=0.6\linewidth]{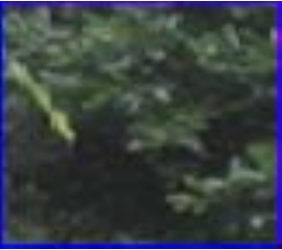}
\end{minipage}
\includegraphics[width=1\linewidth,height=0.6\linewidth]{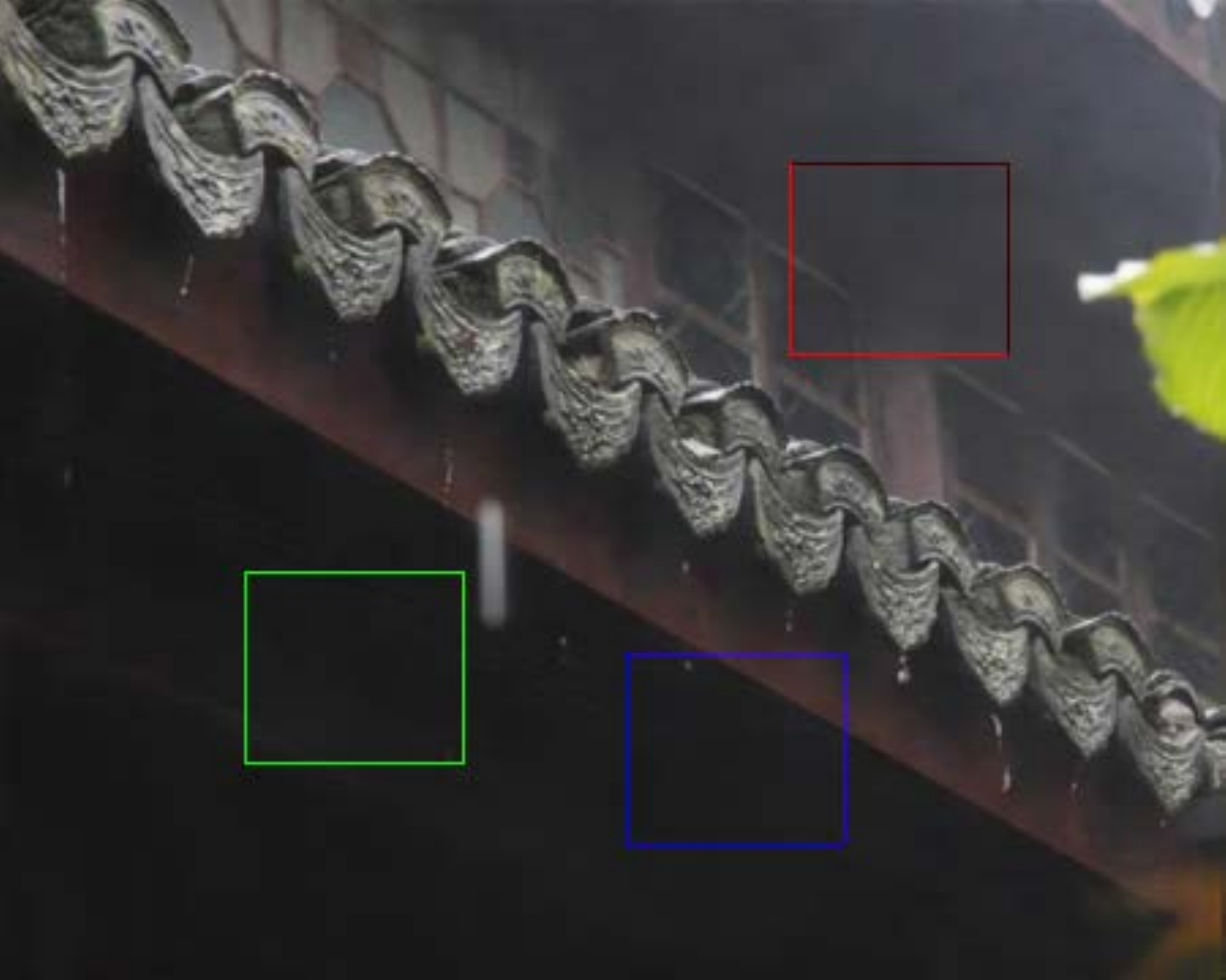}
\begin{minipage}[b]{0.33\linewidth}
\includegraphics[width=1\linewidth,height=0.6\linewidth]{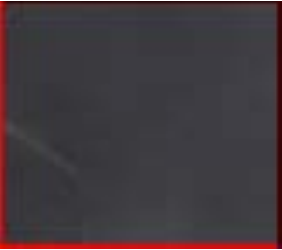}
\end{minipage}
\hspace{-0.15cm}
\begin{minipage}[b]{0.33\linewidth}
\includegraphics[width=1\linewidth,height=0.6\linewidth]{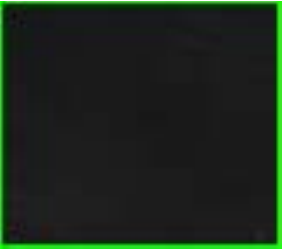}
\end{minipage}
\hspace{-0.15cm}
\begin{minipage}[b]{0.33\linewidth}
\includegraphics[width=1\linewidth,height=0.6\linewidth]{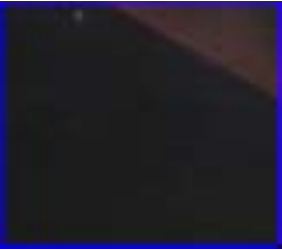}
\end{minipage}
\end{minipage}}
\caption{Visual quality comparisons on sample images from real-world datasets. Zoom in to see the details.}
\label{fig9}
\vspace{-0.4cm}
\end{figure*}

\begin{figure}
\begin{center}
\begin{minipage}[b]{0.116\linewidth}
\includegraphics[width=1\linewidth,height=0.6\linewidth]{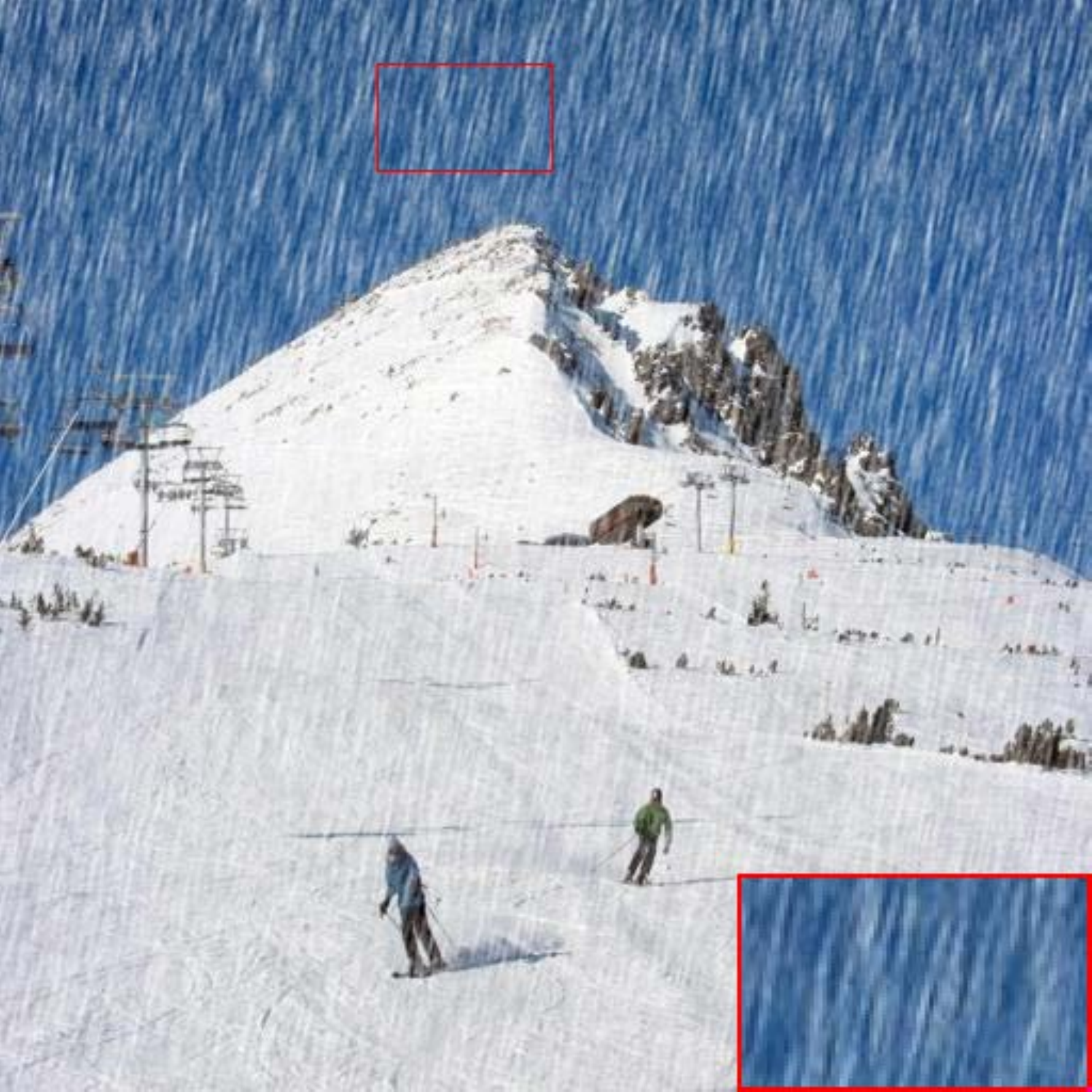}
\end{minipage}
%\vspace{-0.1cm}
\begin{minipage}[b]{0.116\linewidth}
\includegraphics[width=1\linewidth,height=0.6\linewidth]{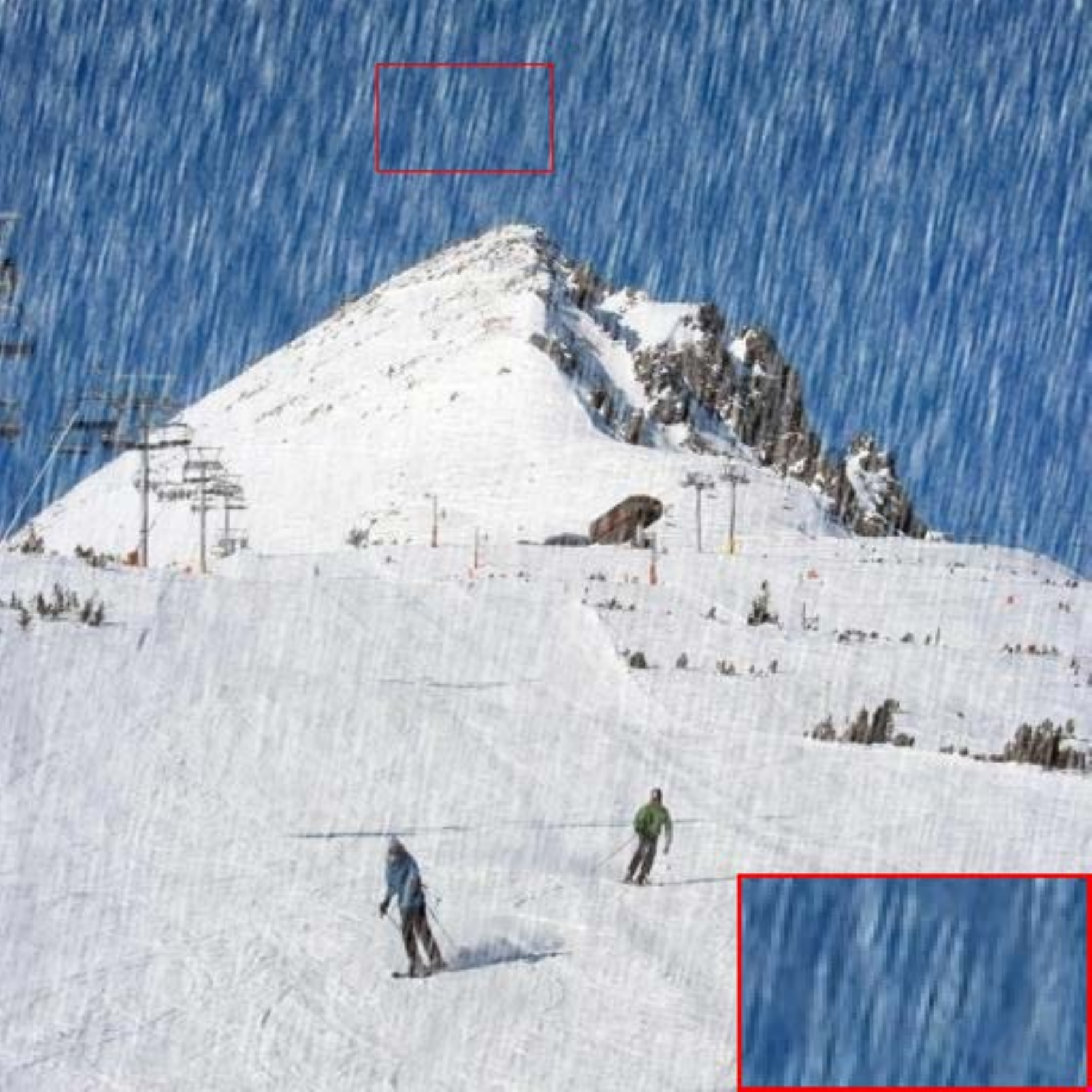}
\end{minipage}
\begin{minipage}[b]{0.116\linewidth}
\includegraphics[width=1\linewidth,height=0.6\linewidth]{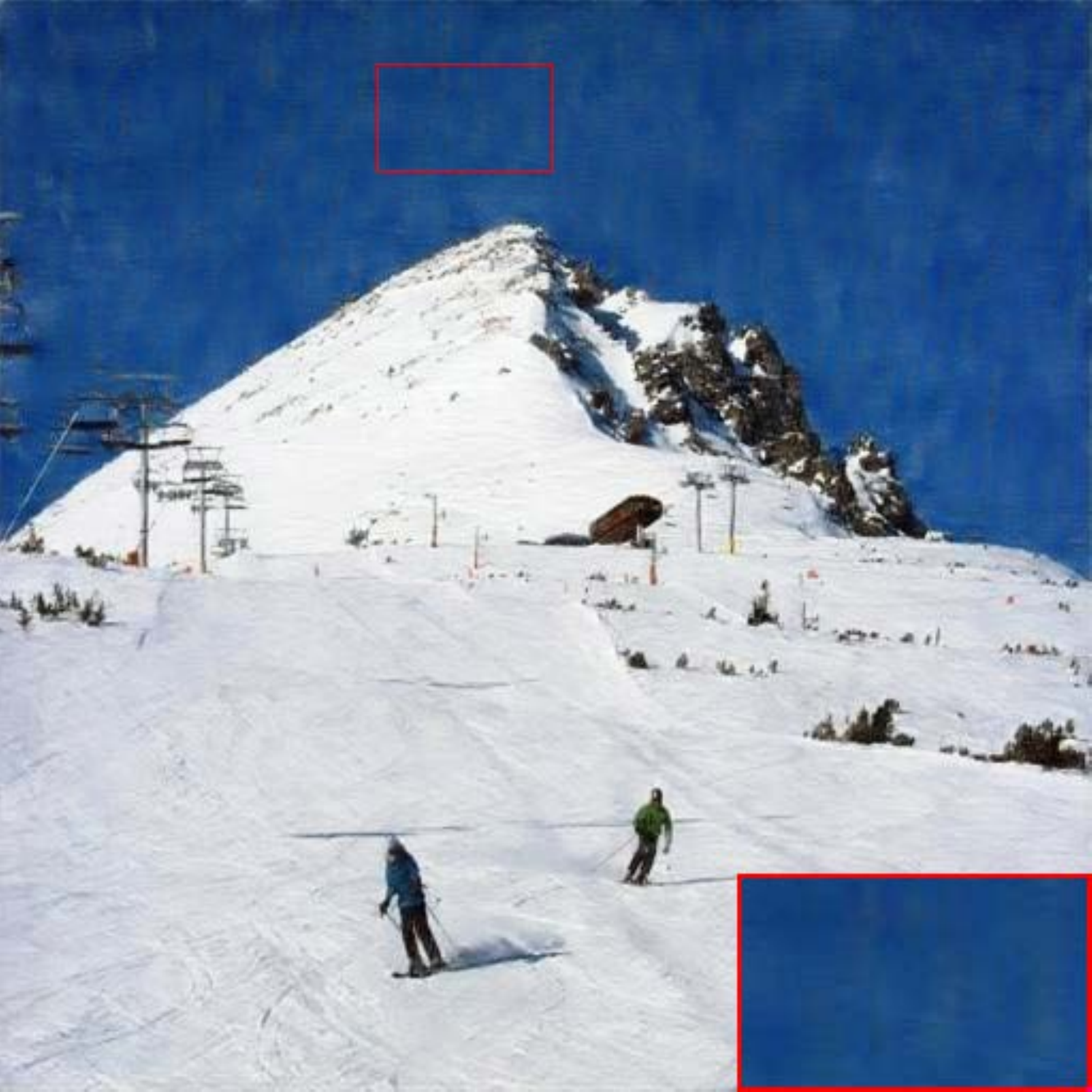}
\end{minipage}
\begin{minipage}[b]{0.116\linewidth}
\includegraphics[width=1\linewidth,height=0.6\linewidth]{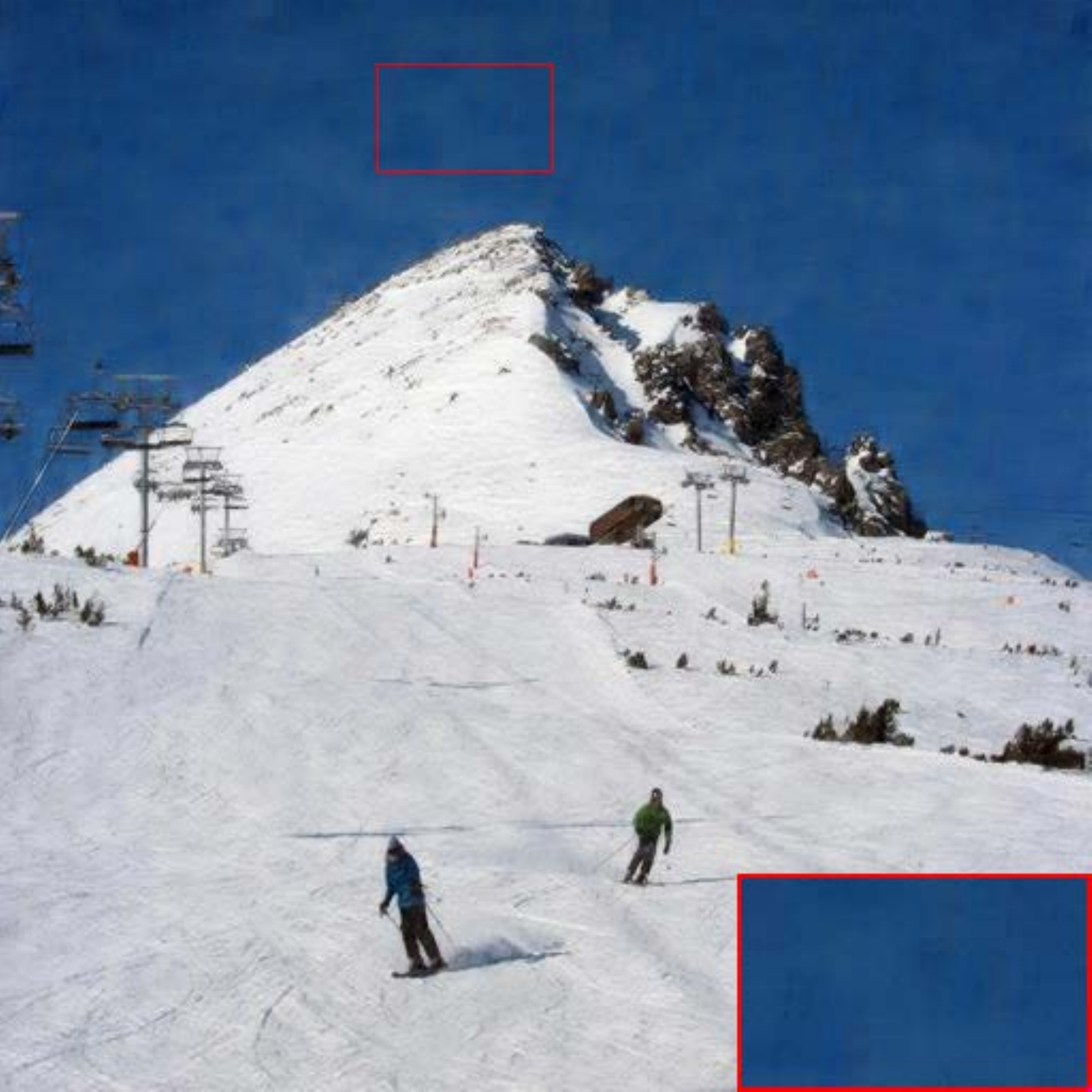}
\end{minipage}
\begin{minipage}[b]{0.116\linewidth}
\includegraphics[width=1\linewidth,height=0.6\linewidth]{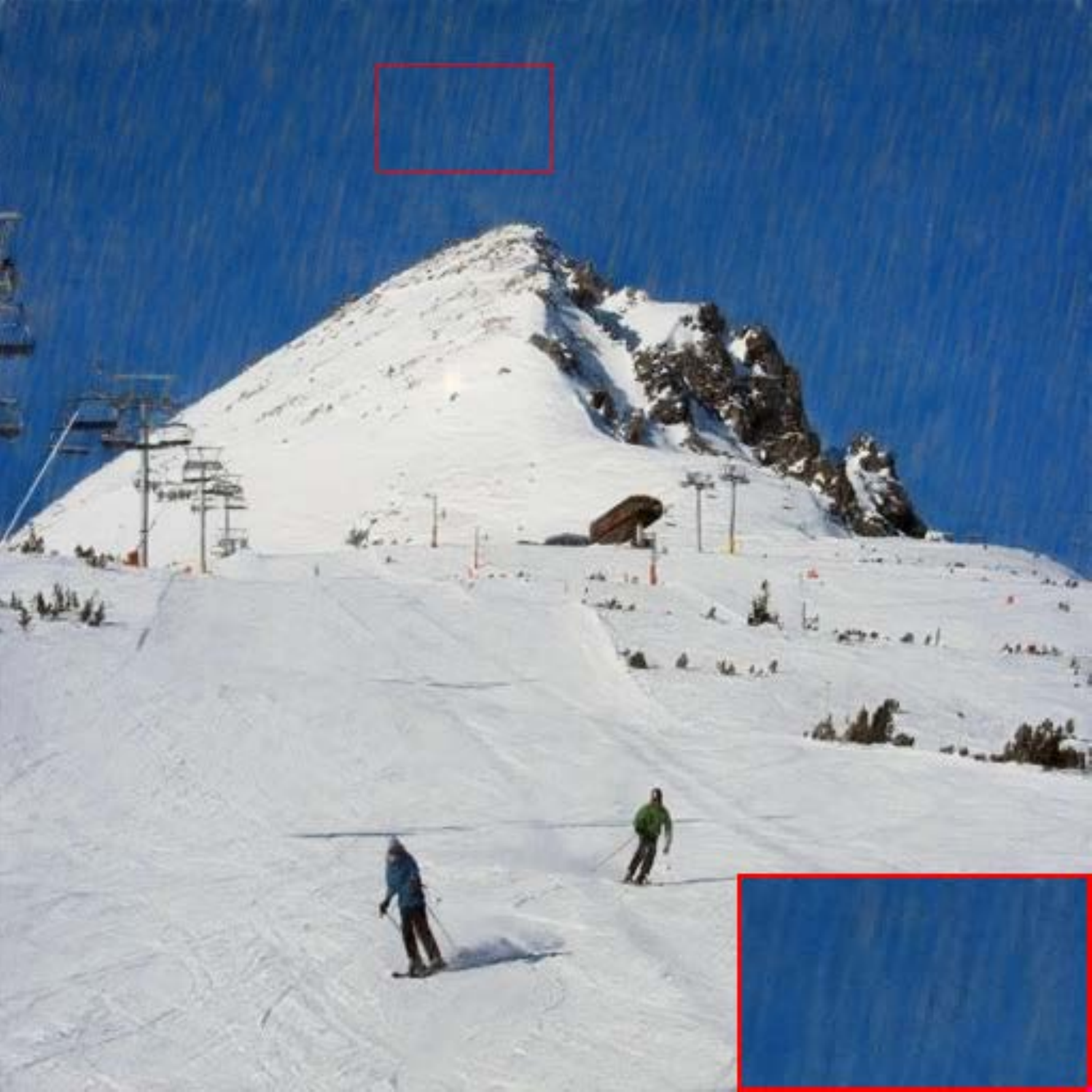}
\end{minipage}
\begin{minipage}[b]{0.116\linewidth}
\includegraphics[width=1\linewidth,height=0.6\linewidth]{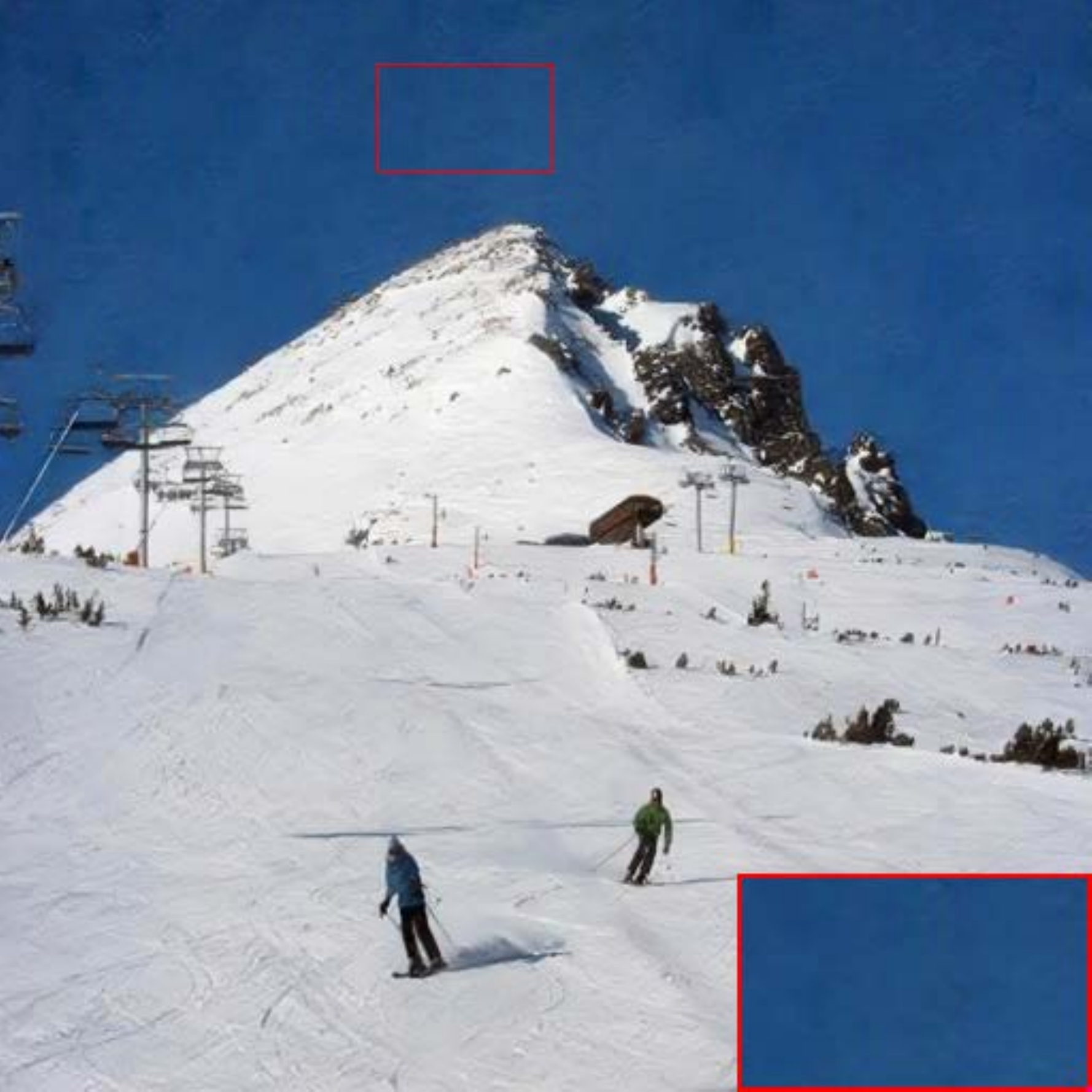}
\end{minipage}
\begin{minipage}[b]{0.116\linewidth}
\includegraphics[width=1\linewidth,height=0.6\linewidth]{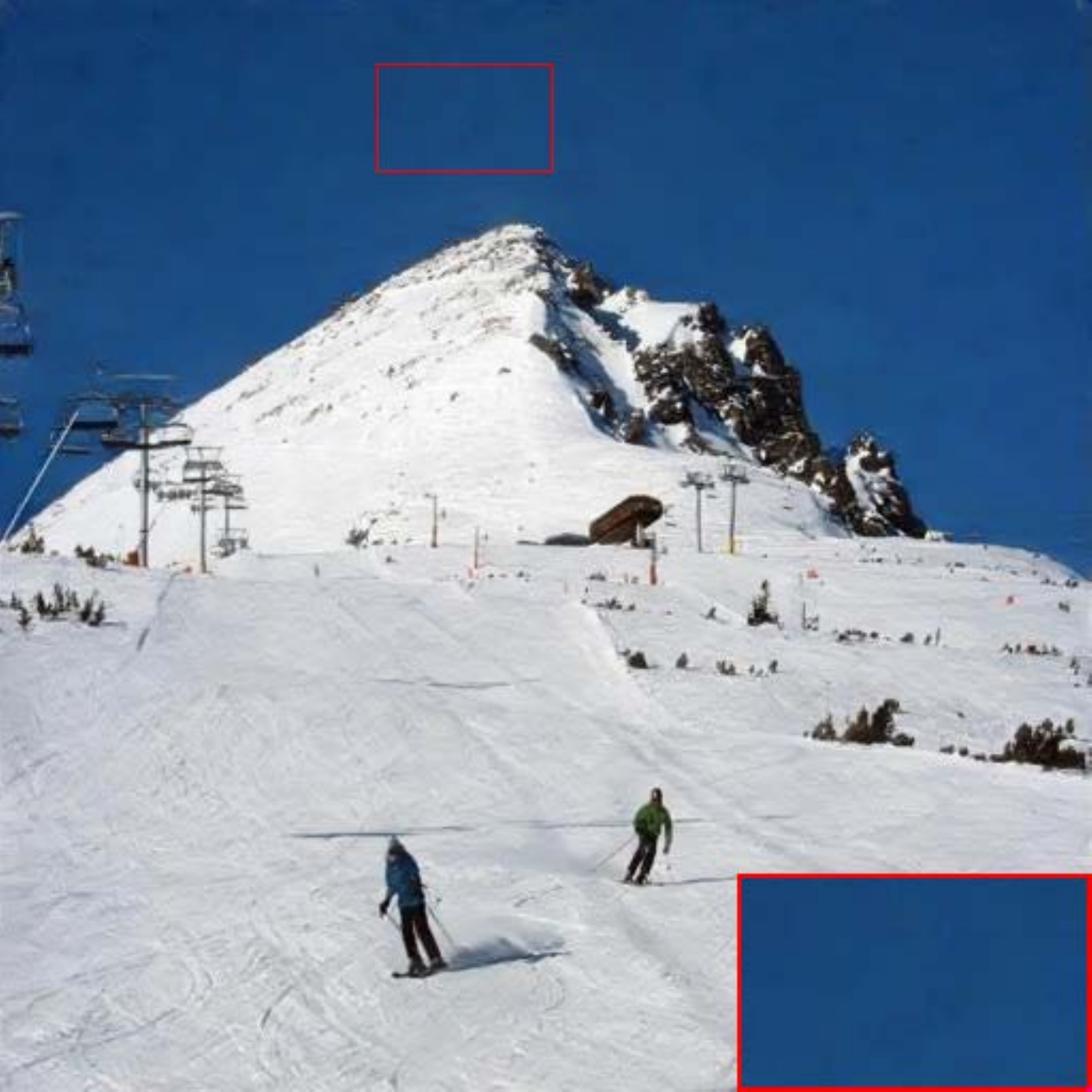}
\end{minipage}
\begin{minipage}[b]{0.116\linewidth}
\includegraphics[width=1\linewidth,height=0.6\linewidth]{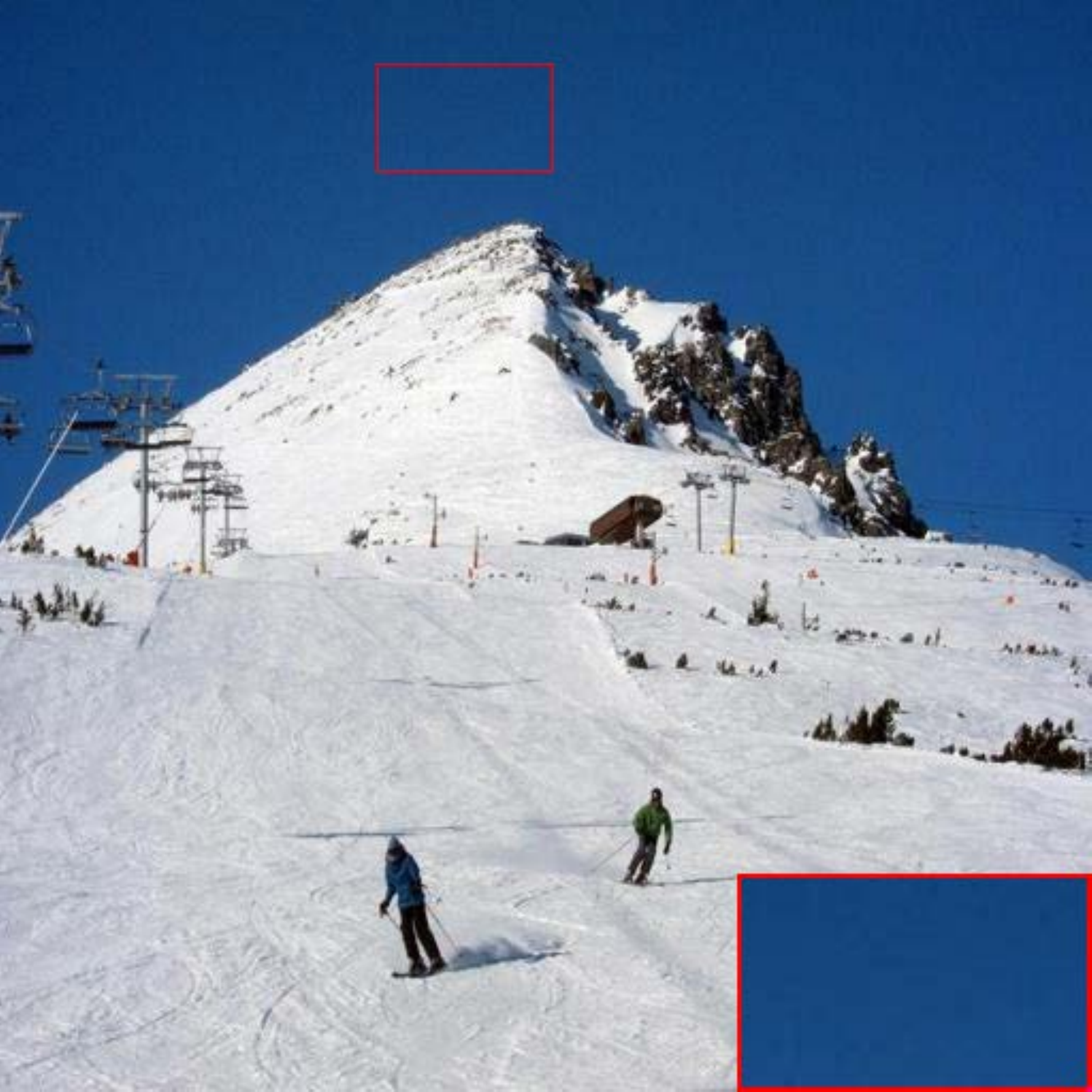}
\end{minipage}
\begin{minipage}[b]{0.116\linewidth}
\includegraphics[width=1\linewidth,height=0.6\linewidth]{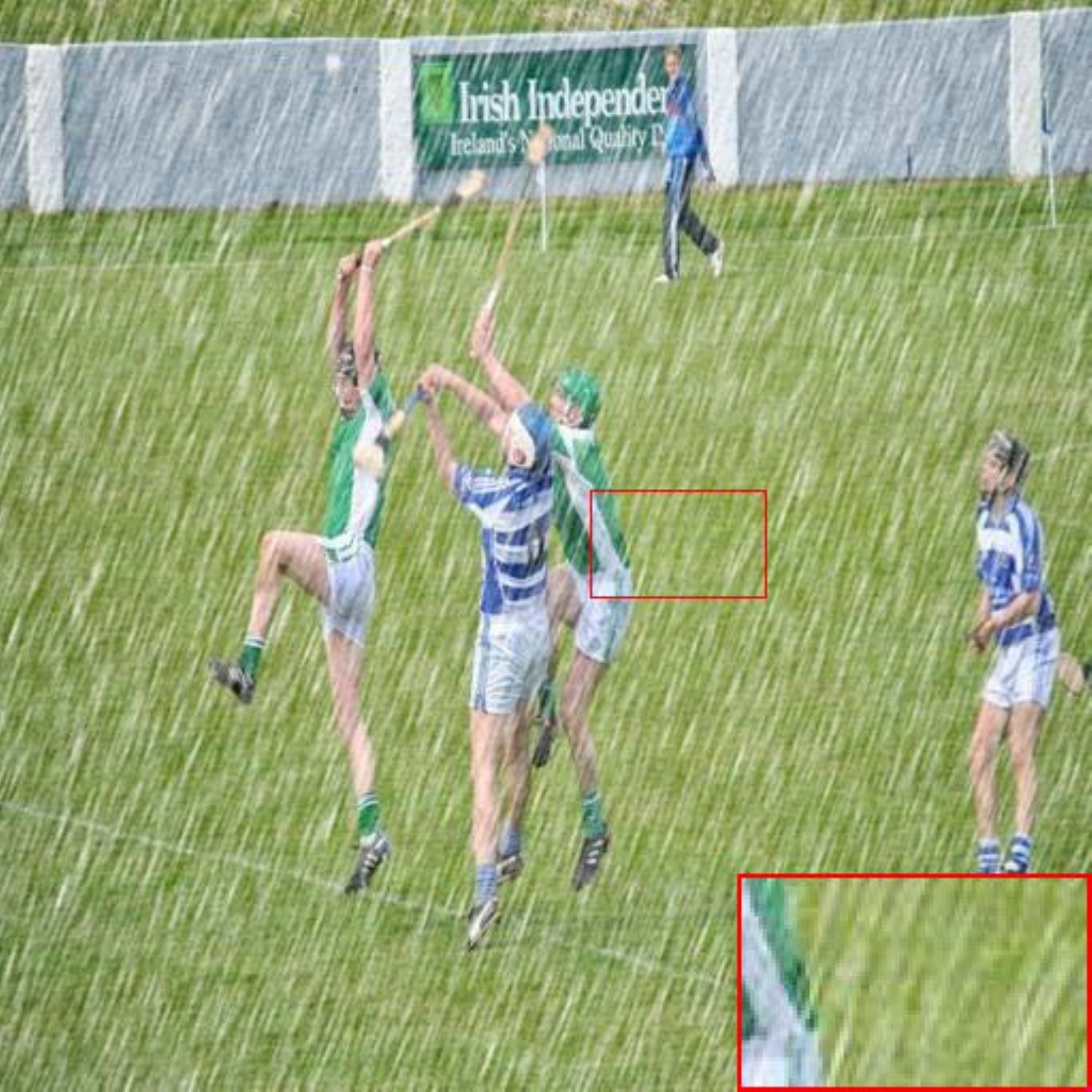}
\end{minipage}
%\vspace{-0.1cm}
\begin{minipage}[b]{0.116\linewidth}
\includegraphics[width=1\linewidth,height=0.6\linewidth]{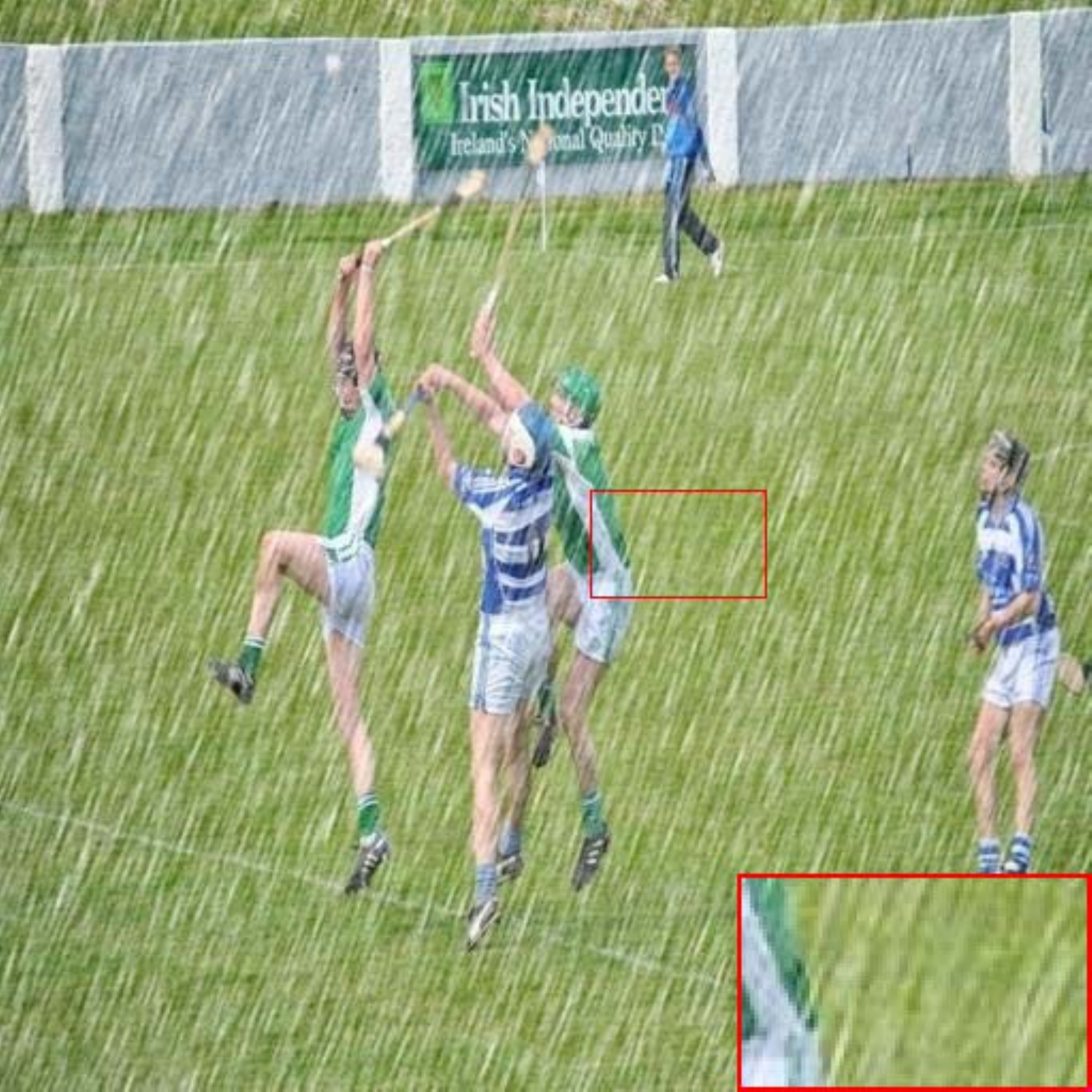}
\end{minipage}
\begin{minipage}[b]{0.116\linewidth}
\includegraphics[width=1\linewidth,height=0.6\linewidth]{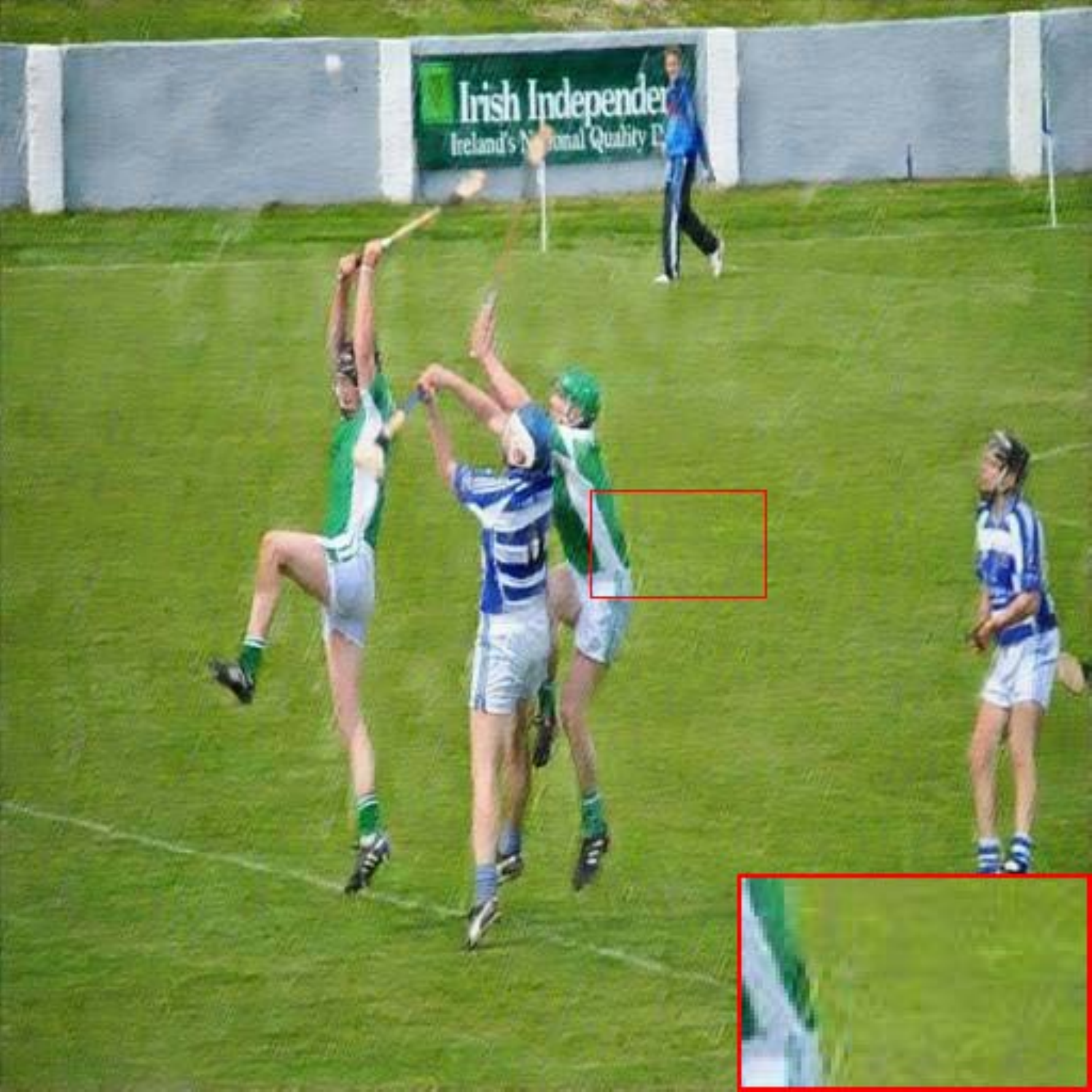}
\end{minipage}
\begin{minipage}[b]{0.116\linewidth}
\includegraphics[width=1\linewidth,height=0.6\linewidth]{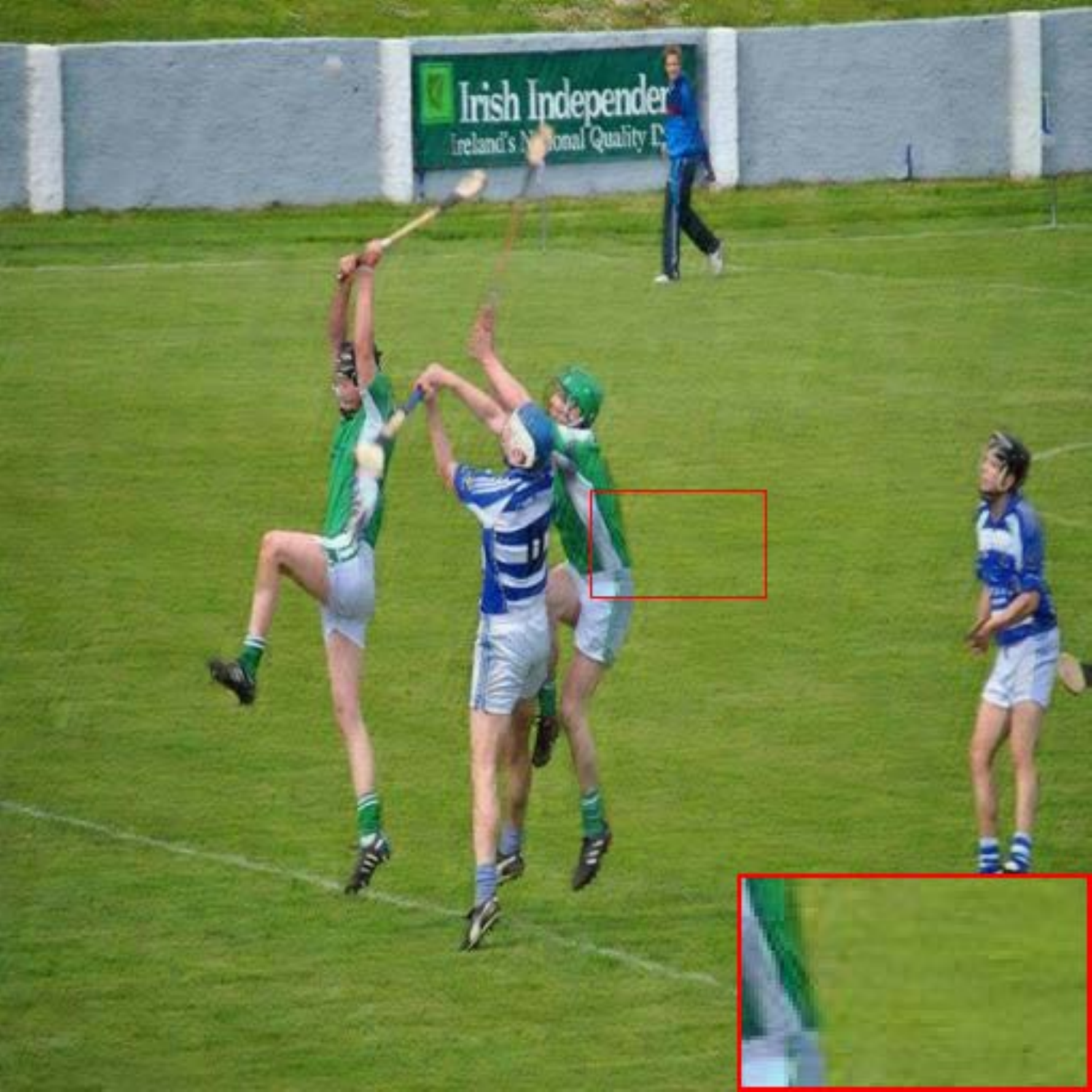}
\end{minipage}
\begin{minipage}[b]{0.116\linewidth}
\includegraphics[width=1\linewidth,height=0.6\linewidth]{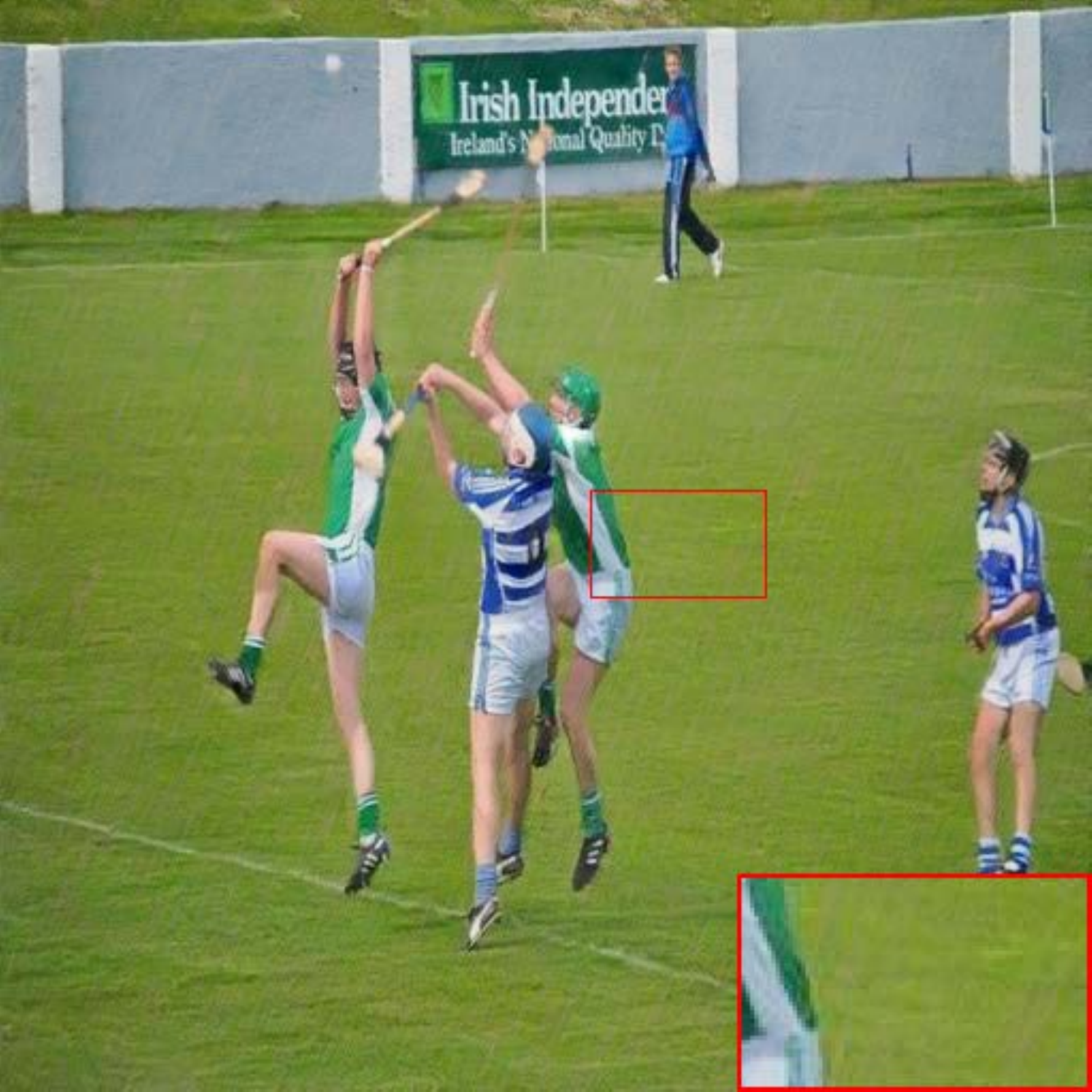}
\end{minipage}
\begin{minipage}[b]{0.116\linewidth}
\includegraphics[width=1\linewidth,height=0.6\linewidth]{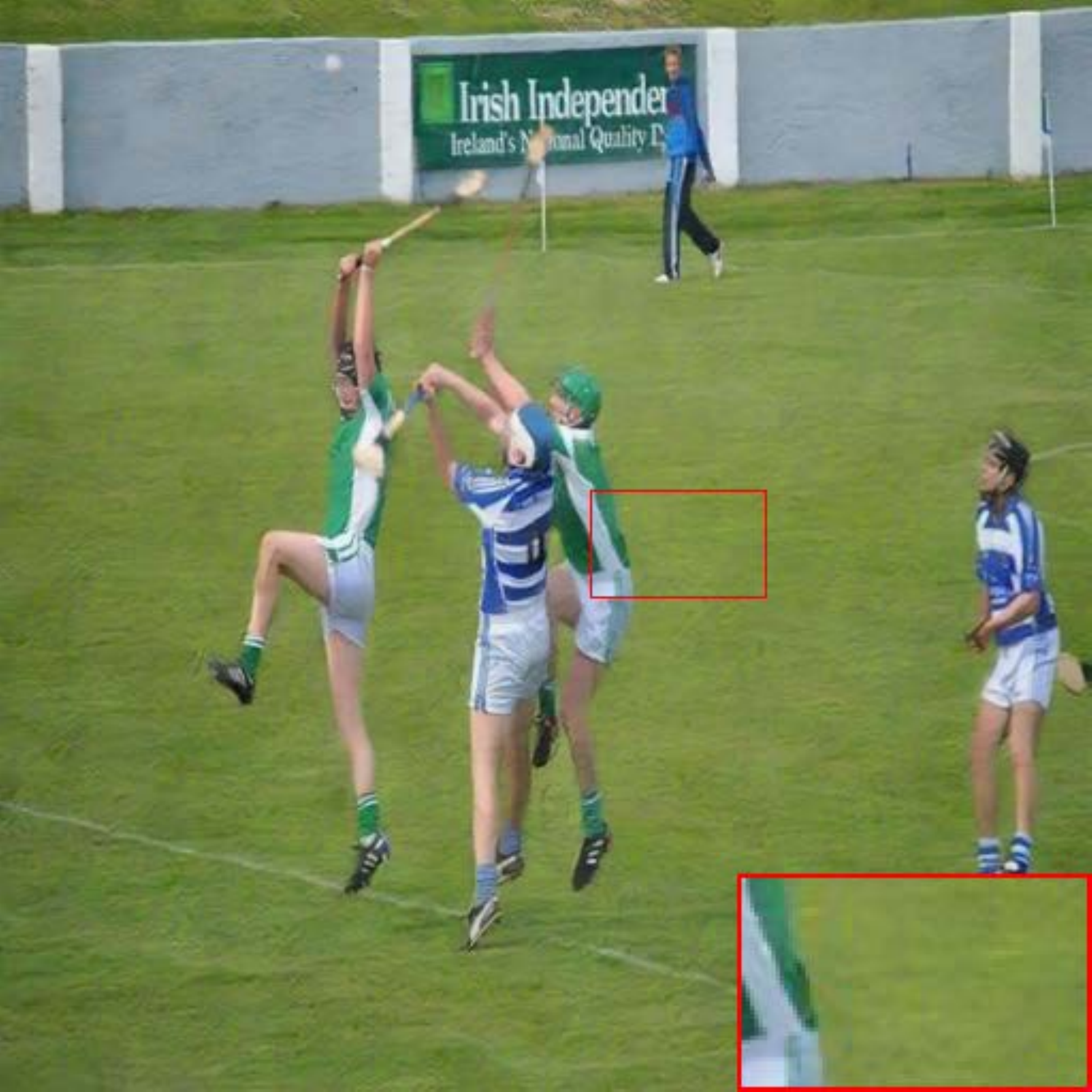}
\end{minipage}
\begin{minipage}[b]{0.116\linewidth}
\includegraphics[width=1\linewidth,height=0.6\linewidth]{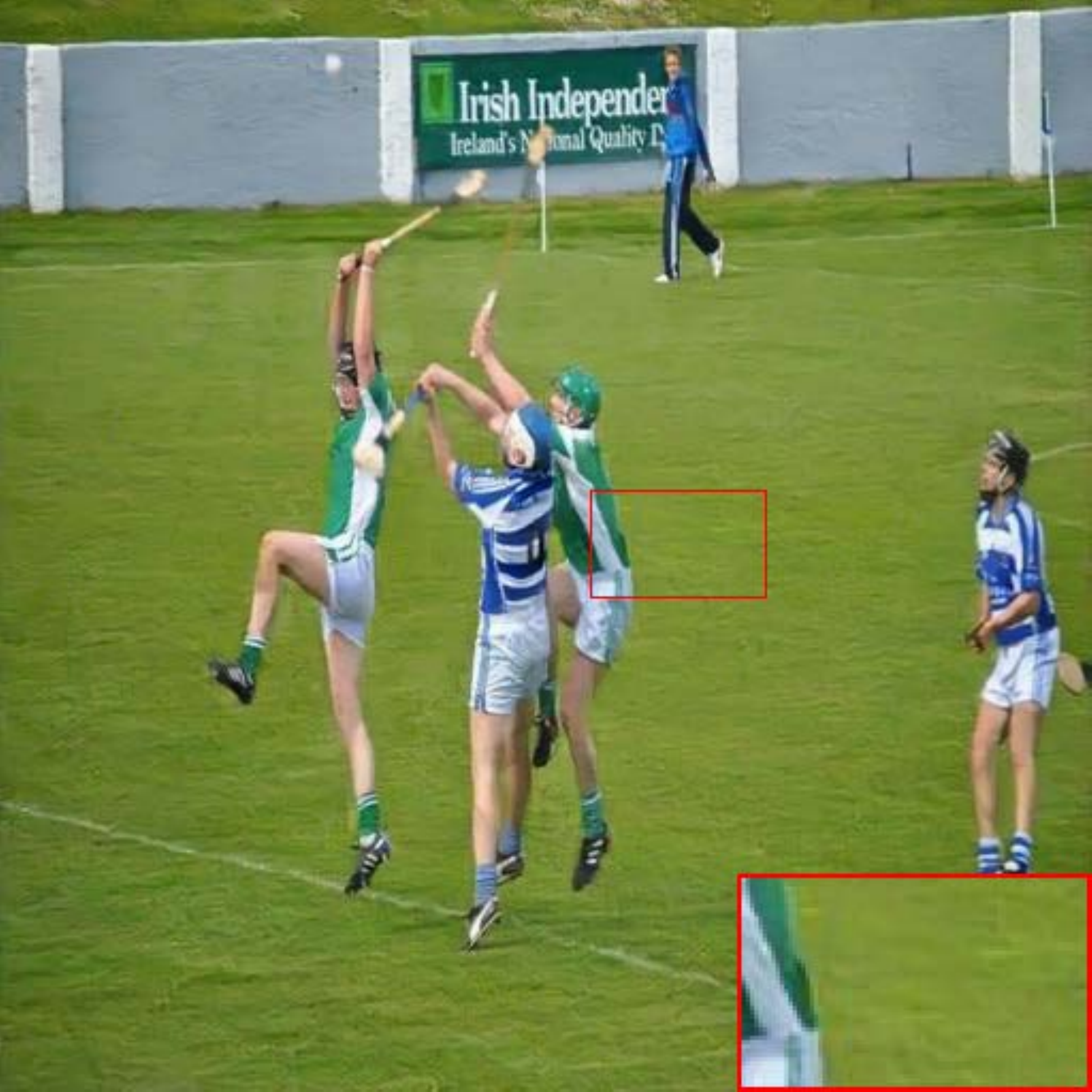}
\end{minipage}
\begin{minipage}[b]{0.116\linewidth}
\includegraphics[width=1\linewidth,height=0.6\linewidth]{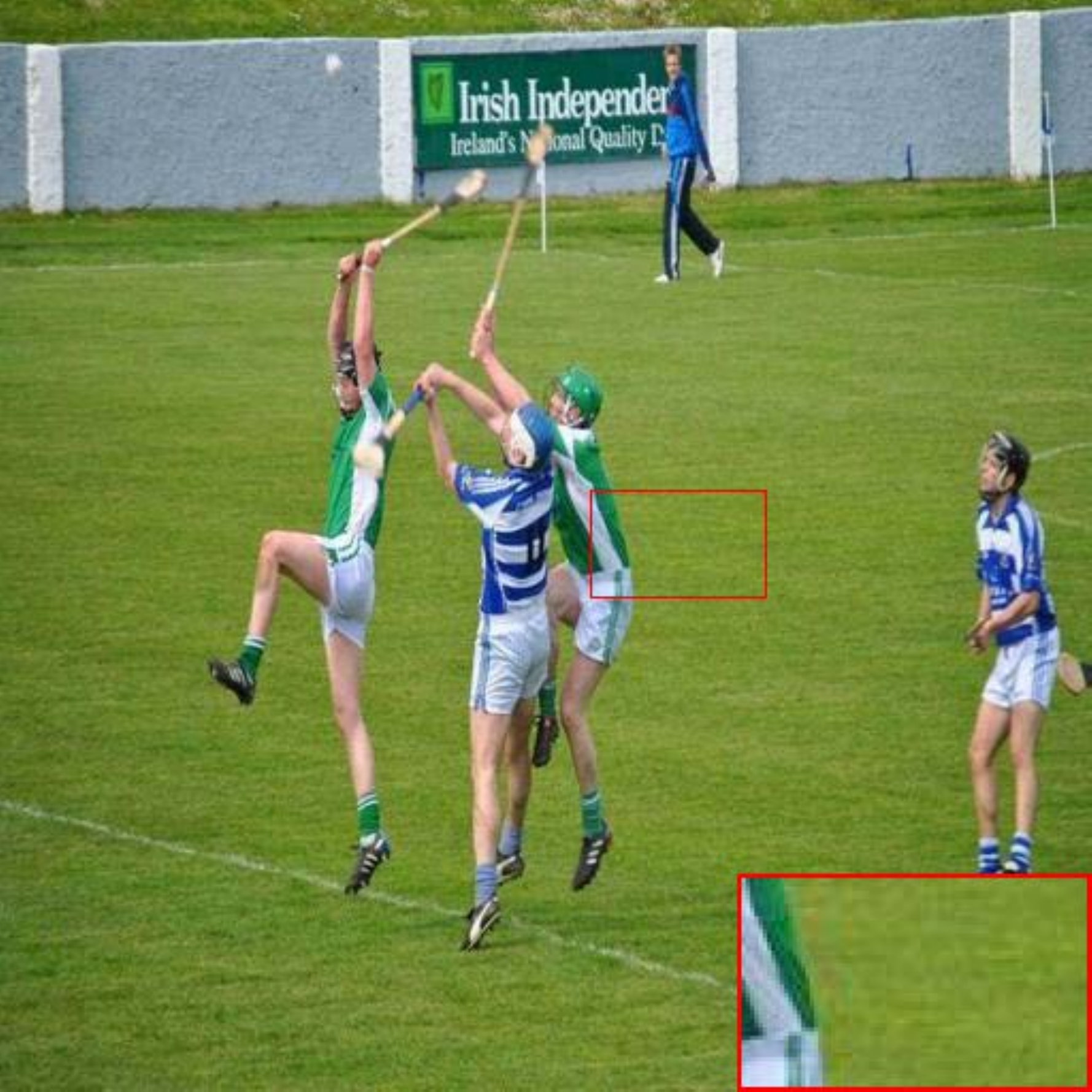}
\end{minipage}
\begin{minipage}[b]{0.116\linewidth}
\includegraphics[width=1\linewidth,height=0.6\linewidth]{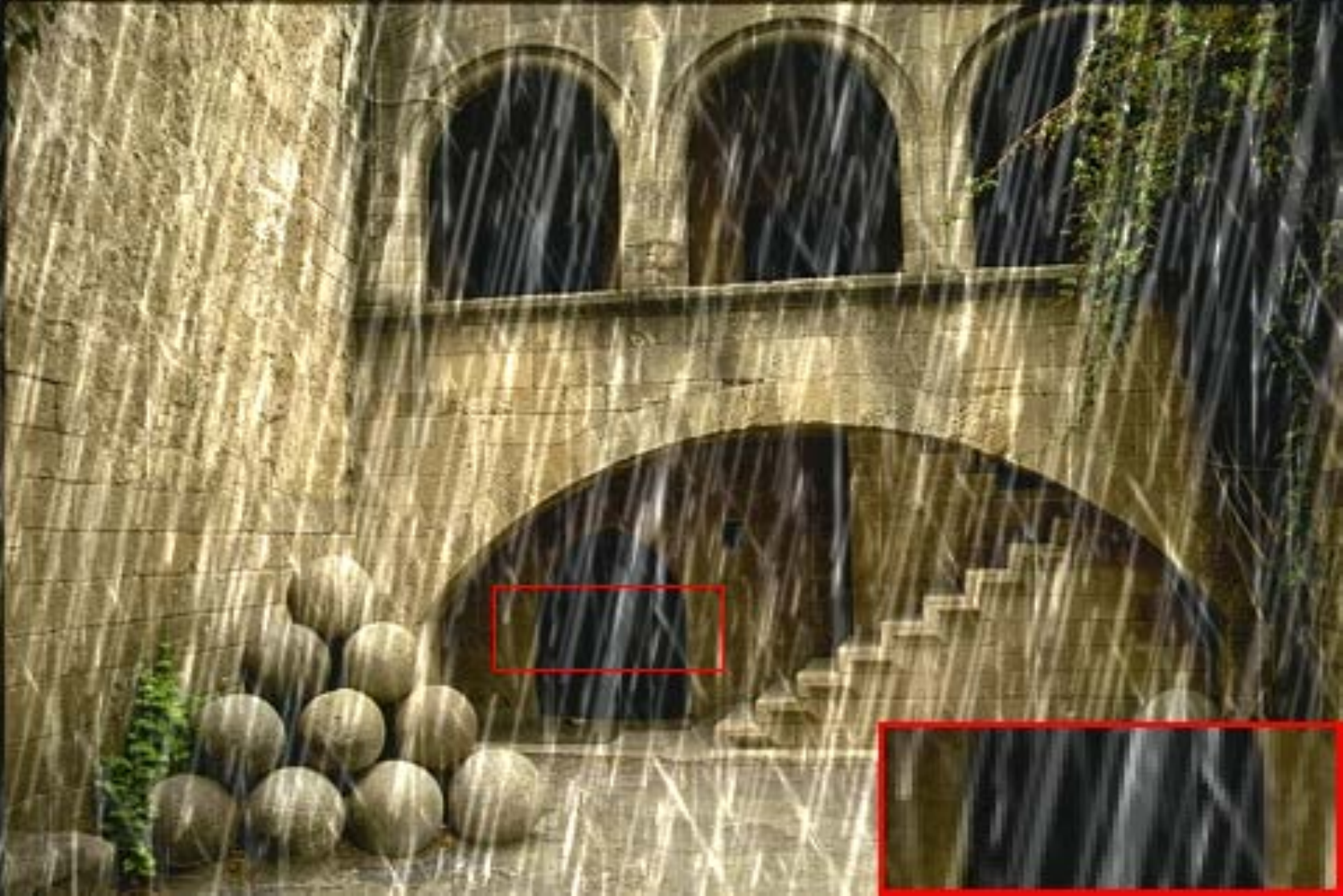}
\end{minipage}
%\vspace{-0.1cm}
\begin{minipage}[b]{0.116\linewidth}
\includegraphics[width=1\linewidth,height=0.6\linewidth]{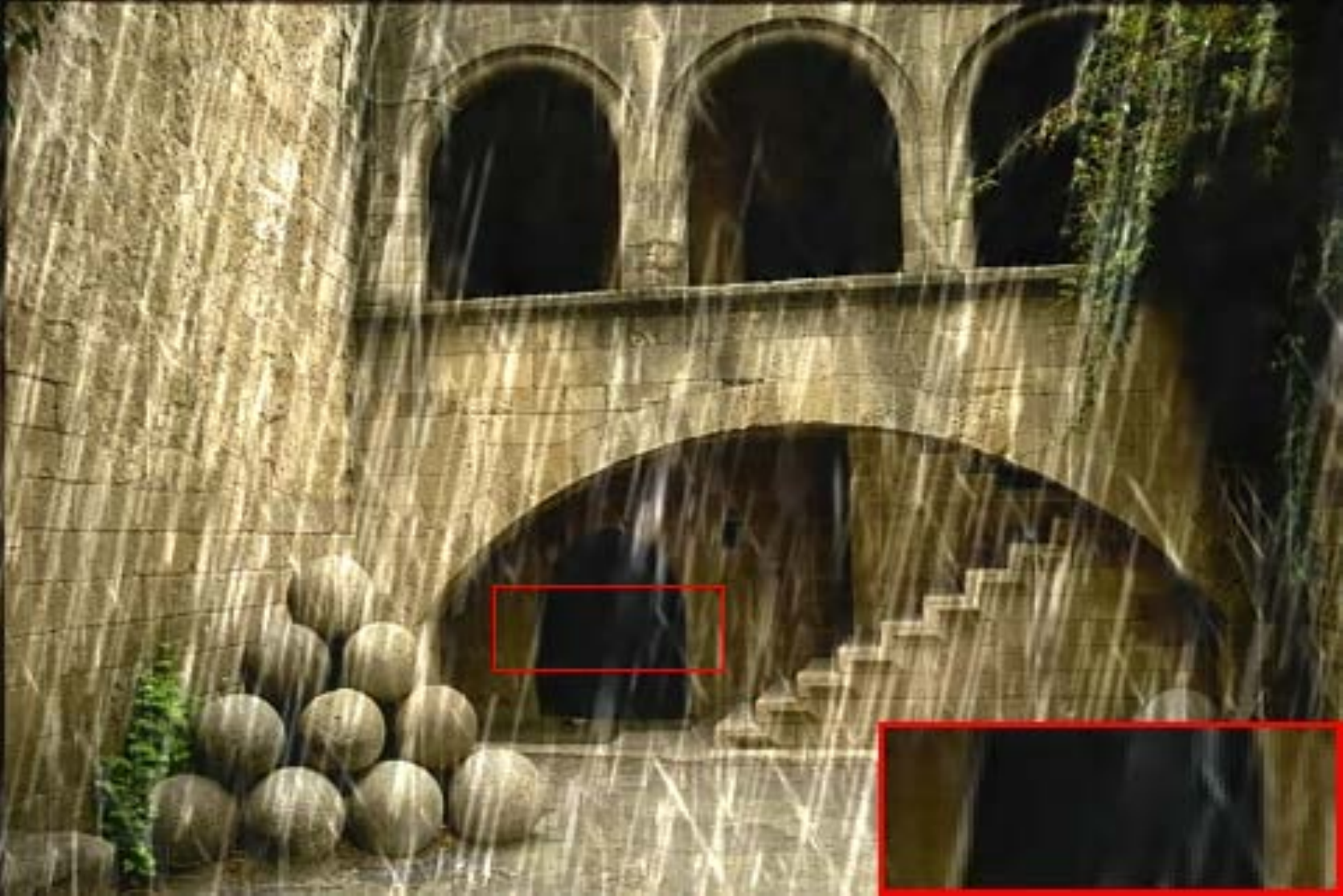}
\end{minipage}
\begin{minipage}[b]{0.116\linewidth}
\includegraphics[width=1\linewidth,height=0.6\linewidth]{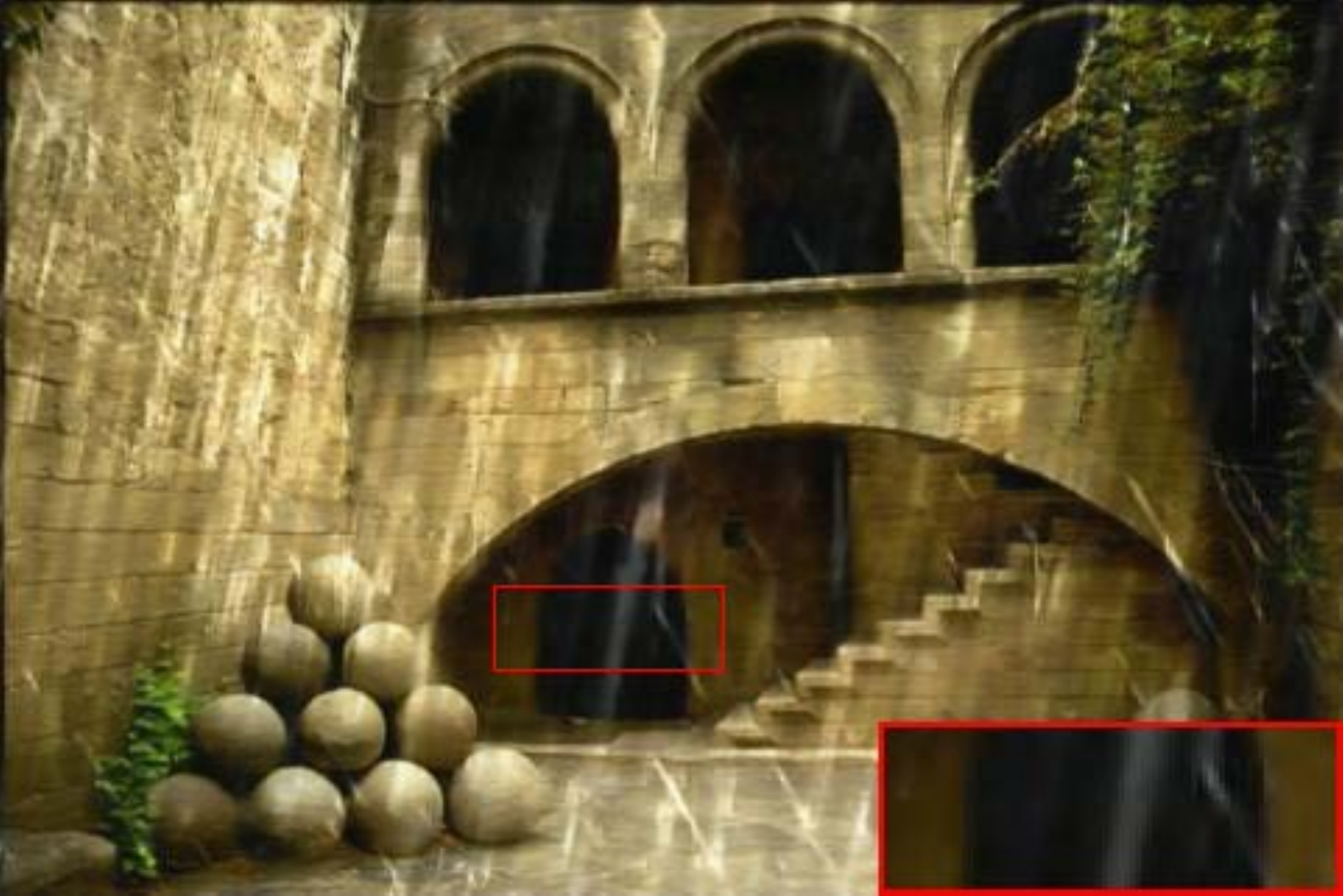}
\end{minipage}
\begin{minipage}[b]{0.116\linewidth}
\includegraphics[width=1\linewidth,height=0.6\linewidth]{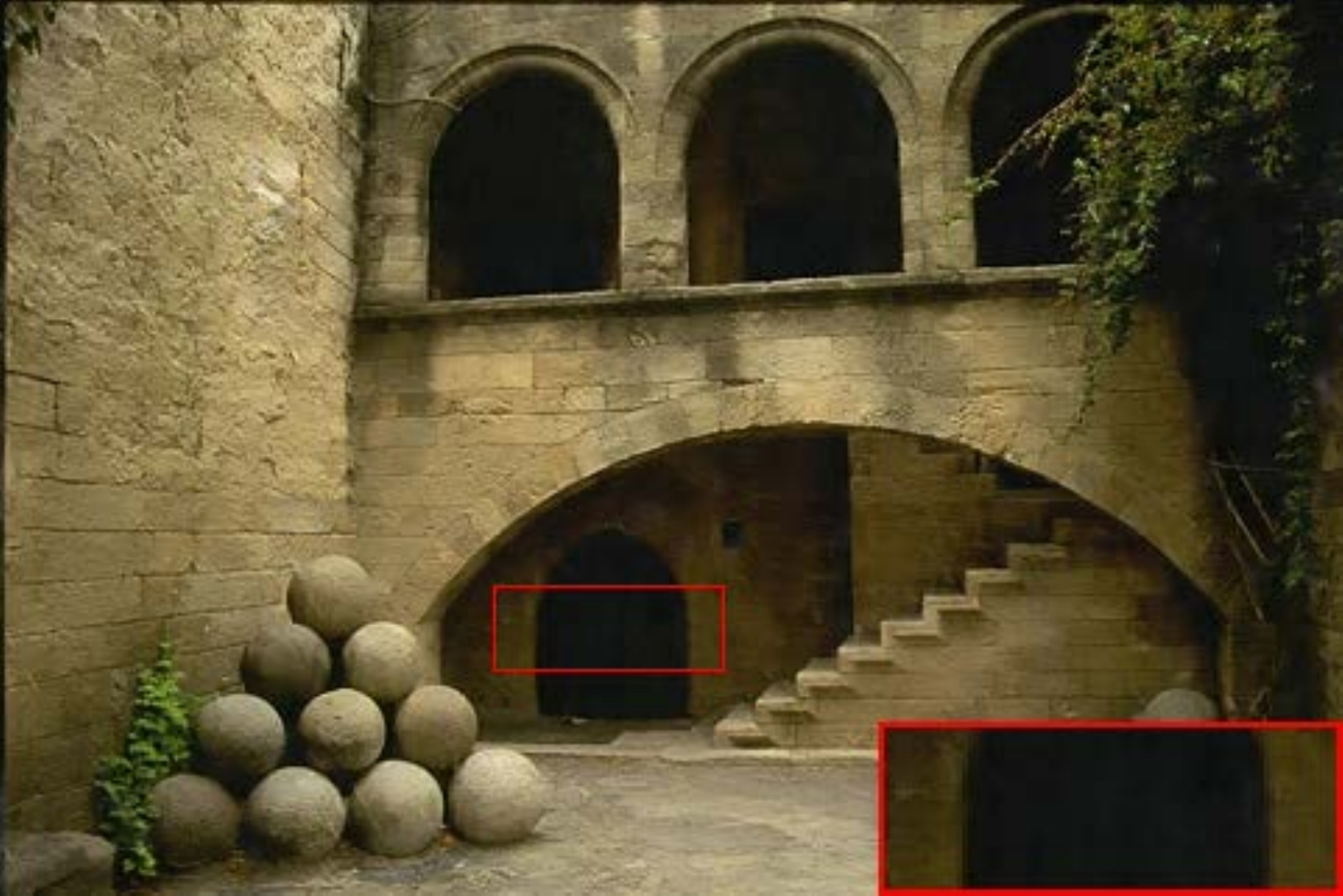}
\end{minipage}
\begin{minipage}[b]{0.116\linewidth}
\includegraphics[width=1\linewidth,height=0.6\linewidth]{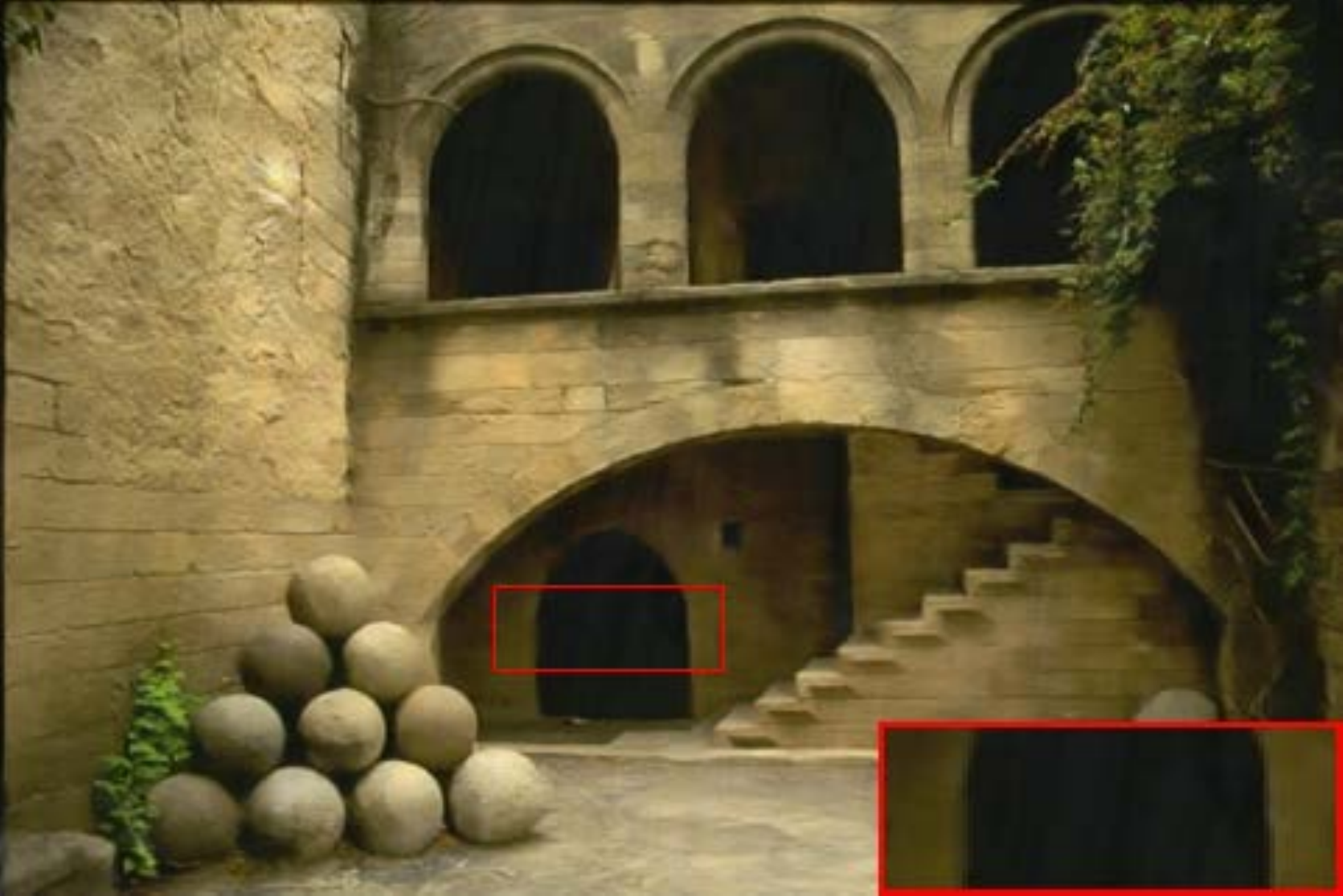}
\end{minipage}
\begin{minipage}[b]{0.116\linewidth}
\includegraphics[width=1\linewidth,height=0.6\linewidth]{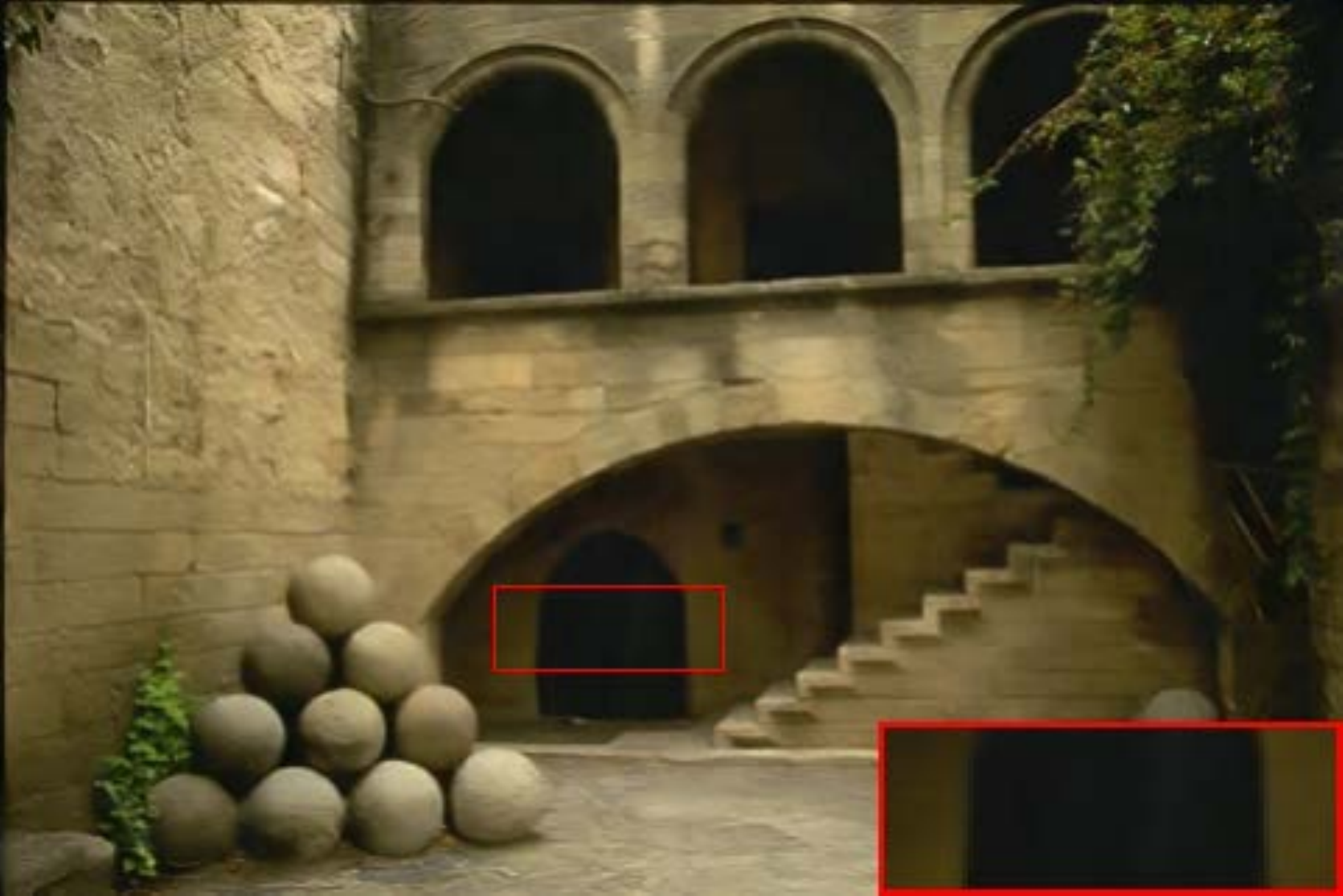}
\end{minipage}
\begin{minipage}[b]{0.116\linewidth}
\includegraphics[width=1\linewidth,height=0.6\linewidth]{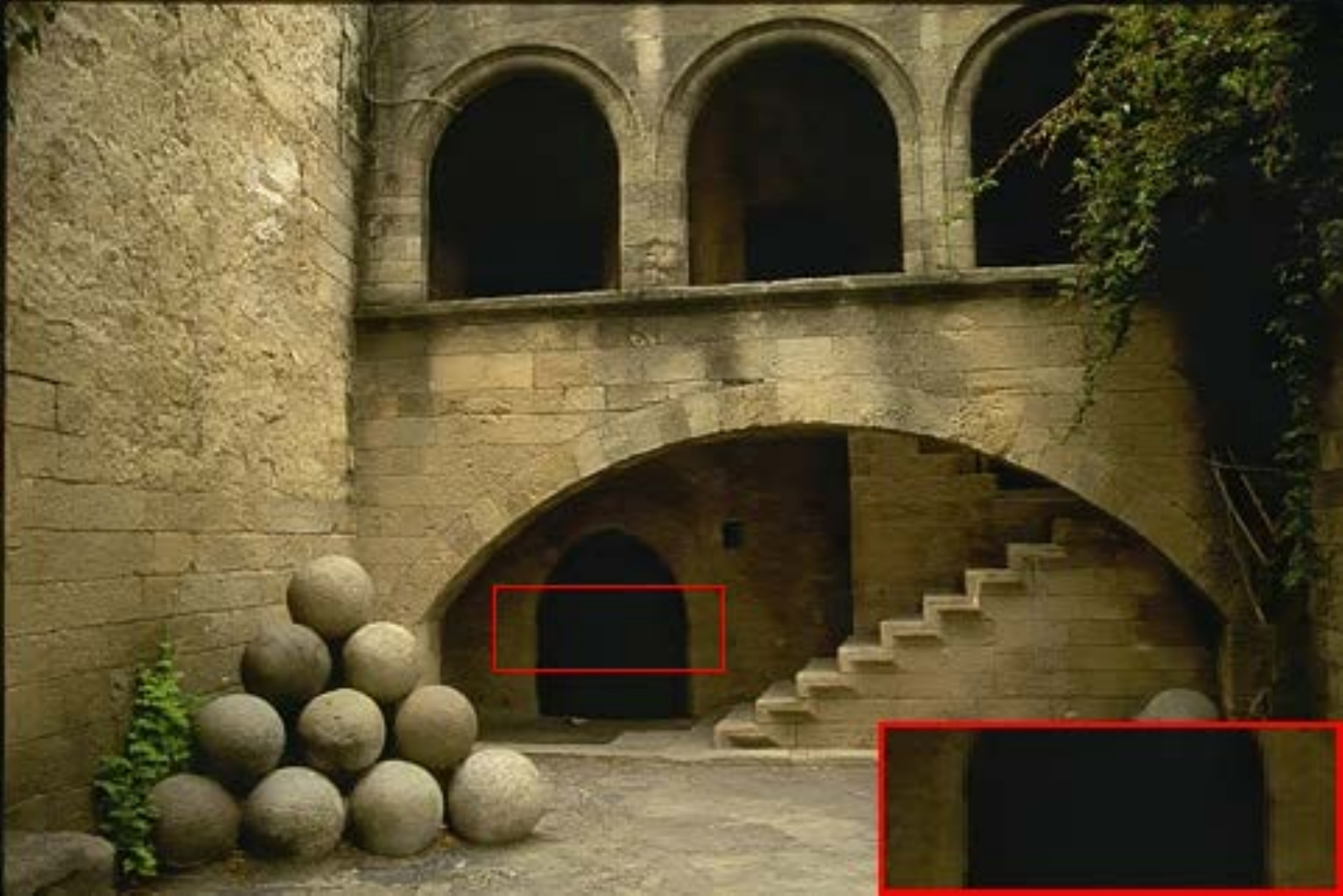}
\end{minipage}
\begin{minipage}[b]{0.116\linewidth}
\includegraphics[width=1\linewidth,height=0.6\linewidth]{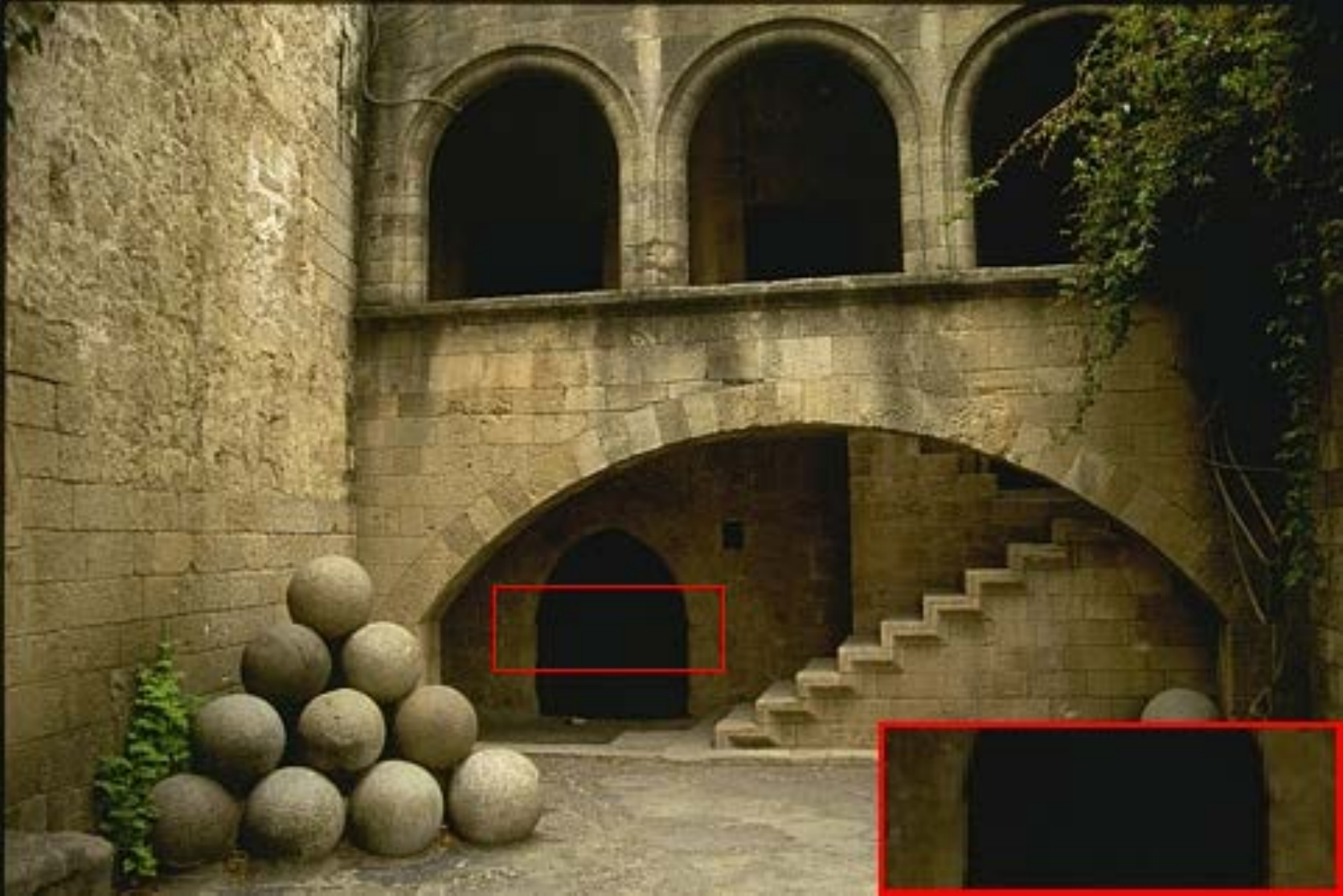}
\end{minipage}
\begin{minipage}[b]{0.116\linewidth}
\includegraphics[width=1\linewidth,height=0.6\linewidth]{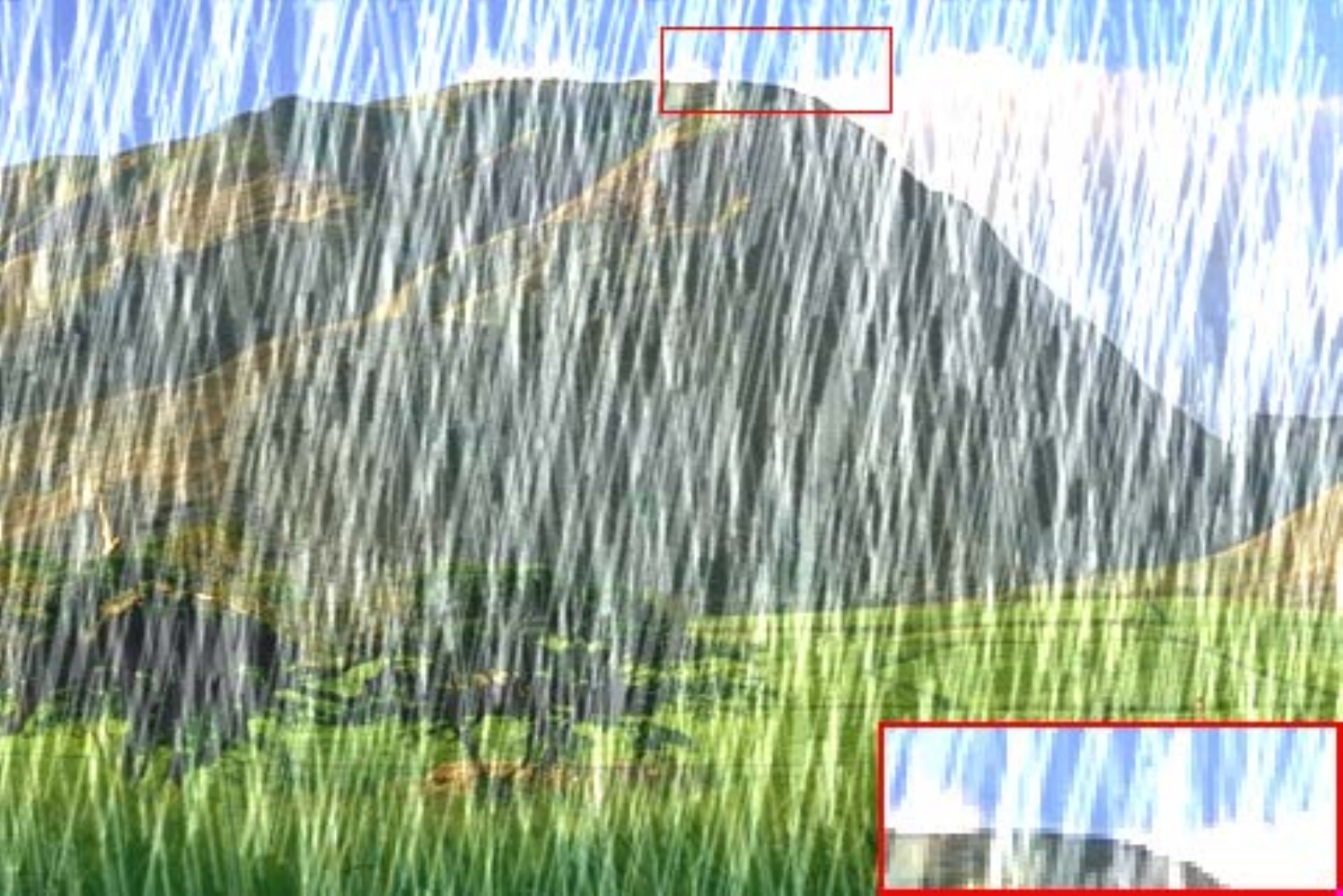}
\end{minipage}
%\vspace{-0.1cm}
\begin{minipage}[b]{0.116\linewidth}
\includegraphics[width=1\linewidth,height=0.6\linewidth]{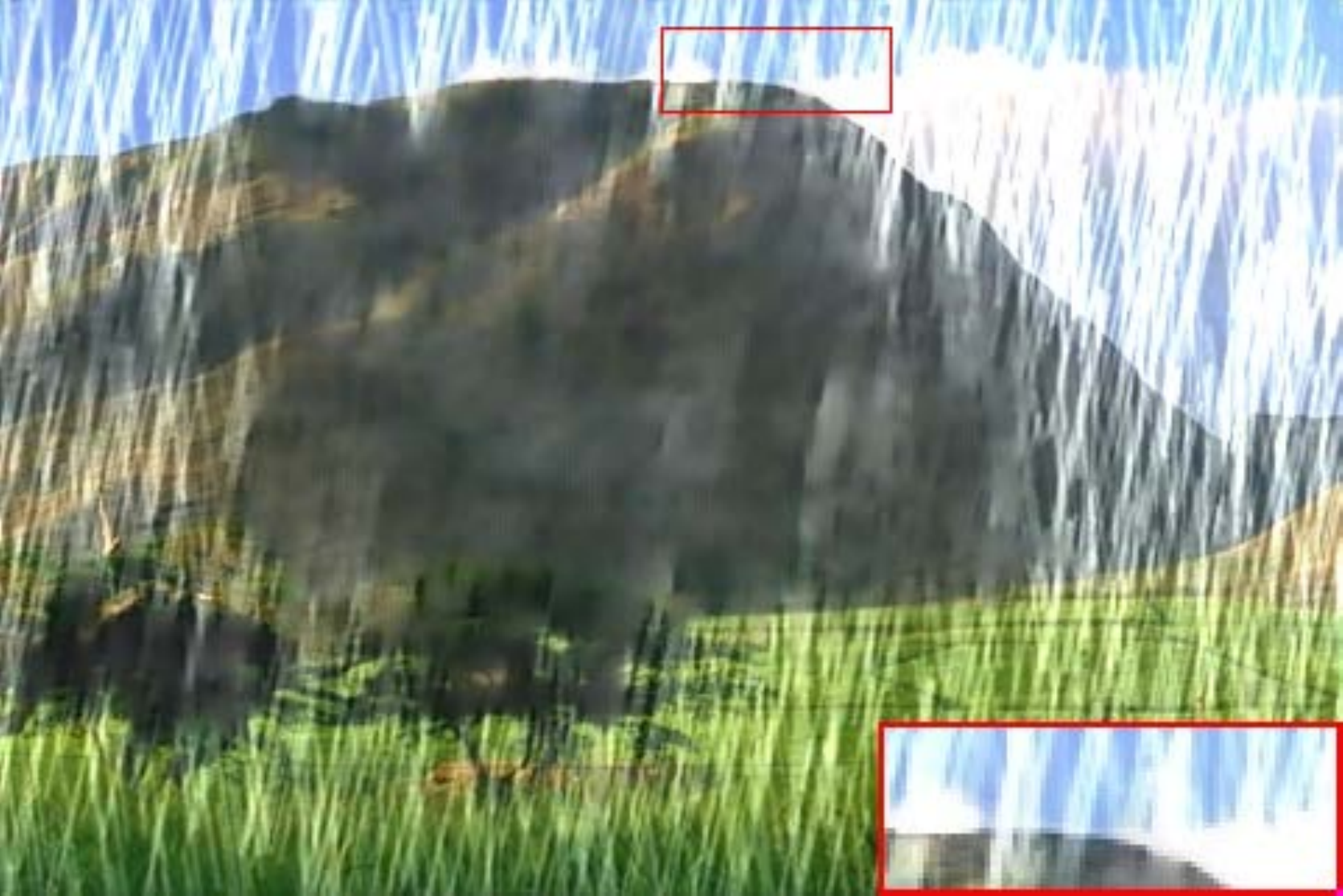}
\end{minipage}
\begin{minipage}[b]{0.116\linewidth}
\includegraphics[width=1\linewidth,height=0.6\linewidth]{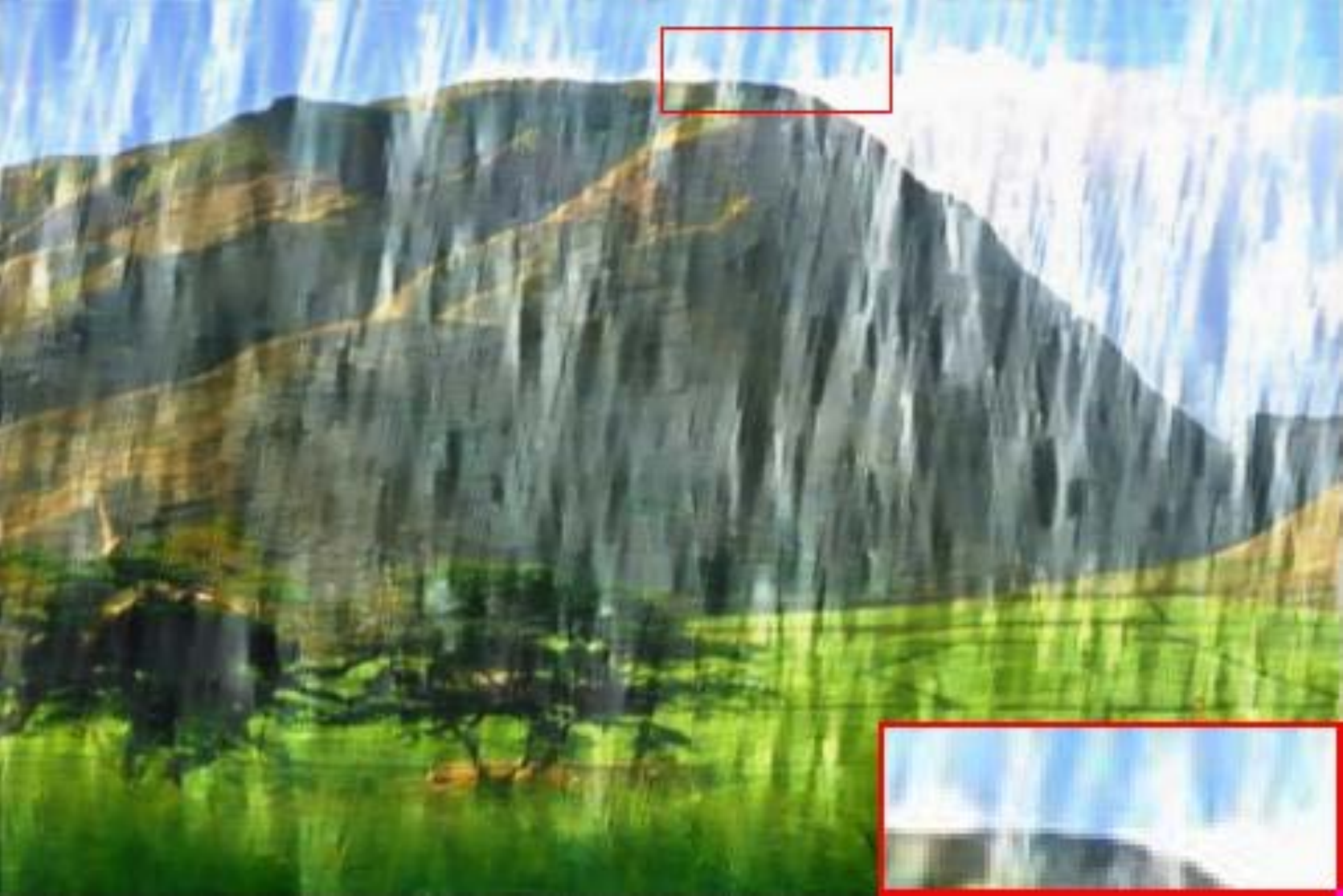}
\end{minipage}
\begin{minipage}[b]{0.116\linewidth}
\includegraphics[width=1\linewidth,height=0.6\linewidth]{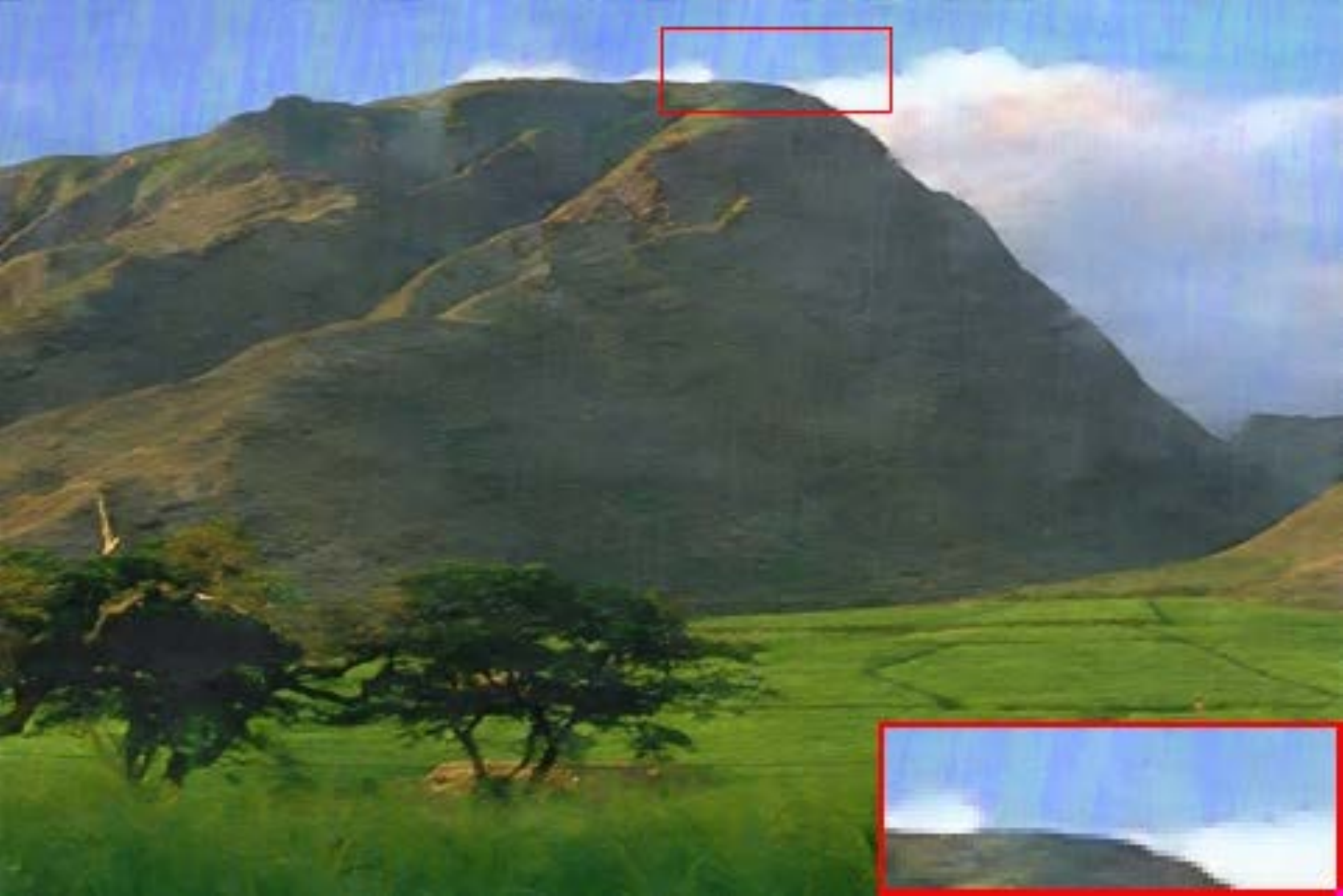}
\end{minipage}
\begin{minipage}[b]{0.116\linewidth}
\includegraphics[width=1\linewidth,height=0.6\linewidth]{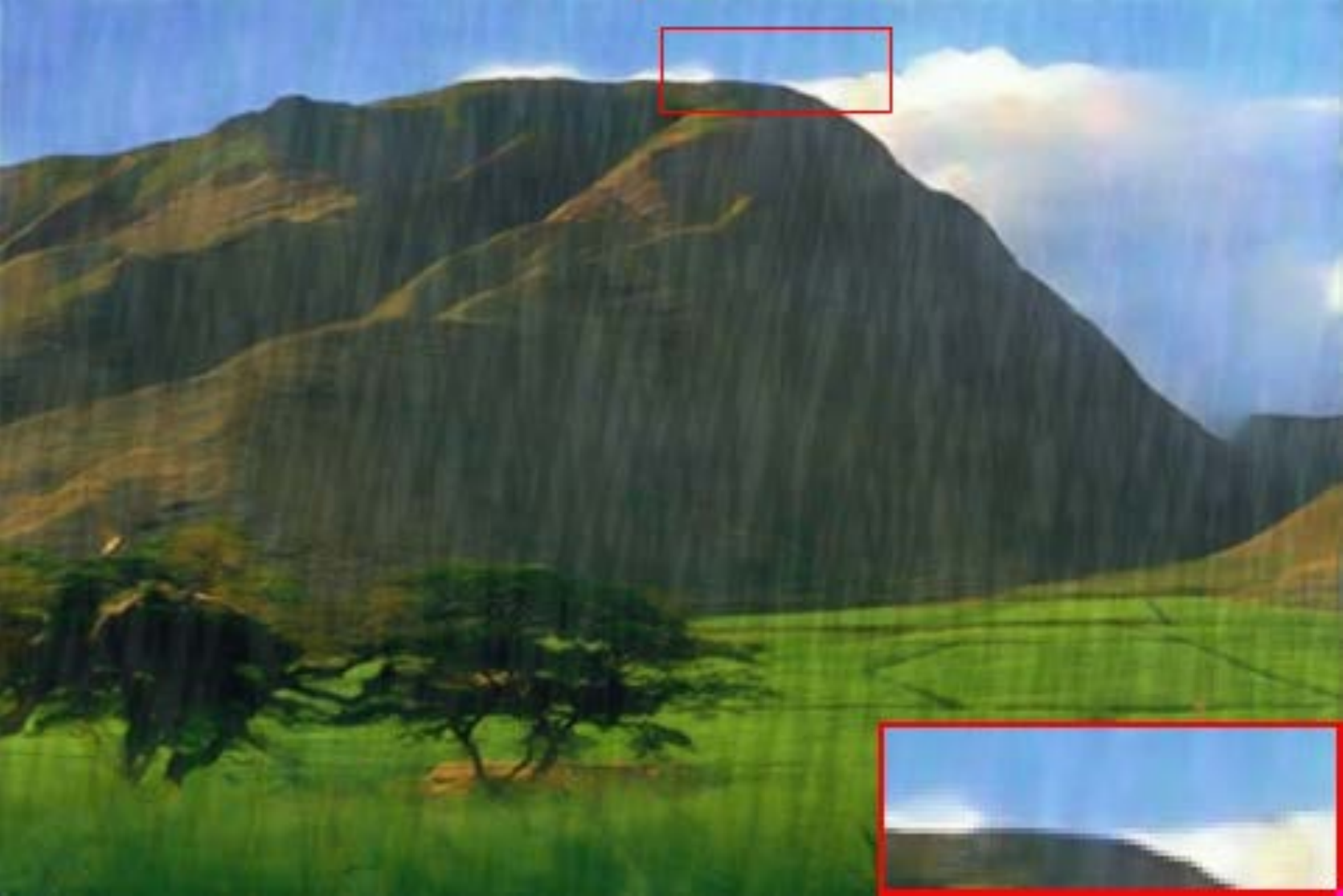}
\end{minipage}
\begin{minipage}[b]{0.116\linewidth}
\includegraphics[width=1\linewidth,height=0.6\linewidth]{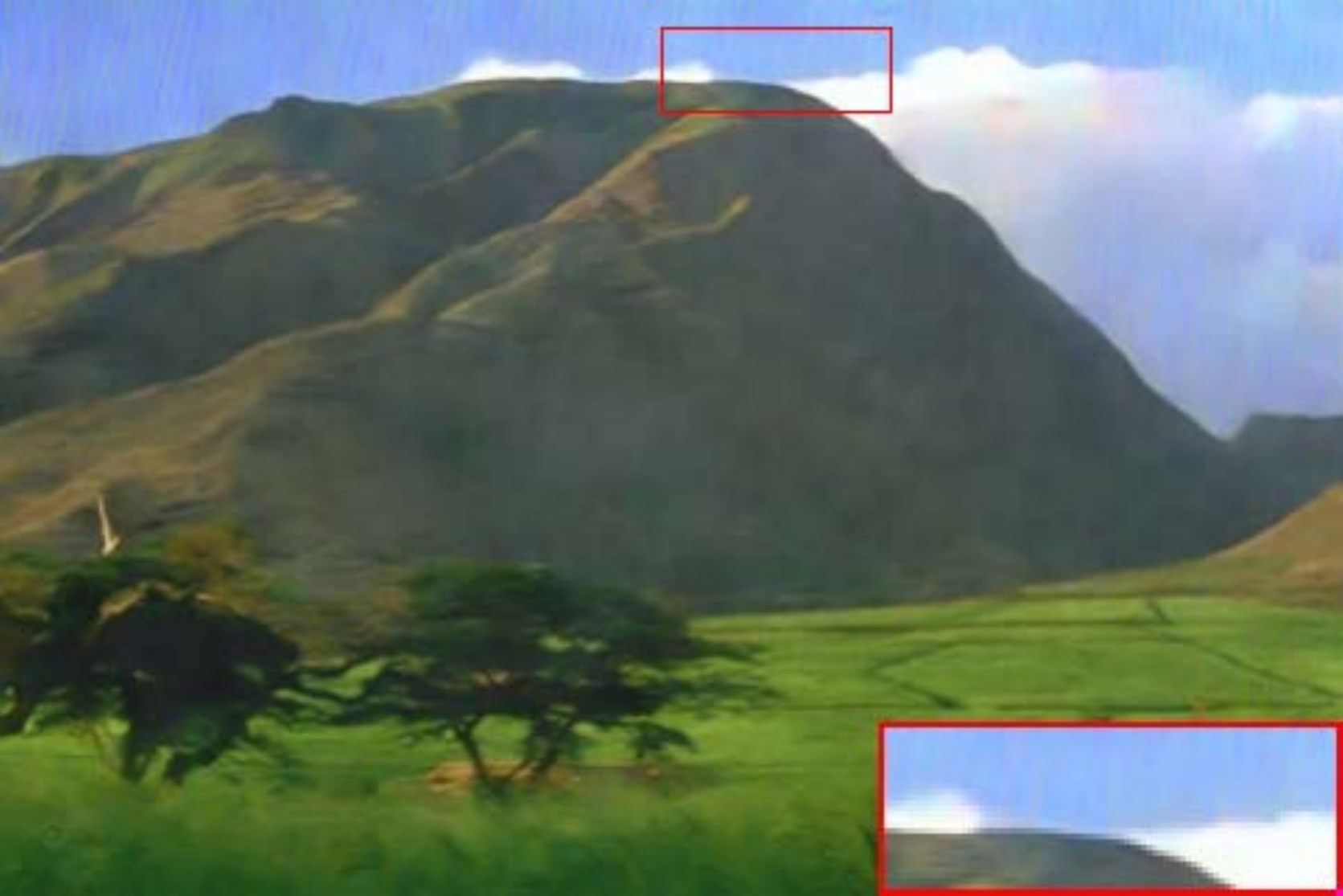}
\end{minipage}
\begin{minipage}[b]{0.116\linewidth}
\includegraphics[width=1\linewidth,height=0.6\linewidth]{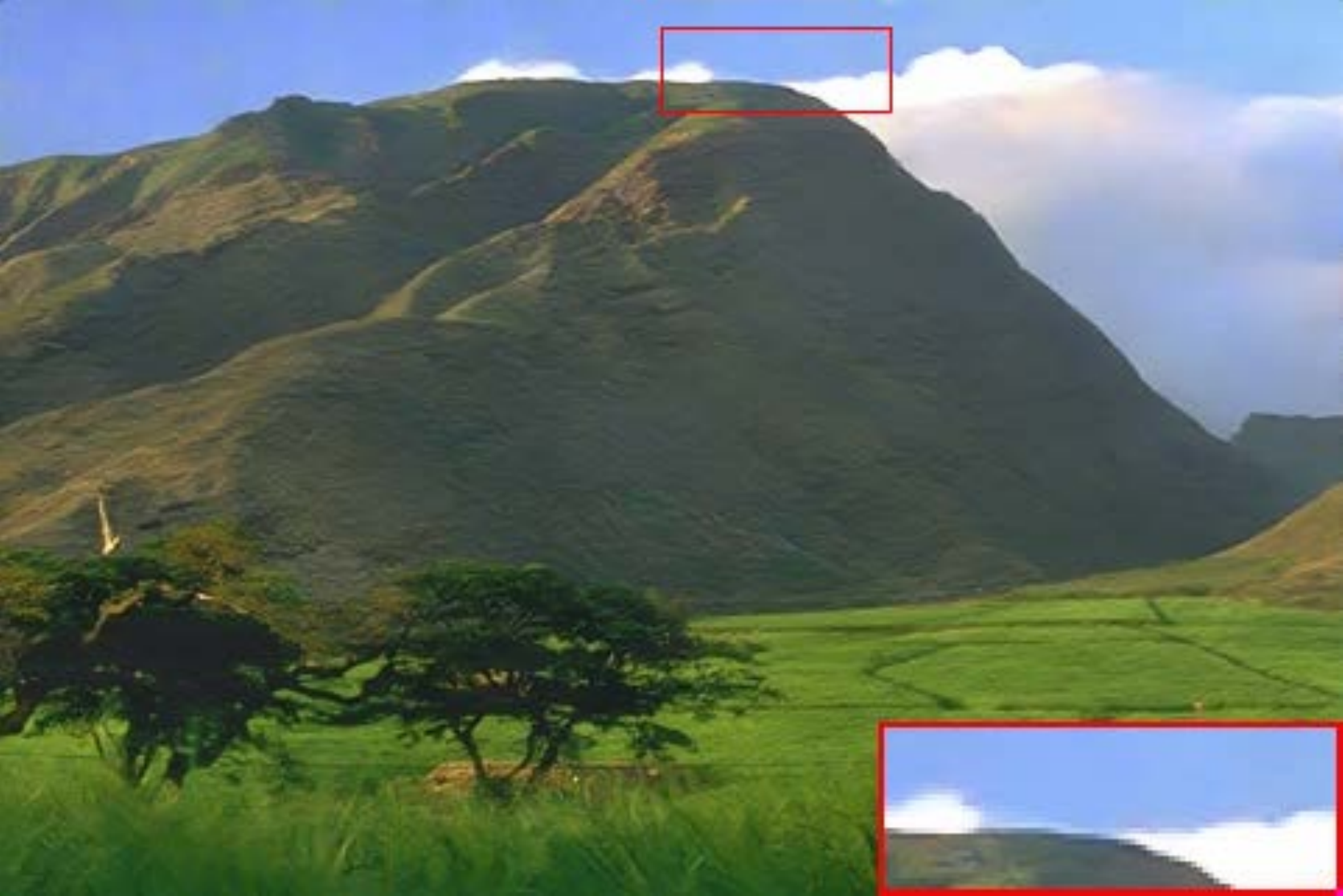}
\end{minipage}
\begin{minipage}[b]{0.116\linewidth}
\includegraphics[width=1\linewidth,height=0.6\linewidth]{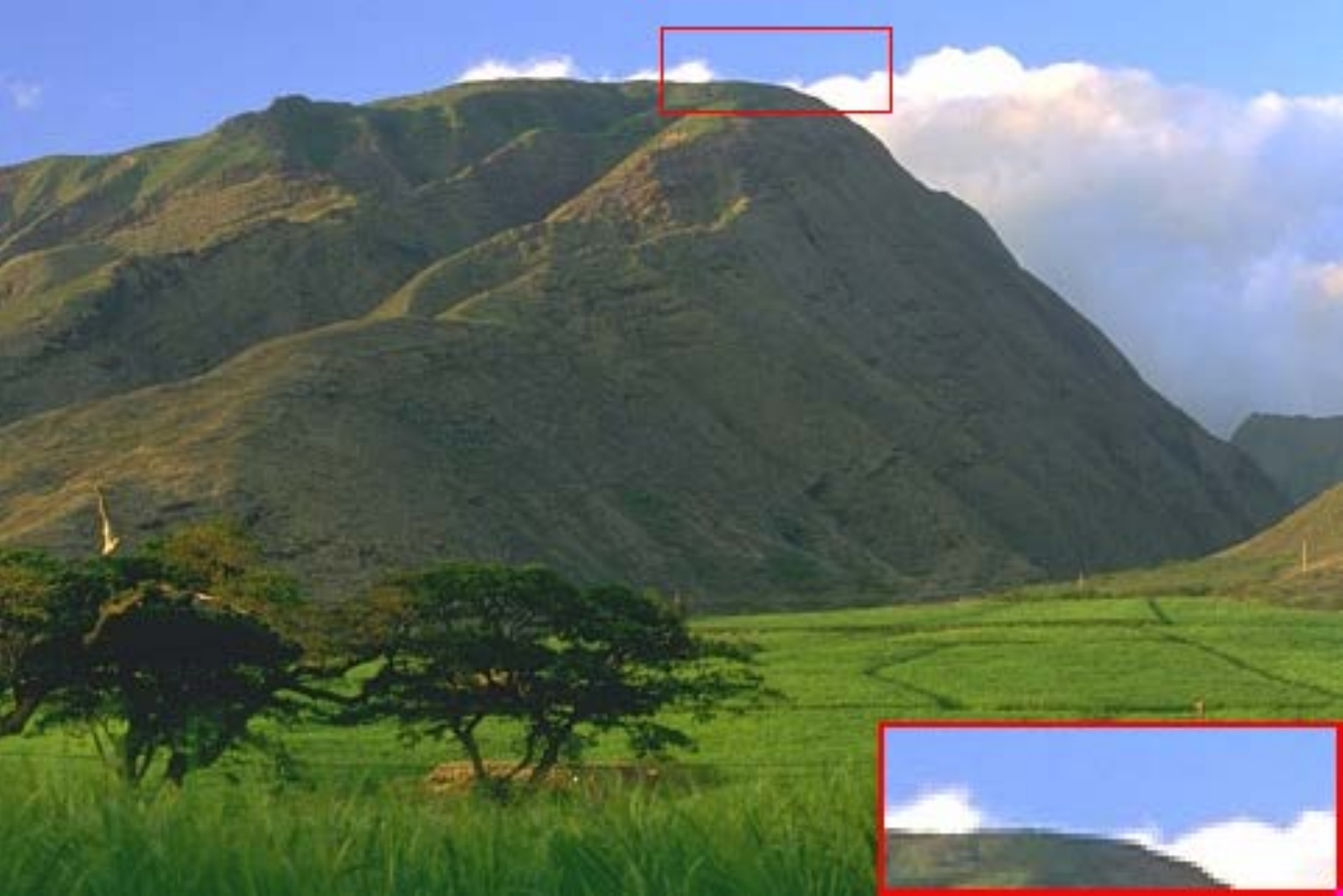}
\end{minipage}
\begin{minipage}[b]{0.116\linewidth}
\includegraphics[width=1\linewidth,height=0.6\linewidth]{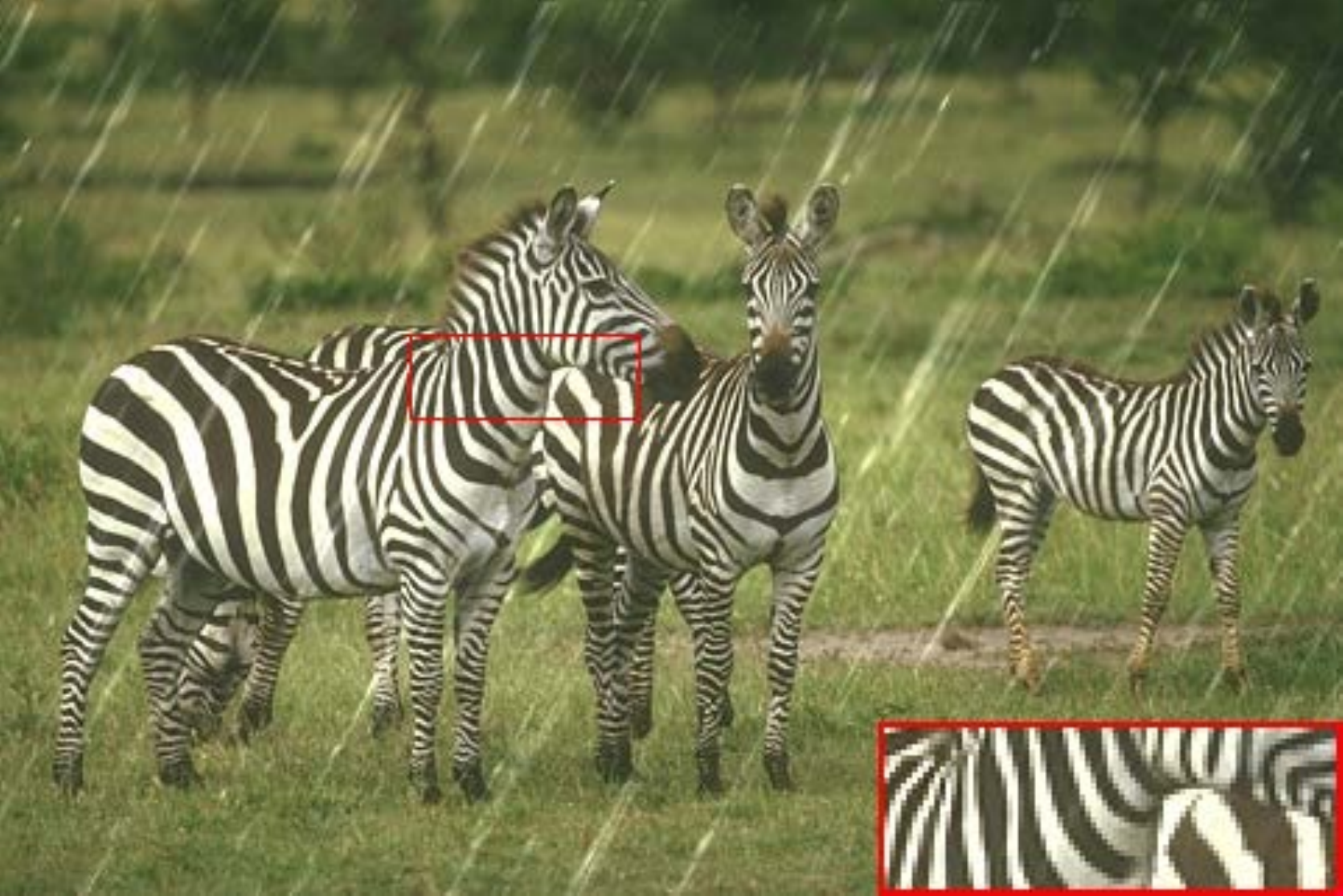}
\end{minipage}
%\vspace{-0.1cm}
\begin{minipage}[b]{0.116\linewidth}
\includegraphics[width=1\linewidth,height=0.6\linewidth]{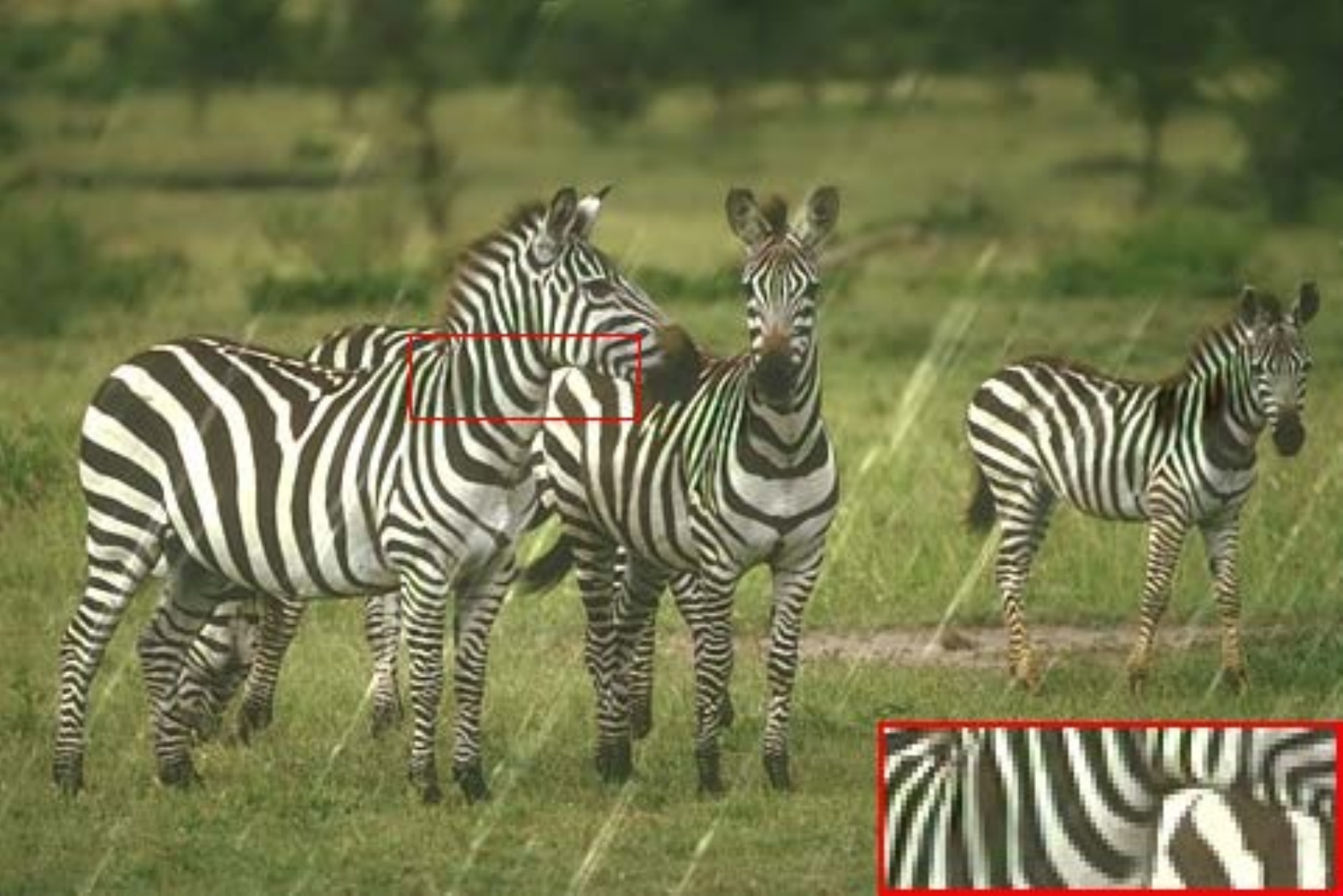}
\end{minipage}
\begin{minipage}[b]{0.116\linewidth}
\includegraphics[width=1\linewidth,height=0.6\linewidth]{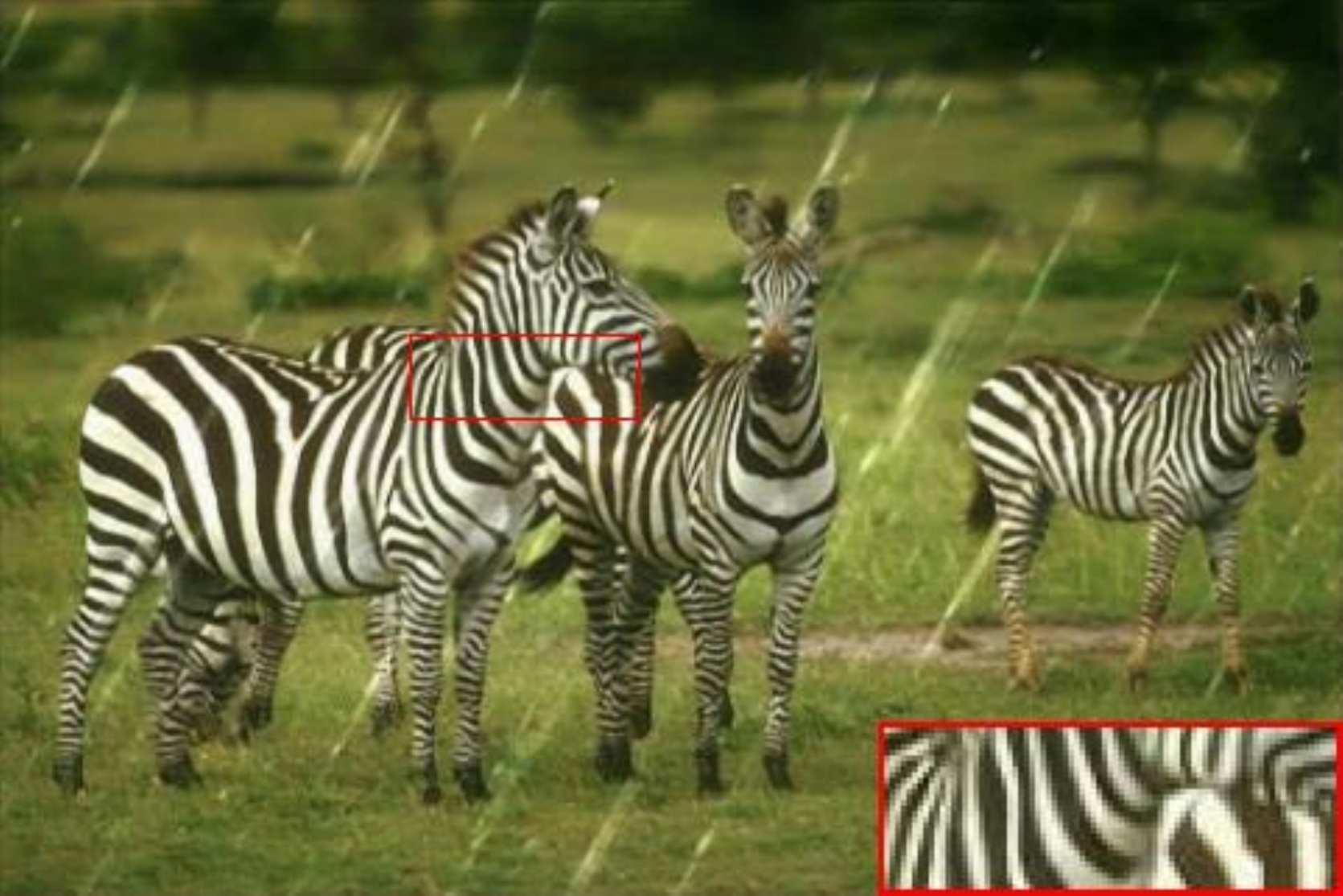}
\end{minipage}
\begin{minipage}[b]{0.116\linewidth}
\includegraphics[width=1\linewidth,height=0.6\linewidth]{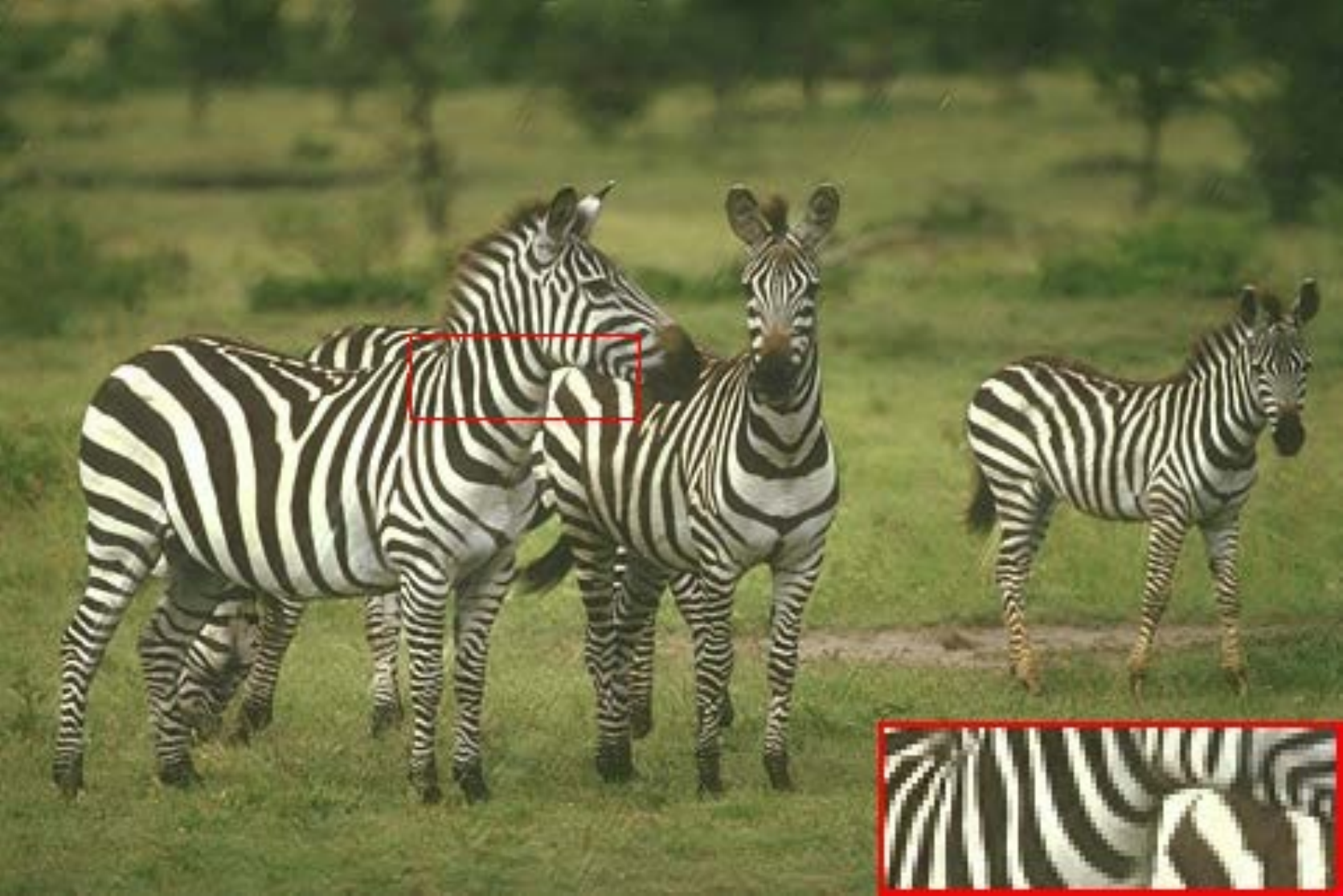}
\end{minipage}
\begin{minipage}[b]{0.116\linewidth}
\includegraphics[width=1\linewidth,height=0.6\linewidth]{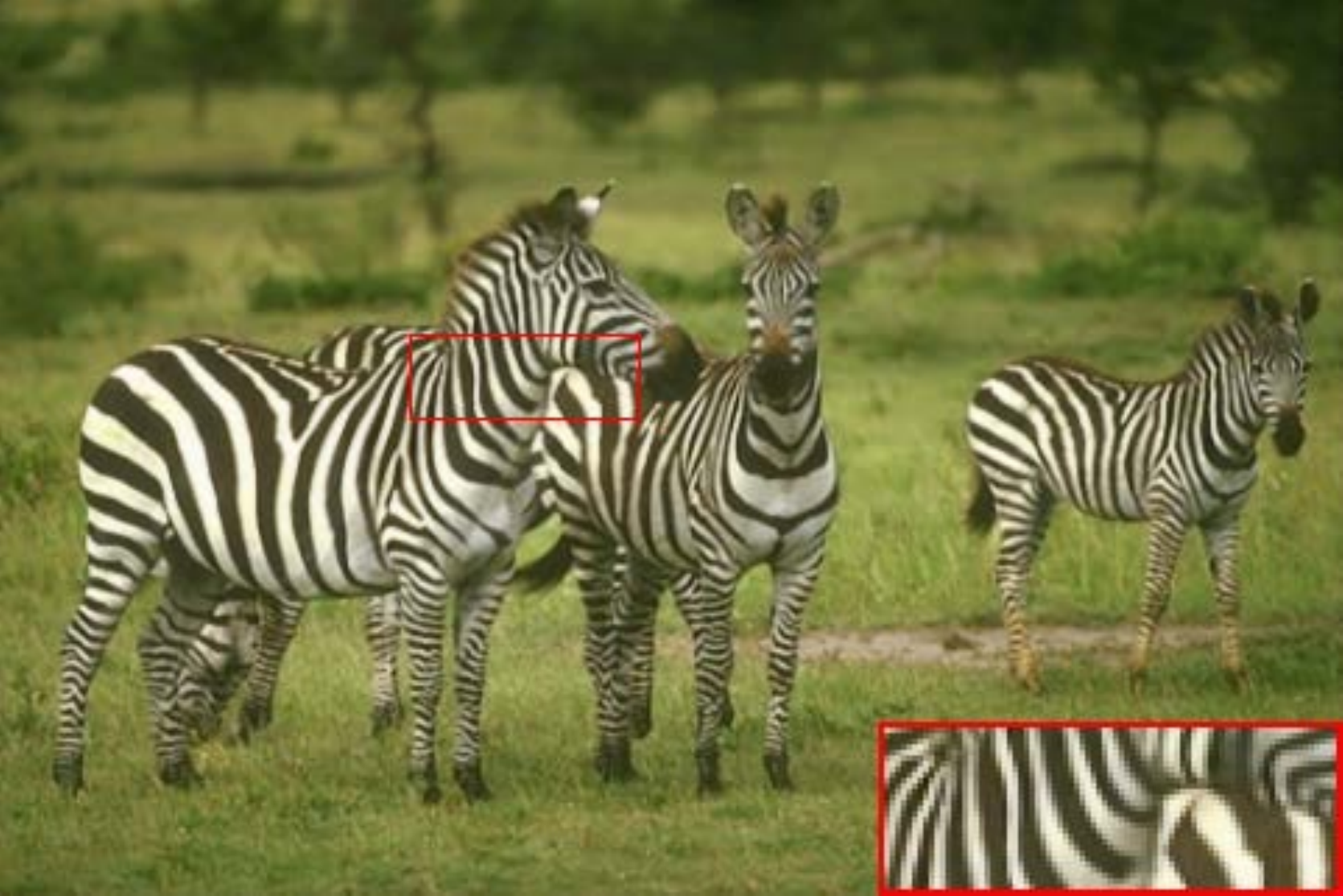}
\end{minipage}
\begin{minipage}[b]{0.116\linewidth}
\includegraphics[width=1\linewidth,height=0.6\linewidth]{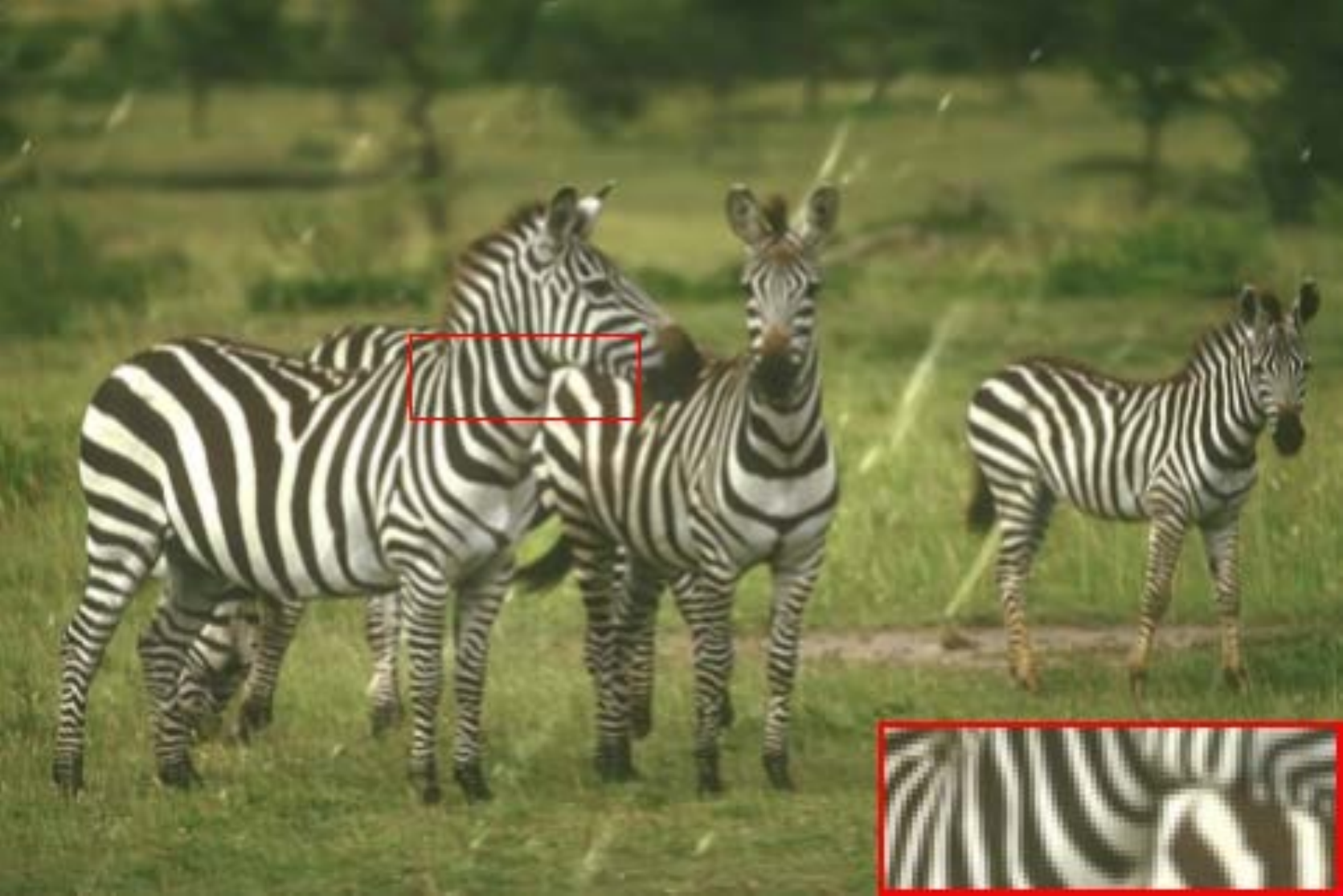}
\end{minipage}
\begin{minipage}[b]{0.116\linewidth}
\includegraphics[width=1\linewidth,height=0.6\linewidth]{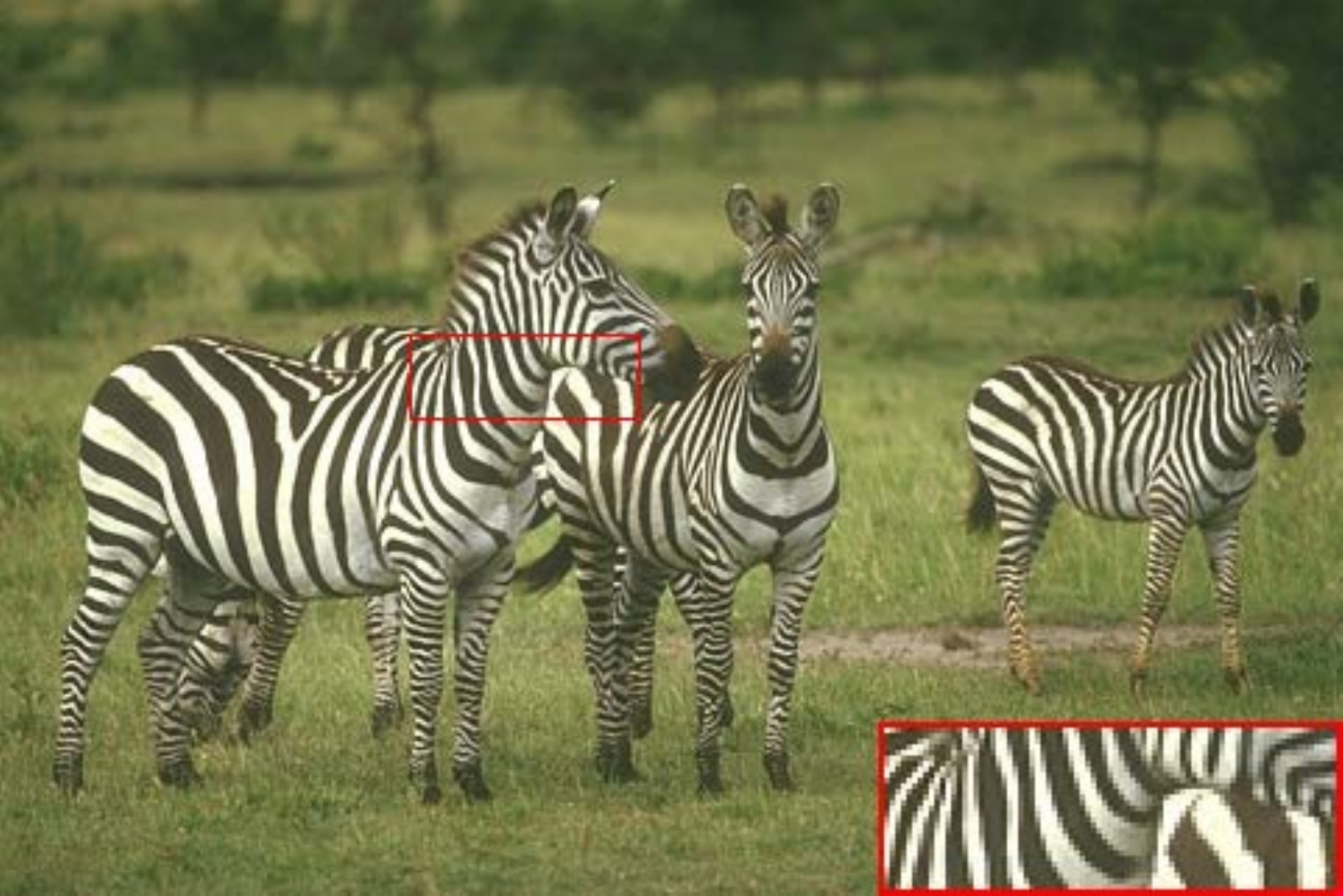}
\end{minipage}
\begin{minipage}[b]{0.116\linewidth}
\includegraphics[width=1\linewidth,height=0.6\linewidth]{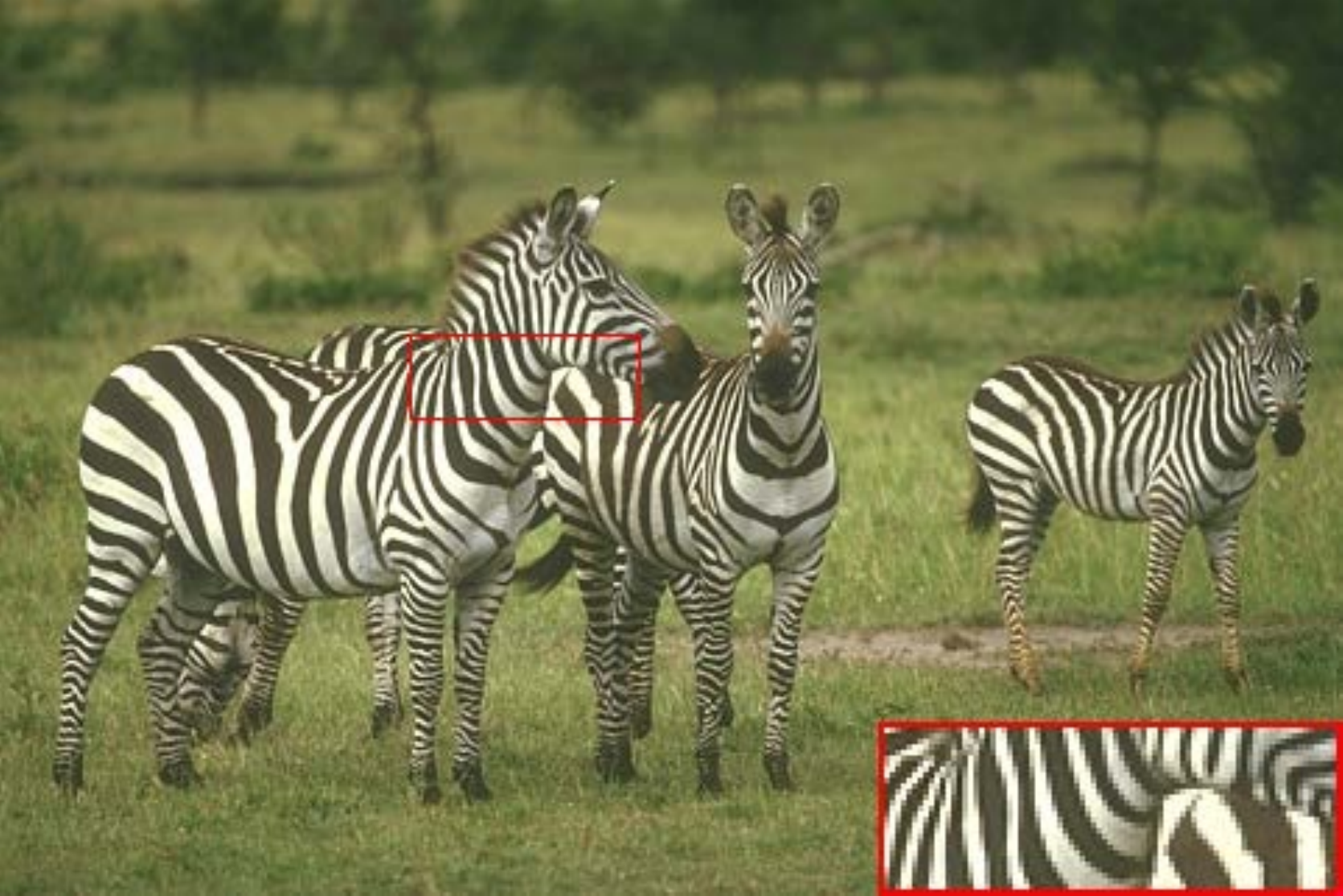}
\end{minipage}
\begin{minipage}[b]{0.116\linewidth}
\includegraphics[width=1\linewidth,height=0.6\linewidth]{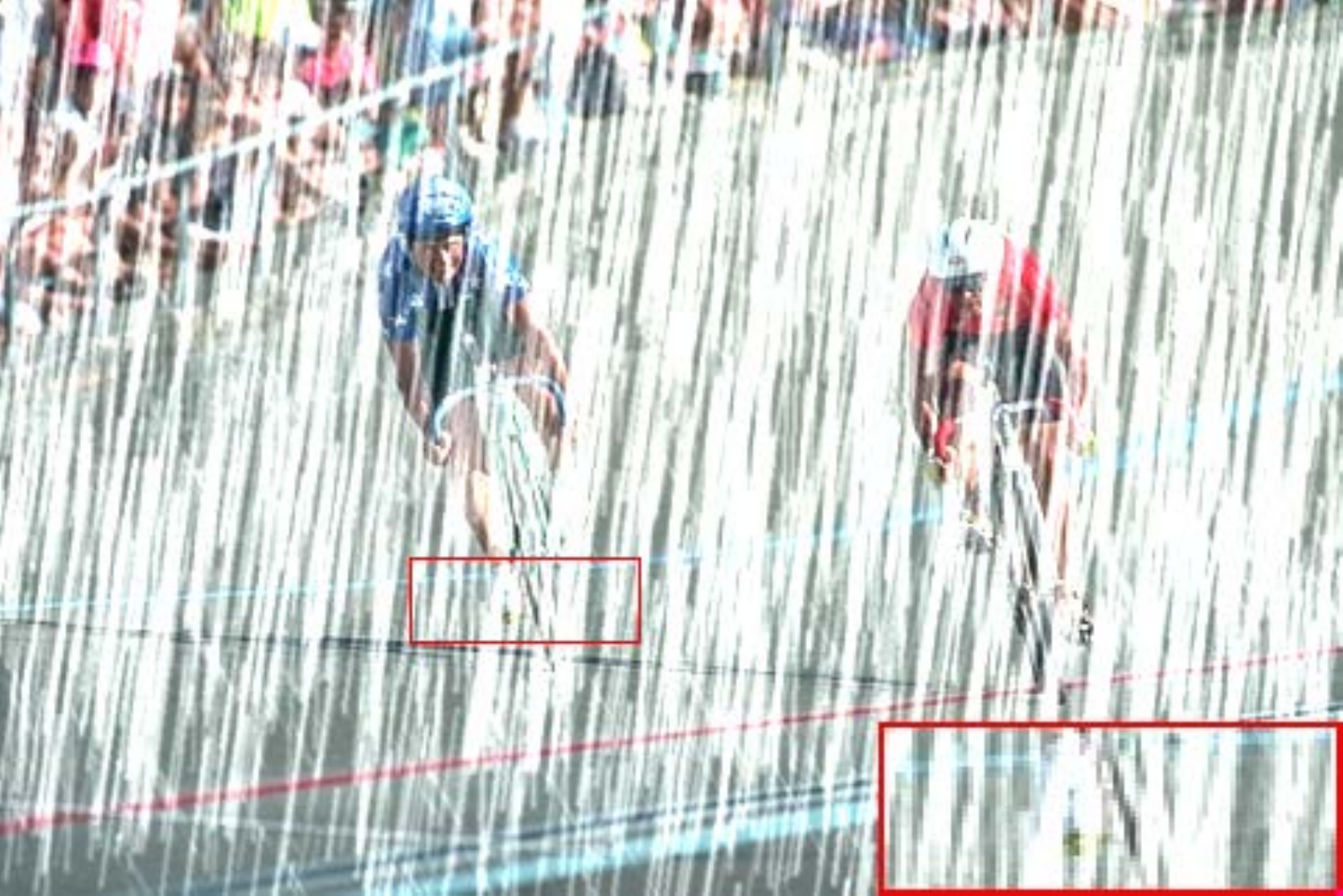}
\end{minipage}
%\vspace{-0.1cm}
\begin{minipage}[b]{0.116\linewidth}
\includegraphics[width=1\linewidth,height=0.6\linewidth]{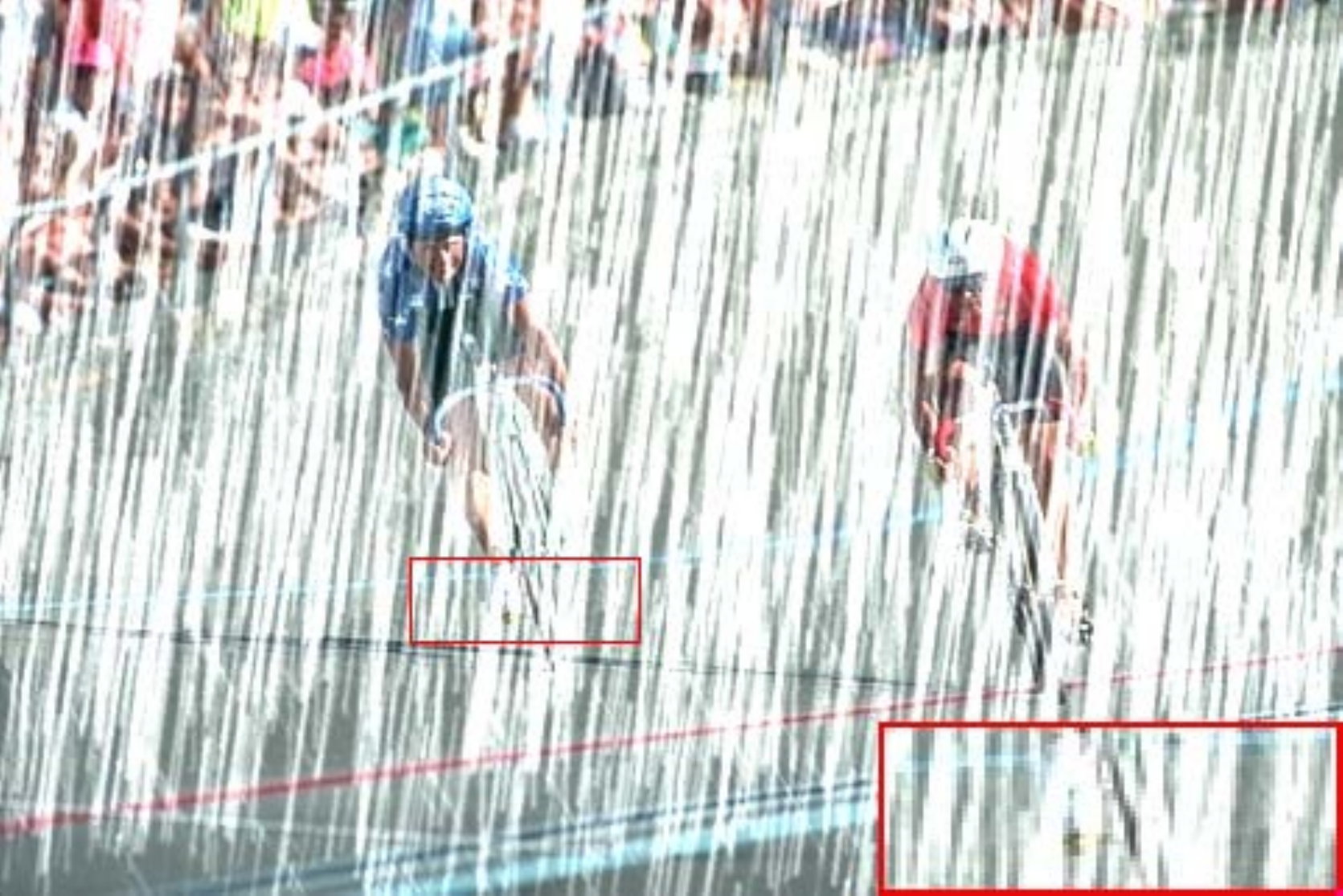}
\end{minipage}
\begin{minipage}[b]{0.116\linewidth}
\includegraphics[width=1\linewidth,height=0.6\linewidth]{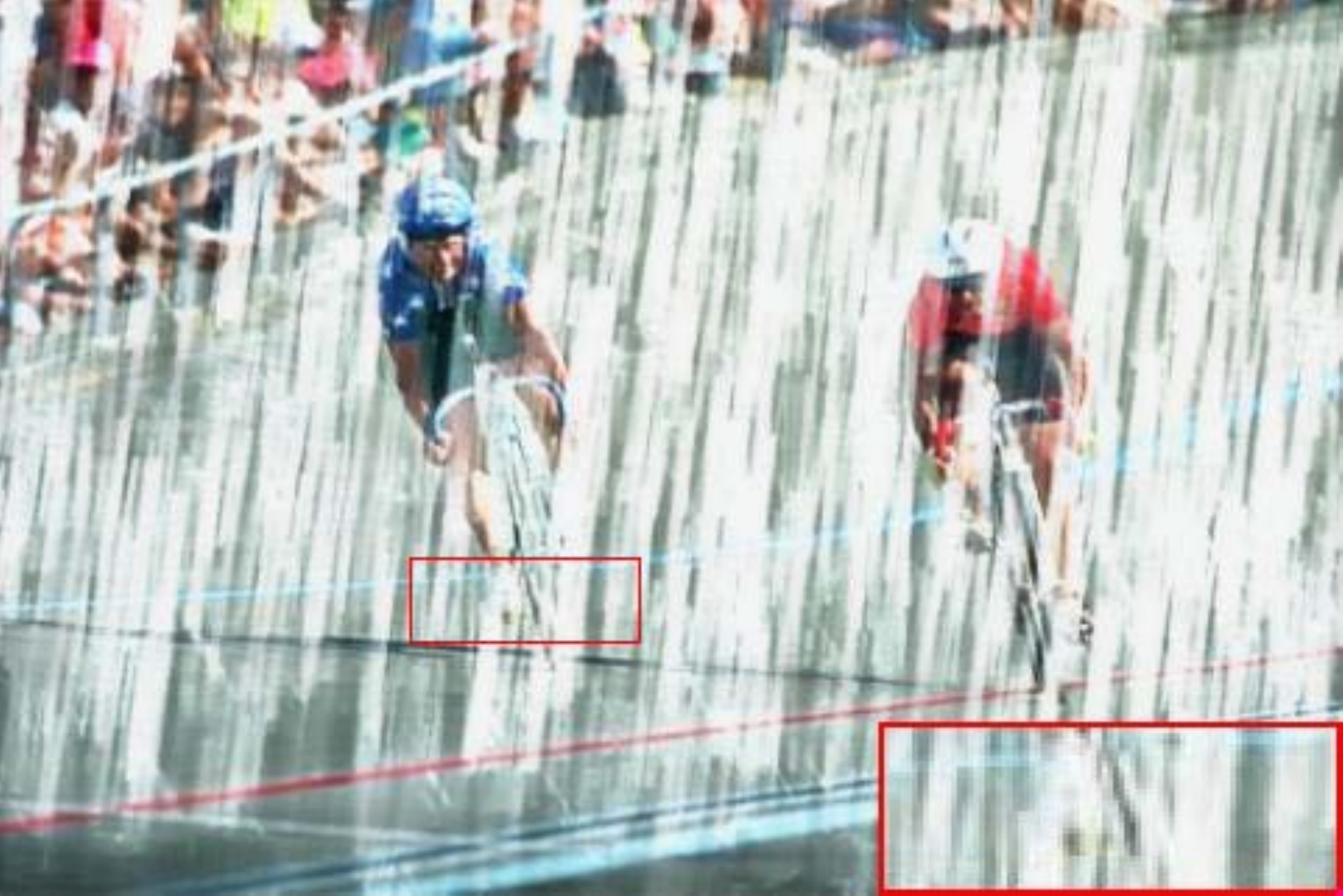}
\end{minipage}
\begin{minipage}[b]{0.116\linewidth}
\includegraphics[width=1\linewidth,height=0.6\linewidth]{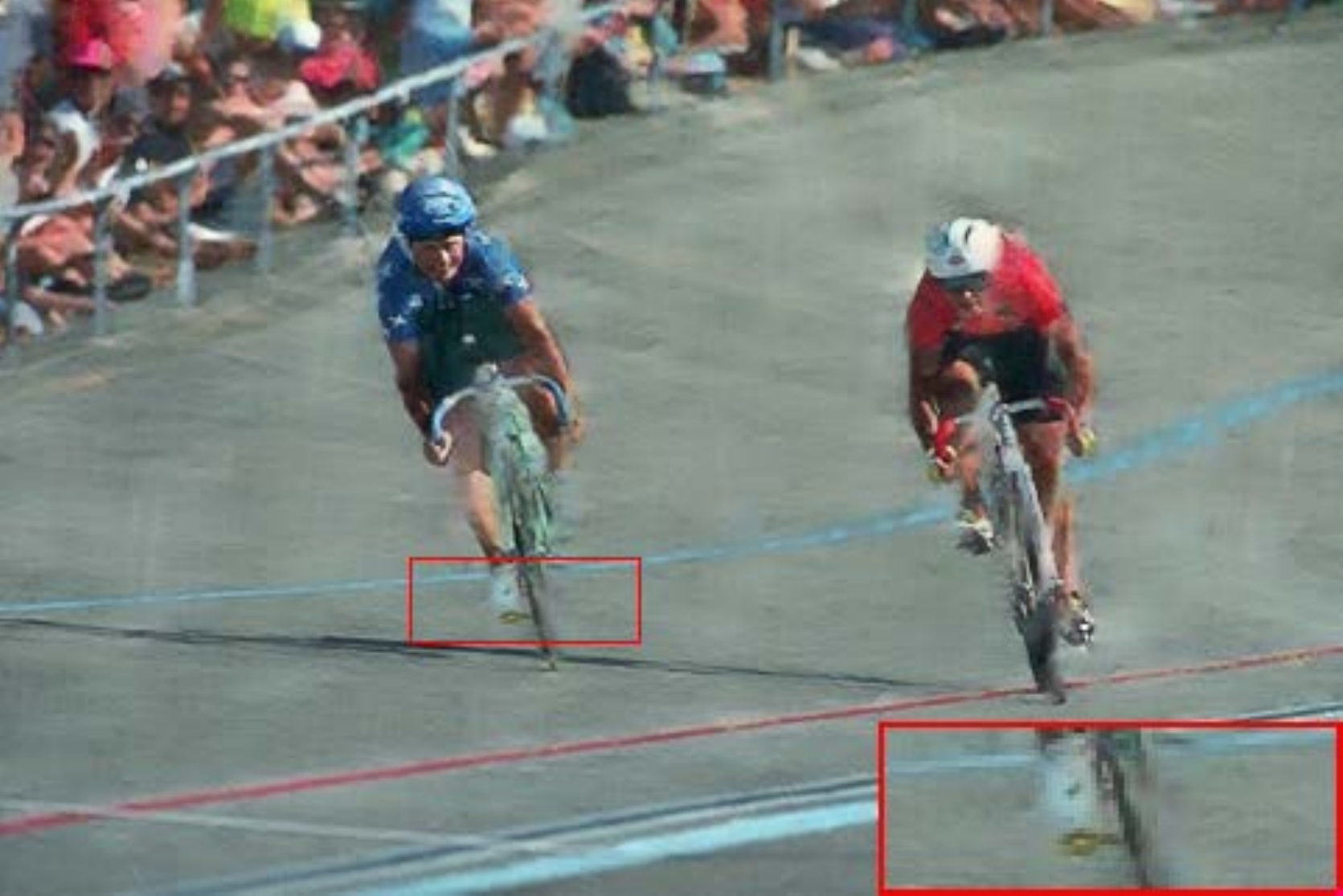}
\end{minipage}
\begin{minipage}[b]{0.116\linewidth}
\includegraphics[width=1\linewidth,height=0.6\linewidth]{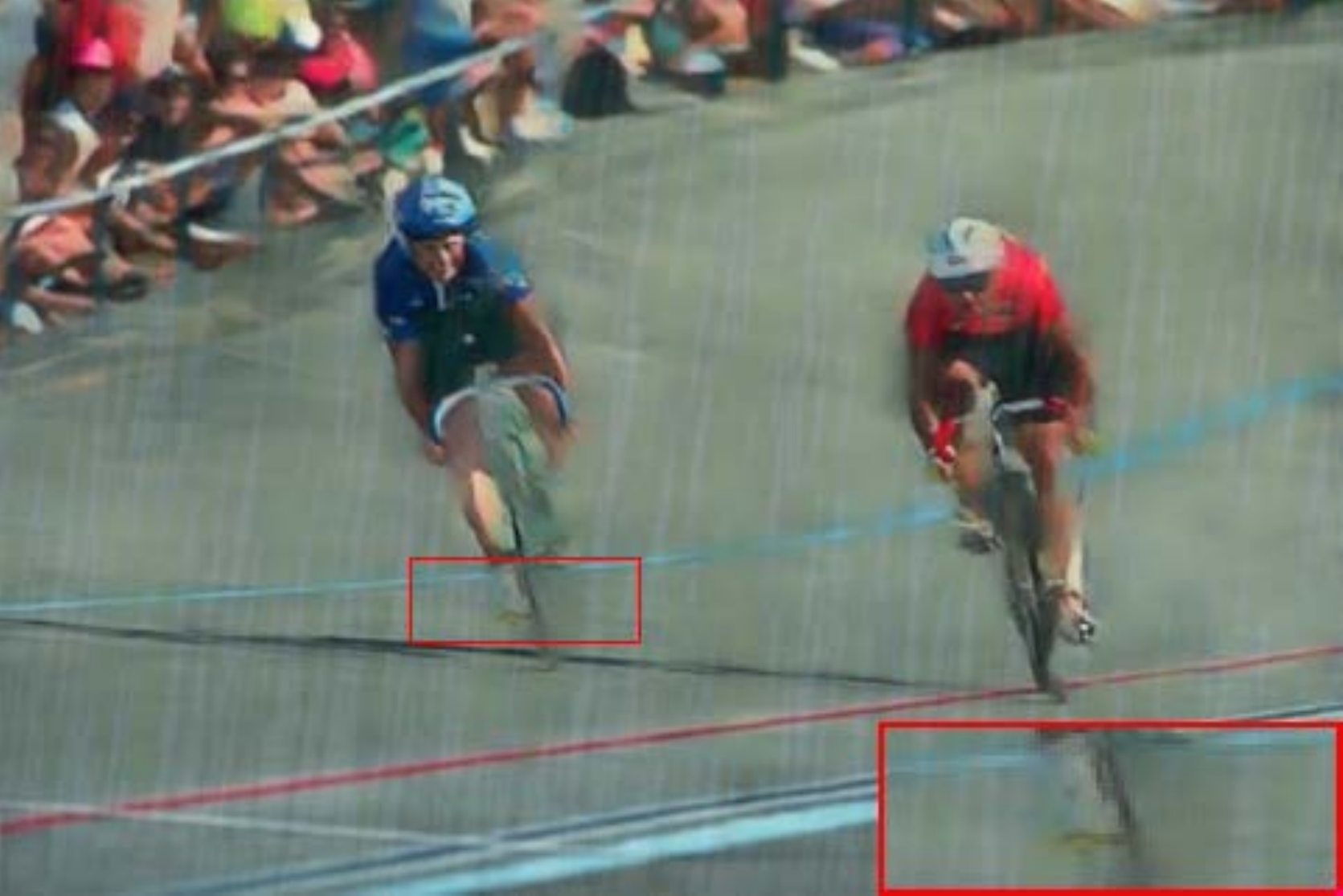}
\end{minipage}
\begin{minipage}[b]{0.116\linewidth}
\includegraphics[width=1\linewidth,height=0.6\linewidth]{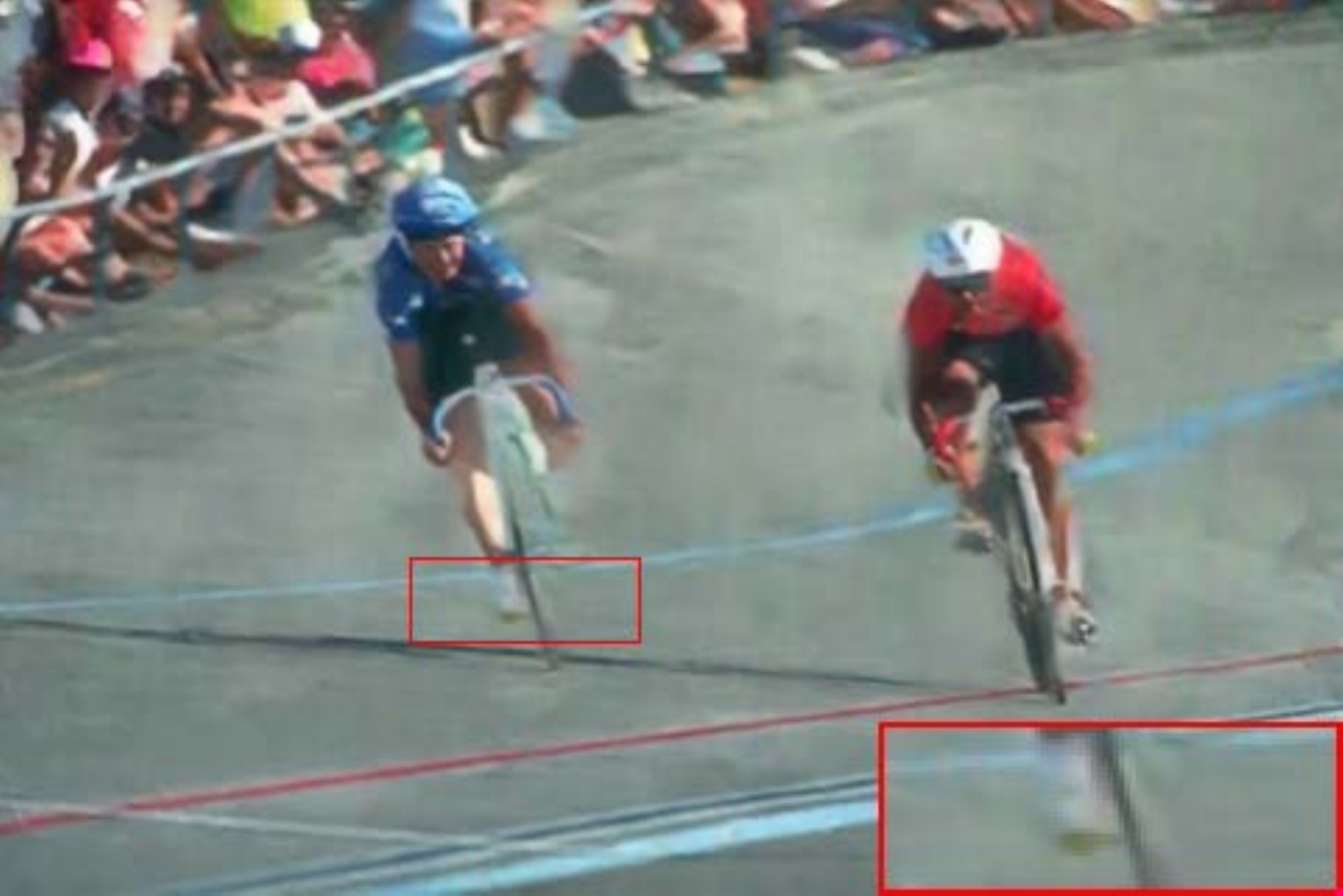}
\end{minipage}
\begin{minipage}[b]{0.116\linewidth}
\includegraphics[width=1\linewidth,height=0.6\linewidth]{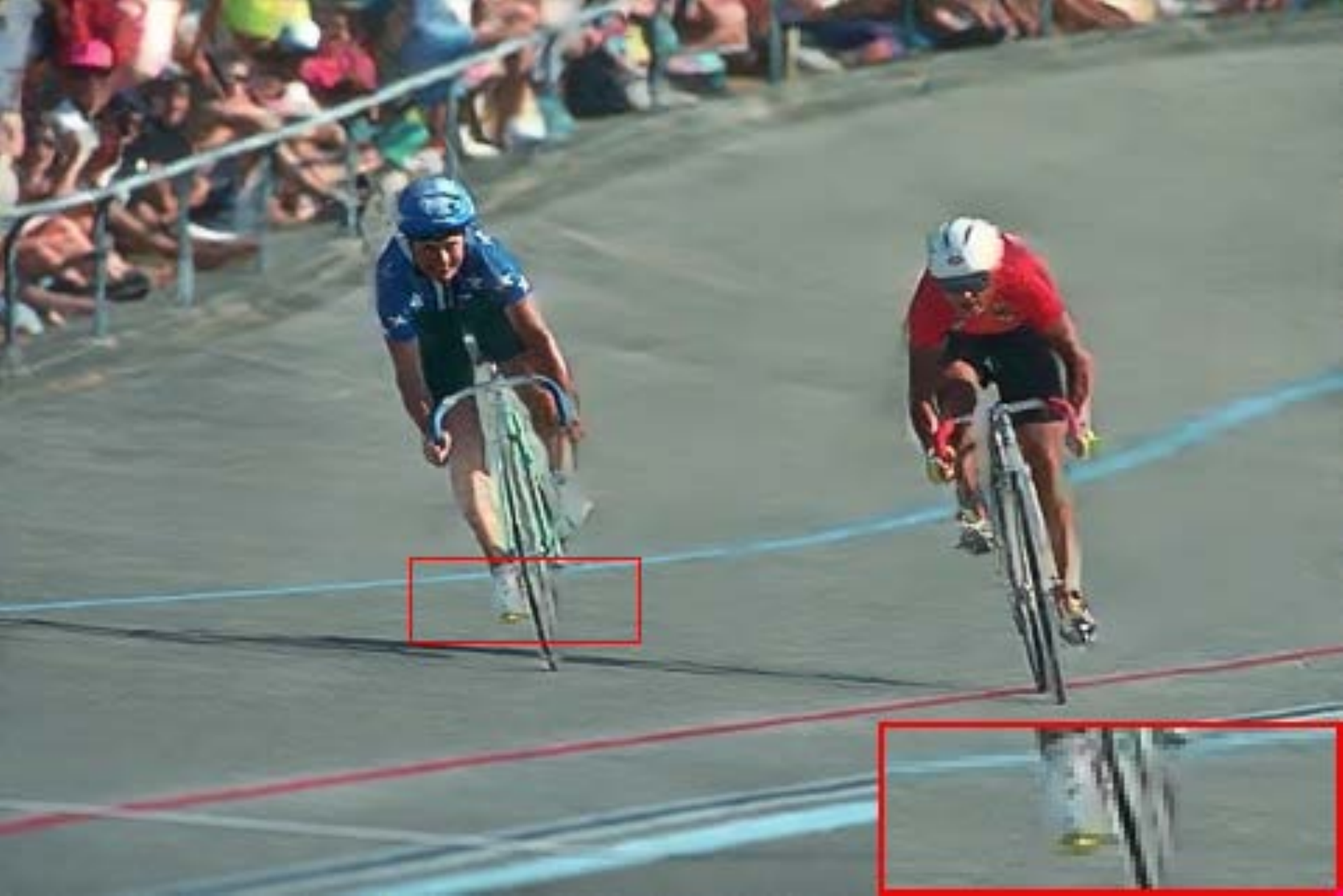}
\end{minipage}
\begin{minipage}[b]{0.116\linewidth}
\includegraphics[width=1\linewidth,height=0.6\linewidth]{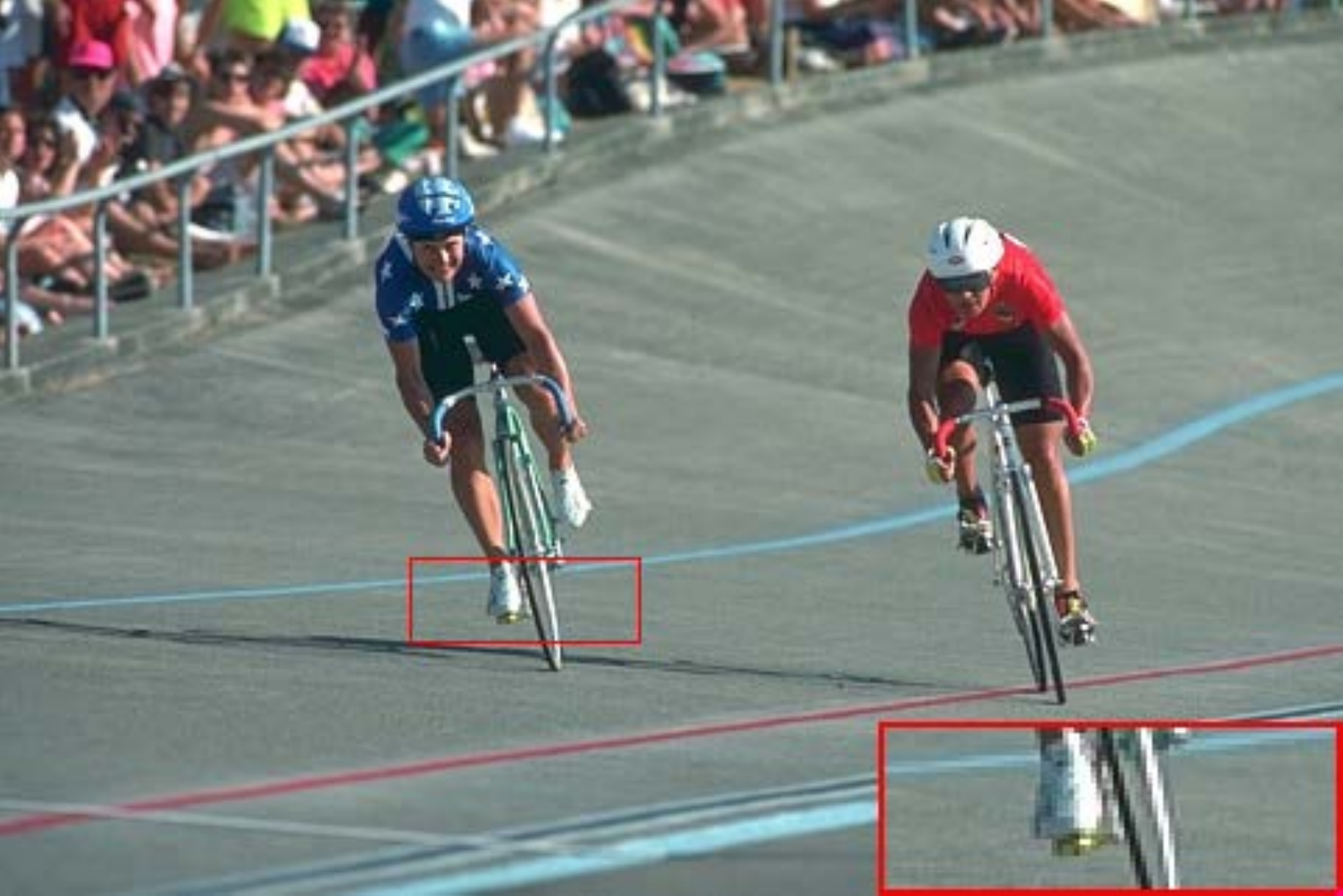}
\end{minipage}
\begin{minipage}[b]{0.116\linewidth}
\includegraphics[width=1\linewidth,height=0.6\linewidth]{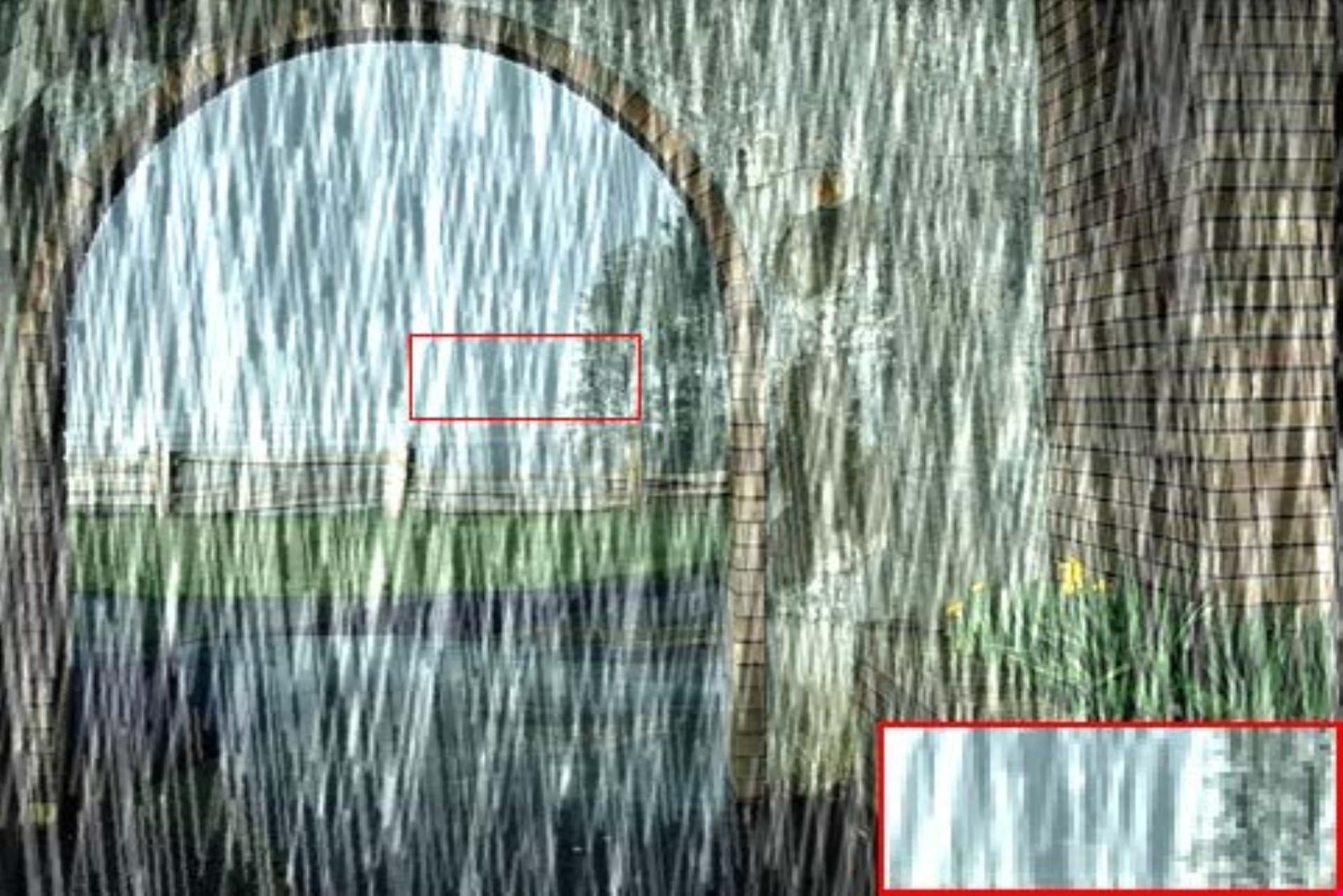}
\end{minipage}
\vspace{-0.2cm}
\begin{minipage}[b]{0.116\linewidth}
\includegraphics[width=1\linewidth,height=0.6\linewidth]{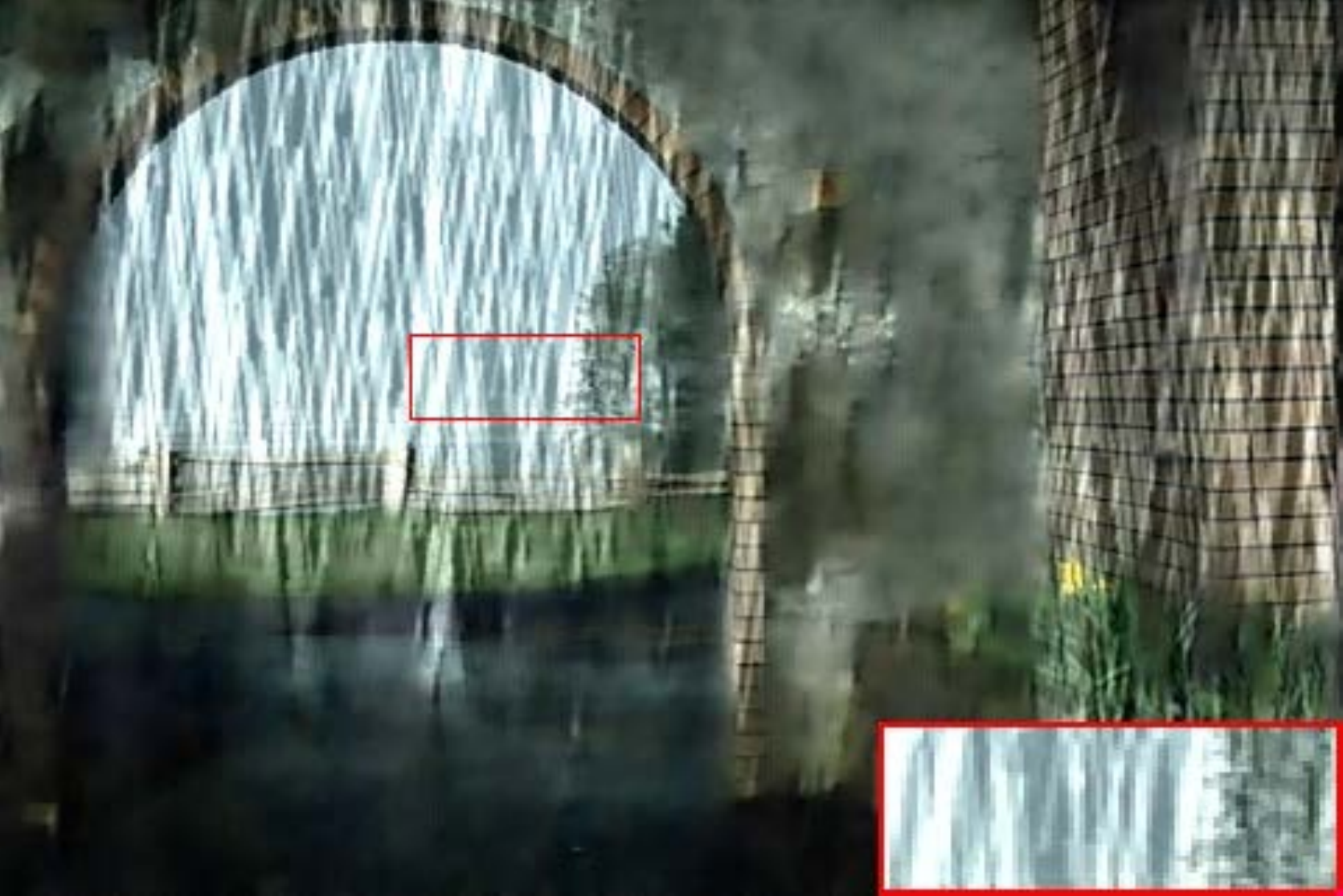}
\end{minipage}
\begin{minipage}[b]{0.116\linewidth}
\includegraphics[width=1\linewidth,height=0.6\linewidth]{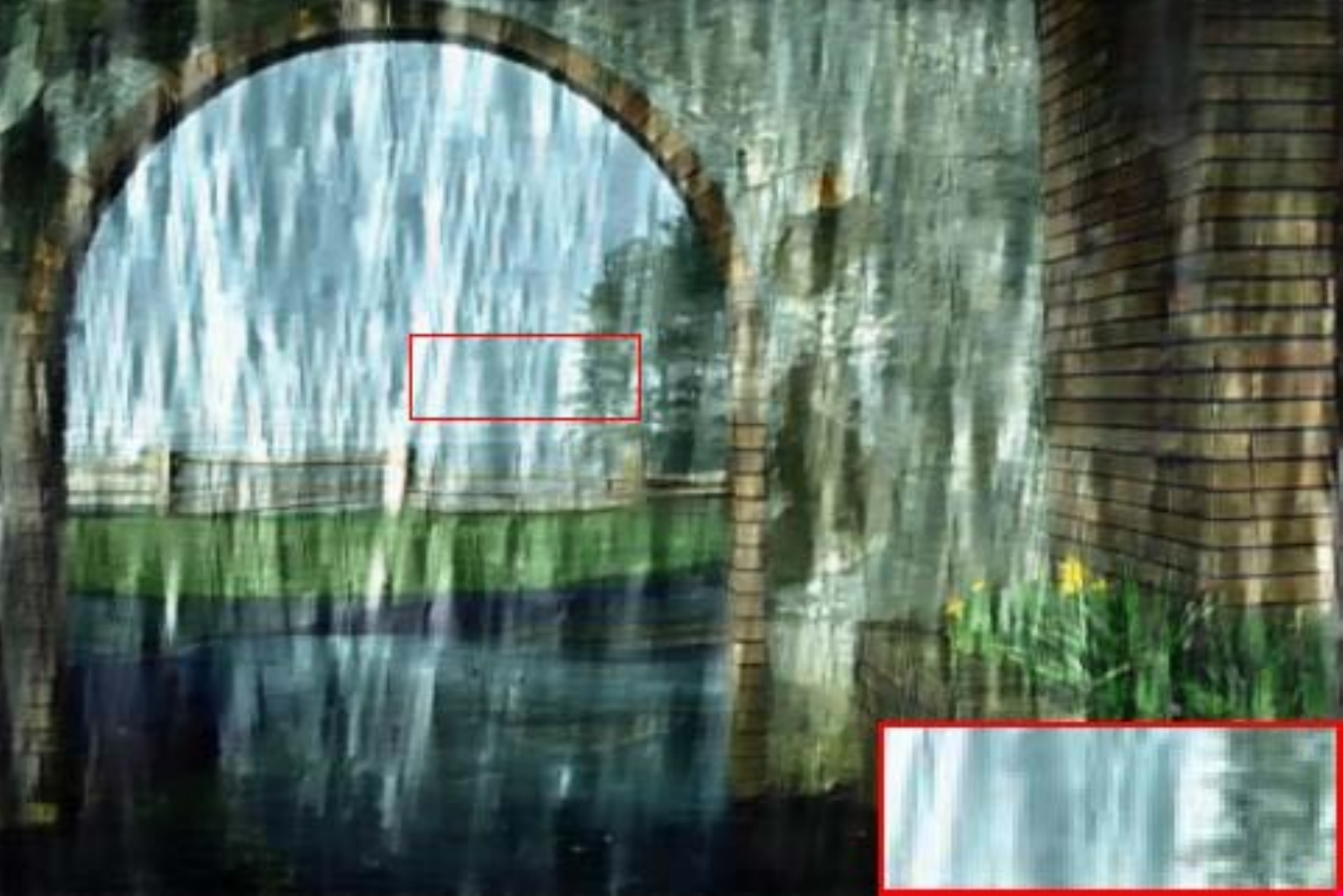}
\end{minipage}
\begin{minipage}[b]{0.116\linewidth}
\includegraphics[width=1\linewidth,height=0.6\linewidth]{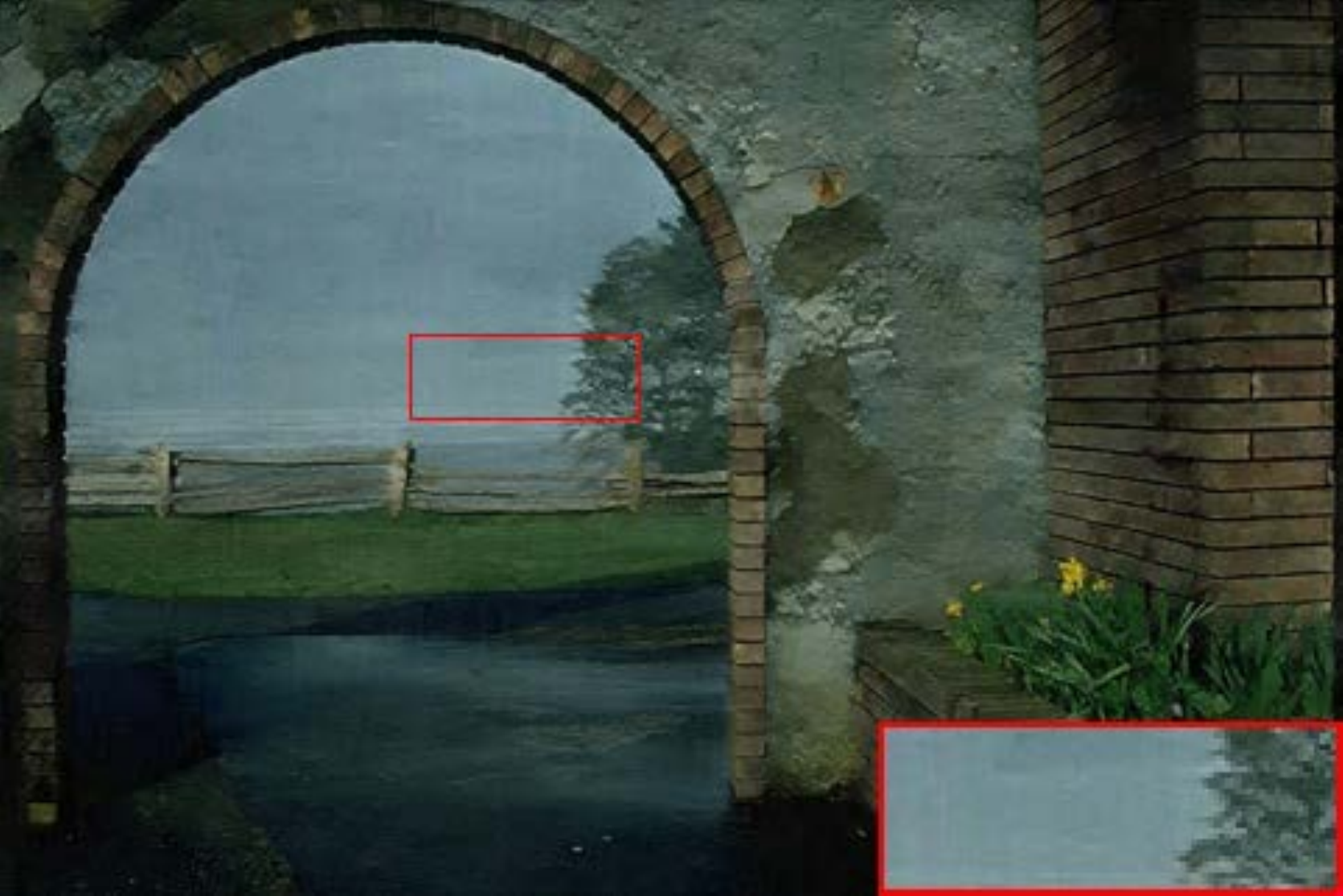}
\end{minipage}
\begin{minipage}[b]{0.116\linewidth}
\includegraphics[width=1\linewidth,height=0.6\linewidth]{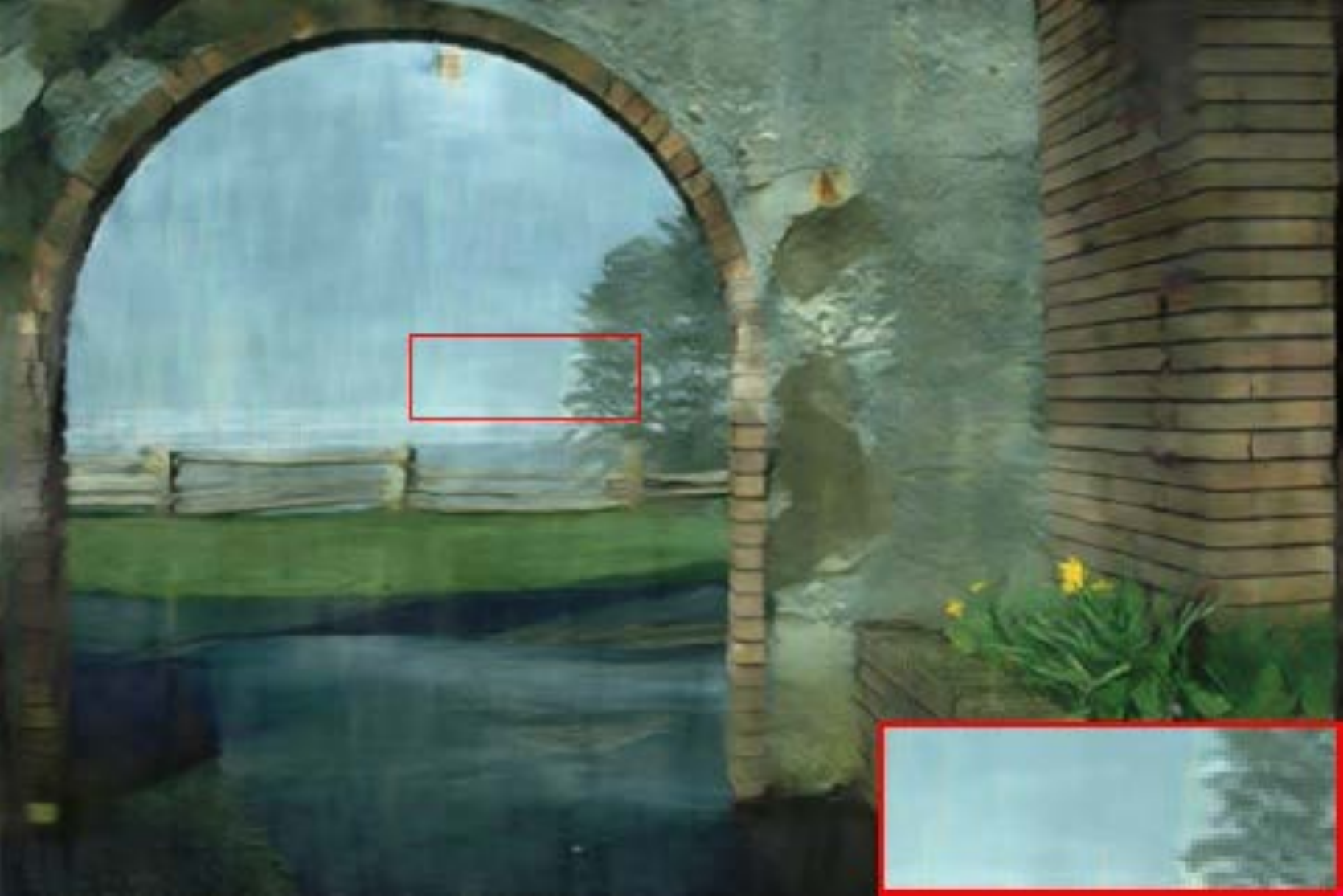}
\end{minipage}
\begin{minipage}[b]{0.116\linewidth}
\includegraphics[width=1\linewidth,height=0.6\linewidth]{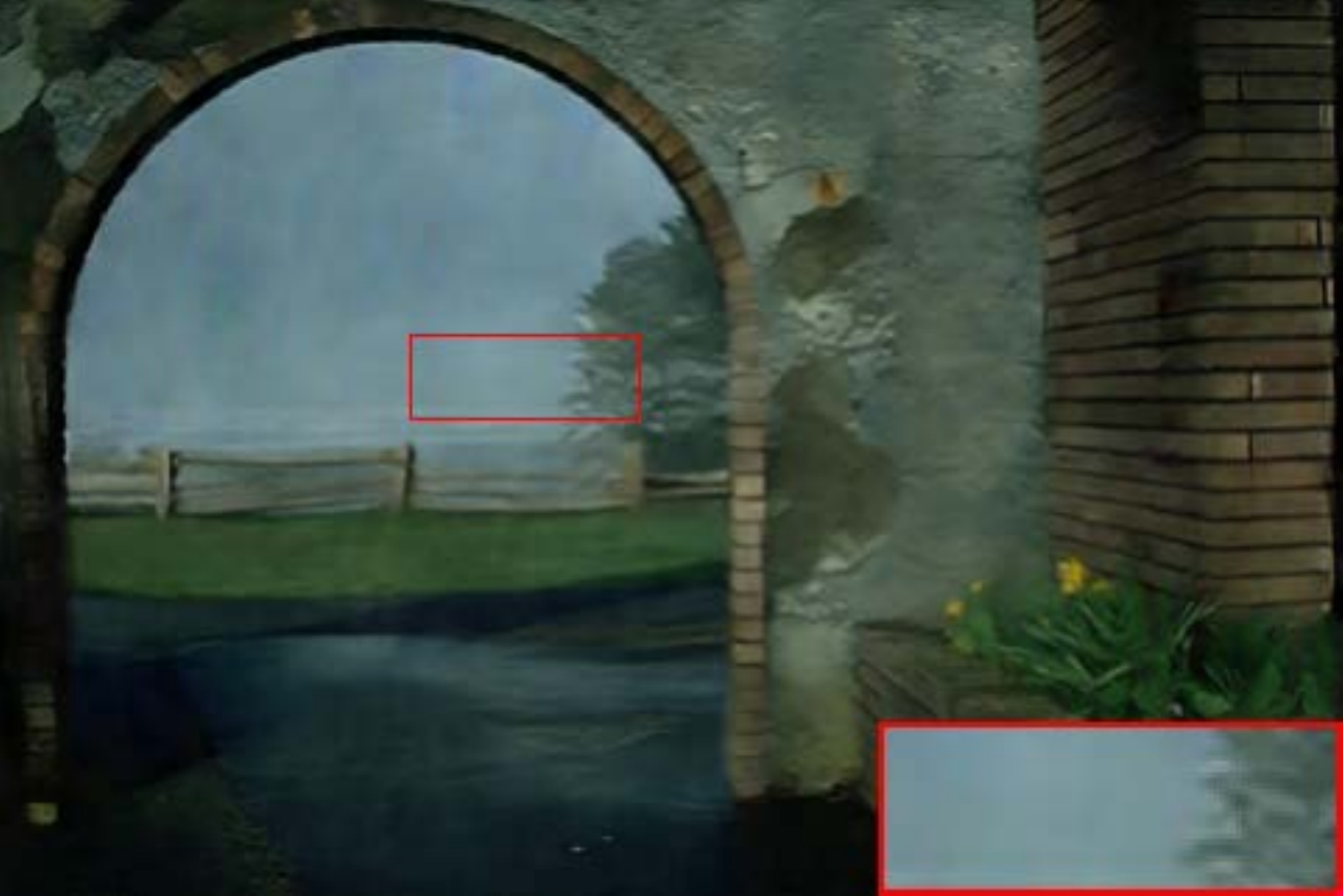}
\end{minipage}
\begin{minipage}[b]{0.116\linewidth}
\includegraphics[width=1\linewidth,height=0.6\linewidth]{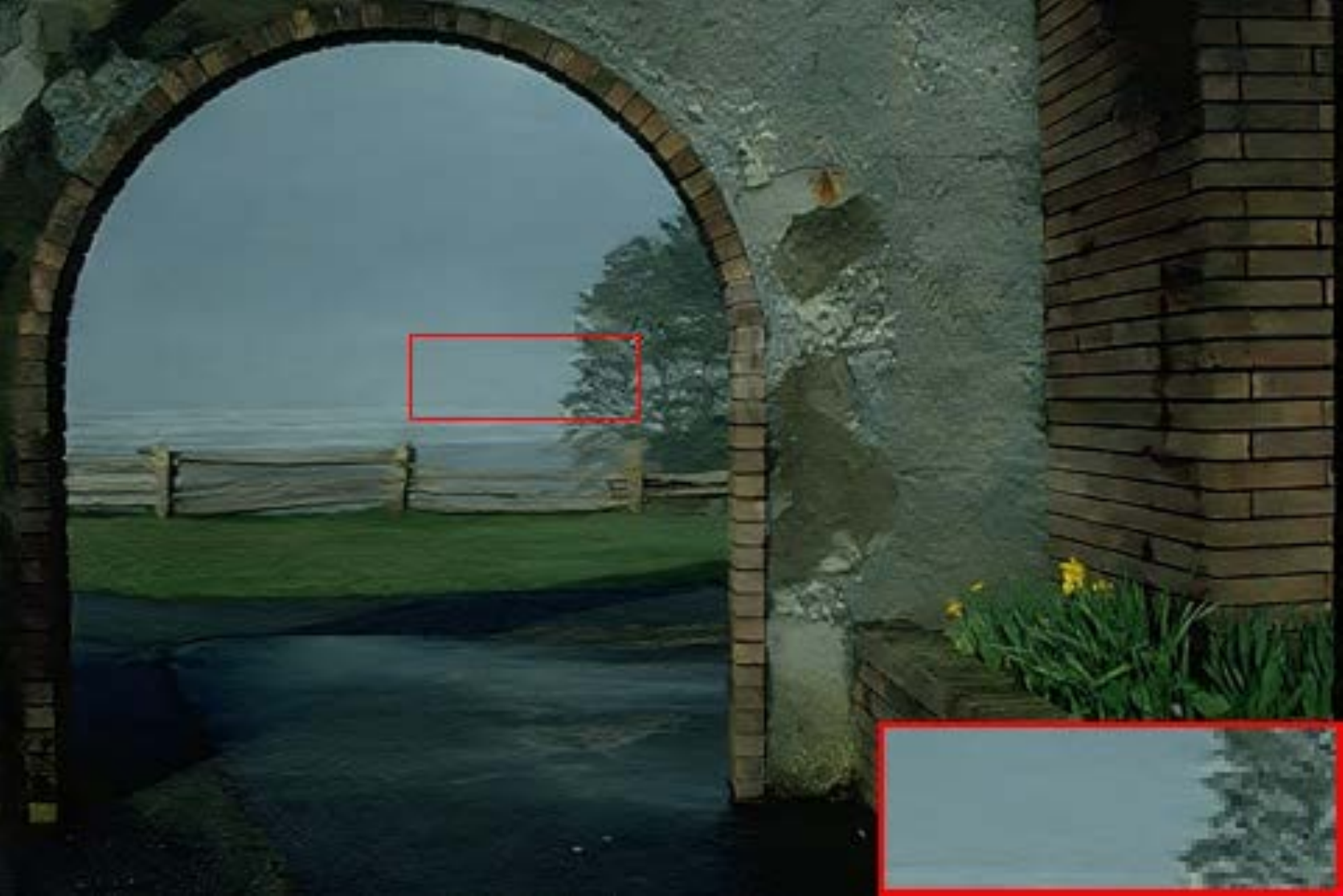}
\end{minipage}
\begin{minipage}[b]{0.116\linewidth}
\includegraphics[width=1\linewidth,height=0.6\linewidth]{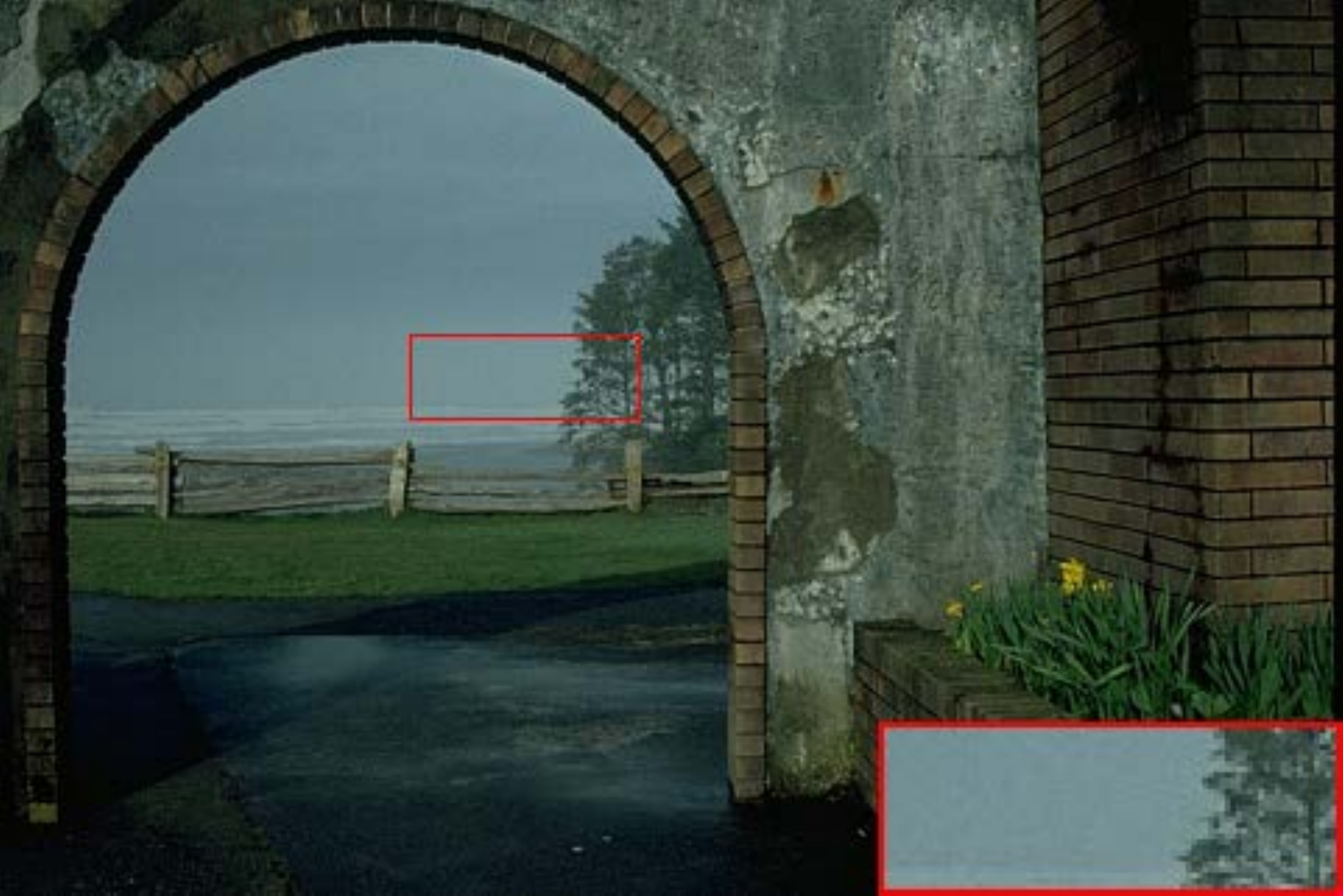}
\end{minipage}
\begin{minipage}[b]{0.116\linewidth}
\centering
\tiny{Rainy images}
\end{minipage}
\begin{minipage}[b]{0.116\linewidth}
\centering
\tiny{SPANet \cite{Alpher37}}
\end{minipage}
\begin{minipage}[b]{0.116\linewidth}
\centering
\tiny{DID \cite{Alpher14}}
\end{minipage}
\begin{minipage}[b]{0.116\linewidth}
\centering
\tiny{Rescan \cite{Alpher26}}
\end{minipage}
\begin{minipage}[b]{0.116\linewidth}
\centering
\tiny{UMRL \cite{Alpher29}}
\end{minipage}
\begin{minipage}[b]{0.116\linewidth}
\centering
\tiny{MSPFN \cite{Alpher41}}
\end{minipage}
\begin{minipage}[b]{0.116\linewidth}
\centering
\tiny{Ours}
\end{minipage}
\begin{minipage}[b]{0.116\linewidth}
\centering
\tiny{Groundtruth}
\end{minipage}
\caption{Visual quality comparisons on sample images from benchmark datasets. Zoom in to see the details.}
\label{fig8}
\vspace{-0.86cm}
\end{center}
\end{figure}
Ten widely adopted methods are compared with our network, including two traditional methods, i.e., {\bf DSC} \cite{Alpher18} and {\bf LP} \cite{Alpher22}, and eight state-of-the-art deep learning based methods, i.e., {\bf DDN} \cite{Alpher13}, {\bf DID} \cite{Alpher14}, {\bf RESCAN} \cite{Alpher26}, {\bf SPANet} \cite{Alpher37}, {\bf UMRL} \cite{Alpher29}, {\bf PReNet} \cite{Alpher15}, {\bf MSPFN} \cite{Alpher41} and {\bf RCDNet} \cite{Alpher45}. Results of our method and other methods are given in Table \ref{table4}. We select seven images from four benchmark datasets to visually validate the qualitative performance of different methods. Specifically, visually one can see in the second and third columns of Fig. \ref{fig8}, though the light rain streaks can be removed by SPANet \cite{Alpher37} and DID \cite{Alpher14}, remaining rain streaks with missing image details usually exist when the rain is heavy. By comparing fourth, sixth and seventh rows of Fig. \ref{fig8}, the UMRL \cite{Alpher29}, RESCAN \cite{Alpher26} and MSPFN \cite{Alpher41} are hard to detect heterogeneously distributed rain streaks in different regions and recover clean image details well. We also test the derained image by DeepLabv3+ \cite{Alpher42} as shown in Fig. \ref{fig10}. In addition, we randomly select 400 image pairs ($512 \times 512$) to evaluate average time and performance of different methods as shown in Table \ref{table6}. The error detector that takes little running time brings the considerable performance improvement. The FLOPs (in $\times10^{11}$) of RESCAN, MSPFN, RCDNet, RLNet$\pm$E, and RLNet are 1.3, 24.2, 7.8, 1.6 and 2.0.
%-------------------------------------------------------------------------

\subsection{Experiments on Real Rainy Images}
Using the real-world datasets \cite{Alpher37} cropped to 481$\times$321, we compare proposed RLNet with other methods as shown in Table. \ref{table7}. In addition, from Fig. \ref{fig9}, it can be observed that RLNet, UMRL \cite{Alpher29} and RESCAN \cite{Alpher26} outperform DID \cite{Alpher14} and MSPFN \cite{Alpher41} in removing heavy rain streaks from real rainy image. Specifically, as can be seen from the Fig. \ref{fig9}, there are more or less visible rain streaks in the results by DID \cite{Alpher14} and MSPFN \cite{Alpher41}, while RLNet, UMRL \cite{Alpher29} and RESCAN \cite{Alpher26} can generate the clean image. For the fourth, fifth and sixth columns of the Fig. \ref{fig9}, the blurs and halo artifacts with missing image details exist in the results by UMRL \cite{Alpher29} and RESCAN \cite{Alpher26}, while the rain streak removal results by RLNet are high-quality and clear.
%\footnote{\vspace{-0.3cm}More results are shown in the supplementary material.}.
\begin{table}
\begin{center}
\caption{Time complexity (in seconds) and performance of different methods. Sizes of testing images are $512 \times 512$. RLNet$-$E means that RLNet does not contain the error detector.}
\scalebox{0.64}{
\begin{tabular}{c|c|c|c|c|c|c|c}
\hline
Method &DSC &UMRL&RCDNet & RESCAN  & MSPFN & RLNet$-$E& RLNet\\
\hline\hline
PSNR &21.90&28.71 & 28.99&27.31 &28.66 &30.19&31.34   \\
\hline
Avg time & 371.13&8.831 &3.023& 0.952 & 1.040 &0.333&0.373  \\
\hline
\end{tabular}\label{table6}}
\end{center}
\vspace{-0.5cm}
\end{table}
\begin{table}
\begin{center}
\caption{NIQE comparisons on real-world datasets, smaller scores indicate better image quality.}
\scalebox{0.68}{
\begin{tabular}{c|c|c|c|c|c|c|c}
\hline
Method &SPANet &DID & MSPFN & UMRL &RCDNet & RESCAN  & RLNet\\
\hline\hline
NIQE$\downarrow$ &5.109&5.068&5.561 &5.257&4.701 &4.631&4.498   \\
\hline
\end{tabular}\label{table7}}
\end{center}
\vspace{-0.7cm}
\end{table}

%-------------------------------------------------------------------------
\section{Conclusion}
We have proposed the RLNet for single image deraining. Based on the CNN in the presence of uncertainty, a new method for error detection and feature compensation is proposed for latent high-quality representation learning. An iterative optimization scheme that unrolls image optimization and error optimization with image priors and correction terms is presented. Experiments demonstrate that the proposed RLNet is robust enough to detect heterogeneous rain streaks and recover details for real rainy images. Taking the portability into account, the proposed method for error detection and feature compensation can be selectively incorporated into learning based image deraining networks.

{\small
\bibliographystyle{ieee_fullname}
\bibliography{egbib}
}
\end{document}